# Coking-Resistant Sub-Nano Dehydrogenation Catalysts: $Pt_nSn_x/SiO_2$ (n = 4, 7)

*Timothy J. Gorey,[a†] Borna Zandkarimi,[b†] Guangjing Li,[a] Eric T. Baxter,[a]*

*Anastassia N. Alexandrova,[b,c*] and Scott L. Anderson[a*]*

[a]Chemistry Department, University of Utah, 315 S. 1400 E., Salt Lake City, UT 84112

[b]Chemistry and Biochemistry, University of California, Los Angeles, and [c]California NanoSystems Institute, Los Angeles, CA 90095

† These authors contributed equally to this work.

*Senior Authors

Corresponding Authors: Scott Anderson, (801) 585-7289, anderson@utah.edu, Anastassia Alexandrova, (310) 825-3769, ana@chem.ucla.edu

## ABSTRACT

We present a combined experimental/theoretical study of $Pt_n/SiO_2$ and $Pt_nSn_x/SiO_2$ (n = 4, 7) model catalysts for the endothermic dehydrogenation of hydrocarbons, using the ethylene intermediate as a model reactant. Mass-selected $Pt_n$ clusters were deposited onto amorphous $SiO_2$/Si(100) to make the $Pt_nSiO_2$ model catalysts. To produce $Pt_nSn_x$ clusters, size-selected $Pt_n$ were used to seed selective deposition of Sn on Pt via a self-limiting $H_2/SnCl_4/H_2$ reaction sequence. Model catalysts were analyzed using $C_2D_4$ and CO temperature programmed desorption (TPD), low energy ion scattering (ISS), X-ray photoelectron spectroscopy (XPS), plane wave density functional theory (DFT) global optimization combined with a statistical mechanical description of the catalytic interface, and a DFT mechanistic study. Supported pure $Pt_n$ clusters are found to be highly active toward dehydrogenation of $C_2D_4$, quickly deactivating due to a combination of carbon deposition and sintering, resulting in loss of accessible Pt sites. Addition of Sn to $Pt_n$ clusters results in the complete suppression of $C_2D_4$ dehydrogenation and carbon deposition, and also stabilizes the clusters against thermal sintering. Theory shows that both systems have thermal access to a multitude of cluster structures and adsorbate configurations that form a statistical ensemble. While $Pt_4/SiO_2$ clusters bind ethylene in both di-σ- and π-bonded configurations, $Pt_4Sn_3/SiO_2$ binds $C_2H_4$ only in the π-mode, with di-σ bonding suppressed by a combination of electronic and geometric features of the PtSn clusters. Dehydrogenation reaction profiles on the accessible cluster isomers were calculated using the climbing image nudged elastic band (CI-NEB) method. Dehydrogenation of di-σ bound ethylene is computed to dominant, and suppressed by Sn addition, in agreement with the experiments. DFT indicates that after Sn alloying, the barrier for ethane-to-ethylene conversion is lower than that for unwanted ethylene dehydrogenation.

## INTRODUCTION

Carbon deposition ("coking") in high temperature reactions under hydrocarbon-rich conditions is a serious catalyst deactivation mechanism, thus understanding the mechanism and developing approaches to suppress carbon deposition are interesting. One reaction of this type is endothermic alkane dehydrogenation, and we have been examining carbon deposition over sub-nano Pt[1] and Pt alloy cluster catalysts,[2-4] with the goal of stabilizing the clusters against both coking and sintering, which is a serious problem for clusters at high temperatures. In addition to maximizing the accessibility of precious metal atoms in the surface layer, sub-nano clusters often have size-dependent properties that provide additional opportunities for catalyst tuning.[5-15] Small clusters are also relatively tractable theoretically, thus these systems allow detailed modeling of the effects of cluster physical and chemical properties on reaction mechanisms.[9, 16-22] Even for a single cluster size, however, cluster reactions can be quite complex, with multiple structural isomers contributing and evolving during reactions due to thermal and adsorbate effects.

Because reducing coking on Pt-based catalysts is important in many applications, there have been many studies of mitigation strategies. Here we focus on use of Pt-Sn alloy catalysts for dehydrogenation and other reactions where coking is problematic. For example, coke formation has been studied on practical Pt catalysts[23-24] and has been addressed by passivating specific metal sites with Sn, ranging from trace to stoichiometric amounts.[25-30] Though successful, the complex nature of practical catalysts makes detailed understanding of the mechanistic origins of the Pt-Sn relationship and its beneficial effects on catalytic dehydrogenation difficult. It is useful, therefore, to consider model catalysts, including ordered surface alloys and planar supported cluster catalysts.

Koel and co-workers studied the branching between alkene desorption and decomposition/dehydrogenation (leading to carbon deposition) on a series of ordered Pt-Sn surface alloys, ranging from pure Pt(111) to an alloy with 2:1 Pt:Sn stoichiometry. With increasing Sn content, the desorption temperature for the alkenes decreased substantially, without any reduction in the saturation coverage. By decreasing the binding energy, desorption of intact alkenes becomes favored over further dehydrogenation. (i.e., coking is suppressed). It was found that the di-σ ethylene binding geometry was preferred for all the alloys, as well as for Pt (111).[31-32] In a comparable study using DFT methods, ethane adsorption to extended Pt and PtSn surfaces was explored by Hook *et al.,* who also found that ethylene binds in a di-σ fashion to both Pt and PtSn extended surfaces. As in the experiments, the ethylene adsorption energy decreased below the barrier for dehydrogenation of ethylene, resulting in suppression of dehydrogenation and carbon deposition.[33] This was rationalized as being due to a combination of binding geometry and electronic effects.

For small PtSn supported nanoparticles, some differences have been noted. For example, Shen *et al.*[30] and Natal-Santiago *et al.*[29] studied adsorption geometries and energies of ethylene to silica-supported Pt and PtSn catalyst particles (diam. = 2-5 nm) using microcalorimetry and IR spectroscopy. The nano Pt catalyst supported both di-σ and π-bound ethylene, and incorporation of Sn resulted in a decrease in the heat of adsorption and an increase in the fraction of π-bound ethylene with increasing Sn content. The appearance of substantial π-bonding for small PtSn nanoparticles differs from the observation of purely di-σ bonding for extended PtSn surface alloys, suggesting significant effects of particle size. Sub-nanometer, or "ultra-dispersed" catalysts would therefore be expected to have additional differences in catalytic behavior. For example, in a study of propane dehydrogenation, Datye *et al*., examined $0.6 \pm 0.2$ nm Pt and PtSn particles on $Al_2O_3$.[34] From a thermal stability perspective, it was proposed that Sn helped maintain the dispersion of Pt particles under reaction and oxidative regeneration conditions, and that this was the major contributor to the prolonged activity for PtSn vs. pure Pt catalysts.

In another study with sub-nanometer particles, Ha and Baxter *et al.*[14] added boron to size-selected $Pt_n/Al_2O_3$ model catalysts to suppress dehydrogenation/coking during ethylene temperature-programmed desorption/reaction (TPD/R). The model catalyst was well-defined enough to allow an accurate DFT model to be built and used to examine changes in energetics and mechanisms for ethylene binding/desorption from $Pt_n/Al_2O_3$[1] and $Pt_nB_x/Al_2O_3$.[3] Addition of boron suppressed di-σ binding of ethylene, and substantially lowered the desorption energy, such that ethylene desorbed rather than undergoing dehydrogenation leading to carbon deposition.

Here we apply size-selection methods to study $PtSn/SiO_2$ model catalysts. Size- and composition-selected clusters were prepared by using mass-selected $Pt_n$ deposition to prepare planar $SiO_2$ supports decorated by a precisely controlled coverage of size-selected clusters. These were then used to seed highly selective Sn deposition on the clusters via a self-limiting reaction process. The composition and morphology were probed using X-ray photoelectron and low-energy ion scattering spectroscopies (XPS, ISS). $Pt_n/SiO_2$ was chosen as the system, because the process used to deposit Sn is quite selective, depositing Sn almost exclusively on the Pt clusters, rather than on the $SiO_2$ support. We initially examined Sn modification of $Pt_n$/alumina/NiAl(110) and $Pt_n$/alumina/Ta(110), but found considerable non-selective Sn deposition on the alumina supports. Those systems have interesting catalytic activity,[35] but for the purpose of studying $Pt_nSn_x$ clusters with well-defined composition, the $SiO_2$ support is better. Additional detail on catalyst preparation is available below in the Methods section, and a complete characterization of stoichiometry, size, and composition has been previously published.[2] Here, we compare catalytic properties of $Pt_4/SiO_2$ and $Pt_7/SiO_2$ to these size- and composition-controlled $Pt_4Sn_{3.3}/SiO_2$ and $Pt_7Sn_{6.3}/SiO_2$ bimetallic cluster catalysts. DFT calculations are used to examine structure and binding properties of the clusters for thermally populated cluster isomer ensembles for $Pt_4/SiO_2$ and $Pt_4Sn_3/SiO_2$. Note that the amorphous structure of the $SiO_2$ native oxide requires a large slab in the calculations. This factor, together with the need to consider the thermally accessible isomer ensembles, makes the DFT calculations quite expensive even for small $Pt_nSn_x/SiO_2$, and infeasible for larger clusters.

## RESULTS AND DISCUSSION

In the following, we present experiments comparing $Pt_n/SiO_2$ and $Pt_nSn_x/SiO_2$, for four clusters: $Pt_4$, $Pt_7$, $Pt_4Sn_{3.3}$, and $Pt_7Sn_{6.3}$. The theory results are focused on $Pt_4/SiO_2$ and $Pt_4Sn_3/SiO_2$, however, the experiments for $Pt_4$ and $Pt_7$ suggest that similar considerations should apply to the $Pt_7$ and $Pt_7Sn_{6.3}$ experiments. The two $Pt_n$ cluster sizes were selected because $Pt_4$ and $Pt_7$ had been studied previously on $Al_2O_3$ supports, with $Pt_4$ found to have all atoms exposed in the surface layer and $Pt_7$ having both single layer and prismatic 3D isomers, the latter of which was found to isomerize to a single layer structure upon adsorption of multiple ethylene molecules.[1] In addition, the low energy structures for $Pt_4Sn_3/SiO_2$ were already available from our paper describing the method for producing size- and composition-selected $Pt_nSn_x/SiO_2$ model catalysts.[2]

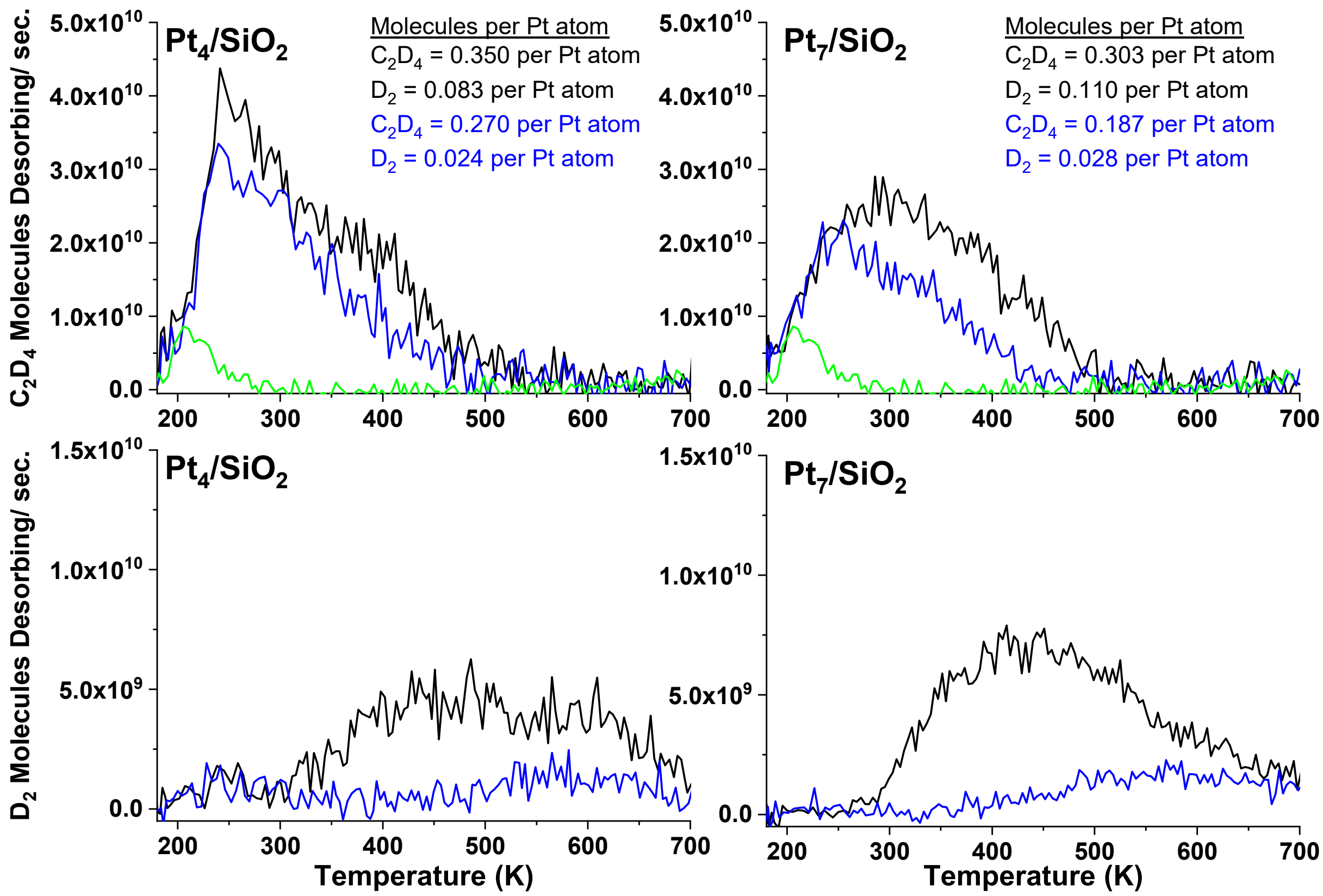

**Figure 1.** Intact desorption of $C_2D_4$ (top) and $D_2$ (bottom) from the first (black) and fourth (blue) $C_2D_4$ TPD. These spectra were collected after a 10 L dose of $C_2D_4$ to $Pt_n/SiO_2$ (n = 4, 7). $C_2D_4$ desorption from $SiO_2$ is also plotted (green), no $D_2$ evolution from $SiO_2$ is observed.

**Ethylene Desorption vs. Dehydrogenation.** $C_2D_4$ and $D_2$ desorption from $Pt_4/SiO_2$ and $Pt_7/SiO_2$ are shown in **Fig. 1**. To examine the effects of repeated $C_2D_4$ exposure and heating, a total of four TPD/R runs were carried out, but the figure shows results only for the 1$^{st}$ (black) and 4$^{th}$ (blue) runs. The data for all four runs are shown in **Fig. S1**. Most experiments were done with $C_2D_4$ to avoid interference with $H_2$ detection from high background at mass 2, however, the background is also high at mass 28, corresponding to $C_2D_2$. Therefore, some experiments were done using $C_2H_4$ to look for acetylene desorption. None was observed. We also looked for ethane signal from possible hydrogenation processes, but the signals were negligible. Because cracking of $C_2D_4$ to $D_2^+$ in the mass spectrometer is weak, this contribution to the $D_2$ signal has not been subtracted in the figure.

The figures also show the $C_2D_4$ desorption observed from a Pt-free $SiO_2$ substrate (green), indicating that for the 180 K $C_2D_4$ exposure temperature chosen, only a small amount of $C_2D_4$ desorbs from the $SiO_2$ substrate, and that $SiO_2$ does not support $D_2$ desorption. We tested the effect of lowering the dose temperature from 180 K to 150 K, observing a large increase in the desorption intensity from the $SiO_2$ support. The integrated $C_2D_4$ TPD signal after 180 K dosing on $SiO_2$ corresponds to ~3.5 x $10^{12}$ molecules/$cm^2$ – well below 1% of a monolayer.

For $Pt_n/SiO_2$ samples, $C_2D_4$ desorption also begins at 180 K, initially matching the desorption from $SiO_2$, but then continuing with much higher intensities before eventually declining and disappearing by ~500 K. The signal between 180 K and ~220 K is, therefore, attributed to desorption mostly from $SiO_2$ sites, while the signal at higher temperatures is attributed to desorption from sites associated with the $Pt_n$ clusters.

The temperature dependence for desorption of intact $C_2D_4$ is different for the two cluster sizes. For $Pt_4$, during the 1st TPD/R run, the desorption appears bimodal, with a strong and relatively sharp peak near 250 K, followed by a shoulder in the 350-400 K range, with a rapid intensity drop above 400 K. For $Pt_7$, $C_2D_4$ desorption grows more slowly at low temperatures, peaking broadly around 300 K, and declining gradually above 400 K. According to the DFT calculations, the ethylene binding energy of the $C_2H_4/Pt_4/SiO_2$ global minimum structure, where the ethylene is bound in the di-σ mode, is roughly 0.14 eV stronger than in the second local minimum structure (23% population at 700 K), where ethylene is π-bonded. More detailed discussion can be found in the DFT analysis section.

For the 1st run on both $Pt_4$ and $Pt_7$, significant $D_2$ desorption starts just below 300 K and peaks broadly around 440 K, declining to baseline above 600 K. The $D_2$ signal is significantly larger for $Pt_7$ than $Pt_4$. In one experiment, the heat ramp was extended to 1000 K to check for additional $C_2D_4$ or $D_2$ desorption above 700 K. None was observed, suggesting that dehydrogenation and desorption go to completion by 700 K. The behavior observed here for $Pt_4/SiO_2$ and $Pt_7/SiO_2$ is similar to that observed for $Pt_n/Al_2O_3$ (n =4, 7), with intact $C_2D_4$ desorption peaking near 300 K, and $D_2$ desorption in the ~300 – 600 K range.[1]

The integrated numbers of $C_2D_4$ and $D_2$ molecules desorbing from each sample are listed in **Table 1**, and because we know the Pt coverage precisely (1.5 x $10^{14}$ atoms/cm$^2$), the integrated desorption numbers are given as molecules *per* Pt atom. To correct the numbers for the contribution of desorption from the $SiO_2$ substrate, ~0.02 $C_2D_4$ molecules/Pt atom should be subtracted. With this correction, we find that ~1.4 and ~2.1 intact $C_2D_4$ molecules desorb per $Pt_4$ and $Pt_7$ cluster, respectively, with the corresponding $D_2$ desorption being ~0.33 and ~0.77 $D_2$ molecules *per* cluster. Given that $D_2$ desorption goes to completion before termination of the heat ramp, the total number of $C_2D_4$ adsorbed *per* cluster can be estimated as #$C_2D_4$ desorbing + 0.5 #$D_2$ desorbing, giving ~1.6 and ~2.5 $C_2D_4$ adsorbed *per* $Pt_4$ and $Pt_7$, respectively. In theoretical modeling of $Pt_4/SiO_2$, we therefore considered coverages of one and two ethylene molecules per cluster. As expected, the number of adsorbed ethylene *per* cluster increases with cluster size, however, on a *per* Pt atom basis, slightly more ethylene adsorbs on $Pt_4$. This effect is consistent with the observation by ISS,[2] that $Pt_7/SiO_2$ has a smaller fraction of its Pt atoms in the surface layer, and thus presents fewer ethylene binding sites.

**Table 1.** Numbers of $C_2D_4$ and $D_2$ molecules desorbing *per* Pt atom from Fig. 1.

| **TPD Run:** | **$Pt_4/SiO_2$** | | **$Pt_7/SiO_2$** | |
|---|---|---|---|---|
| **1st** | 0.350 $C_2D_4$/Pt | 0.083 $D_2$/Pt | 0.303 $C_2D_4$/Pt | 0.110 $D_2$/Pt |
| **2nd** | 0.340 $C_2D_4$/Pt | 0.076 $D_2$/Pt | 0.282 $C_2D_4$/Pt | 0.066 $D_2$/Pt |
| **3rd** | 0.270 $C_2D_4$/Pt | 0.046 $D_2$/Pt | 0.226 $C_2D_4$/Pt | 0.050 $D_2$/Pt |
| **4th** | 0.270 $C_2D_4$/Pt | 0.024 $D_2$/Pt | 0.187 $C_2D_4$/Pt | 0.028 $D_2$/Pt |

**Fig. 1** also shows the $4^{th}$ TPD/R run for each sample. Two significant changes are observed: the amount of ethylene desorbing is lower than in the $1^{st}$ run, and the desorption occurs at lower temperatures. In addition, $D_2$ desorption is substantially weaker, compared to the $1^{st}$ TPD/R run. The changes are qualitatively similar for the two cluster sizes, but more dramatic for $Pt_7$, where more dehydrogenation occurred in the $1^{st}$ run. These changes imply that substantially less ethylene adorbs at 180 K, and that the decrease primarily affects the more stable binding sites responsible for desorption at higher temperatures. As will be shown shortly, stronger binding typically corresponds to the di-σ mode of ethylene attachment, which is a precursor for dehydrogenation, which then poisons those sites by carbon deposits. As shown in **Fig. S1**, the $2^{nd}$, $3^{rd}$, and $4^{th}$ TPD/R runs are quite similar, implying that most of the changes to the binding site distribution occur in the $1^{st}$ TPD/R run. In the $4^{th}$ TPD run, the differences between $Pt_4/SiO_2$ and $Pt_7$/SiO2 are weaker than in the first, but the $Pt_4$ sample still had significantly more ethylene desorbing, and with sharper temperature dependence, compared to the $Pt_7$ sample. Given that both samples had the same total number of Pt atoms, the samples clearly retained at least some memory of the deposited cluster size after repeated TPD/R runs.

Run-to-run losses in the number of $C_2D_4$-Pt sites selectively affects the most-strongly bound $C_2D_4$-Pt sites, and because these are most likely to catalyze dehydrogenation, a run-to-run reduction in $D_2$ desorption is expected, and observed. Given the observation of $D_2$ desorption, the loss of $C_2D_4$ binding sites in repeated $C_2D_4$ TPD runs is at least partially attributable to poisoning by carbon deposition. We show below using XPS that repeated ethylene TPD runs lead to carbon deposition, although the precise amount was difficult to measure because the sensitivity for carbon is relatively low, and the Pt coverage (where C is depositing) is also very low.[1] DFT calculations on $Pt_8/Al_2O_3$ and $Pt_8/Al_2O_3$ with a carbon atom deposited were done to examine changes in cluster morphology and energetics, showing that deposition of electrophilic C causes some rearrangement in cluster geometry.[1] Note, however, that supported $Pt_n$ clusters are expected to be quite dynamic under reaction conditions, sintering is also likely.

Because we have ~3 x $10^{-11}$ Torr of CO background in our UHV system, we always monitor CO desorption during TPD/R experiments. Here, ~0.2 CO molecules desorbed *per* Pt atom during the 1st run, with substantially less in subsequent runs. This CO contamination level is ~20 times larger than would be expected if CO sticks only when impinging directly on a Pt site, and shows that for small, well dispersed clusters, substrate-mediated adsorption (reverse spillover) substantially amplifies the contamination rate. Assuming that CO competes with $C_2D_4$ for binding sites, there probably would have been additional $C_2D_4$ and $D_2$ desorption observed in the 1st TPD/R runs in absences of CO, and the difference between the 1st and subsequent runs would have been larger.

Analogous $C_2D_4$ TPD/R experiments for PtSn alloy clusters are shown in **Fig. 2,** with all four TPDs shown in **Fig. S2**. In this case, the 1$^{st}$ run was done just after completing the $H_2$/$SnCl_4$/$H_2$ exposure process used to deposit Sn, and prior to any heating. At this point there would have been H as well as a small amount of Cl adsorbed on the clusters.[2] Therefore, it is not surprising that the amount of $C_2D_4$ desorbing in this 1$^{st}$ run is substantially smaller than in the 1$^{st}$ run for $Pt_n/SiO_2$, because many $C_2D_4$ binding sites would have been blocked during the 180 K $C_2D_4$ exposure. During this 1$^{st}$ TPD/R run, significant HCl desorption was observed (**Fig. S3.**), and no Cl was detectable by TPD/R, ISS, or XPS after the 1st run.  $H_2$ desorption was not monitored due to large mass 2 background, but 700 K is well above the temperature where hydrogen desorbs from $Pt_n$.[1]

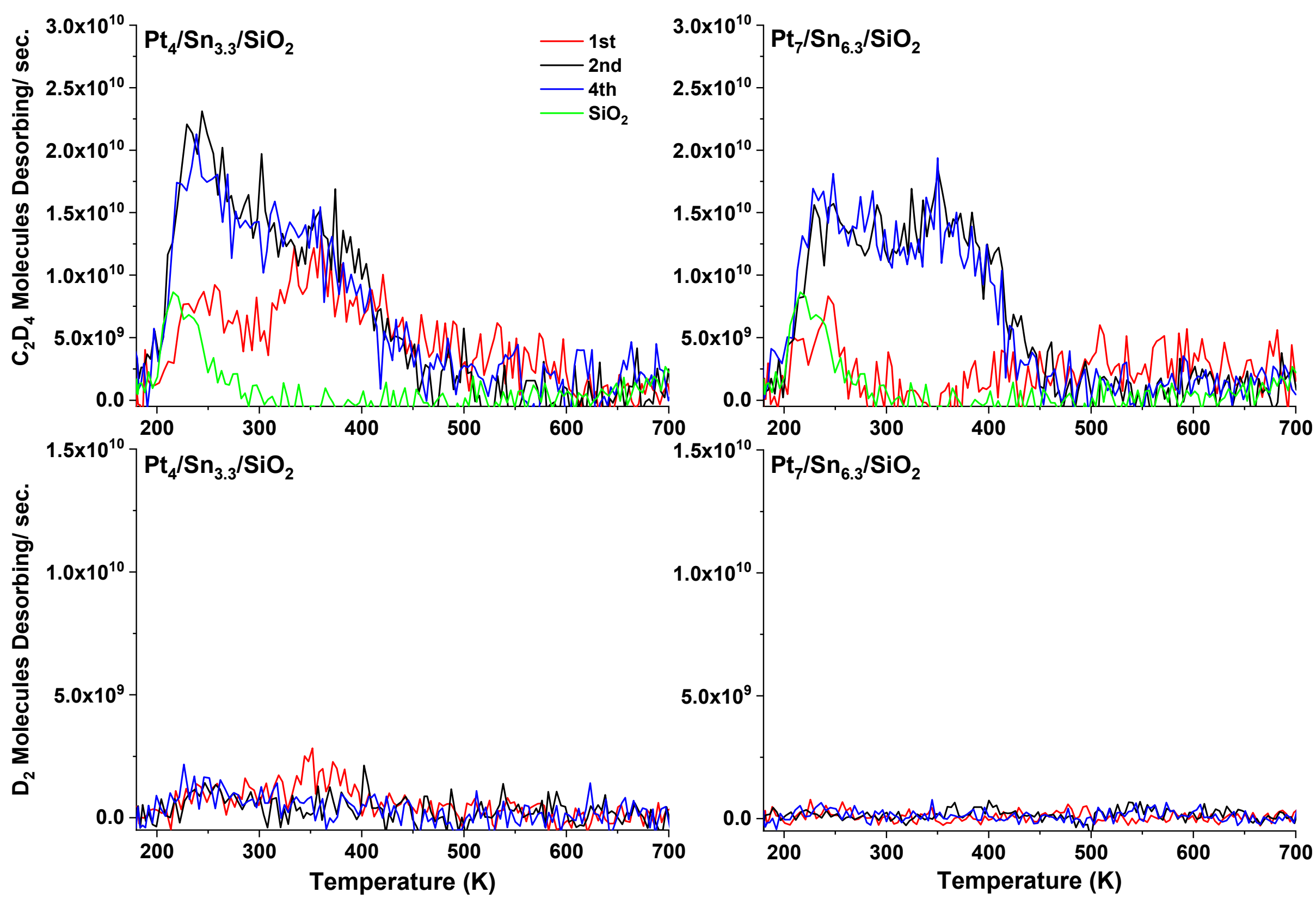

**Figure 2.** Desorption of $C_2D_4$ (top) and $D_2$ (bottom) from the first (red), second (black), and fourth (blue) $C_2D_4$ TPD/R run. Each spectra was collected after a 10 L dose of $C_2D_4$ to $Pt_n/SiO_2$ (n = 4, 7) at 180 K. $C_2D_4$ desorption from bare $SiO_2$ treated with 1 ALD cycle is also plotted (green) – no $D_2$ was observed in that experiment.

In the 2nd TPD/R run on each sample, the amount of $C_2D_4$ desorbing increased substantially, indicating that desorption of HCl and $H_2$ during the 1st run left more binding sites accessible to $C_2D_4$.  Therefore, we feel that the correct comparison is of the 2nd TPD/R run for $Pt_nSn_x/SiO_2$, with the 1st run for the $Pt_n/SiO_2$ samples.  As before, the number of the $C_2D_4$ molecules desorbing *per* Pt atom in the 2nd, 3rd, and 4th TPD/R runs is given in **Table 2.**

**Table 2.** Numbers of $C_2D_4$ molecules desorbing per Pt atom from Fig. 2 and Fig. S2.

| **TPD Run:** | **$Pt_4Sn_{3.3}/SiO_2$** | **$Pt_7Sn_{6.3}/SiO_2$** |
|---|---|---|
| **2nd** | 0.190 $C_2D_4$/Pt | 0.176 $C_2D_4$/Pt |
| **3rd** | 0.210 $C_2D_4$/Pt | 0.170 $C_2D_4$/Pt |
| **4th** | 0.190 $C_2D_4$/Pt | 0.174 $C_2D_4$/Pt |

There are several important points of comparison with the results for pure $Pt_n/SiO_2$.  First, there was no significant $D_2$ desorption in any run on the $Pt_nSn_x/SiO_2$ samples, i.e., all ethylene desorbed intact.  The absence of $D_2$ desorption implies that there should be little if any carbon deposition. **Fig. S4** summarizes an XPS experiment that shows substantial carbon deposition in repeated TPD/R runs on $Pt_n/SiO_2$, and the complete absence of carbon deposition for $Pt_nSn_x/SiO_2$ under identical conditions.

Also consistent with the absence of carbon deposition, is the observation that once the H and Cl were desorbed during the 1st run, there was little change in the intensity or temperature dependence of $C_2D_4$ desorption in TPD runs two through four, compared to the substantial loss of intensity and shifts to lower temperatures observed for pure $Pt_n/SiO_2$.  Additionally, the number of $C_2D_4$ molecules adsorbing during the 180 K dose was ~50% smaller for the $Pt_nSn_x$, compared to the pure $Pt_n$, presumably due to site blocking by Sn atoms (note difference in vertical scales). Finally, for the $Pt_nSn_x/SiO_2$ samples there are cluster size effects on the desorption temperature dependence, weaker than the size effects observed for $Pt_n/SiO_2$, but which are quite persistent in repeated runs. From the perspective of improving the stability of sub-nano Pt cluster catalysts under high temperature hydrocarbon rich conditions, the absence of dehydrogenation (i.e., of carbon deposition) and the stability of the samples in repeated TPD/R runs are both important.

The $C_2D_4$ thermal desorption spectra were fit to extract desorption energy distributions, $\theta(E_{des})$, as described in the supporting information. An example fit is shown in **Fig. S5**, and the $\theta(E_{des})$ distributions for the 1$^{st}$ $C_2D_4$ TPD on $Pt_4/SiO_2$ and the 2$^{nd}$ $C_2D_4$ TPD on $Pt_4Sn_{3.3}/SiO_2$ are compared in **Fig. 3**. $\theta(E_{des})$ distributions for $Pt_7/SiO_2$ and $Pt_7Sn_{6.3}/SiO_2$ are shown in **Fig. S6**. Three features were included in the $\theta(E_{des})$ distributions to fit the experimental temperature dependences: a small feature with $E_{des} < 0.6$ eV attributed to desorption from $SiO_2$ sites, a larger feature with $0.6 < E_{des} < 1.2$ eV attributed to desorption from cluster sites, and a second feature with $0.8 < E_{des} < 1.6$ eV also attributed to desorption from cluster sites. Note that there is little difference in the range of $E_{des}$ observed for $Pt_n$ and $Pt_nSn_x$, consistent with the modest differences in desorption temperatures.

The effects of Sn-alloying on the $Pt_n/SiO_2$ system should be compared to the effects of Sn alloying on ordered PtSn surface alloys studied by Koel and co-workers.[31-32] In both cases, ethylene dehydrogenation, leading to carbon deposition, is strongly suppressed, compared to pure $Pt_n/SiO_2$ or to Pt(111), but the mechanism appears to be different. For the PtSn surface alloys, the saturation ethylene coverage was Sn-independent, whereas for the clusters, Sn significantly reduced the coverage. For the surface alloys, the ethylene desorption temperature dropped with increasing Sn coverage from ~285 K for pure Pt(111)[36] to 184 K for the $\sqrt{3}\,x\,\sqrt{3}\,R30°$ alloy which has 1:2 Sn:Pt ratio in the surface layer. Hook *et al.* used DFT to examine ethane dehydrogenation over PtSn surface alloys, finding, consistent with the experiments of Koel and co-workers, that Sn depresses the alkene desorption energy below the barrier for dehydrogenation, due to a combination of geometric and electronic effects.[33] Thus in the surface alloys, the suppression of carbon deposition appears to be largely due to the reduction in ethylene binding energy. This was also the conclusion in our study of the effects of boron addition to $Pt_n$/alumina model catalysts,[4] where boration shifted the ethylene desorption temperatures into the cryogenic range, thereby suppressing dehydrogenation. In contrast, for the $Pt_n$ vs. $Pt_nSn_x$ clusters here, there is no significant difference between the ethylene desorption temperatures or the extracted $E_{des}$ distributions, suggesting that suppression of dehydrogenation/carbon deposition must result from a different mechanism.

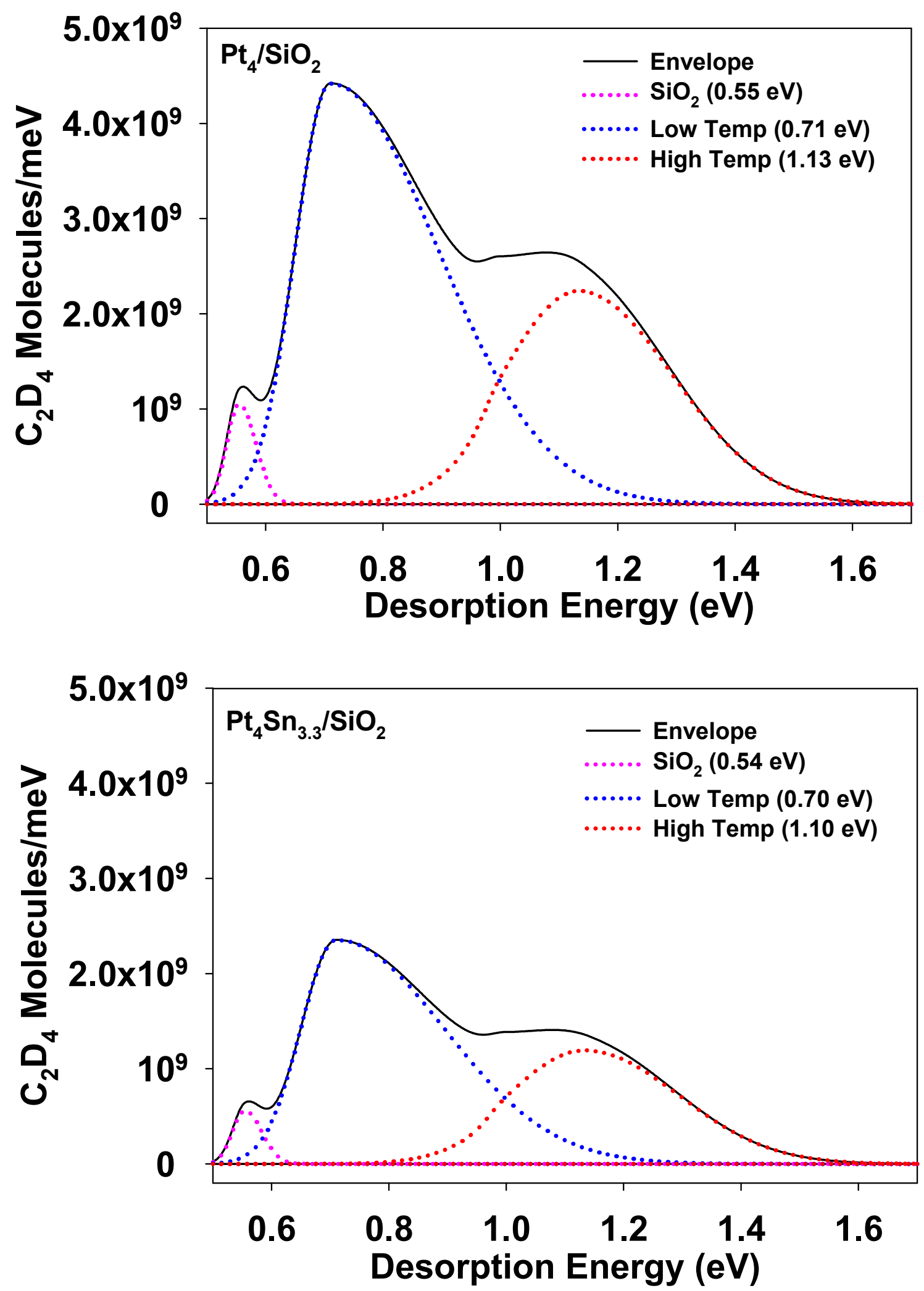

**Figure 3.** Energy desorption ($E_{des}$) profiles for $Pt_4/SiO_2$ (top) and $Pt_4Sn_{3.3}/SiO_2$ (bottom).

**ISS probing of the effects of TPD/R.** ISS experiments were carried out to probe differences in the structure of the $Pt_n$ and $Pt_nSn_x$ samples, and to provide insight regarding the changes in $C_2D_4$ and $D_2$ desorption behavior that occur in multiple TPD runs for $Pt_n/SiO_2$. Peaks in ISS data result mostly from events in which $He^+$ scatters from a single atom in the surface layer, such that the retained energy, $E_k/E_{k0}$, depends on the target atom mass.[37] Multiple scattering, and scattering from atoms deeper in the surface result in low ion survival probability (ISP), and contribute mostly to the featureless background at low $E_k/E_{k0}$. The usefulness of ISS here is that it is sensitive both to the cluster morphology and to the presence of adsorbates on the cluster surface, due to shadowing, blocking, and ISP effects.[38-40]

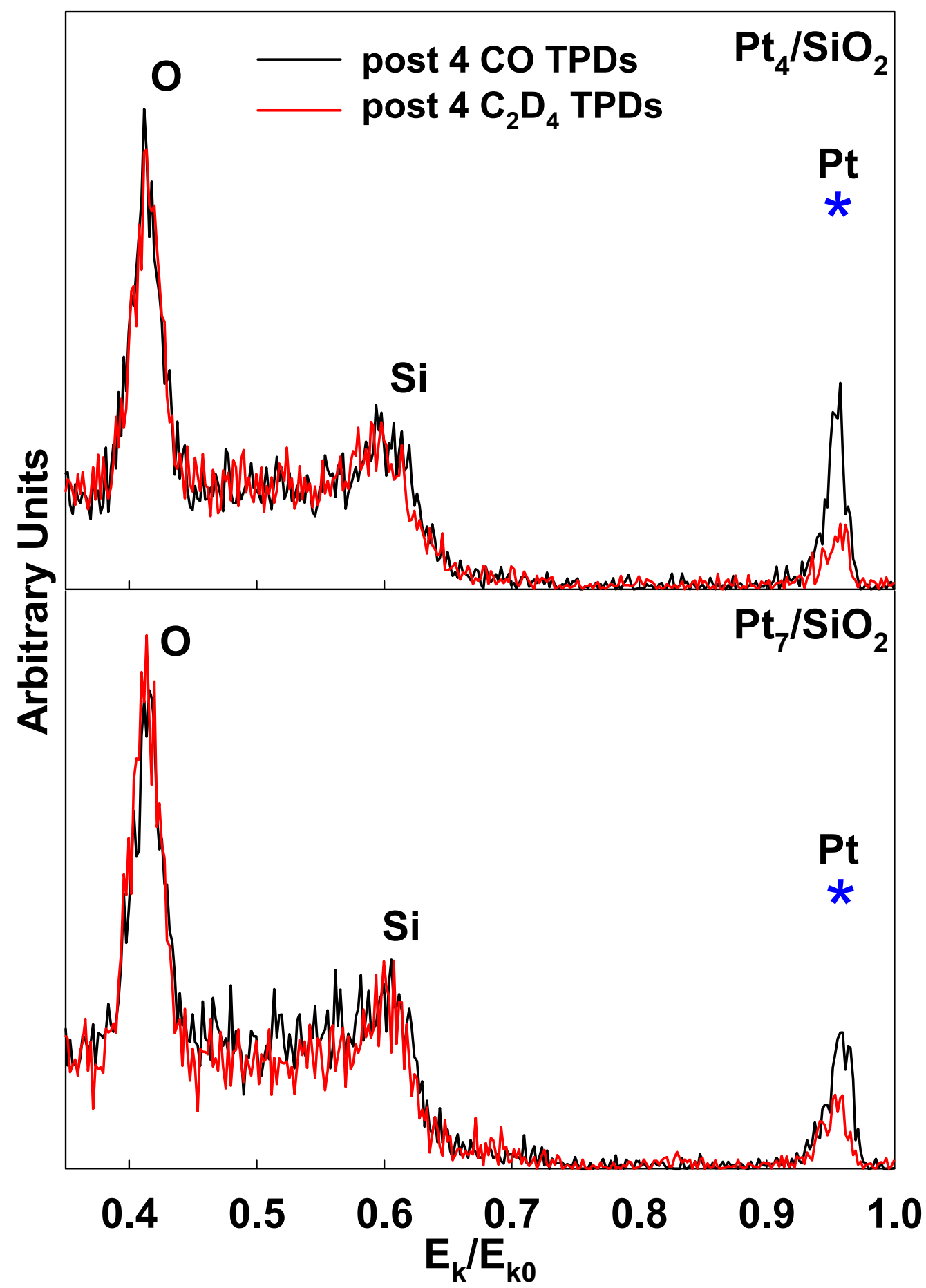

**Figure 4.** Raw ISS spectra for $Pt_n/SiO_2$ samples after they have undergone four CO TPDs to 700 K (black) and four $C_2D_4$ TPDs (red).

**Fig. 4** summarizes ISS data for $Pt_4$ and $Pt_7$ samples after four cycles of $C_2D_4$ TPD/R. The results are compared to ISS (on separate samples) after four cycles of CO TPD under identical conditions, to show the effects of TPD with a non-coking adsorbate. Peaks for O, Si, and Pt are observed. For comparison, the intensities of the Pt peak for as-deposited, unheated $Pt_n/SiO_2$ samples are indicated in **Fig. 4** by blue stars. The O and Si intensities for the as-deposited samples are not shown because they are essentially identical to the post-TPD/R results in the figure. All the samples were prepared with the same total number of Pt atoms, therefore, if the same fraction of Pt atoms are exposed in the surface layer, the Pt ISS intensities should be identical. It can be seen that the as-deposited Pt intensity is highest for $Pt_4$, and ~28% lower for $Pt_7$, indicating that some of the Pt atoms in $Pt_7/SiO_2$ are not in the surface layer, as would be the case for a 3D cluster structure.

After four CO TPD runs, the Pt intensities for both sizes decrease significantly. This decrease suggests significant sintering of the clusters, although there may also be some contribution from isomerization to structures with fewer atoms in the surface layer.[1-2, 39, 41-42] To test for the possibility that CO adsorption might affect the sintering/isomerization process, a separate experiment was performed in which a $Pt_4/SiO_2$ sample was simply flashed to 700 K four times in the absence of any deliberate adsorbate exposure. The resulting Pt ISS intensities were identical to those observed for clusters having undergone four CO TPDs, i.e., the effects are thermal, with no significant effect of small CO exposures.

The decreases in Pt ISS signal after 4 $C_2D_4$ TPD/R experiments are substantially larger. This could be taken as evidence that $C_2D_4$ enhances sintering, however, we know from the observation of $D_2$ desorption and the appearance of carbon in post-TPD/R XPS, that there is carbon deposition on these samples. The ISS results are, therefore, confirmation that this carbon is deposited on, and blocks access to Pt sites, as opposed to depositing on the $SiO_2$ support. Given that the clusters may also be sintering during TPD/R, it is most useful to compare the carbon deposition on a *per* Pt atom, rather than a *per* cluster basis. From the 1$^{st}$ TPD/R run data in (**Fig. 1**), we can see that $Pt_4$ has a higher density of $C_2D_4$ binding sites (0.43/Pt atom), than $Pt_7$ (0.36/Pt atom), and a lower $D_2/C_2D_4$ desorption ratio (0.23), compared to 0.36 for $Pt_7$. If we assume each $D_2$ desorption results in deposition on one C atom, then for $Pt_4$ the 1$^{st}$ TPD/R run deposits ~0.1 C/Pt atom, while for $Pt_7$, it should deposit slightly more – 0.13 C/Pt. As can be seen from **Fig. S1** the differences are smaller in subsequent TPD/R runs. After four runs, the final Pt ISS intensities are similar for the two clusters.

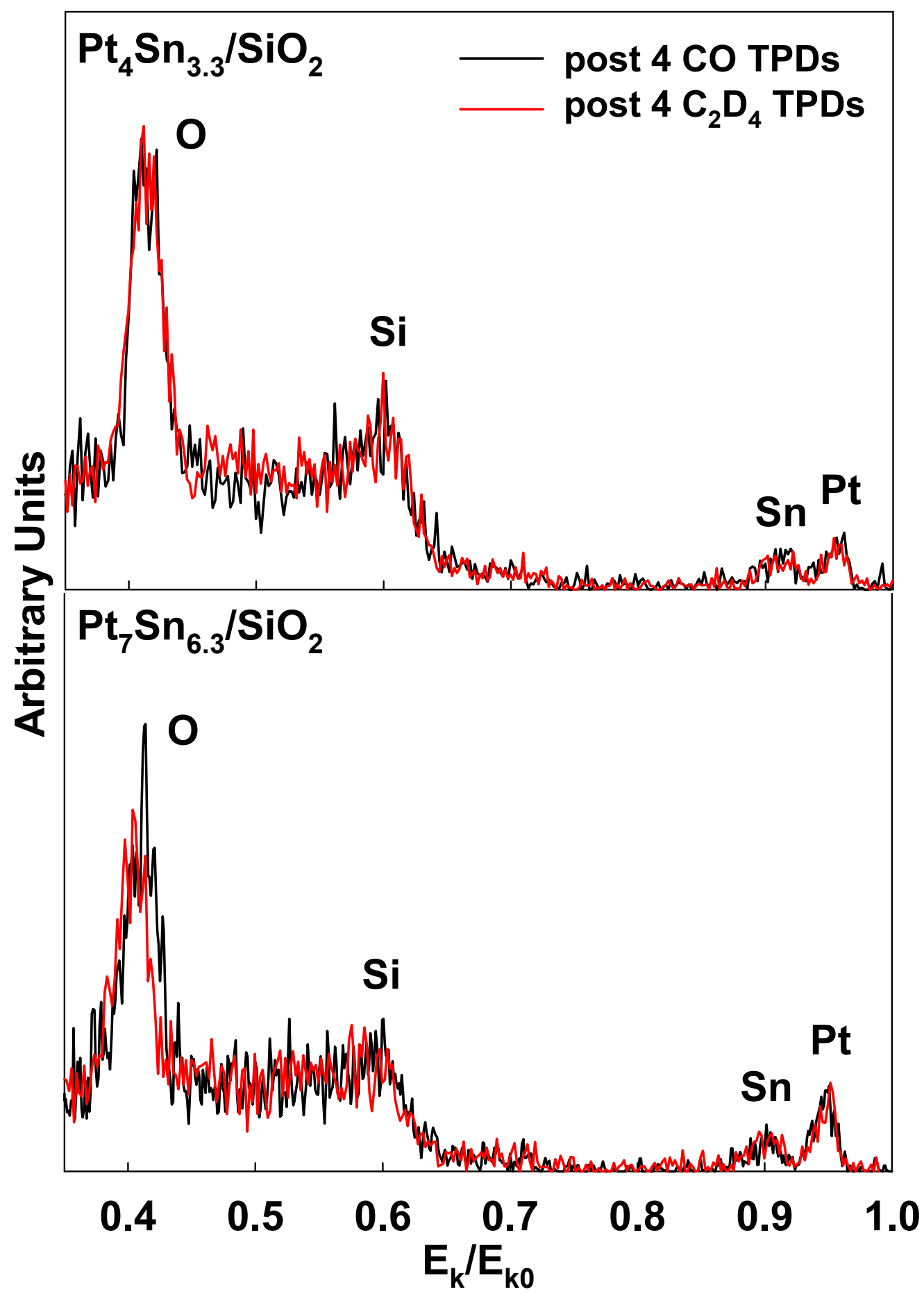

**Figure 5.** Raw ISS Scans for $Pt_4Sn_{3.3}/SiO_2$ and $Pt_7Sn_{6.3}/SiO_2$ after four $C_2D_4$ TPDs (red) and 4 CO TPDs (black)

Analogous ISS experiments for $Pt_nSn_x/SiO_2$ are shown in **Fig. 5**. Sets of $Pt_nSn_x/SiO_2$ samples were prepared and exposed to either four $C_2D_4$ TPD/R runs, or four CO TPD runs, then examined by ISS. It can be seen that the O and Si peaks are similar to those in the $Pt_n/SiO_2$ samples, as expected because Sn deposits primarily on the Pt clusters,[2] but that the Pt peak is substantially less intense, and a peak for Sn is observed.

It is also interesting to compare the post TPD ISS to ISS measured just after the $H_2/SnCl_2/H_2$ treatment (prior to any heating), and after the initial heating to drive of H and Cl. For the treated, unheated sample, the Sn peak is larger and the Pt peak is smaller (Sn:Pt ratio ≈ 2) compared to those in **Fig. 5**, as might be expected for Sn initially deposited on top of the Pt clusters. After initial heating to drive off H and Cl, the spectrum is essentially identical to the post-TPD spectra shown in **Fig. 5**. i.e., neither of the repeated TPD/R experiments has any significant effect on the ISS. The stability with respect to repeated $C_2D_4$ TPD/R runs is a consequence of the absence of carbon deposition on the clusters. Consistent with the TPD/R data in **Fig. 2**, ISS shows that Sn alloying significantly reduces the number accessible Pt binding sites, but that the sites present are quite stable.

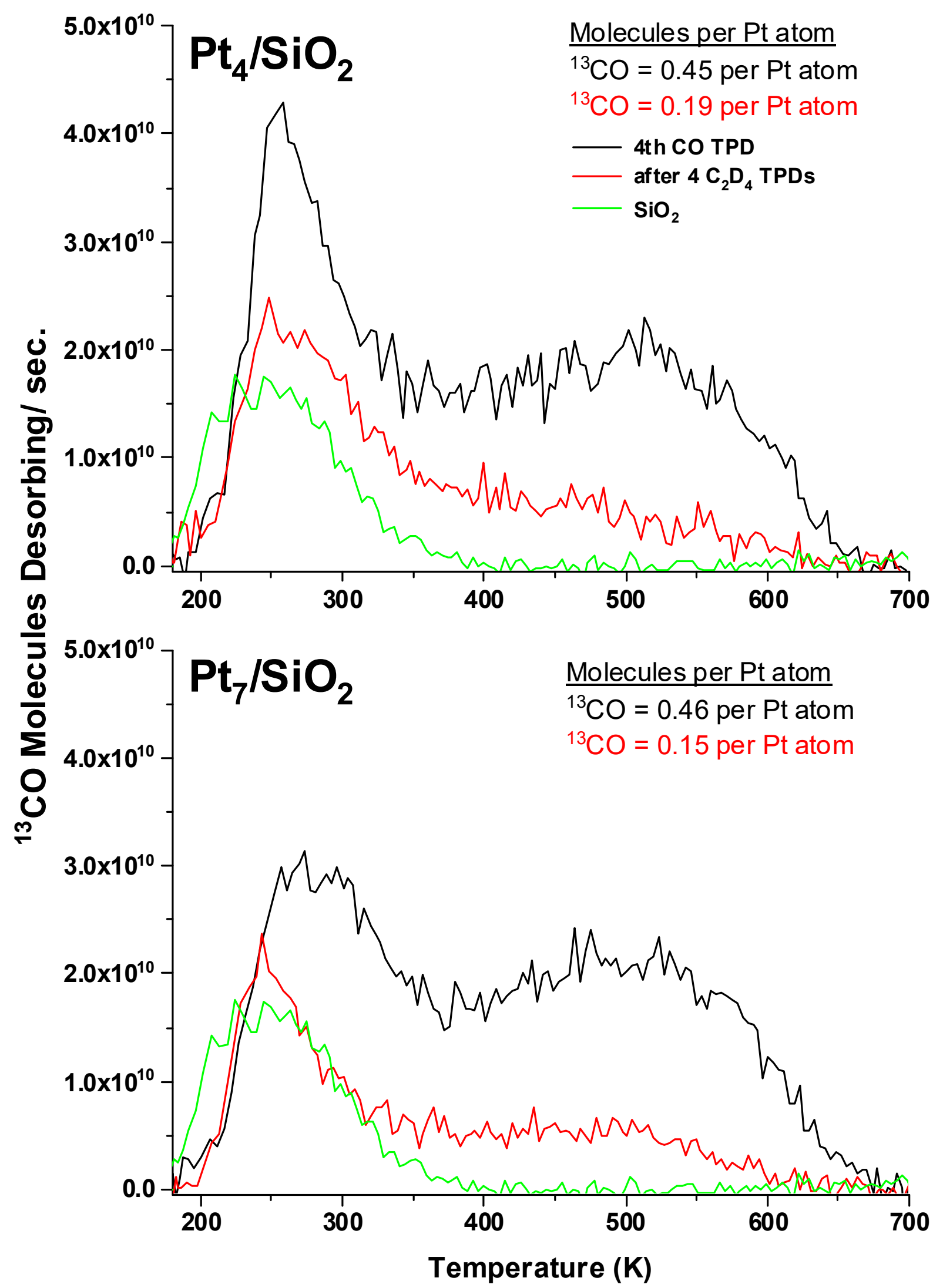

**Figure 6.** $^{13}CO$ TPD after four CO TPDs (black line) and a $^{13}CO$ TPD after four $C_2D_4$ TPDs for $Pt_n/SiO_2$ (red). CO desorption from $SiO_2$ is also shown (green)

**CO desorption as a probe of accessible Pt sites.** The final approach used to probe the effects of repeated heating and $C_2D_4$ exposure on the clusters was to carry out $^{13}CO$ TPD after the $C_2D_4$ TPD/R sequence (on different samples than those used for ISS characterization). **Fig. 6** compares CO TPD measured for a sample first subjected to four $C_2D_4$ TPD/R runs (red) to the CO desorption measured in the 4$^{th}$ in a series of sequential CO TPDs (black). For comparison, CO desorption from the $SiO_2$ substrate is also shown (green). The ion signals were converted to numbers of CO molecules using the process described above, and the integrated number of CO molecules indicated in the text on each frame of the figure have been corrected for CO desorbing from the $SiO_2$ substrate, i.e., the numbers correspond to the CO desorbing from Pt sites. For $Pt_4$ and $Pt_7$, repeated $C_2D_4$ TPD/R results in far larger attenuation of Pt-associated CO binding sites, and particularly of the sites with high $E_{des}$, compared to repeated CO TPD runs. This mirrors the ISS results, where $C_2D_4$ TPD/R caused a much larger Pt intensity attenuation.

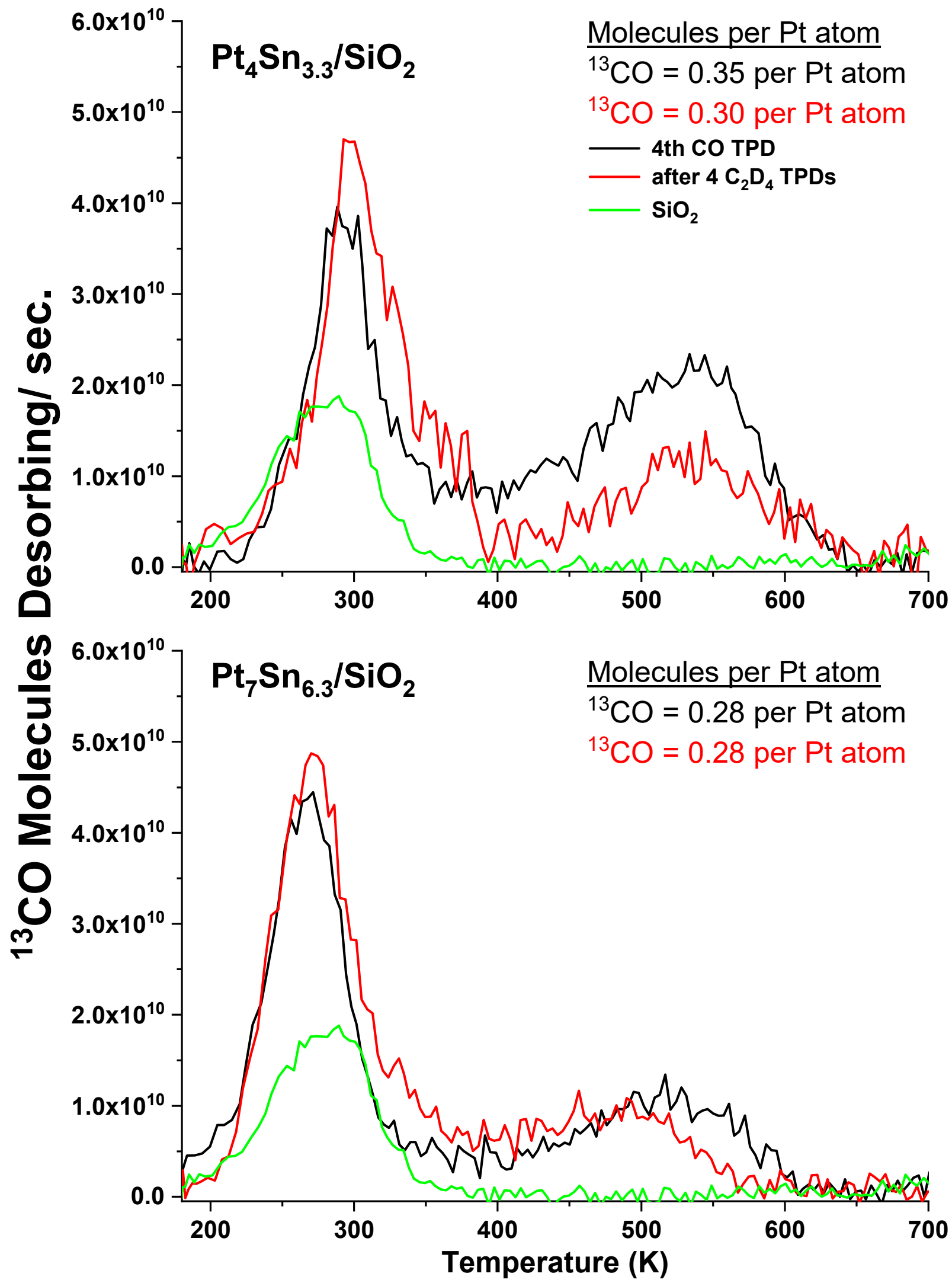

**Figure 7.** CO TPD spectra of $Pt_nSn_x/SiO_2$ after four CO TPDs (black) and four $C_2D_4$ TPDs (red). The 2$^{nd}$ $^{13}CO$ TPD from $H_2/SnCl_4/H_2 – SiO_2$ is also shown (green).

**Fig. 7** shows the analogous results for Sn ALD-treated samples. In this case, the desorption shown for the substrate ("$H_2/SnCl_4/H_2$ – $SiO_2$") is for $SiO_2$ subjected to the Sn ALD treatment, then heated to drive off any residual Cl or H, before being probed by CO TPD. For the Sn-treated samples, the differences in CO desorption for samples after repeated CO TPDs vs. repeated $C_2D_4$ TPD/R runs are much smaller than for the Sn-free $Pt_n/SiO_2$ samples. For $Pt_4Sn_{3.3}$, repeated $C_2D_4$ TPD/R runs resulted in a shift in intensity from the higher to the lower temperature desorption feature, with a ~14% decrease in the integrated number of CO molecules desorbing, compared to the effects of four CO TPD runs. For $Pt_7Sn_{6.3}$, similar, but smaller changes in the desorption temperature dependence were observed, and the integrated number of CO molecules desorbing was identical for the samples subjected to multiple CO or $C_2D_4$ TPD experiments. It is not clear why the difference between the effects of $C_2D_4$ and CO TPD/R was larger for the smaller cluster size, however, for such a small cluster it would not be surprising if the degree of sintering occurring during the TPD heat ramps is affected by whether the clusters are saturated by CO or by $C_2D_4$. Note that in the analogous experiments on pure $Pt_4$, both the low and high temperature desorption features decreased significantly more after $C_2D_4$ TPD/R, such that the decrease in total number of CO binding sites was almost 60% bigger than in multiple CO TPDs. Clearly, the carbon deposition occurring during $C_2D_4$ TPD on pure $Pt_n$ has a large effect on the site availability.

**DFT Analysis of the Effects of Sn Alloying on Ethylene Binding and Dehydrogenation.** Supported cluster catalysts can have many structural isomers accessible in the ≤700 K temperature regime of interest here.[43-46] These isomers can have similar energies, despite having different morphologies, with different morphologies favoring different binding sites and chemical character. It is important to consider these different isomers in the mechanism, because they are expected to be thermally-populated due to high fluxionality of the clusters in this temperature regime.[22,47] Due to the large amount of sampling needed for the reliable exploration of the potential energy surfaces, we focus the theory on $Pt_4/SiO_2$ and $Pt_4Sn_3/SiO_2$.

From the experiments, an average of ~1.4 $C_2D_4$ molecules are adsorbed *per* $Pt_4$ cluster; therefore, each isomer has also been optimized with either one or two $C_2H_4$ molecules adsorbed. Structures obtained from the global optimization are shown in **Figs. 8 and S7**. Note that $C_2H_4$ is able to bind in several configurations depending on the isomer. For instance, $C_2H_4$ in the global minimum structure of $C_2H_4/Pt_4/SiO_2$ preferentially binds in the di-σ binding mode, whereas in the global minimum of $(C_2H_4)_2/Pt_4/SiO_2$ there is one di-σ bound ethylene, and one π-bound ethylene. Note that other higher energy isomers, which are thermally accessible at higher temperatures, can have different $C_2H_4$ binding preferences than the global minimum, as shown in **Fig. 8**. Previously, it was shown that the stronger di-σ interaction favors dehydrogenation.[1]

Referencing the literature augments our interpretations, as ethylene dehydrogenation has been thoroughly studied via DFT and various surface analysis methods over extended Pt surfaces. For example, for temperatures around 100 K, LEED showed ethylene adsorption having a clear preference for di-σ adsorption to Pt(100)[48] and Pt(111).[49] In contrast, EELS analysis of stepped/lower-coordinated Pt sites (Pt(210) and Pt(110)) showed π-binding of ethylene.[50] The di-σ binding was probed using TPD of mixed $C_2D_4$ and $C_2H_4$ adsorbed on Pt(111) by Janssens *et al.*, who reported recombinative desorption of isotope-scrambled ethylene at ~285 K, indicating dissociative adsorption.[51] Just above the 285 K desorption peak, hydrogen evolution begins, with the first H loss resulting in a surface-bound ethylidyne, although this is largely considered to be a spectator between the first and remaining H dissociation events.[52-54] As the sample temperature continues to increase, additional dehydrogenation occurs, indicated by additional $H_2$ desorption until 700 K. Results for the π-bound molecules on low-coordinated stepped surfaces were quite different. For example Yagasaki *et al*. studied $C_2H_4$ decomposition on Pt(210) and reported that some desorbs intact around 250 K, but the residual π-adsorbed $C_2H_4$ undergoes dehydrogenation around 300 K, as shown by desorption of $H_2$ and detection of surface ethylidyne with both C atoms coordinated to the surface. Further heating the Pt(210) surface to 700 K drives dehydrogenation to completion.

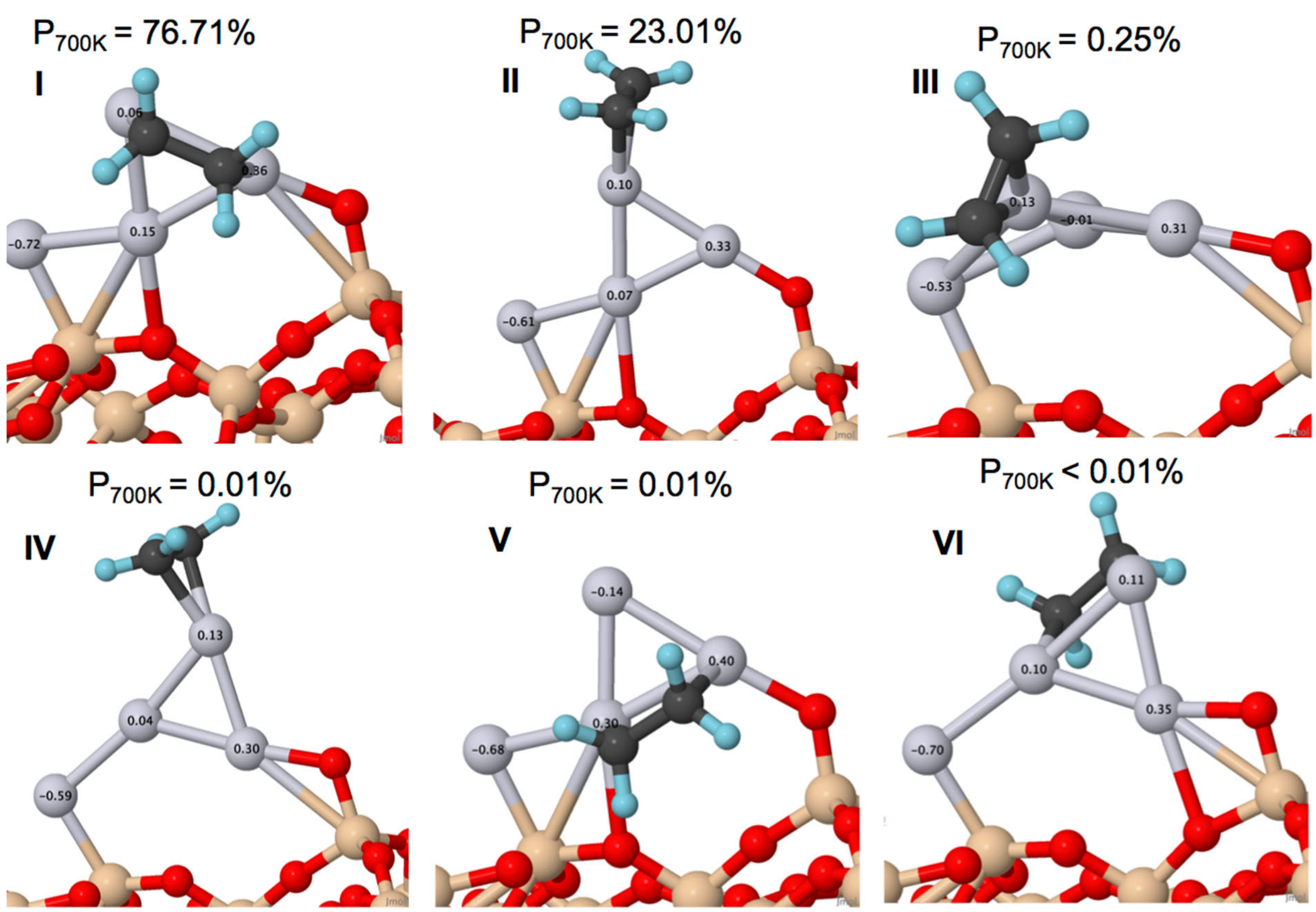

**Figure 8.** Thermally-accessible geometries of $C_2H_4/Pt_4/SiO_2$ obtained from global optimization calculations. The geometries of $(C_2H_4)_2/Pt_4/SiO_2$ are shown in **Fig. S7**.

Pt(110) showed different behavior, as reported by Yagasaki *et al*. Here, the some of the π-adsorbed ethylene is isomerized to the di-σ mode by 160 K. From 270 K–330 K, it undergoes decomposition to both C atoms and ethylidyne surface species, while evolving methane and hydrogen. The strongly bound ethylidyne then completes dehydrogenation by 450 K. For comparison, an IR absorption study[55] was done on $Al_2O_3$-supported Pt nanoparticles around 180 K that resulted three ethylene adsorption modes: π-adsorption, di-σ adsorption, and spontaneous ethylidyne formation. Interestingly, all π-adsorbed ethylene was found to desorb intact by room temperature, while the di-σ bound ethylene is converted to ethylidyne and hydrogen.

Interpretation of our DFT and experimental results in light of these results allows conclusions to be made about the mechanism in our $Pt_n/SiO_2$ experiments. The three structures I, II, and III shown in **Fig. 8** and structure I in **Fig. S7** each have significant contributions to the cluster ensemble and we thus expect surface-bound ethylene species to adsorb in both the π and di-σ modes. We consider the comment by Mohsin *et al.*, that ethylidyne can spontaneously form at this temperature (as a result of dissociation of surface ethylene). In a $D_2$ thermal desorption study for Pt(100), Pasteur *et al.* showed recombinative desorption between 250 K and 500 K.[56] Spontaneous $D_2$ desorption is therefore not expected; rather we would expect these species to exist as surface-bound.

The calculated structures for ethylene on bare $Pt_4/SiO_2$ offer some explanation for the ISS experiments, which were conducted at 150 K, where only the lowest energy isomers are accessible, and all Pt atoms are in the surface layer. This results in the highest scattering intensity for as-deposited $Pt_4$ clusters, when compared to larger, multilayered clusters.[1] At higher temperatures, the catalytic activity observed experimentally is a result of a linear combination of the populated structures and the mechanisms they individually promote. With a combination of binding modes, diverse chemical behavior is expected, where both intact desorption of ethylene and dehydrogenation of ethylene would be expected, as observed.

In bulk, calorimetric experiments by Anres *et al.* showed that PtSn has a highly negative enthalpy of formation.[57] Similarly, Liu and Ascencio reported DFT simulations for clusters of a few hundred atoms that found PtSn clusters to be energetically favorable compared to pure Pt clusters.[58] For sub-nano clusters, DFT shows that Sn incorporation to size-selected Pt clusters results in substantial electron transfer from Sn to Pt atoms, and reduces the fluxionality of the clusters, such that one isomer dominates the ensemble.[2] The dominant isomer is an intermixed $Pt_4Sn_3$ structure, with each atom showing a relatively high degree of coordination. From the chemistry perspective, one important result was that adding Sn quenches all unpaired spins on Pt, while in the pure $Pt_4/SiO_2$ system there are low energy isomers with unpaired electrons available for adsorbate binding.

To examine the effects of ethylene adsorption on PtSn clusters, DFT calculations were performed for ethylene adsorbed on both the global minimum $Pt_4Sn_3/SiO_2$ structure, and on several of the lowest energy isomers. The lowest energy isomer changes when ethylene is adsorbed, thus we expect the isomer ensemble to change as the clusters are heated with adsorbed ethylene. The most stable isomers for $C_2H_4/Pt_4Sn_3/SiO_2$ are shown in **Fig. 9**, along with Boltzmann population and Bader charge data. Structures with two adsorbed ethylene molecules are shown in **Fig. S8.** As shown in **Fig. 9**, the most stable isomers for a single ethylene have it π-bonded to Pt atoms in the cluster, however, as shown **Fig. S8**, one isomer was found for $(C_2D_4)_2/Pt_4Sn_3$ in which one of ethylene molecules is bound with one C atom bound to Sn, and one to an adjacent Pt atom. Note that for a single ethylene, di-σ binding has been completely eliminated by incorporation of Sn into the $Pt_4$ clusters. Structure I is the preferred $C_2H_4/Pt_4Sn_3/SiO_2$ isomer, with greater than 69% of the population at 700 K. The only structure with di-σ $C_2H_4$ is structure II for adsorption of two ethylene molecules (**Fig. S8**). While there is still some diversity in the lowest energy isomer distribution, $C_2H_4$ is only able to adsorb via π interaction, which is expected to limit the catalytic branching toward the products of deeper dehydrogenation on the $Pt_4Sn_{3.3}/SiO_2$ catalyst.

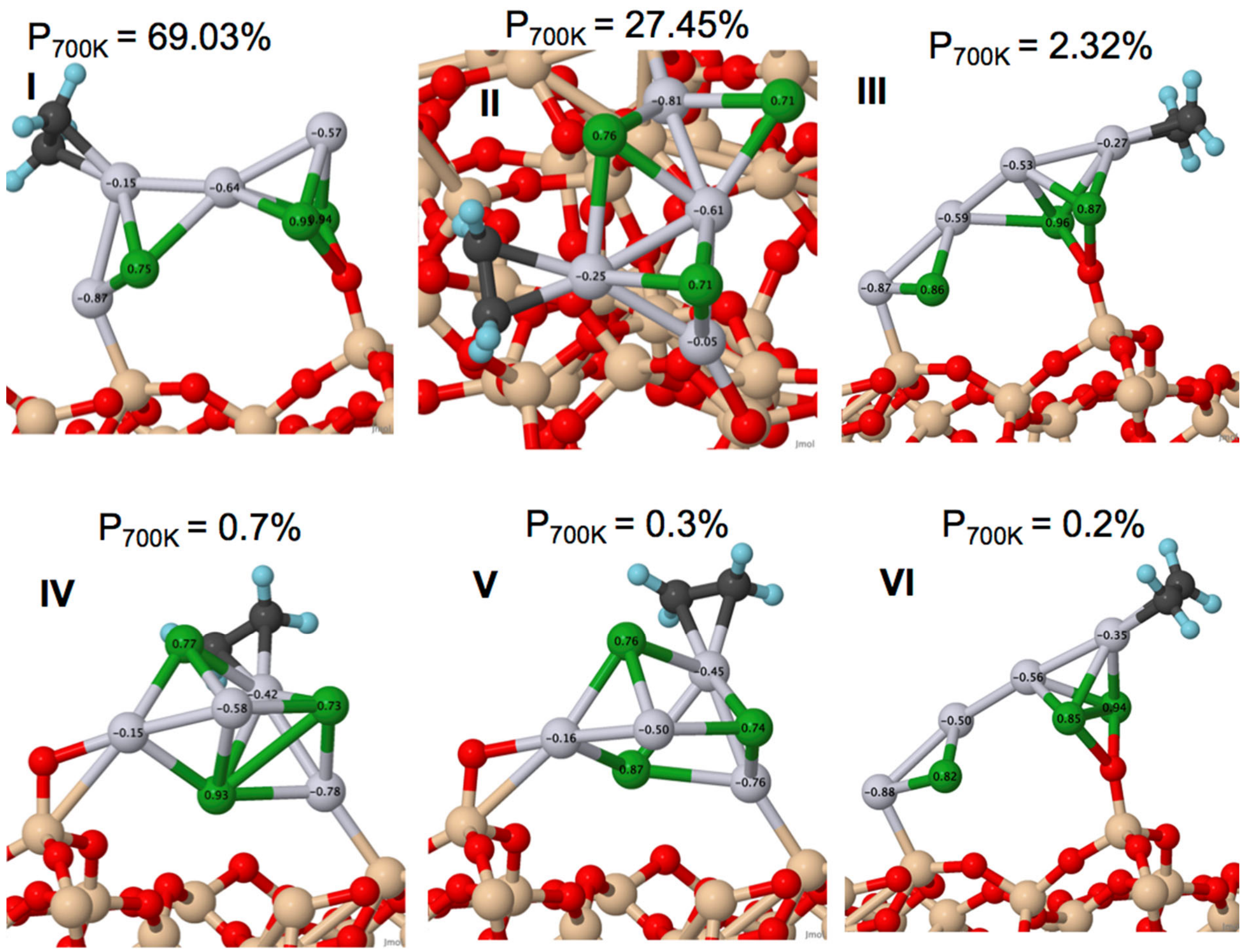

**Figure 9.** Thermally-accessible geometries of $C_2H_4/Pt_4Sn_3/SiO_2$ obtained from global optimization calculations. Note that the ensemble is dominated by $C_2H_4$ π-binding mode. The geometries of $(C_2H_4)_2/Pt_4Sn_3/SiO_2$ are shown in **Fig. S8**.

In order to compare the barriers of $C_2H_4$ dehydrogenation on $Pt_4/SiO_2$ vs. $Pt_4Sn_3/SiO_2$, CI-NEB calculations were done on the three structures of $Pt_4/SiO_2$ and $Pt_4Sn_3/SiO_2$ that are most populated at 700 K. For every structure, the barrier to break each of the four C-H bonds is considered. The minimum energy pathway for each isomer along with the structures of reactants, transition states, and products are shown in **Fig. 10**. Additional CI-NEB reaction paths for other isomers and ethylene binding sites, along with the structures involved, are summarized in **Figs S9 - S13**.

The lowest reaction barrier of the most populated $C_2H_4/Pt_4/SiO_2$ structure is 0.95 eV, whereas this value for $C_2H_4/Pt_4Sn_3/SiO_2$ is 1.29 eV. Isomer II of $C_2H_4/Pt_4Sn_3/SiO_2$ has a more accessible low-energy path (1.00 eV) which makes it the most active of the three lowest-energy isomers of the system. This confirms our working hypothesis that the less stable isomers tend to be more reactive, and they must be considered in order to accurately describe the catalytic properties of dynamic catalysts. Of course, the contributions of the isomers II and III to the reaction is smaller because of their smaller presence in the ensemble (see **Figs. 8 and 9**).

Additionally, **Fig. 11** gives the values for the 700 K ensemble-average rate constants ($k_{ens}$) for $C_2H_4/Pt_4/SiO_2$ and $C_2H_4/Pt_4Sn_3/SiO_2$ along with the rate constant calculated for each isomer, and the contribution of that isomer to the ensemble. The results show that the $Pt_4/SiO_2$ ensemble is more active at 700 K toward ethylene dehydrogenation than the $Pt_4Sn_3/SiO_2$ ensemble. At lower temperatures, such as the ~300 K onset temperature for dehydrogenation, the difference would be larger. These results confirm the experimental finding that introducing Sn to the cluster makes it harder for $C_2H_4$ to undergo dehydrogenation.

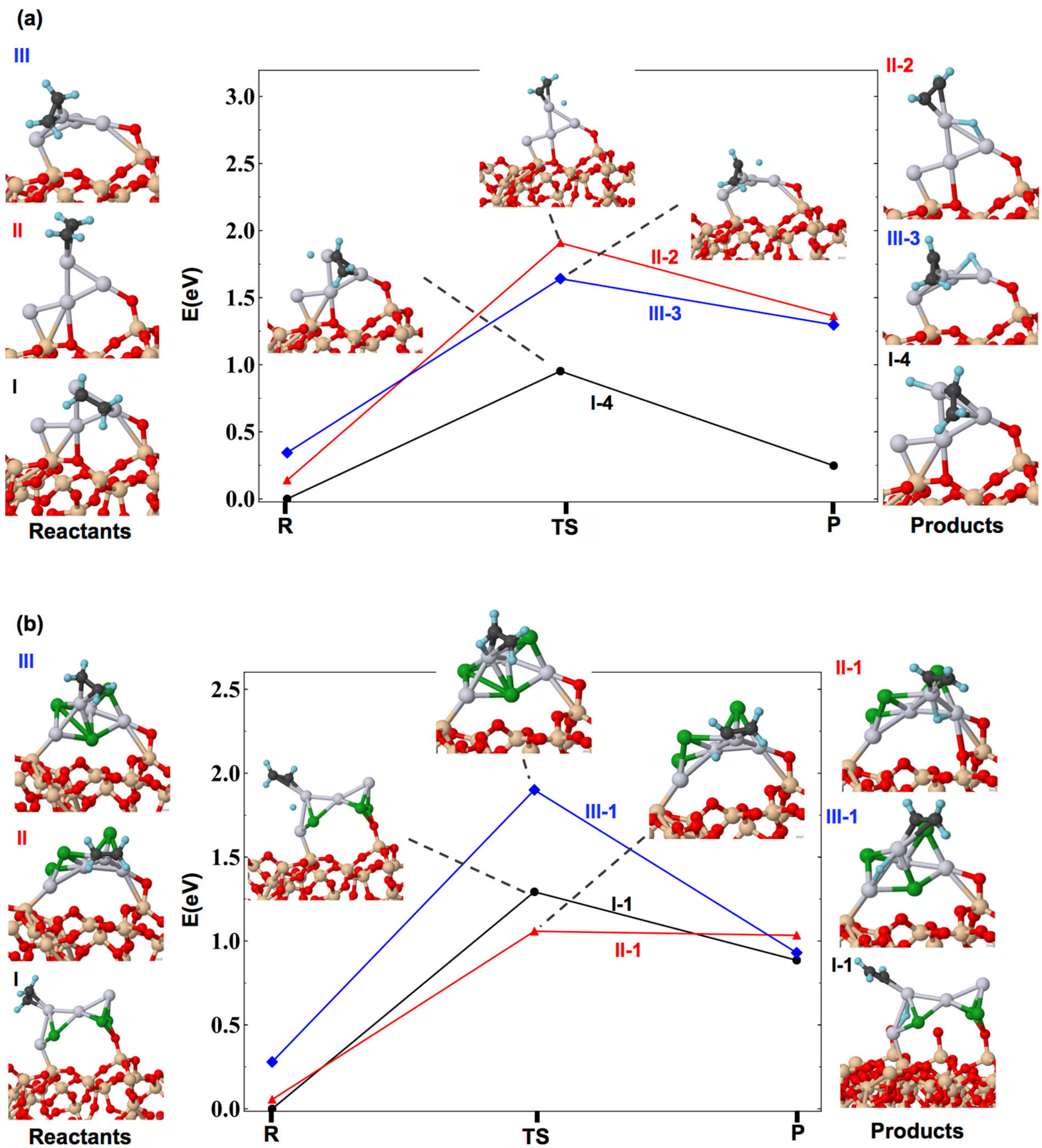

**Figure 10.** Lowest energy reaction profiles of breaking C-H bond obtained from CI-NEB calculations for each isomer of (a) $C_2H_4/Pt_4/SiO_2$ and (b) $C_2H_4/Pt_4Sn_3/SiO_2$ along with the structures of reactants, transition states, and products.

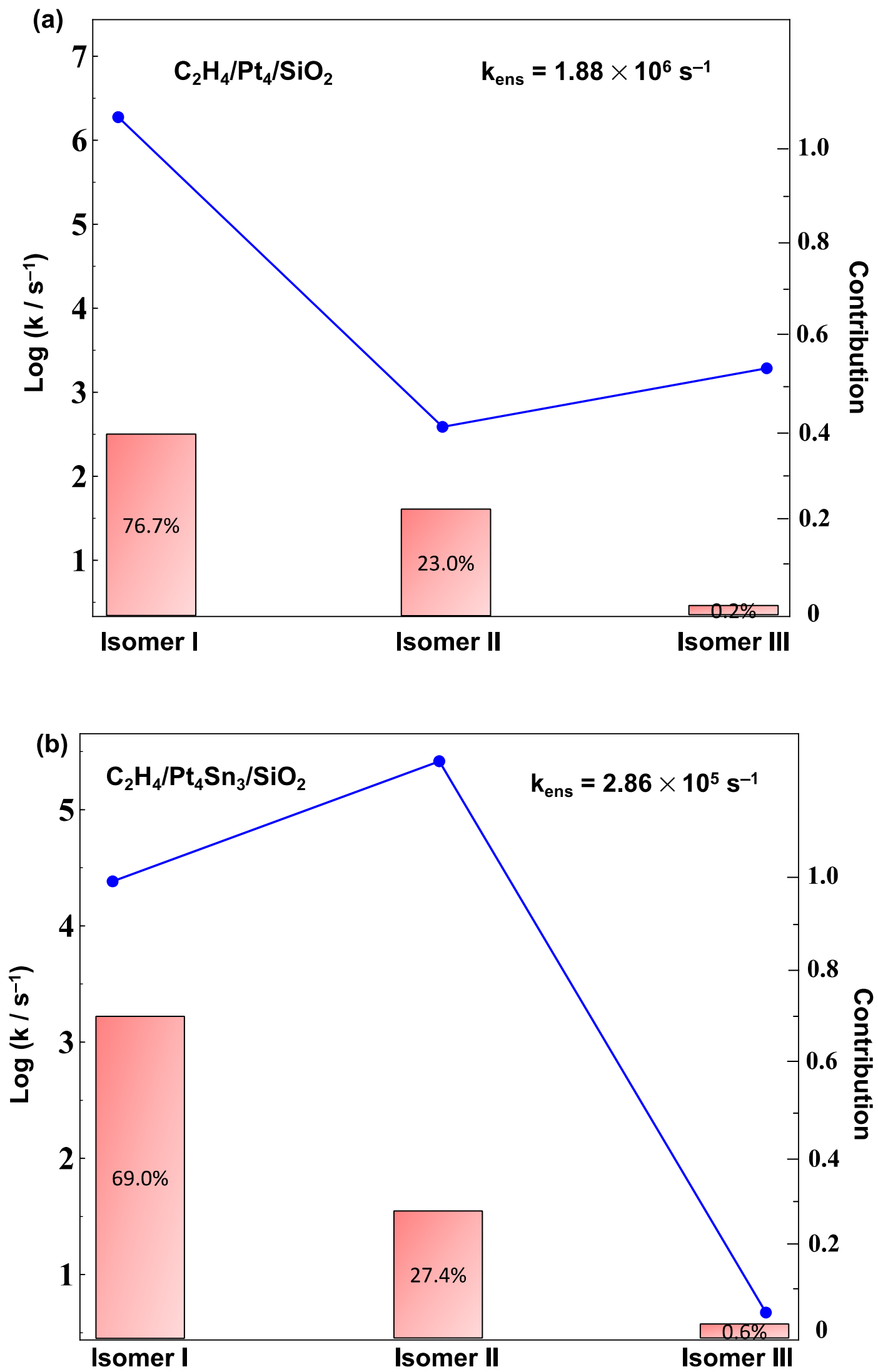

**Figure 11.** (a) Calculated rate constants along with their corresponding contribution to the $k_{ens}$ at 700 K for each isomer of (a) $C_2H_4/Pt_4/SiO_2$ and (b) $C_2H_4/Pt_4Sn_3/SiO_2$. As expected, $Pt_4/SiO_2$ is more active than $Pt_4Sn_3/SiO_2$ toward ethylene dehydrogenation.

Another interesting result is that on many cluster isomers, two of the hydrogens of ethylene can be removed significantly more easily than the other two. For instance, according to the $C_2H_4/Pt_4/SiO_2$ isomer I reaction pathways in **Fig. S9**, the barriers corresponding to breaking C-$H_3$ and C-$H_4$, which are plotted as I-3 and I-4, are almost 2 eV lower than those of C-$H_1$ and C-$H_2$ (I-1 and I-2). This can be understood by structure analysis (**Fig. S10**): the Pt atoms on the top of the cluster are more undercoordinated and active than the Pt atoms attached to the surface; thus, placing the detached H atoms on the top Pt sites is favorable (on either atop or bridge sites).

Note that all reaction profiles obtained from CI-NEB calculations along with the structures can be found in **Figs. S9–13**, and the corresponding numerical data is summarized in **Table S1**. By comparing these structures, we find that on $Pt_4Sn_3/SiO_2$, in the product state, the Pt-C bonds between $C_2H_3$ and the catalyst are elongated, compared to those on $Pt_4/SiO_2$. This suggests that the product might more easily leave $Pt_4Sn_3/SiO_2$ (or recombine and leave), whereas it is more likely to undergo further dehydrogenation on $Pt_4/SiO_2$.

Additionally, the properties of the ensemble of clusters with one and two adsorbed $C_2H_4$ are summarized in **Tables S2** and **S3**. As can be seen, the $Pt_4$ cluster has negative charge due to electron transfer from the support, while the charge of $Pt_4Sn_3$ is positive, as expected from the more electropositive nature of Sn. Moreover, the ethylene binding energy of $Pt_4Sn_3/SiO_2$ is ~0.7–0.8 eV weaker than that of $Pt_4/SiO_2$ as is clear from **Table S3**; this makes it easier for ethylene to desorb from the cluster rather than undergo further dehydrogenation.

Next, MD simulations were run for 10 ps at 700 K to allow sampling of the adsorbate and cluster motion. The average C-C bond distance in $C_2H_4/Pt_4Sn_3/SiO_2$ is 0.1 Å shorter than that in $C_2H_4/Pt_4/SiO_2$, consistent with its looser binding to the PtSn clusters, and indicating the bond being closer to double in character (**Figs. S14 and S15**). Furthermore, the average ∠PtCC bond angle in $C_2H_4/Pt_4/SiO_2$ is very close to the tetrahedral angle (104.94 °) showing the σ-bonding interaction between ethylene and the cluster. On the other hand, the average ∠HCC bond angle in $C_2H_4/Pt_4Sn_3/SiO_2$ is 120.47 ° showing a significant C-C double bond nature and π interaction between the adsorbate and the cluster. Note that during the MD simulations of $C_2H_4/Pt_4Sn_3/SiO_2$, ethylene becomes partially detached from the cluster at 5.059 ps, which is shown in **Fig. S17**, whereas on $C_2H_4/Pt_4/SiO_2$ ethylene is always attached to the cluster. This again suggests that ethylene is more likely to desorb from the $Pt_4Sn_3/SiO_2$ surface, rather than remaining on the surface and undergoing dehydrogenation.

Ethane-to-ethylene conversion is the ultimate target for these catalysts, but in UHV experiments, it is infeasible to probe $C_2D_6$ dehydrogenation over $Pt_n/SiO_2$ and $Pt_nSn_x/SiO_2$, because ethane does not stick at relevant temperatures. However, DFT can be useful to bridge the gap. We performed calculations of $C_2H_6$ dehydrogenation on the pure and Sn-alloyed Pt clusters, to show that PtSn clusters can actually dehydrogenate $C_2H_6$ and form the desired $C_2H_4$ product. **Fig. 12** shows the barriers corresponding to ethane dehydrogenation on the $Pt_4/SiO_2$ and $Pt_4Sn_3/SiO_2$ global minimum structures. The reaction barriers are 0.22 eV for $Pt_4/SiO_2$, and 0.58 eV for $Pt_4Sn_3/SiO_2$, i.e. lower than the ones obtained for $C_2H_4$ dehydrogenation (0.96 eV and 1.30 eV, for $Pt_4/SiO_2$ and $Pt_4Sn_3/SiO_2$, respectively). The reason PtSn can dehydrogenate ethane relatively easily, but not ethylene, has to do with the type of sites required for the two reactions. Whereas di-σ ethylene activation requires the binding to two Pt atoms, ethane binds to a single Pt atom with both the H and the C atoms of the activated C-H bond. Such sites are available on both pure and Sn-alloyed Pt clusters. These results suggest that both catalysts should successfully dehydrogenate alkanes, whereas dehydrogenation of alkenes is harder, and further hindered by alloying of the cluster catalyst with Sn. This justifies the use of $C_2H_4$ as a model for selectivity of dehydrogenation catalysis.

In the past, we proposed that Si and Ge as new highly competitive alloying elements for Pt.[59-60] Both Si and Ge have very similar effect on Pt clusters to the effect of Sn: mixing, Pt site separations, and electronic spin quenching. Both Ge and Si were shown to suppress dehydrogenation beyond ethylene. However, for PtGe, the step of ethane dehydrogenation was actually accelerated compared to the process on pure Pt, and that is different from the effect of Sn, which makes ethane dehydrogenation less facile (though still quite accessible).

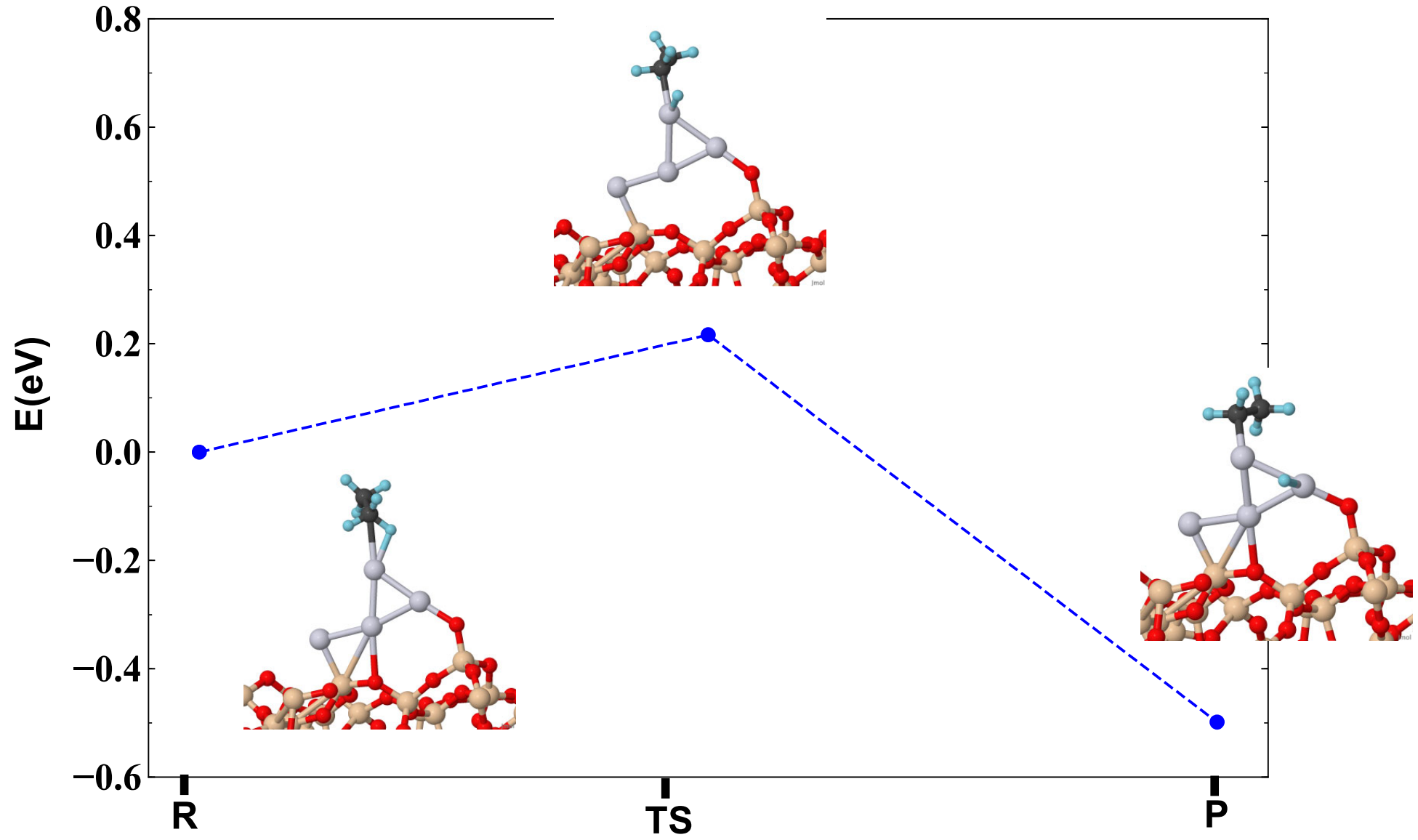

(a)

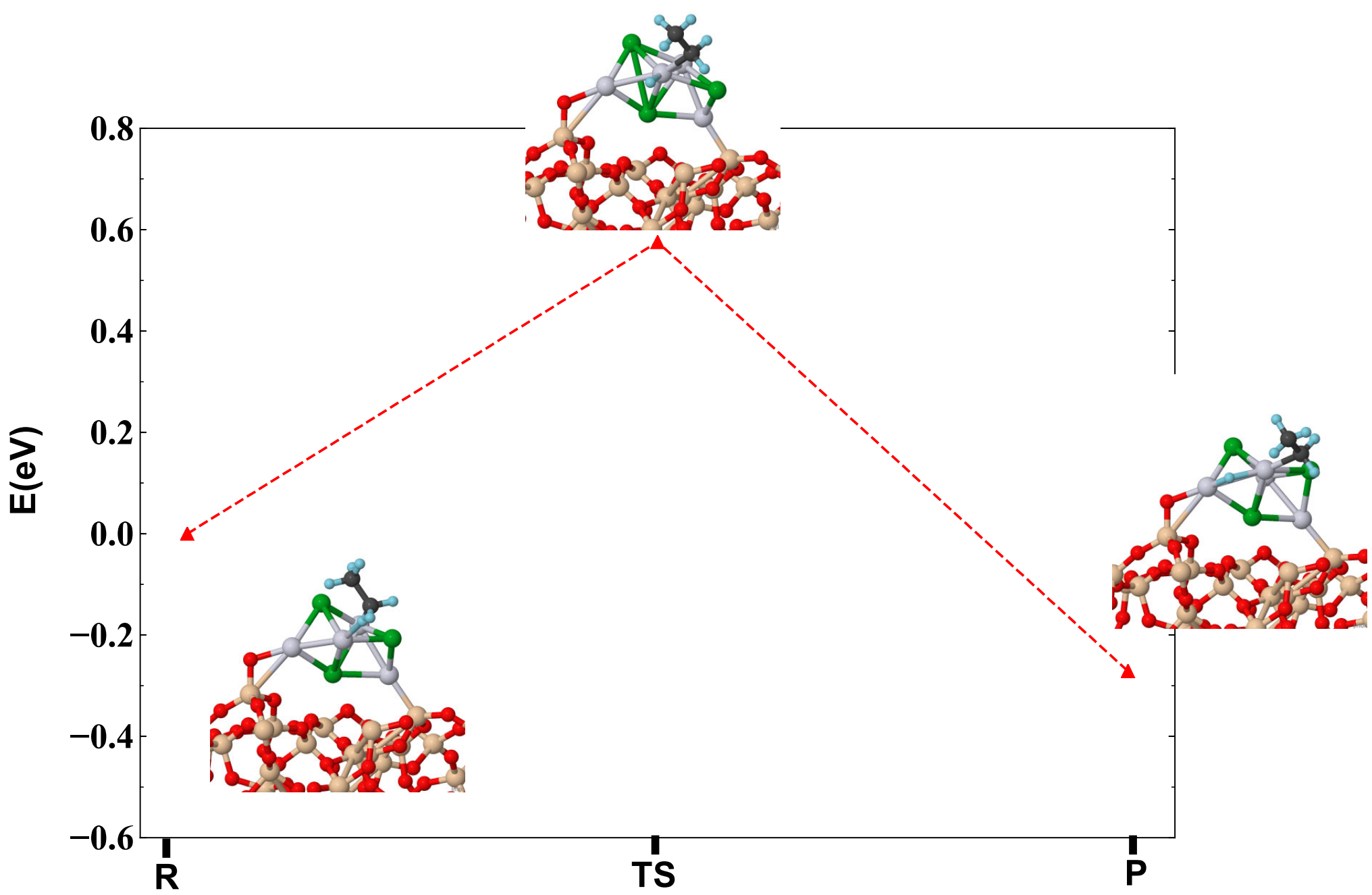

(b)

**Figure 12.** Lowest energy reaction profiles of breaking C-H bond obtained from CI-NEB calculations for the global minimum isomer of (a) $C_2H_6/Pt_4/SiO_2$ and (b) $C_2H_6/Pt_4Sn_3/SiO_2$ along with the structures of reactants, transition states, and products.

**CONCLUSIONS**

We have presented a study comparing the activity and selectivity of $Pt_n/SiO_2$ and $Pt_nSn_x/SiO_2$ for ethylene desorption vs. dehydrogenation and carbon deposition. Alloying of sub-nano $Pt_n$ clusters with Sn resulted in considerable improvement in selectivity by blocking the di-σ binding mode of ethylene, while preserving the π-mode. The origin of this selectivity is electronic and geometric. Adding Sn quenches all unpaired spins on Pt, which are required for the di-σ ethylene binding. Also, Sn intermixes with Pt and separates Pt atoms from each other, thus again preventing di-σ ethylene binding. This resulted in suppression of the dehydrogenation mechanism, allowing ethylene to desorb from the catalytic surface intact before decomposition into high binding energy coke precursor surface species. The improvement in catalyst selectivity thus results in an improvement in run-to-run stability, suggesting that sub-nano PtSn alloy clusters could be a promising candidate for high temperature dehydrogenation catalysts.

On the mechanistic side, we find that many different pathways can feasibly contribute to the reaction mechanism, due to the co-existence of multiple cluster isomers, multiple binding geometries of ethylene to the clusters, and multiple accessible C-H bond dissociation paths. Metastable cluster isomers with attached ethylene can have lower barriers to C-H dissociation than the corresponding global minima of the same stoichiometry, and thus contribute substantially to the overall ensemble-average rate of the reaction. It is indeed important that the selectivity of the cluster catalyst is an ensemble property. Pure Pt cluster catalysts usually bind ethylene in the di-σ mode and efficiently dehydrogenate it, but they can also bind ethylene in the π-mode that discourages dehydrogenation and have high reaction barriers that are within thermal reach to the population. PtSn clusters, on the other hand, mostly favor the π-mode of ethylene binding, favoring desorption instead of dehydrogenation, however, to the extent that some higher energy cluster isomers contribute, they may have thermally-accessible dehydrogenation pathways. DFT calculations combined with statistical analysis confirm that adding Sn to Pt clusters significantly slows down ethylene dehydrogenation.

## METHODS

**Experimental**. Experiments were performed using an instrument described in detail elsewhere,[3, 5, 61-62] which allows sample creation by size-selected cluster deposition in ultra-high vacuum (UHV), with characterization of physical and chemical properties in *in situ*. The cluster deposition beamline includes a laser vaporization cluster source, ion guides for transport through several differential pumping stages, a quadrupole mass filter for cluster size selection, a valve/lens that isolates the cluster beamline from the UHV system, and a final ion guide that transports ions to the sample, where they are deposited through a 2 mm diameter mask. Samples are mounted via heater wires to a liquid $N_2$ cryostat that is mounted to a precision manipulator on a rotating "lid" assembly that allows samples to be positioned for sample exchange, cleaning, film growth, cluster deposition, and characterization. Temperature can be controlled in the range from ~120 K to 2100 K.

The UHV section (base pressure 1.0 x $10^{-10}$ Torr) includes capabilities for X-ray and UV photoelectron spectroscopy, low energy $He^+$ ion scattering spectroscopy, and ion neutralization electron spectroscopy, with analysis area set to 1.1 mm diameter, i.e., smaller than the 2.0 mm cluster spot. The instrument is also equipped with a differentially pumped mass spectrometer that views the sample through a 2.5 mm diameter aperture in the end of skimmer cone. The cone is surrounded by dosing tubes that are connected to either leak valves or pulsed valves, allowing calibrated continuous or pulsed gas doses with local pressure at the sample position roughly an order of magnitude higher than the increase in background pressure.

A small UHV-compatible antechamber chamber is attached to the main UHV section, isolated by a gate valve. When the valve is opened, the sample can be inserted into the antechamber, and when in position, the two chambers are isolated by a triple differentially pumped seal that mates to the cryostat. This allows the antechamber to be vented for sample exchange, or used for "gassy" processes such as annealing in $O_2$ or Sn deposition, without exposing the main UHV system to high gas loads.

Model $Pt_n/SiO_2/Si(100)$ and $Pt_nSn_x/SiO_2/Si(100)$ catalysts were prepared on substrates consisting of 10 mm x 14 mm pieces of oxidized Si(100), using a fresh substrate for each experiment. The $SiO_2/Si(100)$ substrates are referred to throughout as "$SiO_2$" substrates. Substrates were cleaned by annealing at 700 K in $5.0 \times 10^{-6}$ Torr $O_2$ for 20 minutes, followed by 2 min of annealing in UHV. After cleaning, the initial adventitious carbon from air exposure was largely removed, but a small C 1s signal remaining, which we roughly estimated to be ~ ~0.1 ML equivalent. We attempted to remove this residual carbon by sputtering, followed by various annealing protocols in both vacuum and $O_2$ at temperatures up to 1100 K, and while sputtering does remove the carbon, it creates a large number of defect sites that do not completely anneal away. These create large desorption signals during temperature-programmed desorption (TPD) experiments, interfering with detection of species desorbing from the low coverage of clusters, therefore, we opted to use just $O_2$ annealing for sample cleaning.

The thickness of the oxidized surface layer was estimated to be 1.1 nm, from modeling[39] of the $Si^{4+}$ and $Si^0$ relative intensities measured by XPS, with photoemission cross sections and asymmetry parameters taken from work by Yeh and Lindau[63] and electron effective attenuation lengths calculated using the NIST EAL database program of Powell and Jablonski.[64] Prior to cluster deposition, the sample cryostat was cooled, and then the sample was flashed to 700 K to drive off any adventitious species that might have adsorbed during cooling. As the sample cooled after the 700 K flash, $Pt_n$ cluster deposition was initiated once the sample reached 300 K and continued as the sample cooled to 180 K. To allow easy exchange of $SiO_2$ substrates, they are mounted using a tungsten clip to clamp the substrates to a molybdenum backing plate, which is spot welded to Ta heater wires that also act as thermal conductors to the sample cryostat. A type C thermocouple is spot welded to back of the backing plate for temperature measurement.

We chose to study two small clusters, $Pt_4$ and $Pt_7$, both deposited at ~1 eV/atom, as measured using retarding potential analysis of the ion beam on the substrate. Deposition was monitored by integrating the $Pt_n^+$ neutralization current, and for all samples here, the coverage was 1.5 x $10^{14}$ Pt atoms/cm$^2$, equivalent to ~10% of a close-packed Pt monolayer. The absolute coverage of clusters is thus $1.5x10^{14}/n$ clusters/cm$^2$, where n = 4 or 7.

The method used to prepare size- and composition-selected PtSn alloy clusters, along with experimental and theoretical characterization of the cluster properties, has been described in detail,[2] and additional information is given in the Supporting Information. Briefly, size-selected $Pt_n$ (n = 4, 7) clusters deposited on $SiO_2$ substrates were used to "seed" selective Sn deposition via a self-limiting reaction sequence, carried out at 300 K. $Pt_n/SiO_2$ samples were first exposed to ~21000 L of $H_2$, which saturates the Pt clusters with H atoms, with essentially no effect on the $SiO_2$ substrate.[5] Next, the samples were exposed to ~24 L of $SnCl_4$ vapor, which reacts with hydrogenated Pt sites, leading to HCl desorption and binding of $SnCl_x$. The samples were then exposed to an additional 21000 L of $H_2$, causing additional HCl desorption, such that <10% of the Cl remained on the cluster surface, along with H and Sn. Finally, the samples were heated briefly to 700 K, desorbing the remaining H and Cl, but with no loss of Sn. ISS showed evidence that heating also allowed the Sn to mix into the Pt cluster, forming an alloy.

Sn deposition was found to be >40 times more efficient on hydrogenated Pt sites than on the $SiO_2$ support. The cluster Pt:Sn stoichiometry estimated by XPS for the $Pt_4$ and $Pt_7$ seed clusters ($Pt_4Sn_{3.3}$, $Pt_7Sn_{6.3}$) was close to 1:1. It is important to note that the Pt:Sn stoichiometry is unaffected by increasing the reactant exposures,[2] i.e., the stoichiometry is controlled by the seed cluster size[2] and the saturation coverage of the ALD reactants on the clusters. Therefore, we expect that the stoichiometry to be reasonably uniform from cluster to cluster, at least as initially prepared.

Desorption and dehydrogenation of ethylene on $Pt_n/SiO_2$ and $Pt_nSn_x/SiO_2$ was characterized by $C_2D_4$ TPD/R. Samples were cooled to 180 K and exposed to 10 L of $C_2D_4$, which is roughly twice the dose needed to saturate cluster-associated binding sites that are stable at this temperature.[1] 180 K was chosen to minimize adsorption on the amorphous $SiO_2$ substrate. After the $C_2D_4$ exposure, samples were positioned 0.5 mm away from the 2.5 mm diameter aperture that allows desorbing molecules to pass into the ion source of the differentially pumped mass spectrometer (UTI 100C with Extrel electronics). The sample temperature was then ramped at 3 K/sec up to 700 K, while monitoring signal for $D_2^+$, $H_2O^+$, $^{12}CO^+$, $^{13}CO^+$, $C_2D_4^+$, $H^{35}Cl^+$, $^{12}CO_2^+$, $^{13}CO_2^+$, and $Cl_2^+$.

CO TPD was used to examine changes in the number and properties of accessible Pt sites accompanying the ethylene TPD experiments. As with the $C_2D_4$ TPD/R experiments, samples were exposed to 10 L of $^{13}CO$ at 180 K, then positioned 0.5 mm away from the 2.5 mm diameter skimmer cone aperture, and the sample temperature was ramped at 3 K/s, to 700 K while monitoring the same masses as above.

After completion of each set of TPD/R experiments, the sample was moved away from the aperture, and the mass spectrometer sensitivity was calibrated by leaking 2.0 x $10^{-8}$ Torr of $C_2D_4$, $D_2$, $^{13}CO$, and Ar into the main UHV chamber background. Taking the ionization gauge sensitivity into account, these background pressures create well-defined fluxes of the molecules of interest through the skimmer cone aperture into the mass spectrometer ionization source. Ar was used as a UHV-friendly calibrant for HCl. The masses are similar and should therefore have similar transmission efficiency, and the difference in ionization cross-sections was taken into account using ionization gauge sensitivity factors reported by the gauge manufacturer. This calibration information was used to convert ion signals for $C_2D_4^+$, $D_2^+$, $CO^+$, and $H^{35}Cl^+$ to the corresponding number of desorbing molecules, based on the assumption that detection efficiency should be similar for molecules desorbing from the surface and effusing from the gas phase. We previously compared calibrations obtained by this method to calibration by desorption of well-defined (2x2)-CO layers from Pd single crystals, and obtained very similar results.[40, 65] We conservatively estimate that the *absolute* uncertainty in the calibrated desorption signals to be ~±50 %, with uncertainty for comparing experiment to experiment of ~±15%.

**Computational.** Global optimization done in this study was performed based on plane wave density functional theory (PW-DFT) calculations with the Vienna Ab initio Simulation Package (VASP)[66-69] using projector augmented wave (PAW) potentials[70] and the PBE[71] functional. Plane waves were chosen based on the kinetic energy cutoff of 400.0 eV. Moreover, the convergence parameters of $10^{-5}$ ($10^{-6}$) eV for geometric(electronic) relaxations and Gaussian smearing with the sigma value of 0.1 eV were used. In order to model the substrate used in the experiment, the chosen $SiO_2$ slab was previously optimized[72] elsewhere at the B3LYP/6-31G(d,p)[73-76] level. The obtained cell parameters used in this study are $a$ = 12.4 Å, $b$ = 13.1 Å, $c$ = 32.0 Å, $\alpha$ = 90°, $\beta$ = 90°, and $\gamma$ = 88° which includes the vacuum gap of 10 Å in the z-direction. Note that the lower half of the slab was kept fixed during the global optimization and only the Γ-point sampling was used to obtain the energy thanks to the fairly large super cell used in this study.

The supported $Pt_4$ and $Pt_4Sn_3$ clusters, without and with the adsorbed ethylene, were globally optimized. Larger $Pt_7$-based clusters were not considered due to the enormous computational expense, additionally given the need for a larger super cell representing the slab. To produce the initial cluster geometries on the surface, we use our in-house code, parallel global optimization and pathway toolkit PGOPT, that automatically generates structures based on the bond length distribution algorithm (BLDA).[43] Initial structures for the global optimization calculations should be created, such that they are less prone to encounter Self-Consistent Field (SCF) convergence problems. This can be done by avoiding chemically unfavorable configurations, which results in the reduction of configuration search space and the computational cost of calculation. By constraining the distance of atoms to their closest and second closest neighbors to follow a normal distribution, one can achieve this goal. That is to say, both closest and second closest distances are fitted to normal distribution based on which the initial structures are generated. Subsequently, each generated structure was optimized using DFT and duplicate structures were removed thereafter. The ethylene molecules were placed on the clusters in a multitude of possible orientations and binding sites. Then each structure was optimized using DFT, and duplicates were filtered out. In this study, 200 unique structures were generated and optimized to find the putative global minimum and the accessible local minima. To take into account the effect of coverage, we also optimized the clusters with two ethylene adsorbed per cluster. A cut-off energy of 0.4 eV was used to select the thermodynamically accessible isomers at relevant temperatures. More detailed discussion about obtaining the vibrational partition functions and corrected Boltzmann population was published before.[43] The Bader charge analysis[77-80] was used to obtain partial atomic charges. Furthermore, Boltzmann populations were calculated with the assumption that the system can sufficiently equilibrate as

$$P_i = \frac{Z_{elec,i} Z_{trans,i} Z_{vib,i} Z_{rot,i}}{\sum_i Z_i} \approx \frac{g_i e^{-\beta E_i}}{\sum_i g_i e^{-\beta E_i}}$$

where $Z_{elec,i}$, $Z_{trans,i}$, $Z_{vib,i}$, and $Z_{rot,i}$ are electronic, translational, vibrational, and rotational partition functions, respectively. For the lowest-energy isomers with adsorbed ethylene, which contributed significantly into the respective thermal ensembles, the reaction profile for ethylene dehydrogenation were calculated using climbing image nudged elastic band (CI-NEB) method.[81] Three isomers of` each cluster with adsorbed ethylene were chosen and the dissociation of every C-H bond within every adsorbed configuration was tested. All transition states were confirmed by phonon calculations. The distance between the structures (images) in CI-NEB calculations used as reaction coordinate is defined as

$$d_{12} = \sqrt{\frac{1}{N}\sum_{i}^{N}(x_{i,1} - x_{i,2})^2 + (y_{i,1} - y_{i,2})^2 + (z_{i,1} - z_{i,2})^2}$$

Rate constants for each pathway were calculated based on harmonic transition state theory.[82] The ensemble-average rate constant were calculated based on the contribution of each isomer using its Boltzmann population at 700 K:

$$k_{ens} = k_1 P_1 + k_2 P_2 + k_3 P_3$$

, where $P_i$ is the Boltzmann population and $k_i$ ($i$ = 1–3) is the sum over all four pathway rate constants obtained from CI-NEB for isomer $i$.

Finally, *ab initio* MD simulations ran for 10 ps at 700 K with the time step of 1 fs starting from $Pt_4/SiO_2$ and $Pt_4Sn_3/SiO_2$ global minimum structures. Nosé–Hoover thermostat was utilized for equilibration during the MD simulations.[83]

ASSOCIATED CONTENT

**Supporting Information Available:**

Contains complete sets of $C_2D_4$ TPD for $Pt_n/SiO_2$ and $Pt_nSn_x/SiO_2$, HCl desorption during the first $C_2D_4$ TPD for $Pt_nSn_x/SiO_2$, an C 1s XPS regional spectra, the remaining Arrhenius fit plots for $Pt_7/SiO_2$ and $Pt_7Sn_{6.3}/SiO_2$, structures of reactants, products, transition states obtained from CI-NEB, Bader charge of all local minima structures along with their Boltzmann populations, Ethylene binding energies obtained from DFT, bond length and bond angle distribution of $C_2H_4/Pt_4/SiO_2$ and $C_2H_4/Pt_4Sn_3/SiO_2$ obtained MD simulations, XYZ coordinates of minimum energy pathways obtained from CI-NEB.

AUTHOR INFORMATION

**Corresponding Authors**

*(Scott L. Anderson, Anastassia N. Alexandrova) E-mails: anderson@chem.utah.edu, ana@chem.uscla.edu

**ORCIDs**

Scott L. Anderson: 0000-0001-9985-8178

Anastassia Alexandrova: 0000-0002-3003-1911

Timothy J. Gorey: 0000-0002-7491-9648

Borna Zandkarimi: 0000-0002-7633-132X

**Notes**

The authors declare no competing financial interest.

ACKNOWLEDGEMENT

This work was supported by the U.S. Air Force Office of Scientific Research under AFOSR Grants FA9550-19-1-0261 and FA9550-16-1-0141.

## TOC GRAPHIC

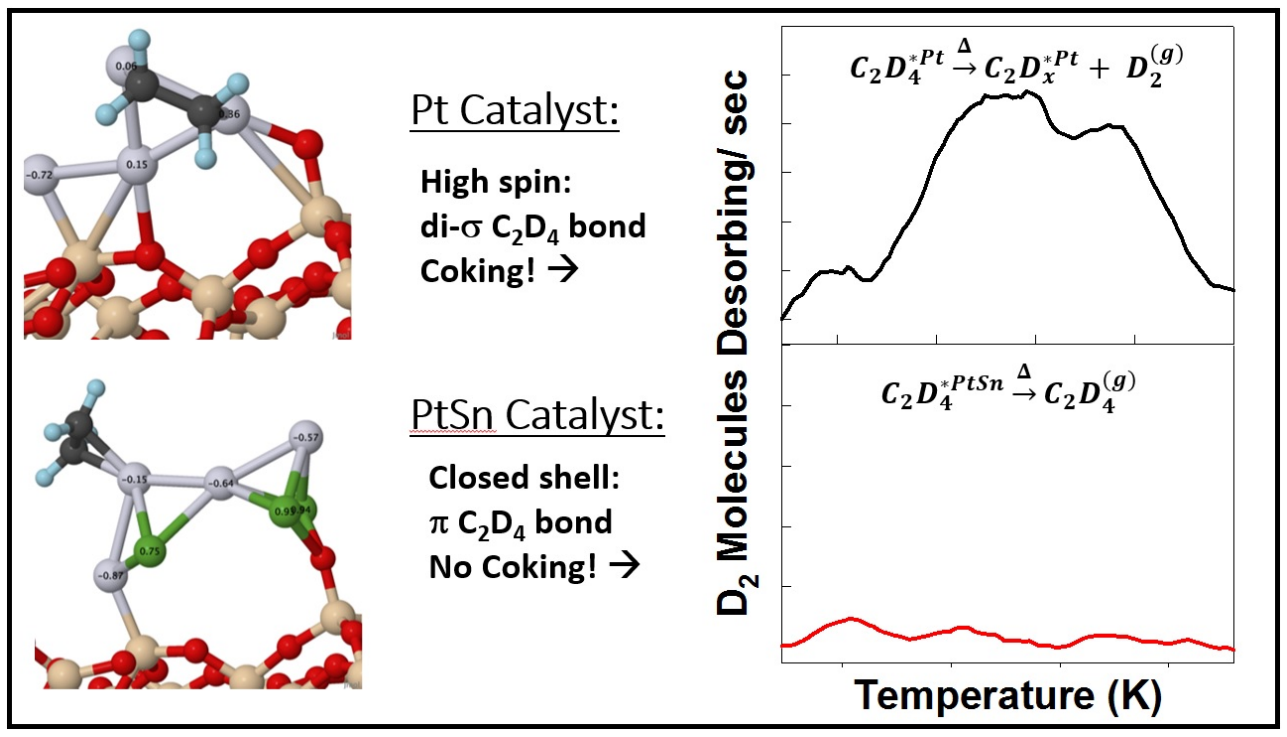

Pt Catalyst:
High spin:
di-σ $C_2D_4$ bond
Coking! →
PtSn Catalyst:
Closed shell:
π $C_2D_4$ bond
No Coking! →
$C_2D_4^{*Pt} \xrightarrow{\Delta} C_2D_x^{*Pt} + D_2^{(g)}$
$C_2D_4^{*PtSn} \xrightarrow{\Delta} C_2D_4^{(g)}$
$D_2$ Molecules Desorbing/ sec
Temperature (K)

# Coking-Resistant Sub-Nano Dehydrogenation Catalysts: $Pt_nSn_x/SiO_2$ (n = 4, 7)

*Timothy J. Gorey,[a†] Borna Zandkarimi,[b†] Guangjing Li,[a] Eric T. Baxter,[a]*

*Anastassia N. Alexandrova,[b,c*] and Scott L. Anderson[a*]*

[a]Chemistry Department, University of Utah, 315 S. 1400 E., Salt Lake City, UT 84112

[b]Chemistry and Biochemistry, University of California, Los Angeles, and [c]California NanoSystems Institute, Los Angeles, CA 90095

## Supporting information

[†] These authors contributed equally to this work.

[*]Senior Authors

Corresponding Authors: Scott Anderson, (801) 585-7289, anderson@utah.edu, Anastassia Alexandrova, (310) 825-3769, ana@chem.ucla.edu

**Fig. S1** shows four sequential $C_2D_4$ TPD runs from $Pt_4/SiO_2$ (top) and $Pt_7/SiO_2$ (bottom). As discussed in the main text and shown in **Table 1**, the overall number of available $C_2D_4$ decreased incrementally with each cycle. Similarly, the amount of liberated $D_2$ incrementally decreases. $C_2D_4$ desorption from clean $SiO_2$ is also shown for each cluster size. There are two distinguishable features for $Pt_4$, and less so for $Pt_7$. $C_2D_4$ desorption begins at the onset of the heat ramp (180 K), peaking around 260 K, and decreasing to baseline by ~500 K. Incremental loss in the number of desorbed $C_2D_4$(intact) and $D_2$ molecules is consistent with coke deposition.

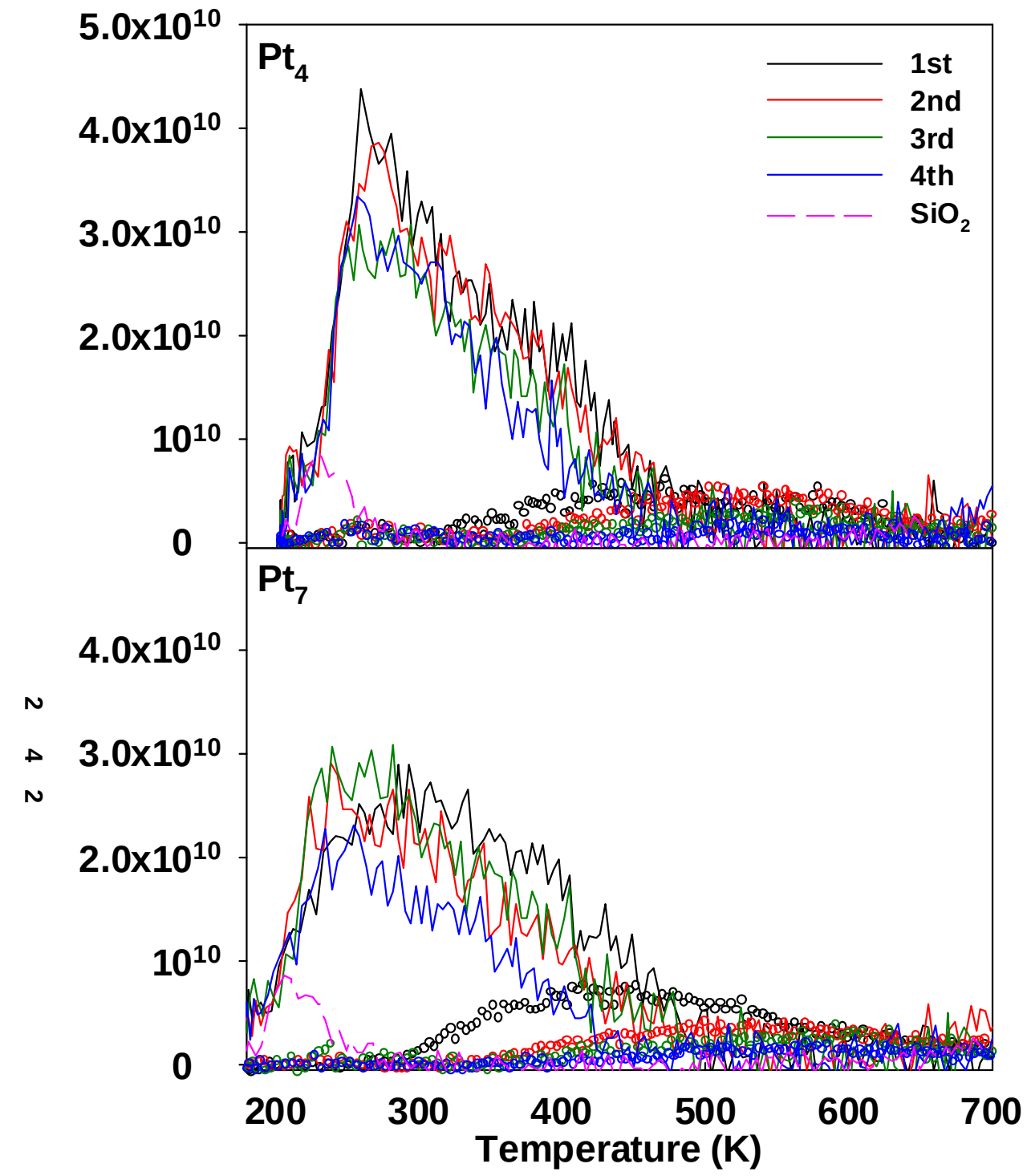

**Figure S1.** Four sequential $C_2D_4$ TPD from $Pt_4/SiO_2$. Intact $C_2D_4$ is shown as solid lines, and $D_2$ desorption is shown at hollow symbols. Intact $C_2D_4$ desorption from clean $SiO_2$ is also shown in each panel as a dashed pink line.

As was done for **Fig. S1**, four sequential $C_2D_4$ TPDs were carried out on the $Pt_4Sn_{3.3}/SiO_2$ catalyst. Numbers of $C_2D_4$ molecules *per* Pt atom are shown in **Table 2** of the manuscript. Results are plotted in in **Fig. S2** along with $C_2D_4$ desorption from a cluster-free $SiO_2$ surface that has been exposed to a $H_2/SnCl_4/H_2$ treatment. All four spectra are shown for $Pt_4Sn_{3.3}/SiO_2$ and $Pt_7Sn_{6.3}/SiO_2$, with the $D_2$ desorption plotted separately on inset plots.

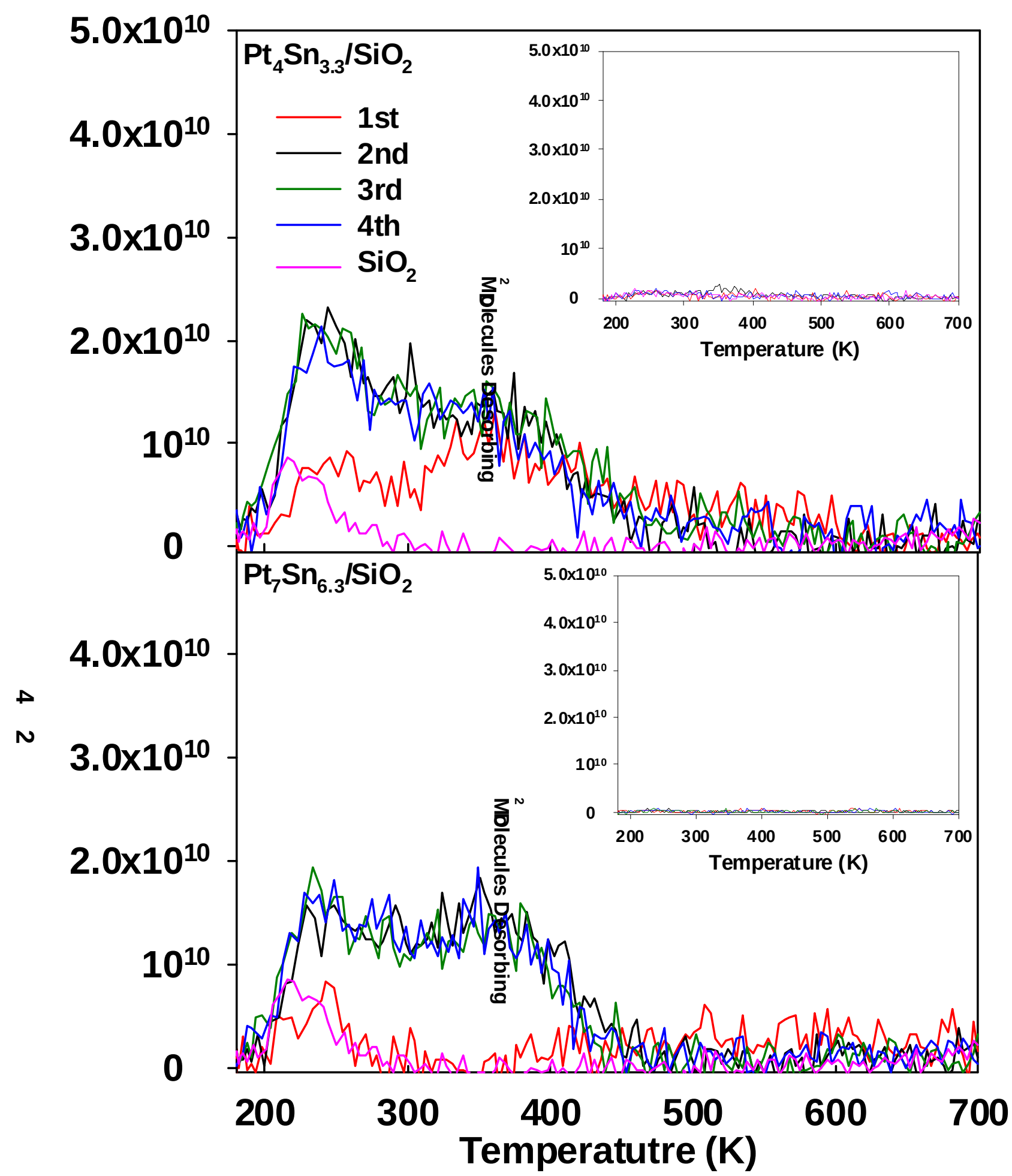

**Figure S2.** Four sequential desorption spectra of $C_2D_4$ (solid lines) and $D_2$ (inset plots) from the first (red) and fourth (blue) $C_2D_4$ TPD. Each spectra was collected after a 10 L dose of $C_2D_4$ to $Pt_nSn_x/SiO_2$ (n = 4, 7). $C_2D_4$ desorption from bare $SiO_2$ treated with 1 ALD cycle is also plotted (pink line), no $D_2$ is observed.

Note the anomalous behavior of the fist ethylene TPD (red) in **Fig. S2**. This is attributed to the surface being covered with residual H and Cl atoms as a result of the Sn deposition process. The removal of these adsorbates has been described in detail previously, and the mass 36 (HCl) desorption during the first $C_2D_4$ TPD is shown in **Fig. S3** for both cluster sizes.

Using a method described previously[1-2], the $C_2D_4$ thermal desorption spectra can be fit to extract desorption energy distributions by using the first order rate equation:

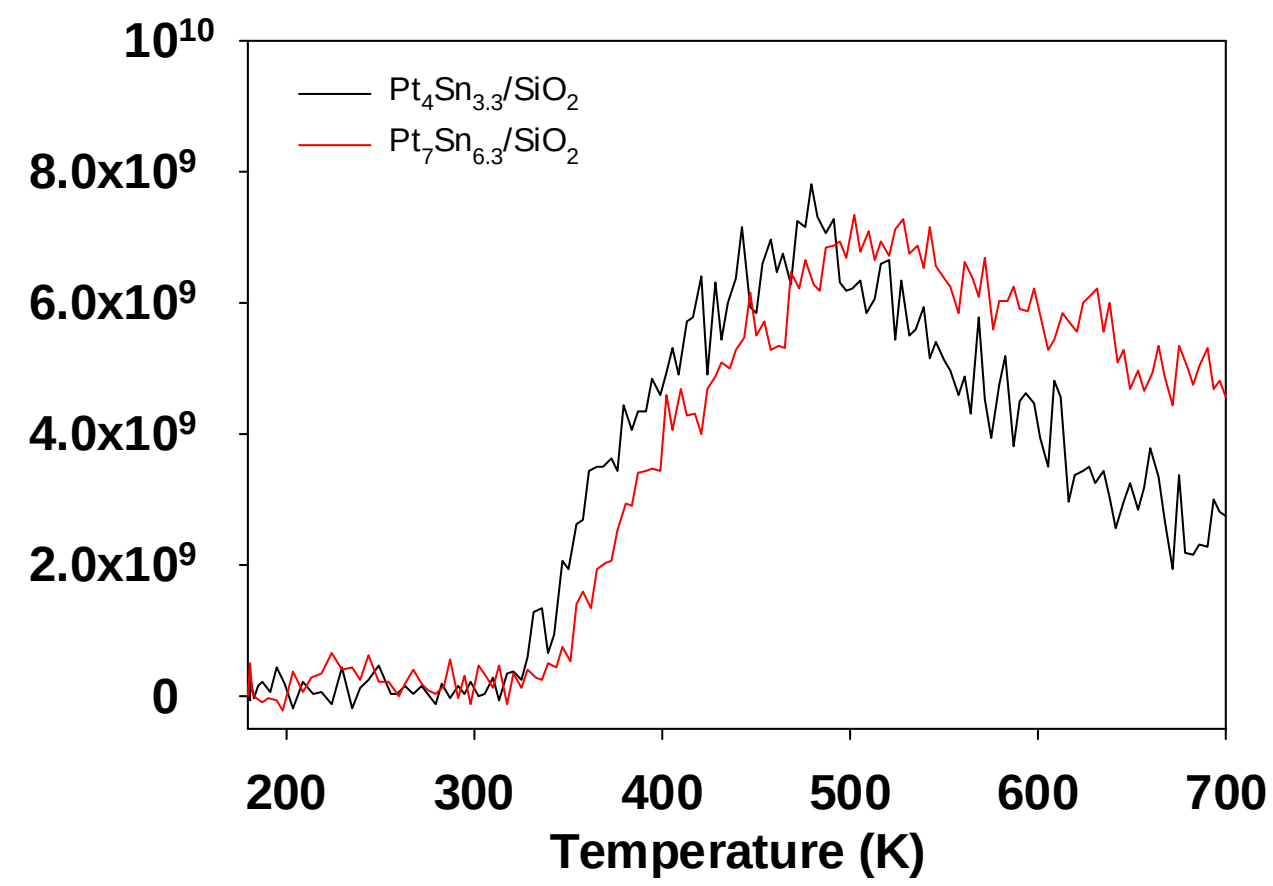

**Figure S3.** Mass 36 desorption (HCl) during the first $C_2D_4$ TPD for $Pt_nSn_x/SiO_2$. No HCl is detected in subsequent runs.

$$I(t) \alpha \frac{-d\theta}{dt} = (\theta(E_{des}) \cdot v) e^{\frac{-E_{des}}{kT(t)}}$$

with I(t) being the time-dependent $C_2D_4$ desorption signal, $E_{des}$ the desorption energy, ν the exponential pre-factor, T(t) is the heat ramp, and $\theta(E_{des})$ is the distribution of occupied binding site energies. For all experiments here, a nominal heat ramp rate of 3 K/s was used, but temperature versus time data is always collected during TPD acquisition in order to account for fluctuations in the heating ramp rate, though no corrections were needed. For the fit, a simulated $\theta(E_{des})$ is calculated for each contributing peak and plotted to match the experimentally measured I(t). The greatest source of uncertainty here is the pre-exponential factor, ν. It is not practicable to carry out the series of studies that would be required to estimate ν from the TPD data, and we simply assumed a value of $10^{14}$ $s^{-1}$, which is in the range of values often used for CO TPD.[3-7] In previous studies[2], the effect of varying ν between $10^{13}$ $s^{-1}$ to $10^{15}$ $s^{-1}$ was tested, and found to shift the $\theta(E_{des})$ distribution by only ~7%.

**XPS Characterization of Carbon Deposition.** Because the C 1s photoemission cross section is quite small,[8] and carbon is deposited only on the clusters, it was difficult to quantify carbon deposition for samples of the sort discussed above. To increase the carbon coverage, we prepared $Pt_4/SiO_2$ and $Pt_4Sn_{3.3}/SiO_2$ samples with twice the usual cluster coverage, and subjected them to eight sequential $C_2D_4$ TPD/R runs. The results are compared in **Fig. S4**. A broad C 1s peak is observed for the blank $SiO_2$ substrate, attributed to adventitious carbon remaining after the substrate was cleaned by annealing in $O_2$. The binding energy is in the range expected for partially oxidized carbon, as might be expected after annealing in $O_2$.

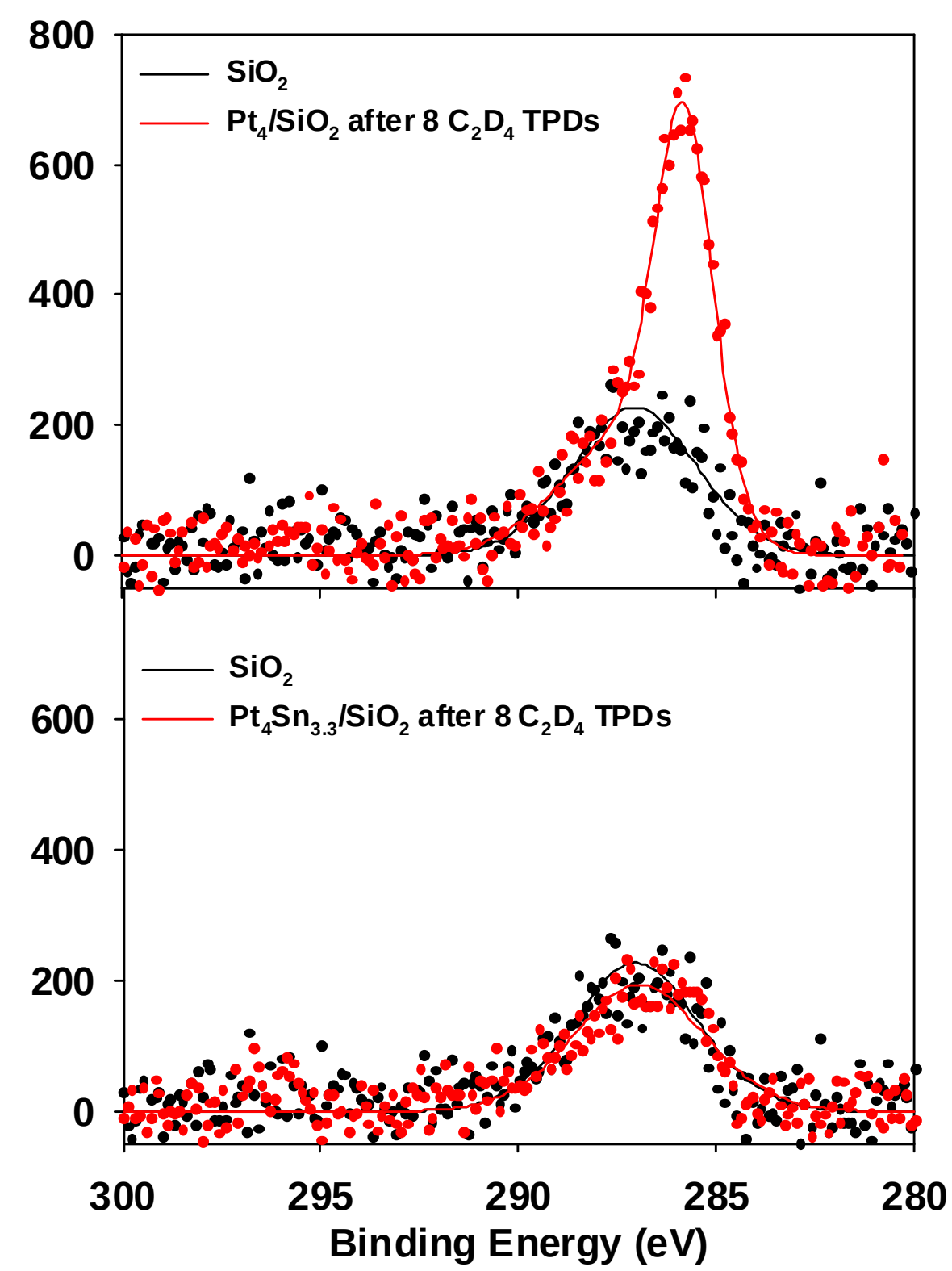

**Figure S4.** Comparison of C 1s XP spectra for an $SiO_2$ substrate, a $Pt_4/SiO_2$ sample, and a $Pt_4Sn_{3.3}/SiO_2$ sample, subjected to eight sequential $C_2D_4$ TPD/R runs.

For $Pt_4/SiO_2$ after $C_2D_4$ TPD/R, this adventitious carbon feature is still present, as expected, but a sharper C 1s feature appears at lower binding energy, which we attribute to carbon deposited on the Pt clusters by $C_2D_4$ dehydrogenation. Subtracting the contribution from adventitious carbon, the intensity of this low binding energy feature corresponds to eight TPD/R runs having deposited ~1.5 C atom/Pt atom.

The lower frame of the figure shows analogous data for $Pt_4Sn_{3.3}/SiO_2$ after eight $C_2D_4$ TPD runs. No increase in carbon intensity was observed, compared to the $SiO_2$ substrate, consistent with the TPD/R results showing no significant $D_2$ desorption from the Sn-alloyed clusters.

**Fig. S5** shows the fit for TPD spectra. Here, we use the first $C_2D_4$ TPD run from $Pt_4/SiO_2$ as a representative example; this method was carried out for all energy desorption profile plots in this paper. Fits were performed using a two component model, representing low temperature and high temperature feature. Using these fits, the desorption profile was computed and plotted as a function of binding energy. This model makes the critical assumption of first order desorption kinetics. While we expect some 2[nd] order behavior for surface-bound $C_2D_4$, i.e., dissociative adsorption → recombinative desorption, the first order fit was carried out simply to approximate the binding energies for comparison to DFT studies.

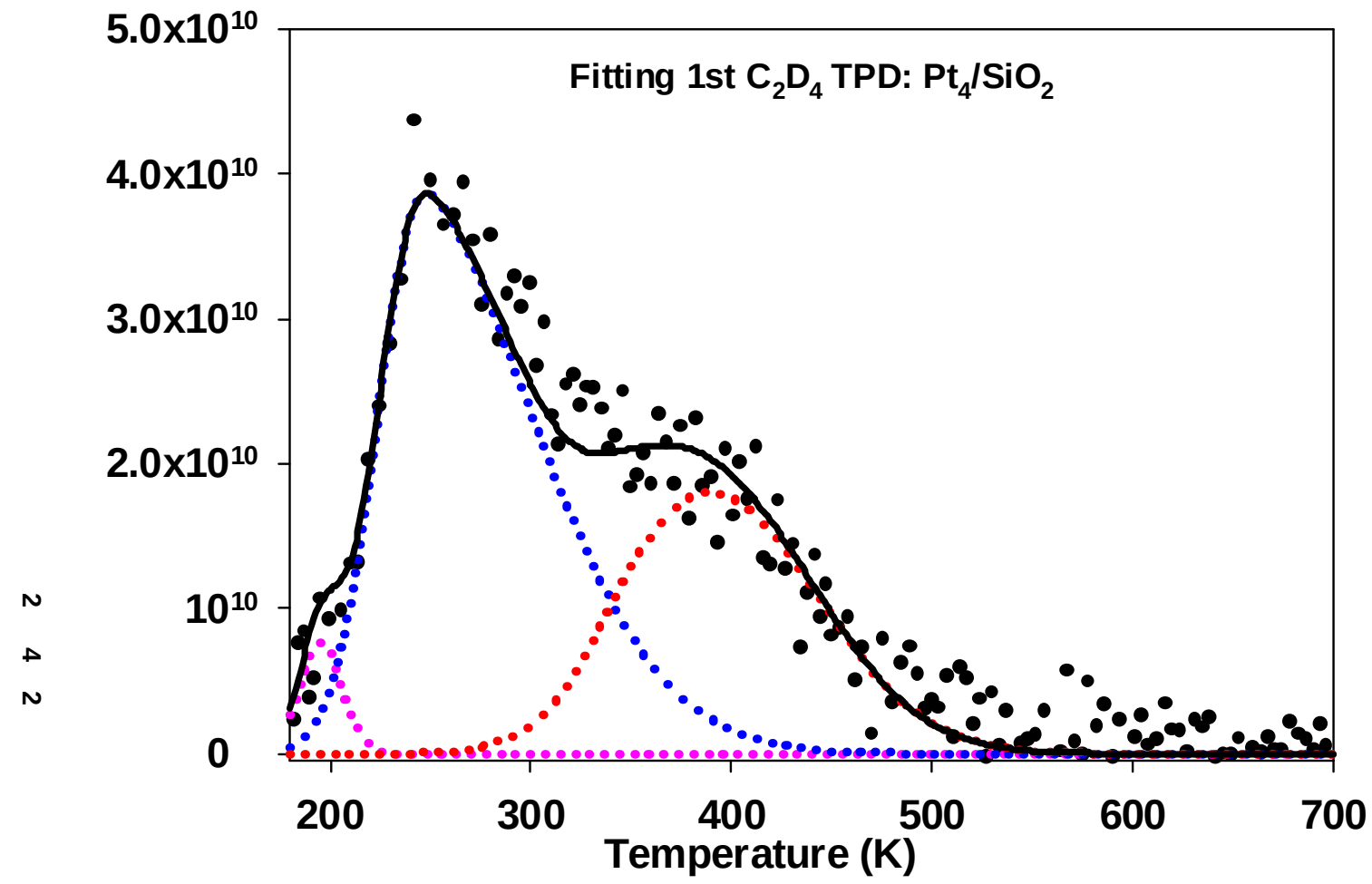

**Figure S5.** 1[st] $C_2D_4$ TPD from $Pt_4/SiO_2$. The spectrum was fit with low(blue) and high(red) temperature components. The fits can be used to calculate energy desorption profiles.

Desorption energy distributions for $Pt_7/SiO_2$ and $Pt_7Sn_{6.3}/SiO_2$ were calculated as described above. Briefly, the first $C_2D_4$ TPD from $Pt_7/SiO_2$ was fit with a low and high temperature component and then modeled using a first order kinetics relation. The same was done for the *second* $C_2D_4$ for the $Pt_7Sn_{6.3}/SiO_2$ sample.

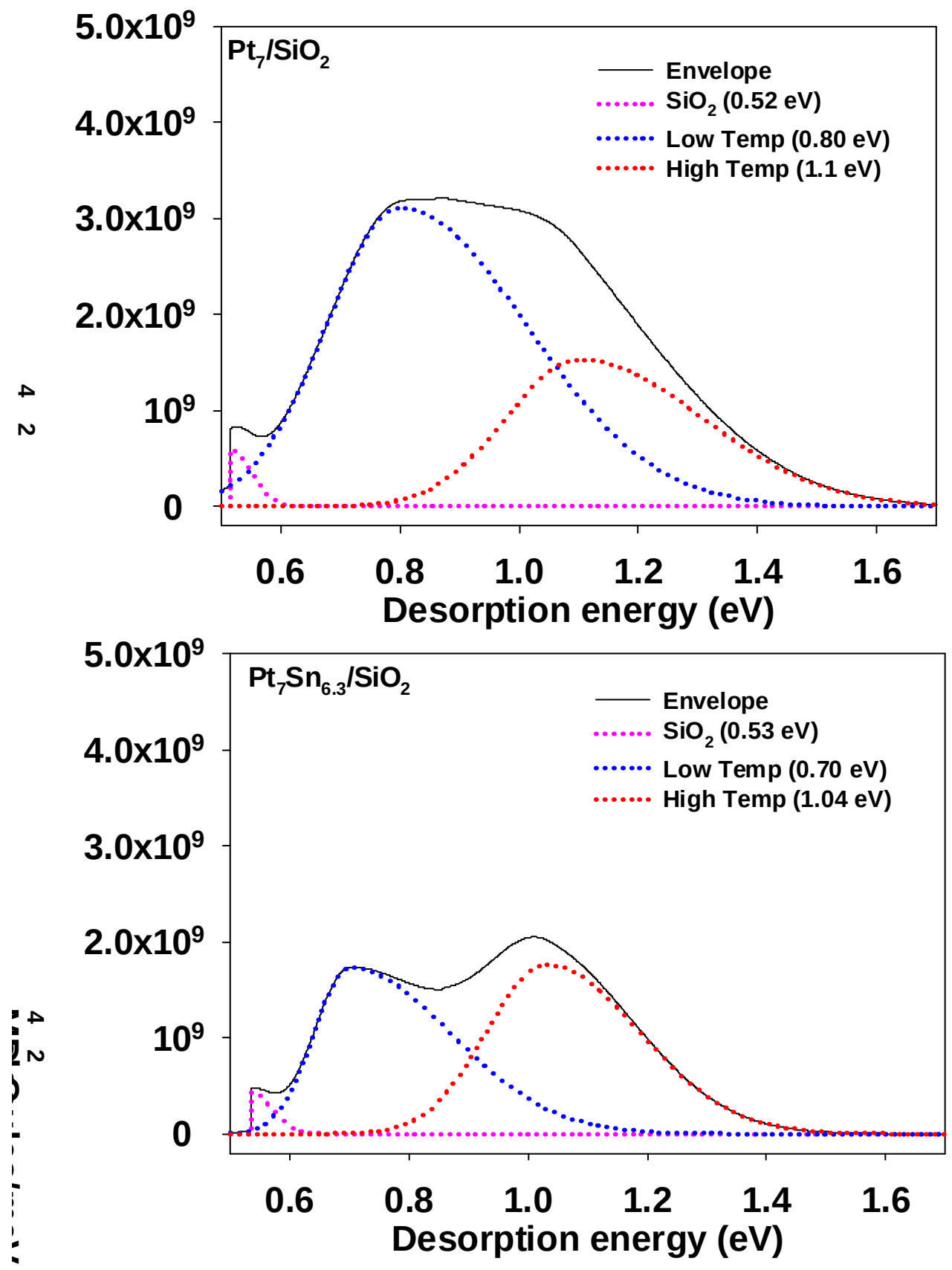

**Figure S6.** Arrhenius fits for the $C_2D_4$ TPD from $Pt_7/SiO_2$ and $Pt_7Sn_{6.3}/SiO_2$ spectra in **Fig. 1**.

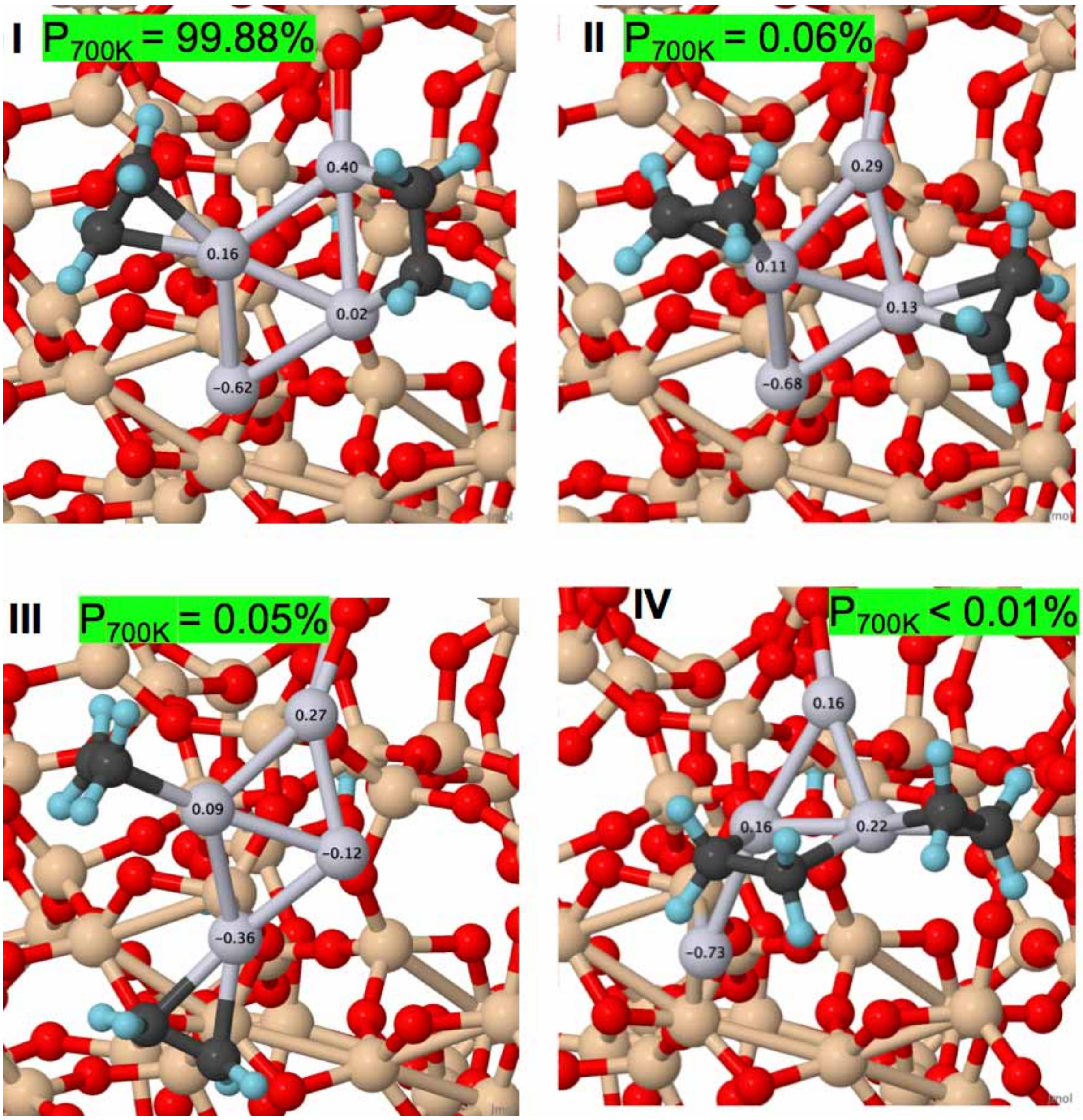

**Figure S7.** Thermally-accessible geometries of $(C_2H_4)_2/Pt_4/SiO_2$ obtained from global optimization calculations along with their Boltzmann populations at 700 K.

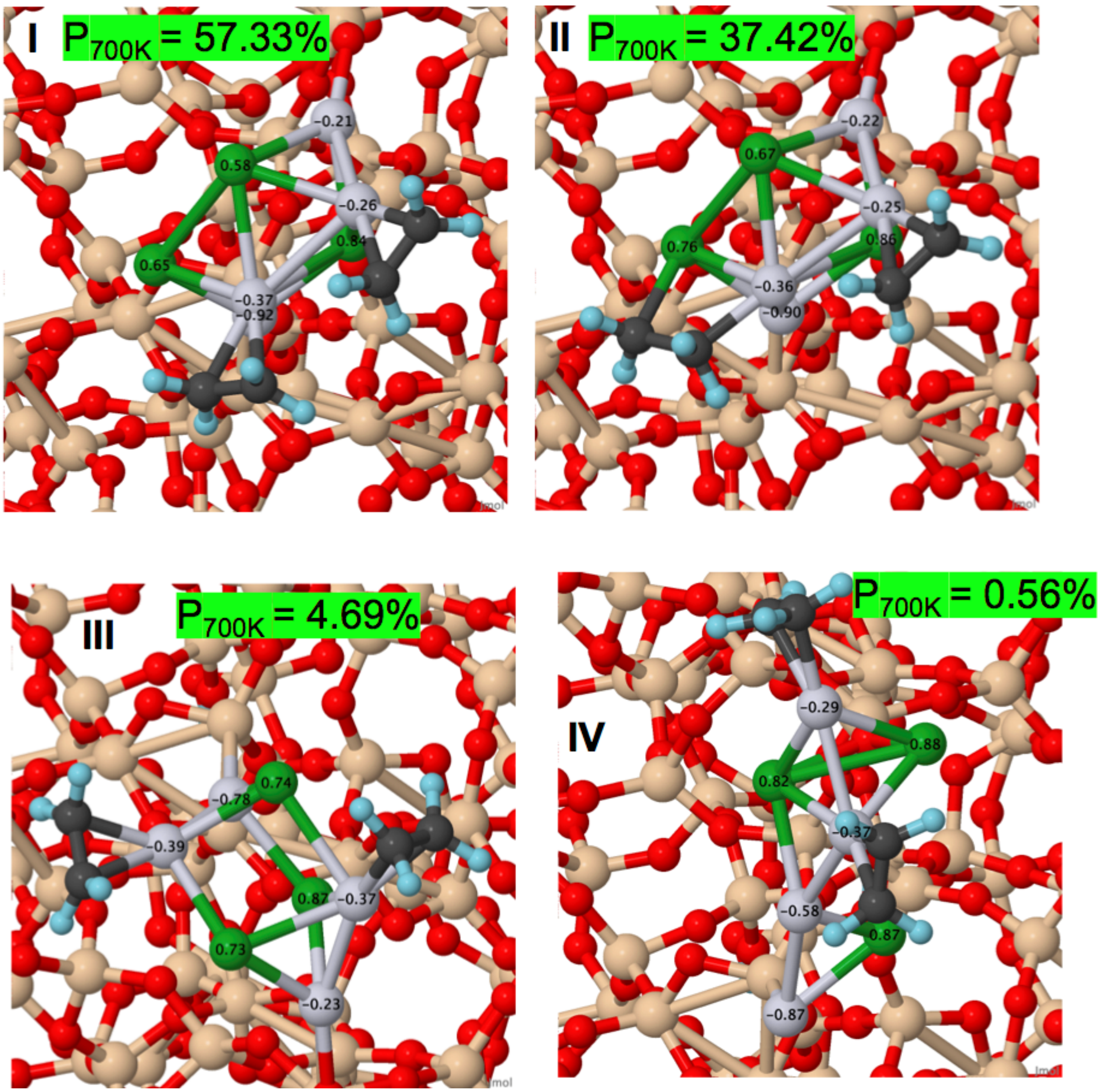

**Figure S8.** Thermally-accessible geometries of $(C_2H_4)_2/Pt_4Sn_3/SiO_2$ obtained from global optimization calculations along with their Boltzmann populations at 700 K. Note that the ensemble is dominated by $C_2H_4$ $\pi$-binding mode.

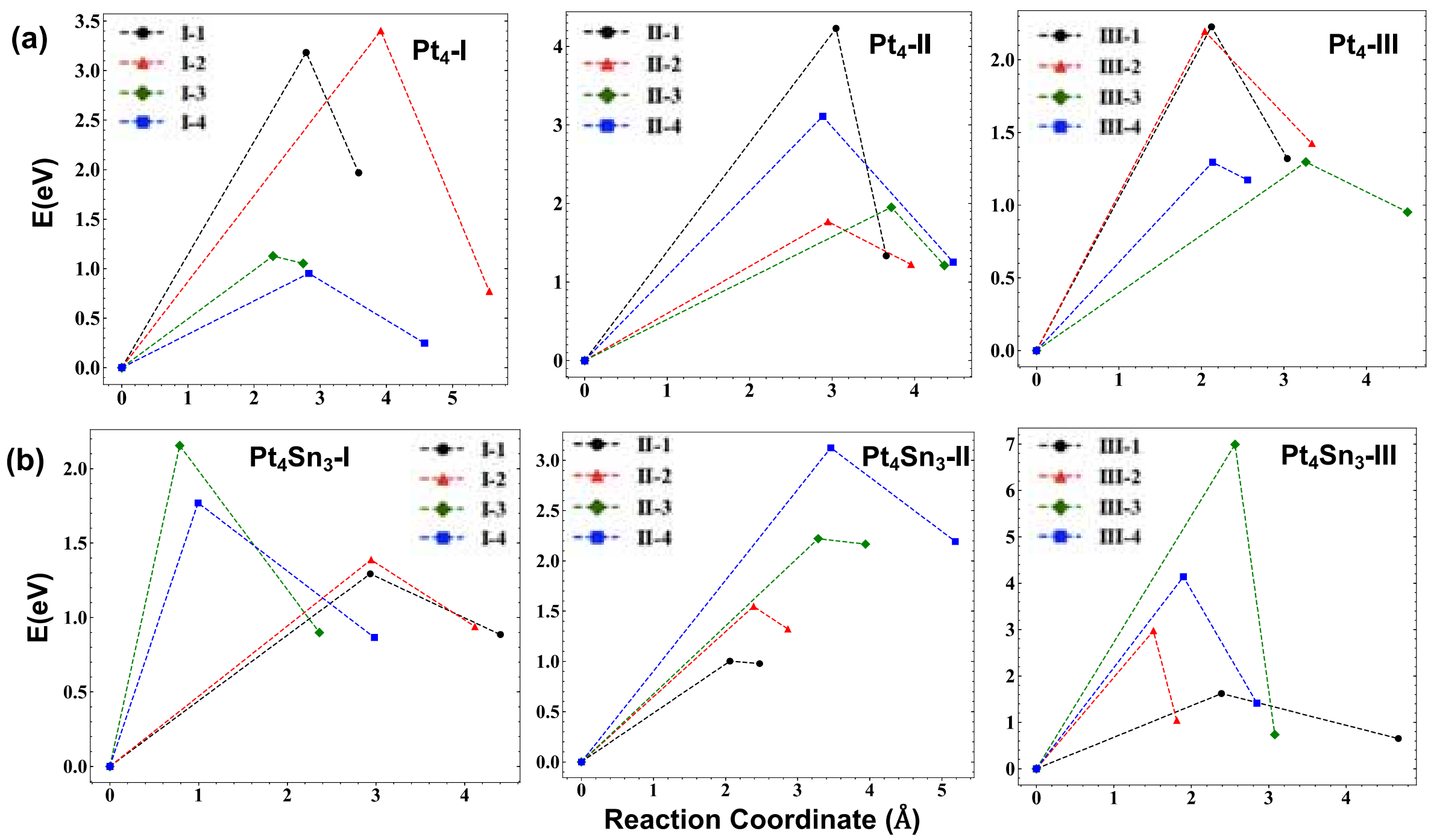

**Figure S9.** Reaction profiles of breaking C-H bond obtained from CI-NEB calculations for (a) $C_2H_4/Pt_4/SiO_2$ and (b) $C_2H_4/Pt_4Sn_3/SiO_2$. The reaction barrier corresponding to the highest populated $C_2H_4/Pt_4Sn_3/SiO_2$ isomer is 0.3 eV higher than that of $C_2H_4/Pt_4/SiO_2$. The reaction coordinate is defined as the root mean square distance from the reactant, which can be found in the computational methods section.

**Table S1.** All reaction barriers of $C_2H_4/Pt_4/SiO_2$ and $C_2H_4/Pt_4Sn_3/SiO_2$ obtained from CI-NEB calculations along with their contribution to the final $k_{ens}$ at 700 K.

| $C_2H_4/Pt_4/SiO_2$ | Barrier (eV) | | | | $P_{700K}$ |
|---|---|---|---|---|---|
| | Path-1 | Path-2 | Path-3 | Path-4 | |
| I | 0.95 | 1.13 | 3.18 | 3.40 | 76.71% |
| II | 1.77 | 1.95 | 3.11 | 4.23 | 23.02% |
| III | 1.29 | 1.30 | 2.20 | 2.23 | 0.01% |
| $C_2H_4/Pt_4Sn_3/SiO_2$ | | | | | |
| I | 1.29 | 1.39 | 1.77 | 2.15 | 69.03% |
| II | 1.00 | 1.55 | 2.22 | 3.12 | 27.45% |
| III | 1.62 | 2.97 | 4.14 | 6.99 | 6.70% |

**Table S2.** Charge on the cluster, ethylene binding mode, and Boltzmann population at 700 K obtained for $C_2H_4/Pt_4/SiO_2$, $(C_2H_4)_2/Pt_4/SiO_2$, $C_2H_4/Pt_4Sn_3/SiO_2$, and $(C_2H_4)_2/Pt_4Sn_3/SiO_2$.

| Isomer | $Q_{cluster}$(e) | | $C_2H_4$ Binding mode | | $P_{700K}$ | |
|---|---|---|---|---|---|---|
| $C_2H_4/Pt_4Sn_{0,3}/SiO_2$ | $Pt_4$ | $Pt_4Sn_3$ | $Pt_4$ | $Pt_4Sn_3$ | $Pt_4$ | $Pt_4Sn_3$ |
| I | –0.15 | 0.39 | di-σ | π | 76.71% | 69.03% |
| II | –0.11 | 0.46 | π | π | 23.01% | 27.45% |
| III | –0.10 | 0.43 | π | π | 0.25% | 2.32% |
| IV | –0.12 | 0.50 | π | π | 0.01% | 0.7% |
| V | –0.12 | 0.50 | di-σ | π | 0.01% | 0.3% |
| VI | –0.14 | 0.32 | di-σ | π | <0.01% | 0.2% |
| $(C_2H_4)_2/Pt_4Sn_{0,3}/SiO_2$ | $Pt_4$ | $Pt_4Sn_3$ | $Pt_4$ | $Pt_4Sn_3$ | $Pt_4$ | $Pt_4Sn_3$ |
| I | –0.04 | 0.31 | π, di-σ | π, π | 99.88% | 57.33% |
| II | –0.16 | 0.56 | π, π | π, di-σ | 0.06% | 37.42% |
| III | –0.11 | 0.58 | π, π | π, π | 0.05% | 4.69% |
| IV | –0.19 | 0.45 | π, di-σ | π, π | <0.01% | 0.56% |

**Table S3.** Ensemble averaged first and second $C_2H_4$ binding energies on $Pt_4/SiO_2$ and $Pt_4Sn_3/SiO_2$ calculated at 700 K.

| Structure | $E_{b1}$(eV) | $E_{b2}$(eV) |
|---|---|---|
| $Pt_4/SiO_2$ | –1.91 | –1.76 |
| $Pt_4Sn_3/SiO_2$ | –1.11 | –1.04 |

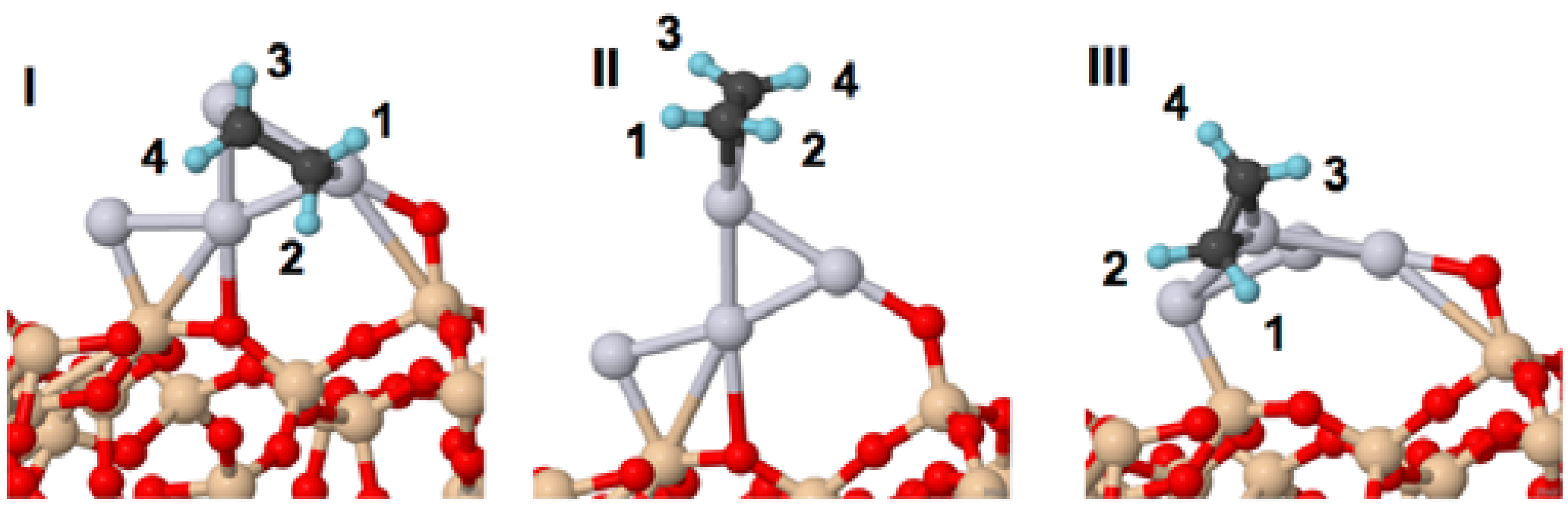

(a)

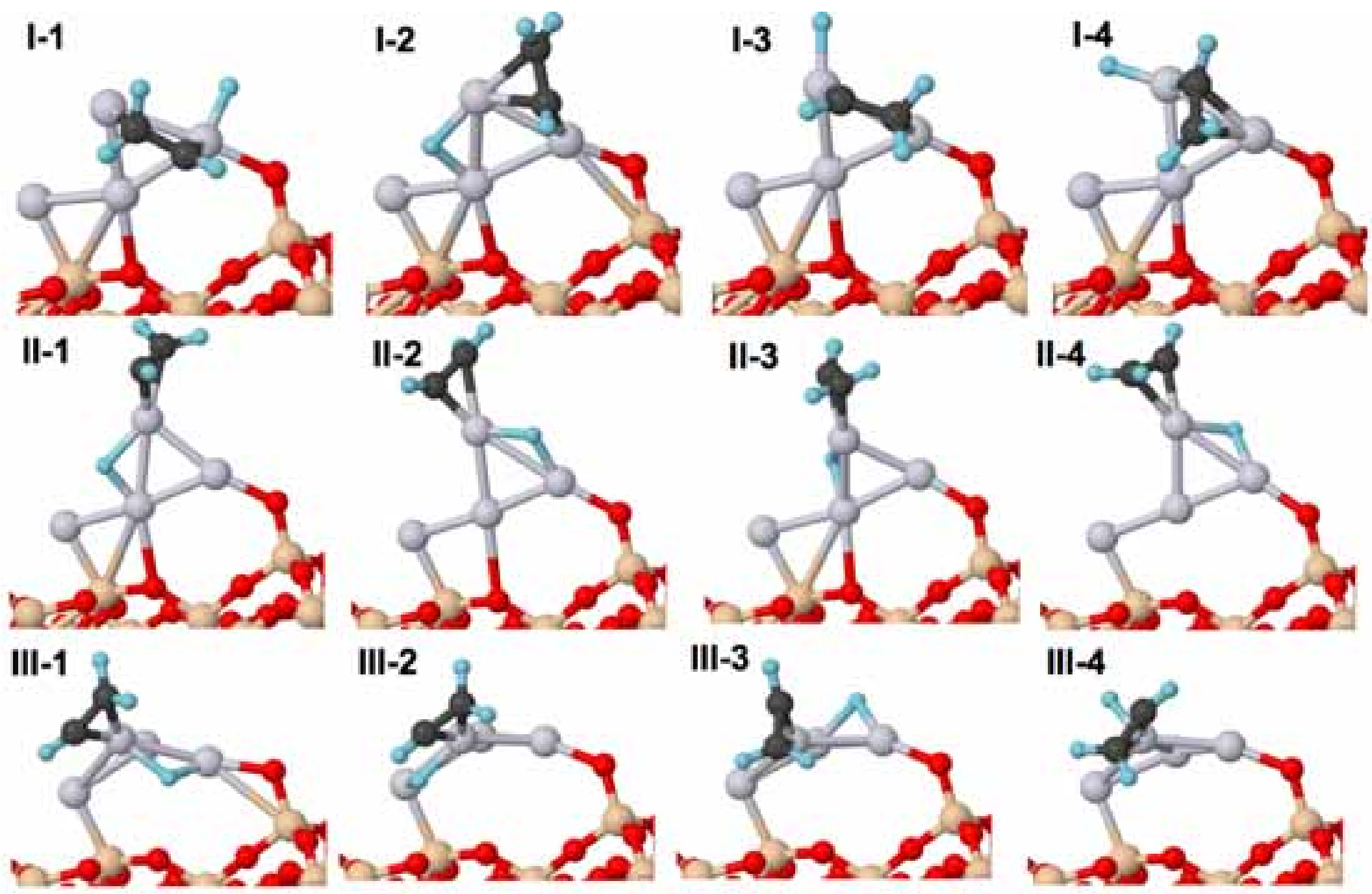

(b)

**Figure S10.** (a) 3 most populated $C_2H_4/Pt_4/SiO_2$ isomers used in CI-NEB calculations and (b) all 12 products (4 for each structure) obtained by cleaving every C-H bond in $C_2H_4$ on $Pt_4/SiO_2$.

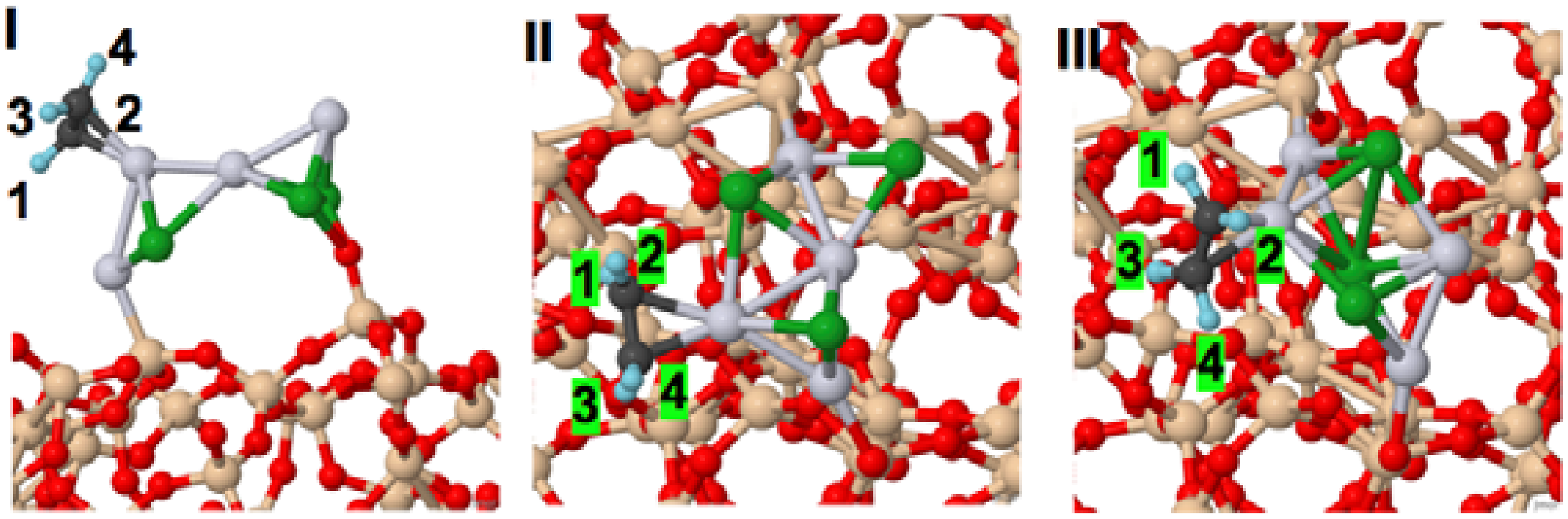

(a)

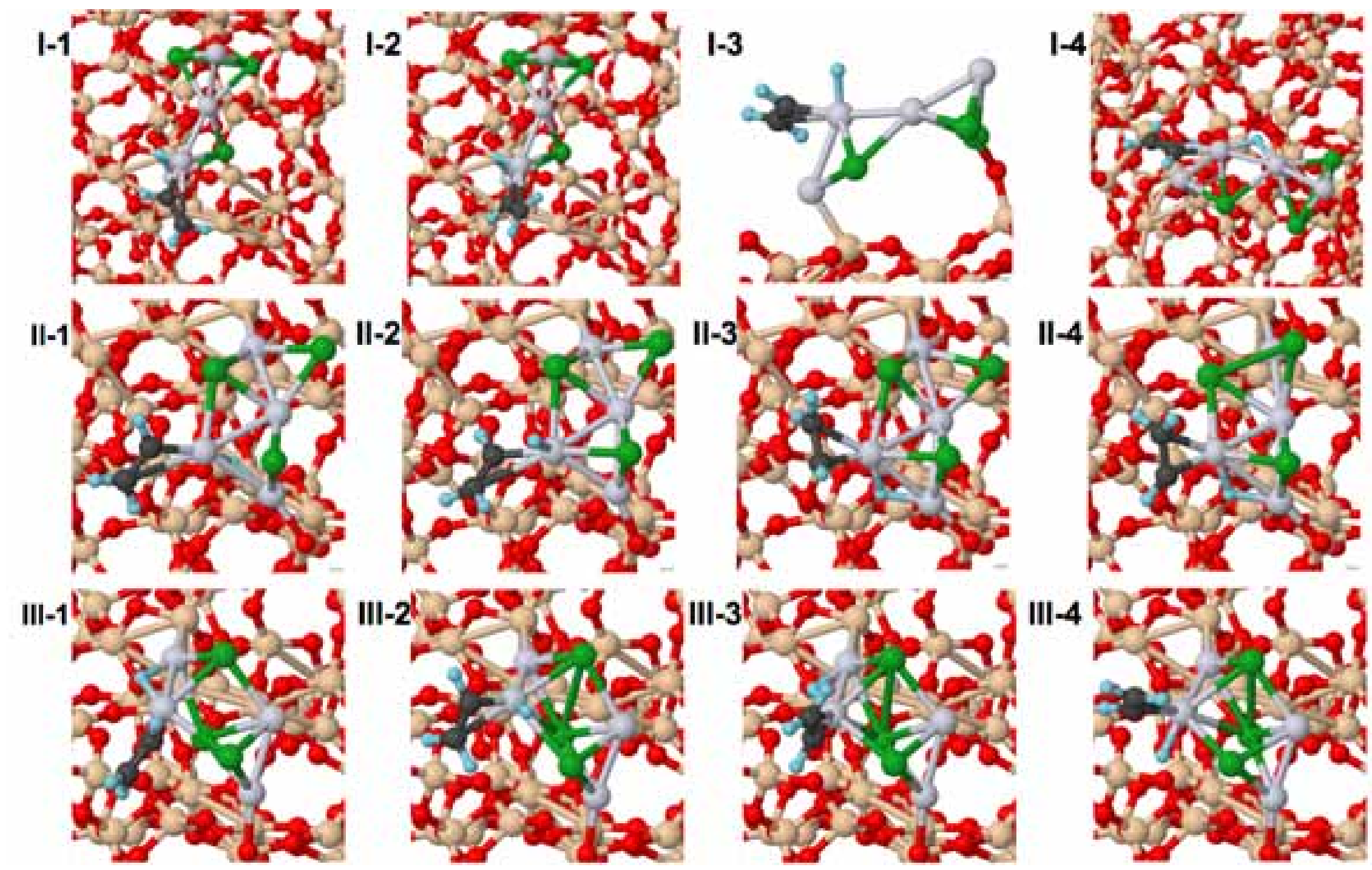

(b)

**Figure S11.** (a) Three most populated $C_2H_4/Pt_4Sn_3/SiO_2$ isomers used in CI-NEB calculations and (b) all 12 products (four for each structure) obtained by cleaving every C-H bond in $C_2H_4$ on $Pt_4Sn_3/SiO_2$.

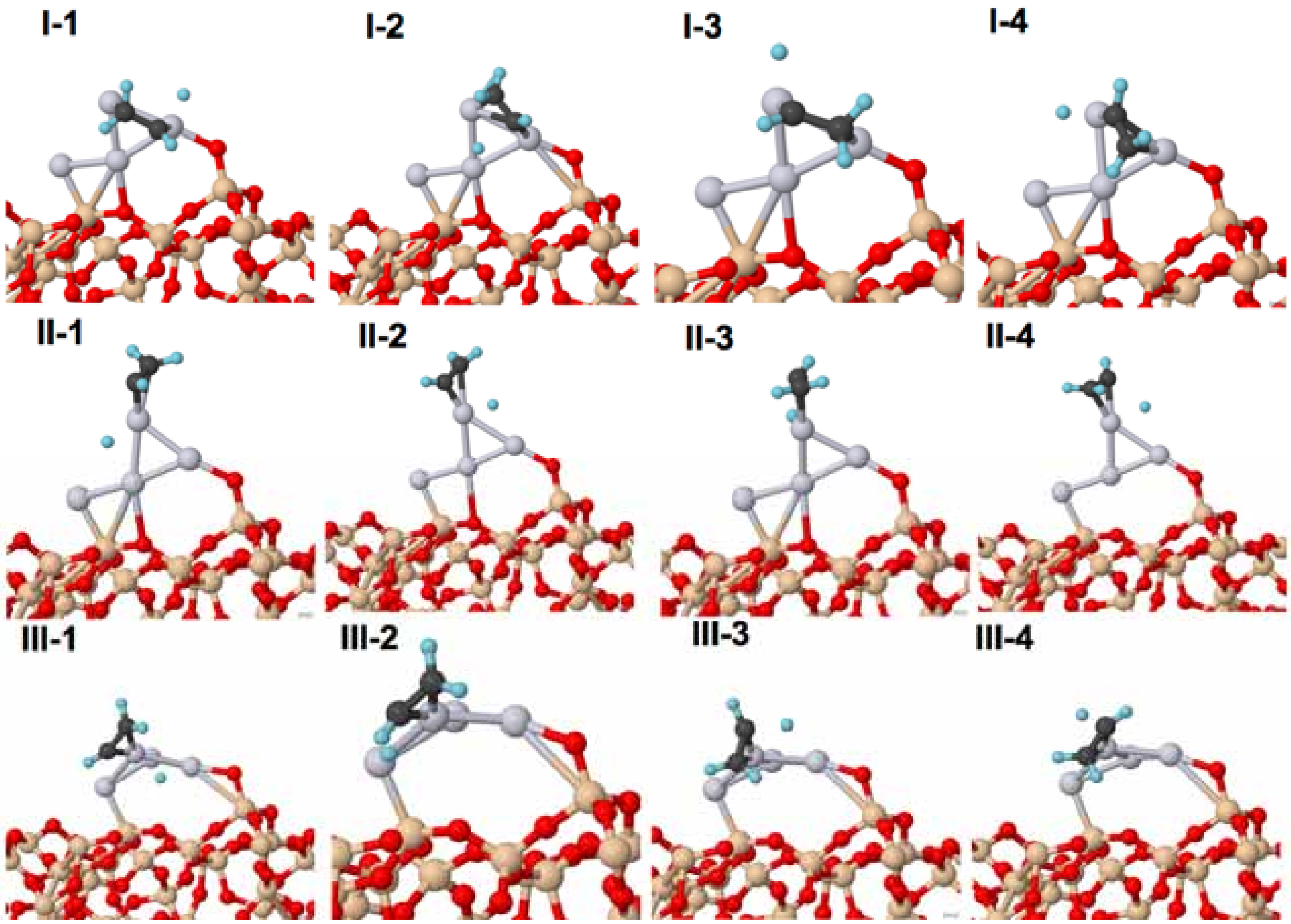

**Figure S12.** Transition state structures corresponding to all 12 different pathways of C-H bond dissociation on $Pt_4/SiO_2$ obtained from CI-NEB calculations.

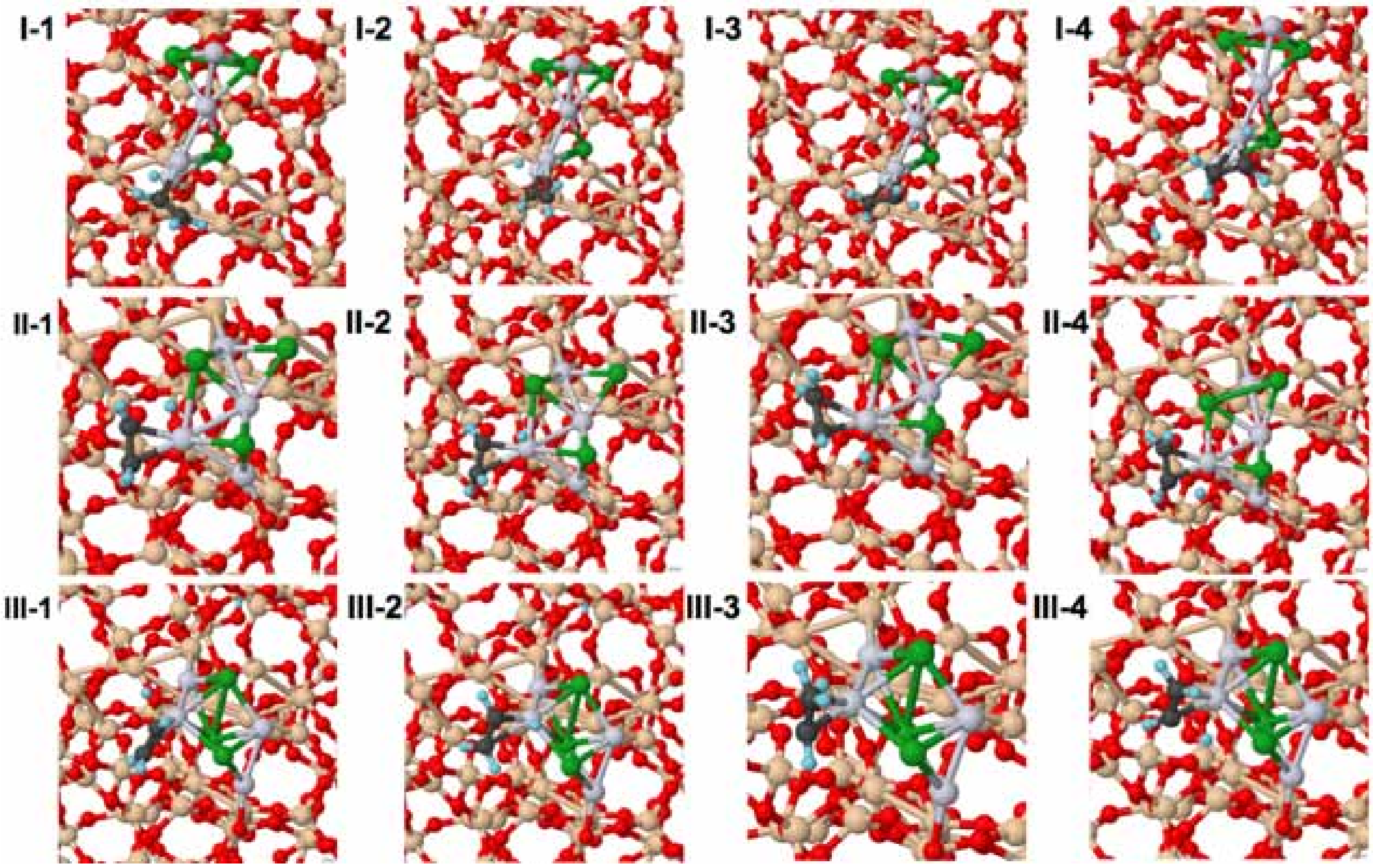

**Figure S13.** Transition state structures corresponding to all 12 different pathways of C-H bond dissociation on $Pt_4Sn_3/SiO_2$ obtained from CI-NEB calculations.

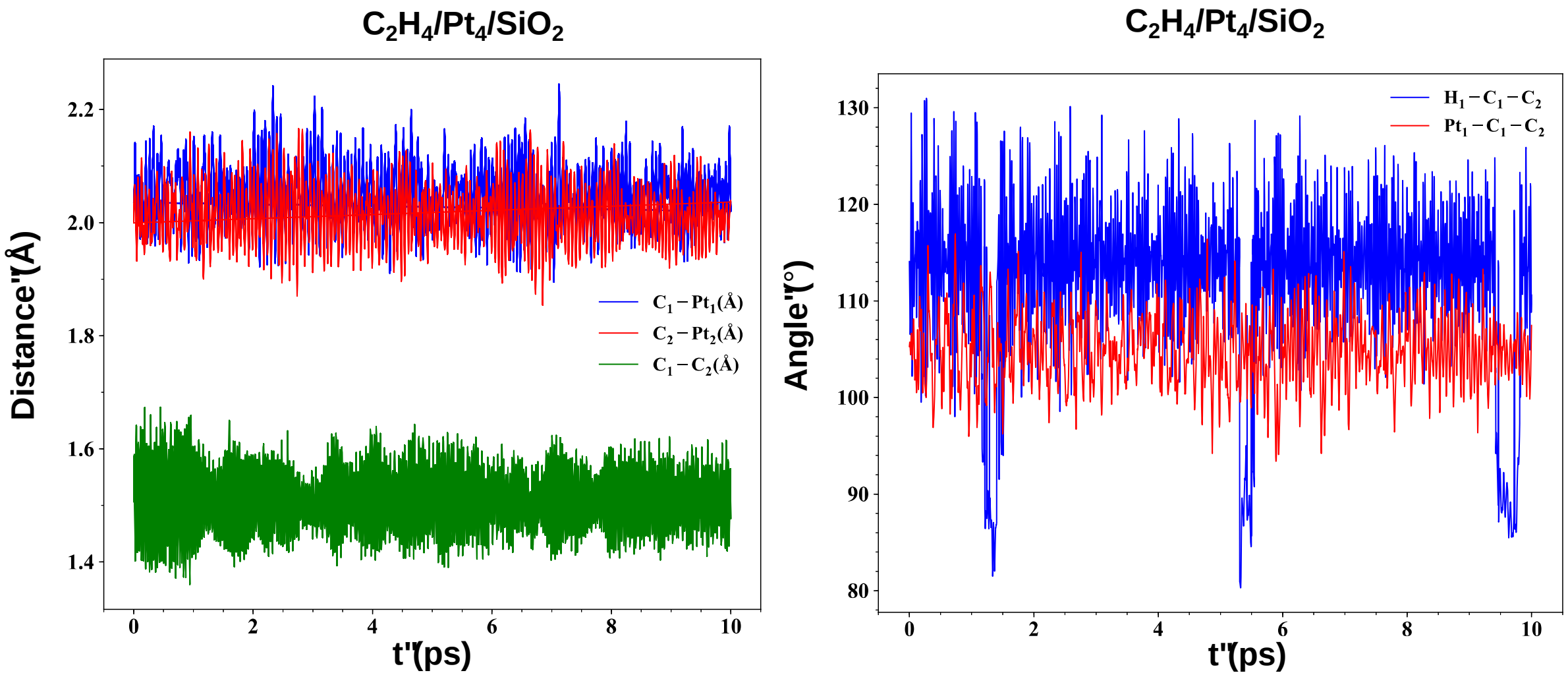

**Figure S14.** C-C and C-Pt Bond distance, and ∠HCC and ∠PtCC bond angle during the MD simulations of $C_2H_4/Pt_4/SiO_2$. The time step is 1 fs and the total run time is 10 ps.

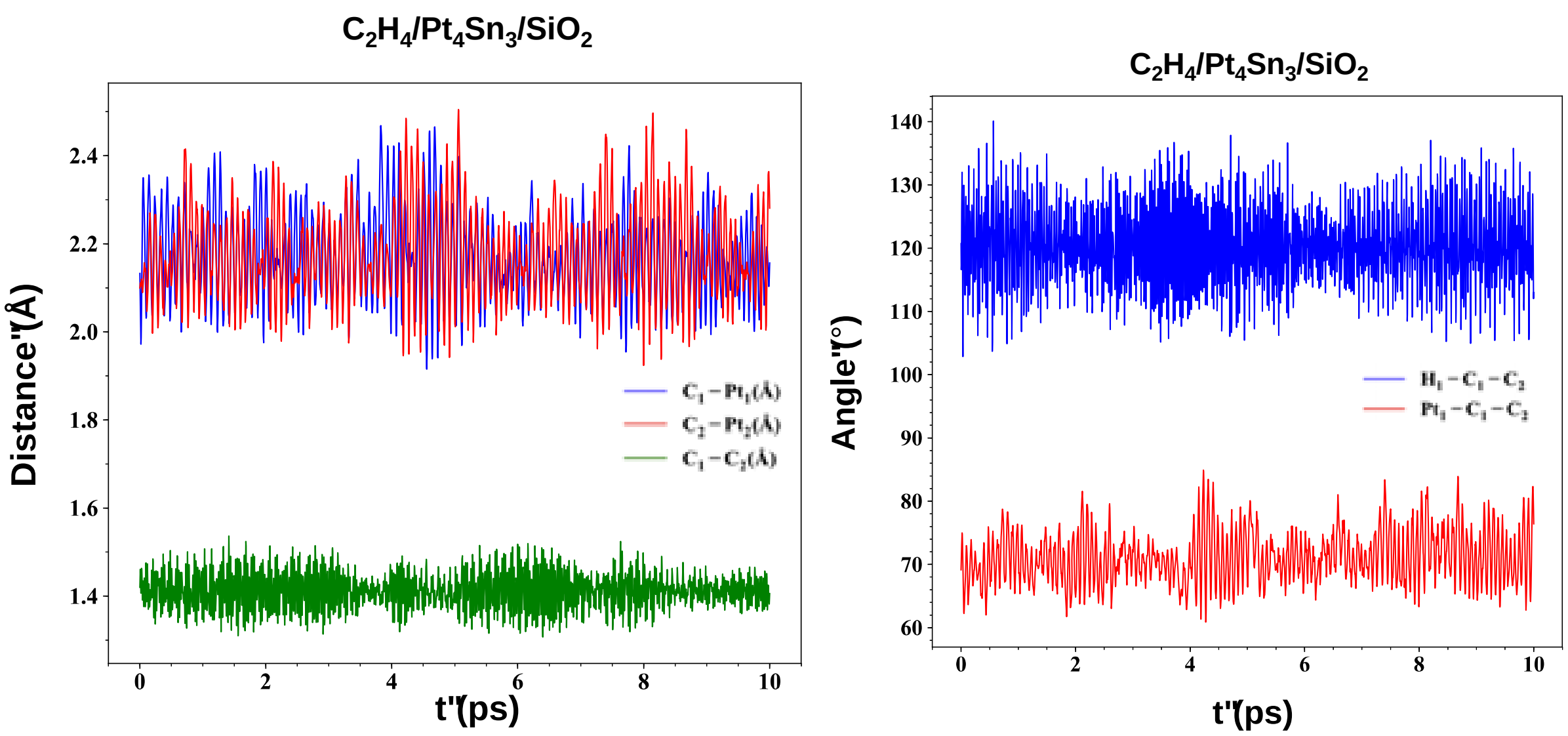

**Figure S15.** C-C and C-Pt Bond distance, and ∠HCC and ∠PtCC bond angle during the MD simulations of $C_2H_4/Pt_4Sn_3/SiO_2$. The time step is 1 fs and the total run time is 10 ps.

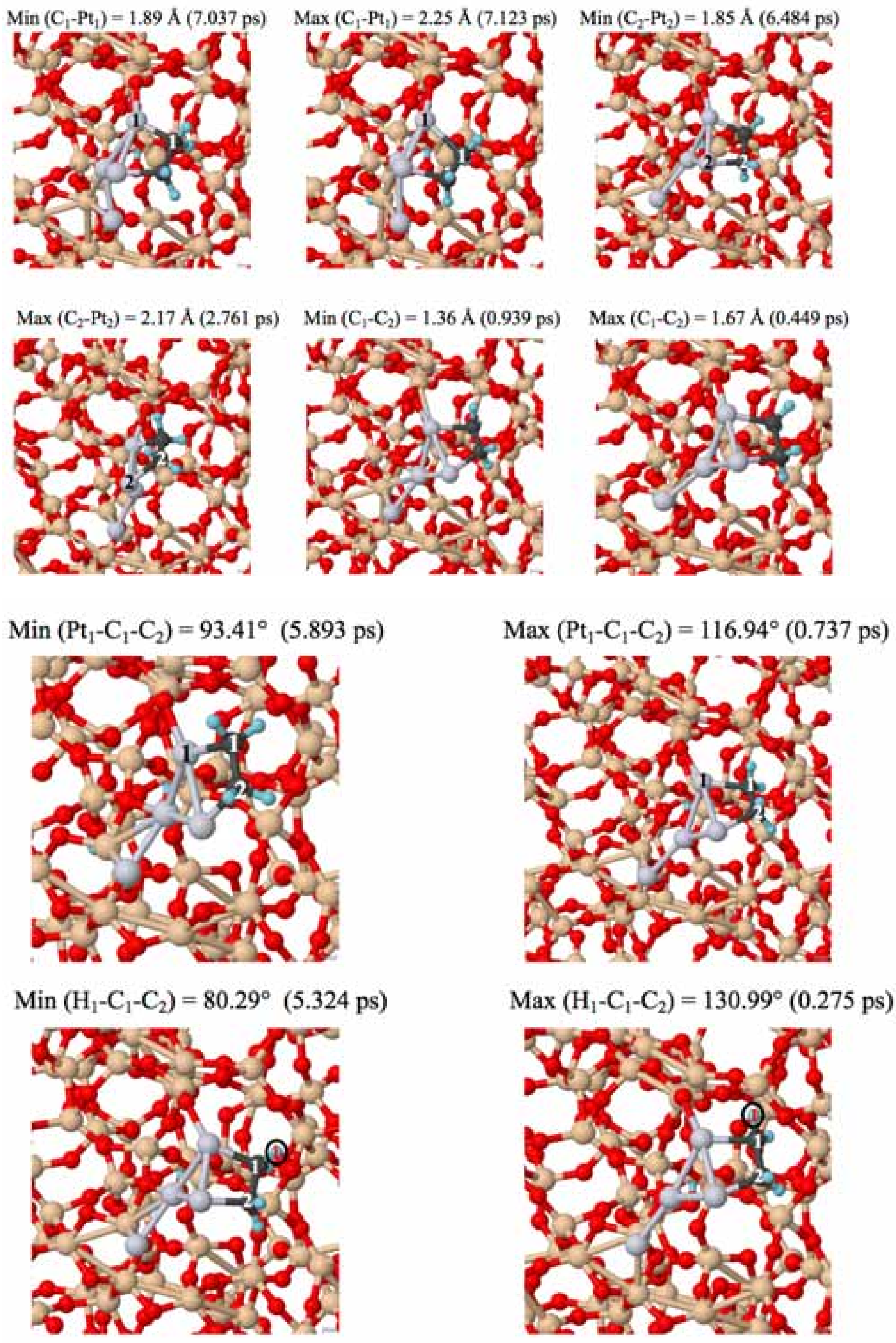

**Figure S16.** All structures corresponding to an extreme in bond distance or bond angle during the MD simulations of $C_2H_4/Pt_4/SiO_2$ with their corresponding value.

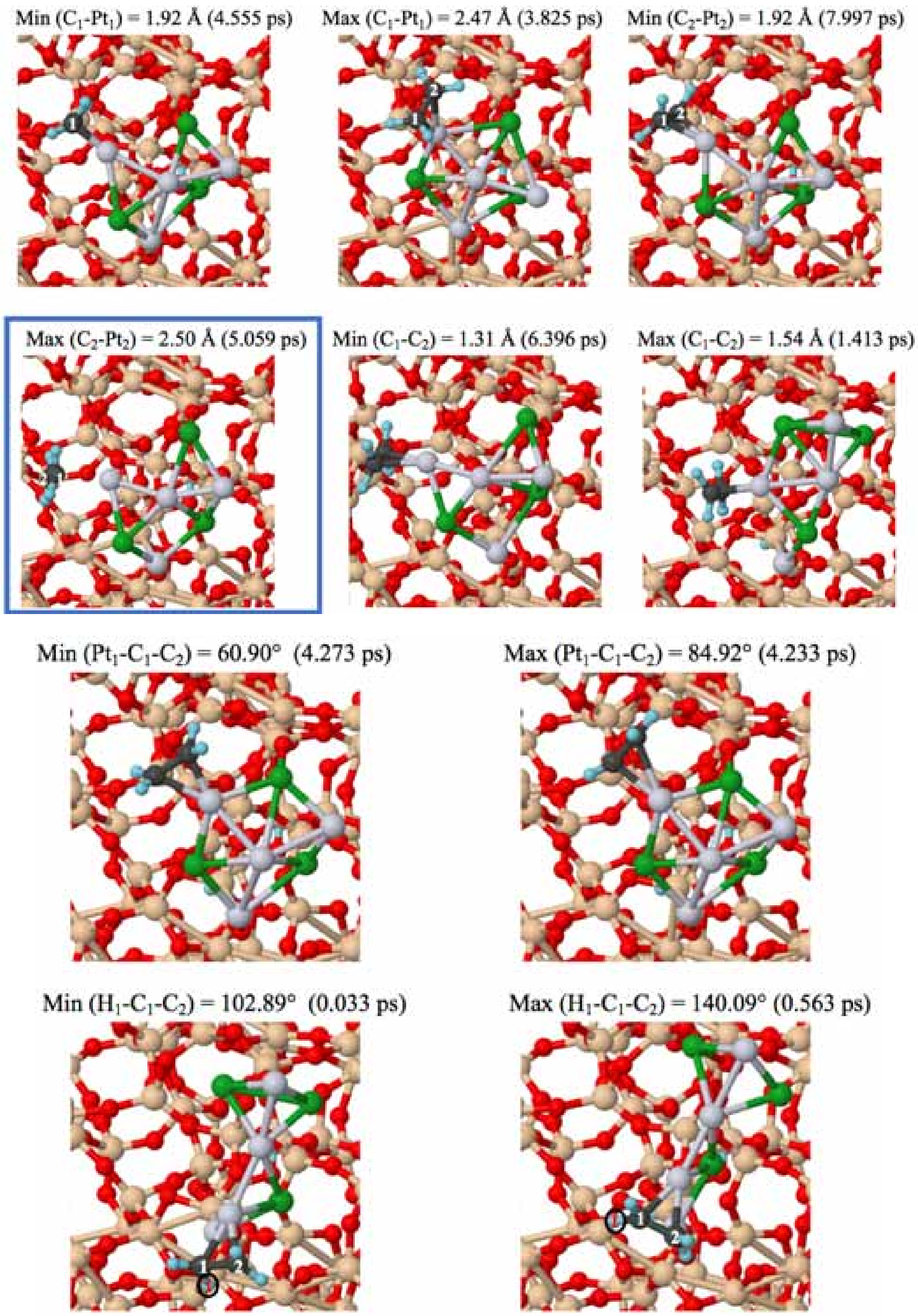

**Figure S17.** All structures corresponding to an extreme in bond distance or bond angle during the MD simulations of $C_2H_4/Pt_4Sn_3/SiO_2$ with their corresponding value. Note that at 5.059 ps $C_2H_4$ is almost detached from the cluster.

| Fig. 1(Top) | 1st TPD(Pt1) | $C_2D_4$ | 1st TPD(Pt1) | $D_2$ | 4th TPD(Pt4) | $C_2D_4$ | 4th TPD (Pt4) | $D_2$ |
|---|---|---|---|---|---|---|---|---|
| | 177.9284 | -988995891.7 | 178.5735 | -1053904532 | 177.0179 | 1606878359 | 176.8743 | -1046569205 |
| | 177.2898 | 1830902237 | 177.4755 | 80770496.23 | 176.8765 | -268256586.8 | 177.0709 | 250164034 |
| | 176.8962 | -1143244080 | 177.1005 | 40380128.71 | 176.8949 | -267944977.9 | 176.7758 | 452857114.6 |
| | 177.1648 | 4106682643 | 177.0129 | -103035.3701 | 177.0179 | 1735246343 | 176.8743 | 736297084.2 |
| | 177.6967 | 3355785322 | 177.4612 | -247767.6593 | 177.7652 | -943272534.7 | 177.3834 | -236043899.9 |
| | 178.4613 | 951999528.6 | 178.1236 | 40049813.15 | 178.9145 | 3078381779 | 178.302 | 208868525.8 |
| | 179.6672 | 2903560574 | 178.9963 | 404371070.7 | 180.4272 | 688818657.5 | 179.5809 | 5540176.024 |
| | 181.6585 | 2451661586 | 180.6015 | -122796024.2 | 182.5031 | 2034987120 | 181.3199 | -76583743.89 |
| | 183.5945 | 7727763084 | 182.7205 | -204503069.6 | 184.8332 | 7249667719 | 183.7877 | 43324204.33 |
| | 186.5118 | 8532805943 | 184.9433 | 888588340.2 | 187.6042 | 5157392537 | 186.0824 | -646666228.8 |
| | 189.0098 | 3947511002 | 187.7298 | 523085650.8 | 190.6397 | 4104646518 | 188.9665 | 363984190.8 |
| | 191.3514 | 5295696462 | 189.9436 | 36233512.66 | 193.9354 | 8576608679 | 192.0704 | 321479381.2 |
| | 194.2818 | 10794475099 | 192.7253 | 156869759 | 197.6211 | 4912707835 | 195.9153 | 440498116.1 |
| | 198.5507 | 9342991614 | 196.4323 | 358230114.6 | 201.9243 | 5675492489 | 199.8722 | 761947099.9 |
| | 204.8242 | 9924630066 | 201.6667 | 437563004.5 | 206.7492 | 10104303223 | 204.3179 | 556573806.7 |
| | 209.5292 | 13173522584 | 207.1249 | 962449538.7 | 211.2661 | 11836756504 | 208.8095 | 675174986.6 |
| | 213.5434 | 13305559188 | 211.5838 | 879987011.3 | 216.0295 | 10951228715 | 213.8094 | 1076951492 |
| | 218.4873 | 20354440683 | 216.1601 | 635440794.2 | 220.7065 | 20690835227 | 218.5865 | 61353974.83 |
| | 223.9512 | 24029381919 | 221.2423 | 593288464.6 | 225.8304 | 26711790343 | 223.1865 | 98884098.42 |
| | 229.3532 | 28324554160 | 226.2656 | 24506359.45 | 230.9546 | 28316501437 | 228.0846 | 1918244240 |
| | 235.079 | 32762853257 | 232.3132 | 1116362884 | 235.6438 | 31362939765 | 233.5325 | 1145216603 |
| | 241.2733 | 43765645200 | 238.4424 | 1803078558 | 239.5939 | 33499833258 | 237.8455 | 1263933133 |
| | 249.3531 | 39613085613 | 245.1594 | 1598352632 | 243.967 | 32814719894 | 241.6653 | 1828473669 |
| | 255.8483 | 36540010279 | 252.5691 | 1069311472 | 249.2605 | 31593134608 | 246.2186 | 813020687.2 |
| | 261.0646 | 37243097490 | 258.7495 | 1918056416 | 254.5395 | 27266130148 | 252.2456 | 282622264.4 |
| | 265.8987 | 39458680302 | 263.7316 | 1430310508 | 258.93 | 28444414549 | 257.0504 | 765525742 |
| | 270.8467 | 35464024174 | 268.4214 | 537544478.2 | 263.6862 | 26247690715 | 261.4999 | 1613163271 |
| | 275.1594 | 31045260950 | 272.813 | 576638024.1 | 267.8407 | 27905012937 | 265.5961 | 557504958.4 |
| | 279.3147 | 35873042742 | 277.0818 | 453725419.4 | 272.0443 | 29770176865 | 269.7478 | 311821024.3 |
| | 283.2962 | 28619159079 | 281.4715 | 411796691.2 | 276.6458 | 27156816222 | 274.5138 | 1118753670 |
| | 287.5097 | 31791313384 | 285.4595 | 1018180808 | 282.4529 | 26633952630 | 279.4637 | 1277559363 |
| | 291.3581 | 33025107088 | 289.29 | 44669318.03 | 287.1822 | 25954859405 | 284.7554 | 1355143385 |

| | | | | | | | | | | | | | | | |
|---|---|---|---|---|---|---|---|---|---|---|---|---|---|---|---|
| 0.598 | 3.31E+08 | 0.598 | 7.78E+08 | 0.598 | 85040.94 | 0.598 | 1.11E+09 | 0.598 | 1.76E+08 | 0.598 | 4.14E+08 | 0.598 | 45273.55 | 0.598 | 5.91E+08 |
| 0.599 | 3.14E+08 | 0.599 | 8.03E+08 | 0.599 | 88288.23 | 0.599 | 1.12E+09 | 0.599 | 1.67E+08 | 0.599 | 4.27E+08 | 0.599 | 47002.33 | 0.599 | 5.94E+08 |
| 0.6 | 2.97E+08 | 0.6 | 8.28E+08 | 0.6 | 91653.18 | 0.6 | 1.12E+09 | 0.6 | 1.58E+08 | 0.6 | 4.41E+08 | 0.6 | 48793.74 | 0.6 | 5.99E+08 |
| 0.601 | 2.8E+08 | 0.601 | 8.53E+08 | 0.601 | 95139.79 | 0.601 | 1.13E+09 | 0.601 | 1.49E+08 | 0.601 | 4.54E+08 | 0.601 | 50649.92 | 0.601 | 6.03E+08 |
| 0.602 | 2.64E+08 | 0.602 | 8.79E+08 | 0.602 | 98752.21 | 0.602 | 1.14E+09 | 0.602 | 1.41E+08 | 0.602 | 4.68E+08 | 0.602 | 52573.07 | 0.602 | 6.09E+08 |
| 0.603 | 2.49E+08 | 0.603 | 9.06E+08 | 0.603 | 102494.7 | 0.603 | 1.16E+09 | 0.603 | 1.33E+08 | 0.603 | 4.82E+08 | 0.603 | 54565.47 | 0.603 | 6.15E+08 |
| 0.604 | 2.35E+08 | 0.604 | 9.33E+08 | 0.604 | 106371.6 | 0.604 | 1.17E+09 | 0.604 | 1.25E+08 | 0.604 | 4.97E+08 | 0.604 | 56629.46 | 0.604 | 6.22E+08 |
| 0.605 | 2.2E+08 | 0.605 | 9.6E+08 | 0.605 | 110387.6 | 0.605 | 1.18E+09 | 0.605 | 1.17E+08 | 0.605 | 5.11E+08 | 0.605 | 58767.45 | 0.605 | 6.29E+08 |
| 0.606 | 2.07E+08 | 0.606 | 9.89E+08 | 0.606 | 114547.3 | 0.606 | 1.2E+09 | 0.606 | 1.1E+08 | 0.606 | 5.26E+08 | 0.606 | 60981.94 | 0.606 | 6.37E+08 |
| 0.607 | 1.94E+08 | 0.607 | 1.02E+09 | 0.607 | 118855.4 | 0.607 | 1.21E+09 | 0.607 | 1.03E+08 | 0.607 | 5.42E+08 | 0.607 | 63275.5 | 0.607 | 6.45E+08 |
| 0.608 | 1.82E+08 | 0.608 | 1.05E+09 | 0.608 | 123317.1 | 0.608 | 1.23E+09 | 0.608 | 96761603 | 0.608 | 5.57E+08 | 0.608 | 65650.77 | 0.608 | 6.54E+08 |
| 0.609 | 1.7E+08 | 0.609 | 1.08E+09 | 0.609 | 127937.4 | 0.609 | 1.25E+09 | 0.609 | 90502298 | 0.609 | 5.73E+08 | 0.609 | 68110.5 | 0.609 | 6.64E+08 |
| 0.61 | 1.59E+08 | 0.61 | 1.11E+09 | 0.61 | 132721.6 | 0.61 | 1.27E+09 | 0.61 | 84542151 | 0.61 | 5.89E+08 | 0.61 | 70657.5 | 0.61 | 6.74E+08 |
| 0.611 | 1.48E+08 | 0.611 | 1.14E+09 | 0.611 | 137675.2 | 0.611 | 1.29E+09 | 0.611 | 78875861 | 0.611 | 6.06E+08 | 0.611 | 73294.66 | 0.611 | 6.85E+08 |
| 0.612 | 1.38E+08 | 0.612 | 1.17E+09 | 0.612 | 142803.8 | 0.612 | 1.31E+09 | 0.612 | 73497415 | 0.612 | 6.22E+08 | 0.612 | 76025 | 0.612 | 6.96E+08 |
| 0.613 | 1.28E+08 | 0.613 | 1.2E+09 | 0.613 | 148113.2 | 0.613 | 1.33E+09 | 0.613 | 68400165 | 0.613 | 6.39E+08 | 0.613 | 78851.58 | 0.613 | 7.08E+08 |
| 0.614 | 1.19E+08 | 0.614 | 1.23E+09 | 0.614 | 153609.4 | 0.614 | 1.35E+09 | 0.614 | 63576902 | 0.614 | 6.57E+08 | 0.614 | 81777.6 | 0.614 | 7.2E+08 |
| 0.615 | 1.11E+08 | 0.615 | 1.27E+09 | 0.615 | 159298.5 | 0.615 | 1.38E+09 | 0.615 | 59019932 | 0.615 | 6.74E+08 | 0.615 | 84806.33 | 0.615 | 7.34E+08 |
| 0.616 | 1.03E+08 | 0.616 | 1.3E+09 | 0.616 | 165186.9 | 0.616 | 1.4E+09 | 0.616 | 54721146 | 0.616 | 6.92E+08 | 0.616 | 87941.14 | 0.616 | 7.47E+08 |
| 0.617 | 95181436 | 0.617 | 1.33E+09 | 0.617 | 171281.1 | 0.617 | 1.43E+09 | 0.617 | 50672087 | 0.617 | 7.1E+08 | 0.617 | 91185.52 | 0.617 | 7.61E+08 |
| 0.618 | 88028438 | 0.618 | 1.37E+09 | 0.618 | 177587.8 | 0.618 | 1.46E+09 | 0.618 | 46864019 | 0.618 | 7.29E+08 | 0.618 | 94543.05 | 0.618 | 7.76E+08 |
| 0.619 | 81311293 | 0.619 | 1.4E+09 | 0.619 | 184114 | 0.619 | 1.49E+09 | 0.619 | 43287989 | 0.619 | 7.48E+08 | 0.619 | 98017.42 | 0.619 | 7.91E+08 |
| 0.62 | 75012886 | 0.62 | 1.44E+09 | 0.62 | 190866.8 | 0.62 | 1.52E+09 | 0.62 | 39934883 | 0.62 | 7.67E+08 | 0.62 | 101612.4 | 0.62 | 8.07E+08 |
| 0.621 | 69115907 | 0.621 | 1.48E+09 | 0.621 | 197853.6 | 0.621 | 1.55E+09 | 0.621 | 36795487 | 0.621 | 7.86E+08 | 0.621 | 105332 | 0.621 | 8.23E+08 |
| 0.622 | 63602953 | 0.622 | 1.51E+09 | 0.622 | 205081.9 | 0.622 | 1.58E+09 | 0.622 | 33860535 | 0.622 | 8.05E+08 | 0.622 | 109180.2 | 0.622 | 8.39E+08 |
| 0.623 | 58456617 | 0.623 | 1.55E+09 | 0.623 | 212559.7 | 0.623 | 1.61E+09 | 0.623 | 31120762 | 0.623 | 8.25E+08 | 0.623 | 113161.2 | 0.623 | 8.56E+08 |
| 0.624 | 53659573 | 0.624 | 1.59E+09 | 0.624 | 220294.8 | 0.624 | 1.64E+09 | 0.624 | 28566942 | 0.624 | 8.45E+08 | 0.624 | 117279.2 | 0.624 | 8.74E+08 |
| 0.625 | 49194650 | 0.625 | 1.63E+09 | 0.625 | 228295.6 | 0.625 | 1.67E+09 | 0.625 | 26189935 | 0.625 | 8.65E+08 | 0.625 | 121538.6 | 0.625 | 8.92E+08 |
| 0.626 | 45044905 | 0.626 | 1.66E+09 | 0.626 | 236570.7 | 0.626 | 1.71E+09 | 0.626 | 23980720 | 0.626 | 8.86E+08 | 0.626 | 125944 | 0.626 | 9.1E+08 |
| 0.627 | 41193681 | 0.627 | 1.7E+09 | 0.627 | 245128.7 | 0.627 | 1.74E+09 | 0.627 | 21930430 | 0.627 | 9.06E+08 | 0.627 | 130500.1 | 0.627 | 9.28E+08 |
| 0.628 | 37624667 | 0.628 | 1.74E+09 | 0.628 | 253978.7 | 0.628 | 1.78E+09 | 0.628 | 20030381 | 0.628 | 9.27E+08 | 0.628 | 135211.6 | 0.628 | 9.47E+08 |
| 0.629 | 34321942 | 0.629 | 1.78E+09 | 0.629 | 263130.1 | 0.629 | 1.82E+09 | 0.629 | 18272097 | 0.629 | 9.48E+08 | 0.629 | 140083.5 | 0.629 | 9.67E+08 |
| 0.63 | 31270021 | 0.63 | 1.82E+09 | 0.63 | 272592.3 | 0.63 | 1.85E+09 | 0.63 | 16647335 | 0.63 | 9.7E+08 | 0.63 | 145120.9 | 0.63 | 9.87E+08 |
| 0.631 | 28453889 | 0.631 | 1.86E+09 | 0.631 | 282375.2 | 0.631 | 1.89E+09 | 0.631 | 15148100 | 0.631 | 9.91E+08 | 0.631 | 150329.1 | 0.631 | 1.01E+09 |
| 0.632 | 25859030 | 0.632 | 1.9E+09 | 0.632 | 292489 | 0.632 | 1.93E+09 | 0.632 | 13766666 | 0.632 | 1.01E+09 | 0.632 | 155713.4 | 0.632 | 1.03E+09 |
| 0.633 | 23471451 | 0.633 | 1.94E+09 | 0.633 | 302944 | 0.633 | 1.97E+09 | 0.633 | 12495582 | 0.633 | 1.04E+09 | 0.633 | 161279.4 | 0.633 | 1.05E+09 |
| 0.634 | 21277705 | 0.634 | 1.99E+09 | 0.634 | 313751.1 | 0.634 | 2.01E+09 | 0.634 | 11327689 | 0.634 | 1.06E+09 | 0.634 | 167032.8 | 0.634 | 1.07E+09 |
| 0.635 | 19264901 | 0.635 | 2.03E+09 | 0.635 | 324921.2 | 0.635 | 2.05E+09 | 0.635 | 10256125 | 0.635 | 1.08E+09 | 0.635 | 172979.5 | 0.635 | 1.09E+09 |
| 0.636 | 17420712 | 0.636 | 2.07E+09 | 0.636 | 336465.7 | 0.636 | 2.09E+09 | 0.636 | 9274328 | 0.636 | 1.1E+09 | 0.636 | 179125.5 | 0.636 | 1.11E+09 |
| 0.637 | 15733384 | 0.637 | 2.11E+09 | 0.637 | 348396.2 | 0.637 | 2.13E+09 | 0.637 | 8376039 | 0.637 | 1.12E+09 | 0.637 | 185477 | 0.637 | 1.13E+09 |
| 0.638 | 14191736 | 0.638 | 2.16E+09 | 0.638 | 360724.9 | 0.638 | 2.17E+09 | 0.638 | 7555306 | 0.638 | 1.15E+09 | 0.638 | 192040.4 | 0.638 | 1.16E+09 |
| 0.639 | 12785156 | 0.639 | 2.2E+09 | 0.639 | 373463.9 | 0.639 | 2.21E+09 | 0.639 | 6806480 | 0.639 | 1.17E+09 | 0.639 | 198822.4 | 0.639 | 1.18E+09 |
| 0.64 | 11503597 | 0.64 | 2.24E+09 | 0.64 | 386626.1 | 0.64 | 2.25E+09 | 0.64 | 6124212 | 0.64 | 1.19E+09 | 0.64 | 205829.6 | 0.64 | 1.2E+09 |
| 0.641 | 10337570 | 0.641 | 2.29E+09 | 0.641 | 400224.5 | 0.641 | 2.3E+09 | 0.641 | 5503449 | 0.641 | 1.22E+09 | 0.641 | 213069 | 0.641 | 1.22E+09 |
| 0.642 | 9278128 | 0.642 | 2.33E+09 | 0.642 | 414272.5 | 0.642 | 2.34E+09 | 0.642 | 4939431 | 0.642 | 1.24E+09 | 0.642 | 220547.7 | 0.642 | 1.25E+09 |
| 0.643 | 8316860 | 0.643 | 2.37E+09 | 0.643 | 428783.8 | 0.643 | 2.38E+09 | 0.643 | 4427677 | 0.643 | 1.26E+09 | 0.643 | 228273.2 | 0.643 | 1.27E+09 |
| 0.644 | 7445872 | 0.644 | 2.42E+09 | 0.644 | 443772.8 | 0.644 | 2.43E+09 | 0.644 | 3963986 | 0.644 | 1.29E+09 | 0.644 | 236253 | 0.644 | 1.29E+09 |
| 0.645 | 6657771 | 0.645 | 2.46E+09 | 0.645 | 459254 | 0.645 | 2.47E+09 | 0.645 | 3544422 | 0.645 | 1.31E+09 | 0.645 | 244494.7 | 0.645 | 1.31E+09 |
| 0.646 | 5945649 | 0.646 | 2.51E+09 | 0.646 | 475242.4 | 0.646 | 2.51E+09 | 0.646 | 3165307 | 0.646 | 1.33E+09 | 0.646 | 253006.5 | 0.646 | 1.34E+09 |
| 0.647 | 5303064 | 0.647 | 2.55E+09 | 0.647 | 491753.3 | 0.647 | 2.56E+09 | 0.647 | 2823211 | 0.647 | 1.36E+09 | 0.647 | 261796.5 | 0.647 | 1.36E+09 |
| 0.648 | 4724018 | 0.648 | 2.59E+09 | 0.648 | 508802.6 | 0.648 | 2.6E+09 | 0.648 | 2514943 | 0.648 | 1.38E+09 | 0.648 | 270873.1 | 0.648 | 1.38E+09 |
| 0.649 | 4202942 | 0.649 | 2.64E+09 | 0.649 | 526406.7 | 0.649 | 2.64E+09 | 0.649 | 2237535 | 0.649 | 1.41E+09 | 0.649 | 280245 | 0.649 | 1.41E+09 |
| 0.65 | 3734671 | 0.65 | 2.68E+09 | 0.65 | 544582.1 | 0.65 | 2.69E+09 | 0.65 | 1988240 | 0.65 | 1.43E+09 | 0.65 | 289921.1 | 0.65 | 1.43E+09 |
| 0.651 | 3314427 | 0.651 | 2.73E+09 | 0.651 | 563346.1 | 0.651 | 2.73E+09 | 0.651 | 1764513 | 0.651 | 1.45E+09 | 0.651 | 299910.6 | 0.651 | 1.45E+09 |
| 0.652 | 2937796 | 0.652 | 2.77E+09 | 0.652 | 582716.3 | 0.652 | 2.78E+09 | 0.652 | 1564005 | 0.652 | 1.48E+09 | 0.652 | 310222.8 | 0.652 | 1.48E+09 |
| 0.653 | 2600711 | 0.653 | 2.82E+09 | 0.653 | 602710.7 | 0.653 | 2.82E+09 | 0.653 | 1384550 | 0.653 | 1.5E+09 | 0.653 | 320867.3 | 0.653 | 1.5E+09 |
| 0.654 | 2299426 | 0.654 | 2.86E+09 | 0.654 | 623348.2 | 0.654 | 2.87E+09 | 0.654 | 1224154 | 0.654 | 1.52E+09 | 0.654 | 331854.1 | 0.654 | 1.53E+09 |
| 0.655 | 2030505 | 0.655 | 2.91E+09 | 0.655 | 644647.6 | 0.655 | 2.91E+09 | 0.655 | 1080987 | 0.655 | 1.55E+09 | 0.655 | 343193.4 | 0.655 | 1.55E+09 |
| 0.656 | 1790795 | 0.656 | 2.95E+09 | 0.656 | 666628.7 | 0.656 | 2.95E+09 | 0.656 | 953372.1 | 0.656 | 1.57E+09 | 0.656 | 354895.6 | 0.656 | 1.57E+09 |
| 0.657 | 1577411 | 0.657 | 2.99E+09 | 0.657 | 689311.6 | 0.657 | 3E+09 | 0.657 | 839771.8 | 0.657 | 1.59E+09 | 0.657 | 366971.3 | 0.657 | 1.6E+09 |
| 0.658 | 1387717 | 0.658 | 3.04E+09 | 0.658 | 712717 | 0.658 | 3.04E+09 | 0.658 | 738783.7 | 0.658 | 1.62E+09 | 0.658 | 379431.8 | 0.658 | 1.62E+09 |
| 0.659 | 1219309 | 0.659 | 3.08E+09 | 0.659 | 736866.2 | 0.659 | 3.08E+09 | 0.659 | 649128.2 | 0.659 | 1.64E+09 | 0.659 | 392288.1 | 0.659 | 1.64E+09 |
| 0.66 | 1070001 | 0.66 | 3.13E+09 | 0.66 | 761780.8 | 0.66 | 3.13E+09 | 0.66 | 569640.4 | 0.66 | 1.66E+09 | 0.66 | 405552 | 0.66 | 1.66E+09 |
| 0.661 | 937803 | 0.661 | 3.17E+09 | 0.661 | 787483.4 | 0.661 | 3.17E+09 | 0.661 | 499261.6 | 0.661 | 1.69E+09 | 0.661 | 419235.4 | 0.661 | 1.69E+09 |
| 0.662 | 820911.2 | 0.662 | 3.21E+09 | 0.662 | 813996.8 | 0.662 | 3.21E+09 | 0.662 | 437031.5 | 0.662 | 1.71E+09 | 0.662 | 433350.4 | 0.662 | 1.71E+09 |
| 0.663 | 717691.6 | 0.663 | 3.25E+09 | 0.663 | 841344.7 | 0.663 | 3.26E+09 | 0.663 | 382080.1 | 0.663 | 1.73E+09 | 0.663 | 447909.7 | 0.663 | 1.73E+09 |
| 0.664 | 626666.7 | 0.664 | 3.3E+09 | 0.664 | 869551.2 | 0.664 | 3.3E+09 | 0.664 | 333620.9 | 0.664 | 1.75E+09 | 0.664 | 462926.1 | 0.664 | 1.76E+09 |
| 0.665 | 546503 | 0.665 | 3.34E+09 | 0.665 | 898641.1 | 0.665 | 3.34E+09 | 0.665 | 290943.8 | 0.665 | 1.78E+09 | 0.665 | 478412.9 | 0.665 | 1.78E+09 |
| 0.666 | 475998.6 | 0.666 | 3.38E+09 | 0.666 | 928640 | 0.666 | 3.38E+09 | 0.666 | 253409.1 | 0.666 | 1.8E+09 | 0.666 | 494383.5 | 0.666 | 1.8E+09 |
| 0.667 | 414072 | 0.667 | 3.42E+09 | 0.667 | 959573.9 | 0.667 | 3.42E+09 | 0.667 | 220441 | 0.667 | 1.82E+09 | 0.667 | 510851.8 | 0.667 | 1.82E+09 |
| 0.668 | 359752 | 0.668 | 3.46E+09 | 0.668 | 991469.6 | 0.668 | 3.46E+09 | 0.668 | 191522.5 | 0.668 | 1.84E+09 | 0.668 | 527832.3 | 0.668 | 1.84E+09 |
| 0.669 | 312167.5 | 0.669 | 3.5E+09 | 0.669 | 1024355 | 0.669 | 3.5E+09 | 0.669 | 166189.7 | 0.669 | 1.86E+09 | 0.669 | 545339.4 | 0.669 | 1.86E+09 |
| 0.67 | 270538.6 | 0.67 | 3.54E+09 | 0.67 | 1058257 | 0.67 | 3.54E+09 | 0.67 | 144027.6 | 0.67 | 1.88E+09 | 0.67 | 563388.2 | 0.67 | 1.89E+09 |
| 0.671 | 234168.2 | 0.671 | 3.58E+09 | 0.671 | 1093206 | 0.671 | 3.58E+09 | 0.671 | 124665 | 0.671 | 1.91E+09 | 0.671 | 581994.1 | 0.671 | 1.91E+09 |
| 0.672 | 202434.2 | 0.672 | 3.62E+09 | 0.672 | 1129231 | 0.672 | 3.62E+09 | 0.672 | 107770.6 | 0.672 | 1.93E+09 | 0.672 | 601172.8 | 0.672 | 1.93E+09 |
| 0.673 | 174782.1 | 0.673 | 3.66E+09 | 0.673 | 1166362 | 0.673 | 3.66E+09 | 0.673 | 93049.36 | 0.673 | 1.95E+09 | 0.673 | 620940.5 | 0.673 | 1.95E+09 |
| 0.674 | 150718.6 | 0.674 | 3.69E+09 | 0.674 | 1204631 | 0.674 | 3.69E+09 | 0.674 | 80238.63 | 0.674 | 1.97E+09 | 0.674 | 641313.9 | 0.674 | 1.97E+09 |
| 0.675 | 129805.8 | 0.675 | 3.73E+09 | 0.675 | 1244070 | 0.675 | 3.73E+09 | 0.675 | 69105.21 | 0.675 | 1.99E+09 | 0.675 | 662309.9 | 0.675 | 1.99E+09 |
| 0.676 | 111655.1 | 0.676 | 3.76E+09 | 0.676 | 1284710 | 0.676 | 3.77E+09 | 0.676 | 59442.24 | 0.676 | 2E+09 | 0.676 | 683946 | 0.676 | 2E+09 |
| 0.677 | 95922.42 | 0.677 | 3.8E+09 | 0.677 | 1326587 | 0.677 | 3.8E+09 | 0.677 | 51066.57 | 0.677 | 2.02E+09 | 0.677 | 706240 | 0.677 | 2.02E+09 |
| 0.678 | 82303.59 | 0.678 | 3.83E+09 | 0.678 | 1369734 | 0.678 | 3.84E+09 | 0.678 | 43816.26 | 0.678 | 2.04E+09 | 0.678 | 729210.2 | 0.678 | 2.04E+09 |
| 0.679 | 70530.11 | 0.679 | 3.87E+09 | 0.679 | 1414186 | 0.679 | 3.87E+09 | 0.679 | 37548.37 | 0.679 | 2.06E+09 | 0.679 | 752875.4 | 0.679 | 2.06E+09 |
| 0.68 | 60365.32 | 0.68 | 3.9E+09 | 0.68 | 1459980 | 0.68 | 3.9E+09 | 0.68 | 32136.9 | 0.68 | 2.08E+09 | 0.68 | 777254.8 | 0.68 | 2.08E+09 |
| 0.681 | 51600.93 | 0.681 | 3.93E+09 | 0.681 | 1507152 | 0.681 | 3.93E+09 | 0.681 | 27470.98 | 0.681 | 2.09E+09 | 0.681 | 802368.1 | 0.681 | 2.09E+09 |
| 0.682 | 44053.94 | 0.682 | 3.96E+09 | 0.682 | 1555741 | 0.682 | 3.96E+09 | 0.682 | 23453.15 | 0.682 | 2.11E+09 | 0.682 | 828235.6 | 0.682 | 2.11E+09 |
| 0.683 | 37563.76 | 0.683 | 3.99E+09 | 0.683 | 1605785 | 0.683 | 4E+09 | 0.683 | 19997.96 | 0.683 | 2.13E+09 | 0.683 | 854877.8 | 0.683 | 2.13E+09 |
| 0.684 | 31989.72 | 0.684 | 4.02E+09 | 0.684 | 1657325 | 0.684 | 4.02E+09 | 0.684 | 17030.49 | 0.684 | 2.14E+09 | 0.684 | 882316 | 0.684 | 2.14E+09 |
| 0.685 | 27208.78 | 0.685 | 4.05E+09 | 0.685 | 1710400 | 0.685 | 4.05E+09 | 0.685 | 14485.24 | 0.685 | 2.16E+09 | 0.685 | 910571.8 | 0.685 | 2.16E+09 |
| 0.686 | 23113.45 | 0.686 | 4.08E+09 | 0.686 | 1765053 | 0.686 | 4.08E+09 | 0.686 | 12304.99 | 0.686 | 2.17E+09 | 0.686 | 939667.4 | 0.686 | 2.17E+09 |
| 0.687 | 19610 | 0.687 | 4.11E+09 | 0.687 | 1821326 | 0.687 | 4.11E+09 | 0.687 | 10439.85 | 0.687 | 2.19E+09 | 0.687 | 969625.7 | 0.687 | 2.19E+09 |
| 0.688 | 16616.81 | 0.688 | 4.13E+09 | 0.688 | 1879263 | 0.688 | 4.13E+09 | 0.688 | 8846.35 | 0.688 | 2.2E+09 | 0.688 | 1000470 | 0.688 | 2.2E+09 |
| 0.689 | 14062.89 | 0.689 | 4.16E+09 | 0.689 | 1938908 | 0.689 | 4.16E+09 | 0.689 | 7486.714 | 0.689 | 2.21E+09 | 0.689 | 1032224 | 0.689 | 2.21E+09 |
| 0.69 | 11886.64 | 0.69 | 4.18E+09 | 0.69 | 2000309 | 0.69 | 4.18E+09 | 0.69 | 6328.131 | 0.69 | 2.23E+09 | 0.69 | 1064912 | 0.69 | 2.23E+09 |
| 0.691 | 10034.61 | 0.691 | 4.2E+09 | 0.691 | 2063511 | 0.691 | 4.2E+09 | 0.691 | 5342.159 | 0.691 | 2.24E+09 | 0.691 | 1098559 | 0.691 | 2.24E+09 |
| 0.692 | 8460.554 | 0.692 | 4.22E+09 | 0.692 | 2128563 | 0.692 | 4.23E+09 | 0.692 | 4504.176 | 0.692 | 2.25E+09 | 0.692 | 1133191 | 0.692 | 2.25E+09 |
| 0.693 | 7124.5 | 0.693 | 4.24E+09 | 0.693 | 2195513 | 0.693 | 4.25E+09 | 0.693 | 3792.896 | 0.693 | 2.26E+09 | 0.693 | 1168833 | 0.693 | 2.26E+09 |
| 0.694 | 5991.936 | 0.694 | 4.26E+09 | 0.694 | 2264413 | 0.694 | 4.27E+09 | 0.694 | 3189.949 | 0.694 | 2.27E+09 | 0.694 | 1205514 | 0.694 | 2.27E+09 |
| 0.695 | 5033.118 | 0.695 | 4.28E+09 | 0.695 | 2335313 | 0.695 | 4.28E+09 | 0.695 | 2679.499 | 0.695 | 2.28E+09 | 0.695 | 1243259 | 0.695 | 2.28E+09 |
| 0.696 | 4222.446 | 0.696 | 4.3E+09 | 0.696 | 2408266 | 0.696 | 4.3E+09 | 0.696 | 2247.919 | 0.696 | 2.29E+09 | 0.696 | 1282097 | 0.696 | 2.29E+09 |
| 0.697 | 3537.922 | 0.697 | 4.32E+09 | 0.697 | 2483327 | 0.697 | 4.32E+09 | 0.697 | 1883.496 | 0.697 | 2.3E+09 | 0.697 | 1322058 | 0.697 | 2.3E+09 |
| 0.698 | 2960.667 | 0.698 | 4.33E+09 | 0.698 | 2560550 | 0.698 | 4.33E+09 | 0.698 | 1576.181 | 0.698 | 2.31E+09 | 0.698 | 1363169 | 0.698 | 2.31E+09 |
| 0.699 | 2474.503 | 0.699 | 4.34E+09 | 0.699 | 2639991 | 0.699 | 4.35E+09 | 0.699 | 1317.36 | 0.699 | 2.31E+09 | 0.699 | 1405462 | 0.699 | 2.31E+09 |
| 0.7 | 2065.587 | 0.7 | 4.36E+09 | 0.7 | 2721709 | 0.7 | 4.36E+09 | 0.7 | 1099.664 | 0.7 | 2.32E+09 | 0.7 | 1448966 | 0.7 | 2.32E+09 |
| 0.701 | 1722.091 | 0.701 | 4.37E+09 | 0.701 | 2805762 | 0.701 | 4.37E+09 | 0.701 | 916.796 | 0.701 | 2.33E+09 | 0.701 | 1493714 | 0.701 | 2.33E+09 |
| 0.702 | 1433.924 | 0.702 | 4.38E+09 | 0.702 | 2892211 | 0.702 | 4.38E+09 | 0.702 | 763.383 | 0.702 | 2.33E+09 | 0.702 | 1539737 | 0.702 | 2.33E+09 |
| 0.703 | 1192.485 | 0.703 | 4.39E+09 | 0.703 | 2981117 | 0.703 | 4.39E+09 | 0.703 | 634.848 | 0.703 | 2.34E+09 | 0.703 | 1587068 | 0.703 | 2.34E+09 |
| 0.704 | 990.46 | 0.704 | 4.4E+09 | 0.704 | 3072544 | 0.704 | 4.4E+09 | 0.704 | 527.295 | 0.704 | 2.34E+09 | 0.704 | 1635741 | 0.704 | 2.34E+09 |
| 0.705 | 821.634 | 0.705 | 4.4E+09 | 0.705 | 3166555 | 0.705 | 4.41E+09 | 0.705 | 437.416 | 0.705 | 2.34E+09 | 0.705 | 1685790 | 0.705 | 2.35E+09 |
| 0.706 | 680.733 | 0.706 | 4.41E+09 | 0.706 | 3263216 | 0.706 | 4.41E+09 | 0.706 | 362.404 | 0.706 | 2.35E+09 | 0.706 | 1737250 | 0.706 | 2.35E+09 |
| 0.707 | 563.29 | 0.707 | 4.41E+09 | 0.707 | 3362596 | 0.707 | 4.42E+09 | 0.707 | 299.881 | 0.707 | 2.35E+09 | 0.707 | 1790158 | 0.707 | 2.35E+09 |
| 0.708 | 465.527 | 0.708 | 4.42E+09 | 0.708 | 3464763 | 0.708 | 4.42E+09 | 0.708 | 247.834 | 0.708 | 2.35E+09 | 0.708 | 1844548 | 0.708 | 2.35E+09 |
| 0.709 | 384.251 | 0.709 | 4.42E+09 | 0.709 | 3569786 | 0.709 | 4.42E+09 | 0.709 | 204.565 | 0.709 | 2.35E+09 | 0.709 | 1900460 | 0.709 | 2.35E+09 |
| 0.71 | 316.768 | 0.71 | 4.42E+09 | 0.71 | 3677739 | 0.71 | 4.42E+09 | 0.71 | 168.639 | 0.71 | 2.35E+09 | 0.71 | 1957931 | 0.71 | 2.35E+09 |
| 0.711 | 260.811 | 0.711 | 4.42E+09 | 0.711 | 3788694 | 0.711 | 4.42E+09 | 0.711 | 138.849 | 0.711 | 2.35E+09 | 0.711 | 2017001 | 0.711 | 2.35E+09 |
| 0.712 | 214.47 | 0.712 | 4.42E+09 | 0.712 | 3902726 | 0.712 | 4.42E+09 | 0.712 | 114.178 | 0.712 | 2.35E+09 | 0.712 | 2077708 | 0.712 | 2.35E+09 |
| 0.713 | 176.143 | 0.713 | 4.42E+09 | 0.713 | 4019912 | 0.713 | 4.42E+09 | 0.713 | 93.774 | 0.713 | 2.35E+09 | 0.713 | 2140095 | 0.713 | 2.35E+09 |
| 0.714 | 144.484 | 0.714 | 4.42E+09 | 0.714 | 4140331 | 0.714 | 4.42E+09 | 0.714 | 76.92 | 0.714 | 2.35E+09 | 0.714 | 2204203 | 0.714 | 2.35E+09 |
| 0.715 | 118.368 | 0.715 | 4.42E+09 | 0.715 | 4264061 | 0.715 | 4.42E+09 | 0.715 | 63.016 | 0.715 | 2.35E+09 | 0.715 | 2270074 | 0.715 | 2.35E+09 |
| 0.716 | 96.851 | 0.716 | 4.42E+09 | 0.716 | 4391185 | 0.716 | 4.42E+09 | 0.716 | 51.561 | 0.716 | 2.35E+09 | 0.716 | 2337751 | 0.716 | 2.35E+09 |
| 0.717 | 79.146 | 0.717 | 4.41E+09 | 0.717 | 4521786 | 0.717 | 4.42E+09 | 0.717 | 42.135 | 0.717 | 2.35E+09 | 0.717 | 2407280 | 0.717 | 2.35E+09 |
| 0.718 | 64.597 | 0.718 | 4.41E+09 | 0.718 | 4655950 | 0.718 | 4.42E+09 | 0.718 | 34.39 | 0.718 | 2.35E+09 | 0.718 | 2478705 | 0.718 | 2.35E+09 |
| 0.719 | 52.657 | 0.719 | 4.41E+09 | 0.719 | 4793762 | 0.719 | 4.42E+09 | 0.719 | 28.033 | 0.719 | 2.35E+09 | 0.719 | 2552072 | 0.719 | 2.35E+09 |
| 0.72 | 42.87 | 0.72 | 4.41E+09 | 0.72 | 4935312 | 0.72 | 4.42E+09 | 0.72 | 22.823 | 0.72 | 2.35E+09 | 0.72 | 2627430 | 0.72 | 2.35E+09 |
| 0.721 | 34.858 | 0.721 | 4.41E+09 | 0.721 | 5080689 | 0.721 | 4.41E+09 | 0.721 | 18.558 | 0.721 | 2.35E+09 | 0.721 | 2704825 | 0.721 | 2.35E+09 |
| 0.722 | 28.309 | 0.722 | 4.41E+09 | 0.722 | 5229988 | 0.722 | 4.41E+09 | 0.722 | 15.071 | 0.722 | 2.35E+09 | 0.722 | 2784307 | 0.722 | 2.35E+09 |
| 0.723 | 22.961 | 0.723 | 4.41E+09 | 0.723 | 5383300 | 0.723 | 4.41E+09 | 0.723 | 12.224 | 0.723 | 2.35E+09 | 0.723 | 2865927 | 0.723 | 2.35E+09 |
| 0.724 | 18.6 | 0.724 | 4.4E+09 | 0.724 | 5540724 | 0.724 | 4.41E+09 | 0.724 | 9.902 | 0.724 | 2.35E+09 | 0.724 | 2949735 | 0.724 | 2.35E+09 |
| 0.725 | 15.049 | 0.725 | 4.4E+09 | 0.725 | 5702356 | 0.725 | 4.41E+09 | 0.725 | 8.012 | 0.725 | 2.34E+09 | 0.725 | 3035784 | 0.725 | 2.35E+09 |
| 0.726 | 12.16 | 0.726 | 4.4E+09 | 0.726 | 5868298 | 0.726 | 4.41E+09 | 0.726 | 6.474 | 0.726 | 2.34E+09 | 0.726 | 3124127 | 0.726 | 2.35E+09 |
| 0.727 | 9.814 | 0.727 | 4.4E+09 | 0.727 | 6038650 | 0.727 | 4.4E+09 | 0.727 | 5.225 | 0.727 | 2.34E+09 | 0.727 | 3214818 | 0.727 | 2.35E+09 |
| 0.728 | 7.91 | 0.728 | 4.4E+09 | 0.728 | 6213518 | 0.728 | 4.4E+09 | 0.728 | 4.211 | 0.728 | 2.34E+09 | 0.728 | 3307913 | 0.728 | 2.34E+09 |
| 0.729 | 6.368 | 0.729 | 4.39E+09 | 0.729 | 6393007 | 0.729 | 4.4E+09 | 0.729 | 3.39 | 0.729 | 2.34E+09 | 0.729 | 3403468 | 0.729 | 2.34E+09 |
| 0.73 | 5.12 | 0.73 | 4.39E+09 | 0.73 | 6577226 | 0.73 | 4.4E+09 | 0.73 | 2.726 | 0.73 | 2.34E+09 | 0.73 | 3501542 | 0.73 | 2.34E+09 |
| 0.731 | 4.112 | 0.731 | 4.39E+09 | 0.731 | 6766285 | 0.731 | 4.4E+09 | 0.731 | 2.189 | 0.731 | 2.34E+09 | 0.731 | 3602192 | 0.731 | 2.34E+09 |
| 0.732 | 3.298 | 0.732 | 4.39E+09 | 0.732 | 6960296 | 0.732 | 4.39E+09 | 0.732 | 1.756 | 0.732 | 2.34E+09 | 0.732 | 3705478 | 0.732 | 2.34E+09 |
| 0.733 | 2.641 | 0.733 | 4.38E+09 | 0.733 | 7159375 | 0.733 | 4.39E+09 | 0.733 | 1.406 | 0.733 | 2.33E+09 | 0.733 | 3811463 | 0.733 | 2.34E+09 |
| 0.734 | 2.113 | 0.734 | 4.38E+09 | 0.734 | 7363639 | 0.734 | 4.39E+09 | 0.734 | 1.125 | 0.734 | 2.33E+09 | 0.734 | 3920207 | 0.734 | 2.34E+09 |
| 0.735 | 1.688 | 0.735 | 4.38E+09 | 0.735 | 7573206 | 0.735 | 4.38E+09 | 0.735 | 0.899 | 0.735 | 2.33E+09 | 0.735 | 4031775 | 0.735 | 2.33E+09 |
| 0.736 | 1.347 | 0.736 | 4.37E+09 | 0.736 | 7788199 | 0.736 | 4.38E+09 | 0.736 | 0.717 | 0.736 | 2.33E+09 | 0.736 | 4146232 | 0.736 | 2.33E+09 |
| 0.737 | 1.074 | 0.737 | 4.37E+09 | 0.737 | 8008740 | 0.737 | 4.38E+09 | 0.737 | 0.572 | 0.737 | 2.33E+09 | 0.737 | 4263642 | 0.737 | 2.33E+09 |
| 0.738 | 0.855 | 0.738 | 4.37E+09 | 0.738 | 8234957 | 0.738 | 4.37E+09 | 0.738 | 0.455 | 0.738 | 2.32E+09 | 0.738 | 4384074 | 0.738 | 2.33E+09 |
| 0.739 | 0.68 | 0.739 | 4.36E+09 | 0.739 | 8466978 | 0.739 | 4.37E+09 | 0.739 | 0.362 | 0.739 | 2.32E+09 | 0.739 | 4507596 | 0.739 | 2.33E+09 |
| 0.74 | 0.54 | 0.74 | 4.36E+09 | 0.74 | 8704933 | 0.74 | 4.37E+09 | 0.74 | 0.287 | 0.74 | 2.32E+09 | 0.74 | 4634277 | 0.74 | 2.33E+09 |
| 0.741 | 0.428 | 0.741 | 4.36E+09 | 0.741 | 8948957 | 0.741 | 4.36E+09 | 0.741 | 0.228 | 0.741 | 2.32E+09 | 0.741 | 4764189 | 0.741 | 2.32E+09 |
| 0.742 | 0.339 | 0.742 | 4.35E+09 | 0.742 | 9199184 | 0.742 | 4.36E+09 | 0.742 | 0.18 | 0.742 | 2.32E+09 | 0.742 | 4897403 | 0.742 | 2.32E+09 |
| 0.743 | 0.268 | 0.743 | 4.35E+09 | 0.743 | 9455754 | 0.743 | 4.36E+09 | 0.743 | 0.143 | 0.743 | 2.31E+09 | 0.743 | 5033994 | 0.743 | 2.32E+09 |
| 0.744 | 0.212 | 0.744 | 4.34E+09 | 0.744 | 9718807 | 0.744 | 4.35E+09 | 0.744 | 0.113 | 0.744 | 2.31E+09 | 0.744 | 5174037 | 0.744 | 2.32E+09 |
| 0.745 | 0.167 | 0.745 | 4.34E+09 | 0.745 | 9988487 | 0.745 | 4.35E+09 | 0.745 | 0.089 | 0.745 | 2.31E+09 | 0.745 | 5317607 | 0.745 | 2.31E+09 |
| 0.746 | 0.132 | 0.746 | 4.33E+09 | 0.746 | 10264940 | 0.746 | 4.34E+09 | 0.746 | 0.07 | 0.746 | 2.31E+09 | 0.746 | 5464783 | 0.746 | 2.31E+09 |
| 0.747 | 0.104 | 0.747 | 4.33E+09 | 0.747 | 10548314 | 0.747 | 4.34E+09 | 0.747 | 0.055 | 0.747 | 2.3E+09 | 0.747 | 5615644 | 0.747 | 2.31E+09 |
| 0.748 | 0.082 | 0.748 | 4.32E+09 | 0.748 | 10838761 | 0.748 | 4.34E+09 | 0.748 | 0.043 | 0.748 | 2.3E+09 | 0.748 | 5770270 | 0.748 | 2.31E+09 |
| 0.749 | 0.064 | 0.749 | 4.32E+09 | 0.749 | 11136434 | 0.749 | 4.33E+09 | 0.749 | 0.034 | 0.749 | 2.3E+09 | 0.749 | 5928744 | 0.749 | 2.31E+09 |
| 0.75 | 0.05 | 0.75 | 4.31E+09 | 0.75 | 11441491 | 0.75 | 4.33E+09 | 0.75 | 0.027 | 0.75 | 2.3E+09 | 0.75 | 6091148 | 0.75 | 2.3E+09 |
| 0.751 | 0.039 | 0.751 | 4.31E+09 | 0.751 | 11754091 | 0.751 | 4.32E+09 | 0.751 | 0.021 | 0.751 | 2.29E+09 | 0.751 | 6257568 | 0.751 | 2.3E+09 |
| 0.752 | 0.031 | 0.752 | 4.3E+09 | 0.752 | 12074396 | 0.752 | 4.32E+09 | 0.752 | 0.016 | 0.752 | 2.29E+09 | 0.752 | 6428090 | 0.752 | 2.3E+09 |
| 0.753 | 0.024 | 0.753 | 4.3E+09 | 0.753 | 12402571 | 0.753 | 4.31E+09 | 0.753 | 0.013 | 0.753 | 2.29E+09 | 0.753 | 6602802 | 0.753 | 2.3E+09 |
| 0.754 | 0.019 | 0.754 | 4.29E+09 | 0.754 | 12738784 | 0.754 | 4.31E+09 | 0.754 | 0.01 | 0.754 | 2.29E+09 | 0.754 | 6781793 | 0.754 | 2.29E+09 |
| 0.755 | 0.015 | 0.755 | 4.29E+09 | 0.755 | 13083205 | 0.755 | 4.3E+09 | 0.755 | 8.00E-03 | 0.755 | 2.28E+09 | 0.755 | 6965154 | 0.755 | 2.29E+09 |
| 0.756 | 0.011 | 0.756 | 4.28E+09 | 0.756 | 13436009 | 0.756 | 4.29E+09 | 0.756 | 6.00E-03 | 0.756 | 2.28E+09 | 0.756 | 7152977 | 0.756 | 2.29E+09 |
| 0.757 | 9.00E-03 | 0.757 | 4.28E+09 | 0.757 | 13797372 | 0.757 | 4.29E+09 | 0.757 | 5.00E-03 | 0.757 | 2.28E+09 | 0.757 | 7345357 | 0.757 | 2.28E+09 |
| 0.758 | 7.00E-03 | 0.758 | 4.27E+09 | 0.758 | 14167474 | 0.758 | 4.28E+09 | 0.758 | 4.00E-03 | 0.758 | 2.27E+09 | 0.758 | 7542389 | 0.758 | 2.28E+09 |
| 0.759 | 5.00E-03 | 0.759 | 4.26E+09 | 0.759 | 14546496 | 0.759 | 4.28E+09 | 0.759 | 3.00E-03 | 0.759 | 2.27E+09 | 0.759 | 7744171 | 0.759 | 2.28E+09 |
| 0.76 | 4.00E-03 | 0.76 | 4.26E+09 | 0.76 | 14934625 | 0.76 | 4.27E+09 | 0.76 | 2.00E-03 | 0.76 | 2.27E+09 | 0.76 | 7950800 | 0.76 | 2.27E+09 |
| 0.761 | 3.00E-03 | 0.761 | 4.25E+09 | 0.761 | 15332048 | 0.761 | 4.27E+09 | 0.761 | 2.00E-03 | 0.761 | 2.26E+09 | 0.761 | 8162378 | 0.761 | 2.27E+09 |
| 0.762 | 2.00E-03 | 0.762 | 4.24E+09 | 0.762 | 15738958 | 0.762 | 4.26E+09 | 0.762 | 1.00E-03 | 0.762 | 2.26E+09 | 0.762 | 8379006 | 0.762 | 2.27E+09 |
| 0.763 | 2.00E-03 | 0.763 | 4.24E+09 | 0.763 | 16155550 | 0.763 | 4.25E+09 | 0.763 | 1.00E-03 | 0.763 | 2.26E+09 | 0.763 | 8600789 | 0.763 | 2.26E+09 |
| 0.764 | 1.00E-03 | 0.764 | 4.23E+09 | 0.764 | 16582020 | 0.764 | 4.25E+09 | 0.764 | 1.00E-03 | 0.764 | 2.25E+09 | 0.764 | 8827830 | 0.764 | 2.26E+09 |
| 0.765 | 1.00E-03 | 0.765 | 4.22E+09 | 0.765 | 17018571 | 0.765 | 4.24E+09 | 0.765 | 1.00E-03 | 0.765 | 2.25E+09 | 0.765 | 9060239 | 0.765 | 2.26E+09 |
| 0.766 | 1.00E-03 | 0.766 | 4.22E+09 | 0.766 | 17465406 | 0.766 | 4.23E+09 | 0.766 | 0 | 0.766 | 2.25E+09 | 0.766 | 9298122 | 0.766 | 2.25E+09 |
| 0.767 | 1.00E-03 | 0.767 | 4.21E+09 | 0.767 | 17922732 | 0.767 | 4.23E+09 | 0.767 | 0 | 0.767 | 2.24E+09 | 0.767 | 9541590 | 0.767 | 2.25E+09 |
| 0.768 | 1.00E-03 | 0.768 | 4.2E+09 | 0.768 | 18390761 | 0.768 | 4.22E+09 | 0.768 | 0 | 0.768 | 2.24E+09 | 0.768 | 9790756 | 0.768 | 2.25E+09 |
| 0.769 | 0 | 0.769 | 4.2E+09 | 0.769 | 18869706 | 0.769 | 4.21E+09 | 0.769 | 0 | 0.769 | 2.23E+09 | 0.769 | 10045734 | 0.769 | 2.24E+09 |
| 0.77 | 0 | 0.77 | 4.19E+09 | 0.77 | 19359784 | 0.77 | 4.21E+09 | 0.77 | 0 | 0.77 | 2.23E+09 | 0.77 | 10306638 | 0.77 | 2.24E+09 |
| 0.771 | 0 | 0.771 | 4.18E+09 | 0.771 | 19861215 | 0.771 | 4.2E+09 | 0.771 | 0 | 0.771 | 2.23E+09 | 0.771 | 10573587 | 0.771 | 2.24E+09 |
| 0.772 | 0 | 0.772 | 4.17E+09 | 0.772 | 20374224 | 0.772 | 4.19E+09 | 0.772 | 0 | 0.772 | 2.22E+09 | 0.772 | 10846700 | 0.772 | 2.23E+09 |
| 0.773 | 0 | 0.773 | 4.17E+09 | 0.773 | 20899037 | 0.773 | 4.19E+09 | 0.773 | 0 | 0.773 | 2.22E+09 | 0.773 | 11126096 | 0.773 | 2.23E+09 |
| 0.774 | 0 | 0.774 | 4.16E+09 | 0.774 | 21435886 | 0.774 | 4.18E+09 | 0.774 | 0 | 0.774 | 2.21E+09 | 0.774 | 11411900 | 0.774 | 2.22E+09 |
| 0.775 | 0 | 0.775 | 4.15E+09 | 0.775 | 21985003 | 0.775 | 4.17E+09 | 0.775 | 0 | 0.775 | 2.21E+09 | 0.775 | 11704236 | 0.775 | 2.22E+09 |
| 0.776 | 0 | 0.776 | 4.14E+09 | 0.776 | 22546627 | 0.776 | 4.16E+09 | 0.776 | 0 | 0.776 | 2.21E+09 | 0.776 | 12003230 | 0.776 | 2.22E+09 |
| 0.777 | 0 | 0.777 | 4.13E+09 | 0.777 | 23120997 | 0.777 | 4.16E+09 | 0.777 | 0 | 0.777 | 2.2E+09 | 0.777 | 12309009 | 0.777 | 2.21E+09 |
| 0.778 | 0 | 0.778 | 4.13E+09 | 0.778 | 23708359 | 0.778 | 4.15E+09 | 0.778 | 0 | 0.778 | 2.2E+09 | 0.778 | 12621705 | 0.778 | 2.21E+09 |
| 0.779 | 0 | 0.779 | 4.12E+09 | 0.779 | 24308959 | 0.779 | 4.14E+09 | 0.779 | 0 | 0.779 | 2.19E+09 | 0.779 | 12941449 | 0.779 | 2.2E+09 |
| 0.78 | 0 | 0.78 | 4.11E+09 | 0.78 | 24923050 | 0.78 | 4.13E+09 | 0.78 | 0 | 0.78 | 2.19E+09 | 0.78 | 13268375 | 0.78 | 2.2E+09 |
| 0.781 | 0 | 0.781 | 4.1E+09 | 0.781 | 25550885 | 0.781 | 4.13E+09 | 0.781 | 0 | 0.781 | 2.18E+09 | 0.781 | 13602618 | 0.781 | 2.2E+09 |
| 0.782 | 0 | 0.782 | 4.09E+09 | 0.782 | 26192724 | 0.782 | 4.12E+09 | 0.782 | 0 | 0.782 | 2.18E+09 | 0.782 | 13944316 | 0.782 | 2.19E+09 |
| 0.783 | 0 | 0.783 | 4.08E+09 | 0.783 | 26848827 | 0.783 | 4.11E+09 | 0.783 | 0 | 0.783 | 2.17E+09 | 0.783 | 14293608 | 0.783 | 2.19E+09 |
| 0.784 | 0 | 0.784 | 4.07E+09 | 0.784 | 27519461 | 0.784 | 4.1E+09 | 0.784 | 0 | 0.784 | 2.17E+09 | 0.784 | 14650635 | 0.784 | 2.18E+09 |
| 0.785 | 0 | 0.785 | 4.06E+09 | 0.785 | 28204894 | 0.785 | 4.09E+09 | 0.785 | 0 | 0.785 | 2.16E+09 | 0.785 | 15015542 | 0.785 | 2.18E+09 |
| 0.786 | 0 | 0.786 | 4.06E+09 | 0.786 | 28905399 | 0.786 | 4.08E+09 | 0.786 | 0 | 0.786 | 2.16E+09 | 0.786 | 15388472 | 0.786 | 2.17E+09 |
| 0.787 | 0 | 0.787 | 4.05E+09 | 0.787 | 29621251 | 0.787 | 4.08E+09 | 0.787 | 0 | 0.787 | 2.15E+09 | 0.787 | 15769573 | 0.787 | 2.17E+09 |
| 0.788 | 0 | 0.788 | 4.04E+09 | 0.788 | 30352731 | 0.788 | 4.07E+09 | 0.788 | 0 | 0.788 | 2.15E+09 | 0.788 | 16158994 | 0.788 | 2.17E+09 |
| 0.789 | 0 | 0.789 | 4.03E+09 | 0.789 | 31100123 | 0.789 | 4.06E+09 | 0.789 | 0 | 0.789 | 2.14E+09 | 0.789 | 16556885 | 0.789 | 2.16E+09 |
| 0.79 | 0 | 0.79 | 4.02E+09 | 0.79 | 31863713 | 0.79 | 4.05E+09 | 0.79 | 0 | 0.79 | 2.14E+09 | 0.79 | 16963401 | 0.79 | 2.16E+09 |
| 0.791 | 0 | 0.791 | 4.01E+09 | 0.791 | 32643791 | 0.791 | 4.04E+09 | 0.791 | 0 | 0.791 | 2.13E+09 | 0.791 | 17378694 | 0.791 | 2.15E+09 |
| 0.792 | 0 | 0.792 | 4E+09 | 0.792 | 33440654 | 0.792 | 4.03E+09 | 0.792 | 0 | 0.792 | 2.13E+09 | 0.792 | 17802922 | 0.792 | 2.15E+09 |
| 0.793 | 0 | 0.793 | 3.99E+09 | 0.793 | 34254597 | 0.793 | 4.02E+09 | 0.793 | 0 | 0.793 | 2.12E+09 | 0.793 | 18236244 | 0.793 | 2.14E+09 |
| 0.794 | 0 | 0.794 | 3.98E+09 | 0.794 | 35085924 | 0.794 | 4.01E+09 | 0.794 | 0 | 0.794 | 2.12E+09 | 0.794 | 18678821 | 0.794 | 2.14E+09 |
| 0.795 | 0 | 0.795 | 3.97E+09 | 0.795 | 35934939 | 0.795 | 4.01E+09 | 0.795 | 0 | 0.795 | 2.11E+09 | 0.795 | 19130814 | 0.795 | 2.13E+09 |
| 0.796 | 0 | 0.796 | 3.96E+09 | 0.796 | 36801952 | 0.796 | 4E+09 | 0.796 | 0 | 0.796 | 2.11E+09 | 0.796 | 19592389 | 0.796 | 2.13E+09 |
| 0.797 | 0 | 0.797 | 3.95E+09 | 0.797 | 37687275 | 0.797 | 3.99E+09 | 0.797 | 0 | 0.797 | 2.1E+09 | 0.797 | 20063712 | 0.797 | 2.12E+09 |
| 0.798 | 0 | 0.798 | 3.94E+09 | 0.798 | 38591226 | 0.798 | 3.98E+09 | 0.798 | 0 | 0.798 | 2.1E+09 | 0.798 | 20544951 | 0.798 | 2.12E+09 |
| 0.799 | 0 | 0.799 | 3.93E+09 | 0.799 | 39514123 | 0.799 | 3.97E+09 | 0.799 | 0 | 0.799 | 2.09E+09 | 0.799 | 21036277 | 0.799 | 2.11E+09 |
| 0.8 | 0 | 0.8 | 3.92E+09 | 0.8 | 40456292 | 0.8 | 3.96E+09 | 0.8 | 0 | 0.8 | 2.09E+09 | 0.8 | 21537863 | 0.8 | 2.11E+09 |
| 0.801 | 0 | 0.801 | 3.91E+09 | 0.801 | 41418059 | 0.801 | 3.95E+09 | 0.801 | 0 | 0.801 | 2.08E+09 | 0.801 | 22049882 | 0.801 | 2.1E+09 |
| 0.802 | 0 | 0.802 | 3.9E+09 | 0.802 | 42399755 | 0.802 | 3.94E+09 | 0.802 | 0 | 0.802 | 2.08E+09 | 0.802 | 22572512 | 0.802 | 2.1E+09 |
| 0.803 | 0 | 0.803 | 3.89E+09 | 0.803 | 43401717 | 0.803 | 3.93E+09 | 0.803 | 0 | 0.803 | 2.07E+09 | 0.803 | 23105930 | 0.803 | 2.09E+09 |
| 0.804 | 0 | 0.804 | 3.88E+09 | 0.804 | 44424281 | 0.804 | 3.92E+09 | 0.804 | 0 | 0.804 | 2.06E+09 | 0.804 | 23650316 | 0.804 | 2.09E+09 |
| 0.805 | 0 | 0.805 | 3.87E+09 | 0.805 | 45467791 | 0.805 | 3.91E+09 | 0.805 | 0 | 0.805 | 2.06E+09 | 0.805 | 24205853 | 0.805 | 2.08E+09 |
| 0.806 | 0 | 0.806 | 3.85E+09 | 0.806 | 46532592 | 0.806 | 3.9E+09 | 0.806 | 0 | 0.806 | 2.05E+09 | 0.806 | 24772725 | 0.806 | 2.08E+09 |
| 0.807 | 0 | 0.807 | 3.84E+09 | 0.807 | 47619034 | 0.807 | 3.89E+09 | 0.807 | 0 | 0.807 | 2.05E+09 | 0.807 | 25351118 | 0.807 | 2.07E+09 |
| 0.808 | 0 | 0.808 | 3.83E+09 | 0.808 | 48727470 | 0.808 | 3.88E+09 | 0.808 | 0 | 0.808 | 2.04E+09 | 0.808 | 25941220 | 0.808 | 2.07E+09 |
| 0.809 | 0 | 0.809 | 3.82E+09 | 0.809 | 49858257 | 0.809 | 3.87E+09 | 0.809 | 0 | 0.809 | 2.03E+09 | 0.809 | 26543222 | 0.809 | 2.06E+09 |
| 0.81 | 0 | 0.81 | 3.81E+09 | 0.81 | 51011755 | 0.81 | 3.86E+09 | 0.81 | 0 | 0.81 | 2.03E+09 | 0.81 | 27157313 | 0.81 | 2.06E+09 |
| 0.811 | 0 | 0.811 | 3.8E+09 | 0.811 | 52188328 | 0.811 | 3.85E+09 | 0.811 | 0 | 0.811 | 2.02E+09 | 0.811 | 27783690 | 0.811 | 2.05E+09 |
| 0.812 | 0 | 0.812 | 3.79E+09 | 0.812 | 53388343 | 0.812 | 3.84E+09 | 0.812 | 0 | 0.812 | 2.02E+09 | 0.812 | 28422546 | 0.812 | 2.04E+09 |
| 0.813 | 0 | 0.813 | 3.78E+09 | 0.813 | 54612172 | 0.813 | 3.83E+09 | 0.813 | 0 | 0.813 | 2.01E+09 | 0.813 | 29074081 | 0.813 | 2.04E+09 |
| 0.814 | 0 | 0.814 | 3.76E+09 | 0.814 | 55860189 | 0.814 | 3.82E+09 | 0.814 | 0 | 0.814 | 2E+09 | 0.814 | 29738492 | 0.814 | 2.03E+09 |
| 0.815 | 0 | 0.815 | 3.75E+09 | 0.815 | 57132773 | 0.815 | 3.81E+09 | 0.815 | 0 | 0.815 | 2E+09 | 0.815 | 30415982 | 0.815 | 2.03E+09 |
| 0.816 | 0 | 0.816 | 3.74E+09 | 0.816 | 58430304 | 0.816 | 3.8E+09 | 0.816 | 0 | 0.816 | 1.99E+09 | 0.816 | 31106753 | 0.816 | 2.02E+09 |
| 0.817 | 0 | 0.817 | 3.73E+09 | 0.817 | 59753167 | 0.817 | 3.79E+09 | 0.817 | 0 | 0.817 | 1.99E+09 | 0.817 | 31811011 | 0.817 | 2.02E+09 |
| 0.818 | 0 | 0.818 | 3.72E+09 | 0.818 | 61101752 | 0.818 | 3.78E+09 | 0.818 | 0 | 0.818 | 1.98E+09 | 0.818 | 32528962 | 0.818 | 2.01E+09 |
| 0.819 | 0 | 0.819 | 3.71E+09 | 0.819 | 62476449 | 0.819 | 3.77E+09 | 0.819 | 0 | 0.819 | 1.97E+09 | 0.819 | 33260814 | 0.819 | 2.01E+09 |
| 0.82 | 0 | 0.82 | 3.69E+09 | 0.82 | 63877655 | 0.82 | 3.76E+09 | 0.82 | 0 | 0.82 | 1.97E+09 | 0.82 | 34006779 | 0.82 | 2E+09 |
| 0.821 | 0 | 0.821 | 3.68E+09 | 0.821 | 65305766 | 0.821 | 3.75E+09 | 0.821 | 0 | 0.821 | 1.96E+09 | 0.821 | 34767068 | 0.821 | 1.99E+09 |
| 0.822 | 0 | 0.822 | 3.67E+09 | 0.822 | 66761186 | 0.822 | 3.74E+09 | 0.822 | 0 | 0.822 | 1.95E+09 | 0.822 | 35541895 | 0.822 | 1.99E+09 |
| 0.823 | 0 | 0.823 | 3.66E+09 | 0.823 | 68244319 | 0.823 | 3.73E+09 | 0.823 | 0 | 0.823 | 1.95E+09 | 0.823 | 36331476 | 0.823 | 1.98E+09 |
| 0.824 | 0 | 0.824 | 3.65E+09 | 0.824 | 69755572 | 0.824 | 3.71E+09 | 0.824 | 0 | 0.824 | 1.94E+09 | 0.824 | 37136027 | 0.824 | 1.98E+09 |
| 0.825 | 0 | 0.825 | 3.63E+09 | 0.825 | 71295358 | 0.825 | 3.7E+09 | 0.825 | 0 | 0.825 | 1.93E+09 | 0.825 | 37955769 | 0.825 | 1.97E+09 |
| 0.826 | 0 | 0.826 | 3.62E+09 | 0.826 | 72864090 | 0.826 | 3.69E+09 | 0.826 | 0 | 0.826 | 1.93E+09 | 0.826 | 38790921 | 0.826 | 1.97E+09 |
| 0.827 | 0 | 0.827 | 3.61E+09 | 0.827 | 74462187 | 0.827 | 3.68E+09 | 0.827 | 0 | 0.827 | 1.92E+09 | 0.827 | 39641705 | 0.827 | 1.96E+09 |
| 0.828 | 0 | 0.828 | 3.6E+09 | 0.828 | 76090068 | 0.828 | 3.67E+09 | 0.828 | 0 | 0.828 | 1.91E+09 | 0.828 | 40508346 | 0.828 | 1.95E+09 |
| 0.829 | 0 | 0.829 | 3.58E+09 | 0.829 | 77748156 | 0.829 | 3.66E+09 | 0.829 | 0 | 0.829 | 1.91E+09 | 0.829 | 41391069 | 0.829 | 1.95E+09 |
| 0.83 | 0 | 0.83 | 3.57E+09 | 0.83 | 79436879 | 0.83 | 3.65E+09 | 0.83 | 0 | 0.83 | 1.9E+09 | 0.83 | 42290100 | 0.83 | 1.94E+09 |
| 0.831 | 0 | 0.831 | 3.56E+09 | 0.831 | 81156666 | 0.831 | 3.64E+09 | 0.831 | 0 | 0.831 | 1.89E+09 | 0.831 | 43205669 | 0.831 | 1.94E+09 |
| 0.832 | 0 | 0.832 | 3.54E+09 | 0.832 | 82907947 | 0.832 | 3.63E+09 | 0.832 | 0 | 0.832 | 1.89E+09 | 0.832 | 44138005 | 0.832 | 1.93E+09 |
| 0.833 | 0 | 0.833 | 3.53E+09 | 0.833 | 84691159 | 0.833 | 3.62E+09 | 0.833 | 0 | 0.833 | 1.88E+09 | 0.833 | 45087340 | 0.833 | 1.93E+09 |
| 0.834 | 0 | 0.834 | 3.52E+09 | 0.834 | 86506737 | 0.834 | 3.61E+09 | 0.834 | 0 | 0.834 | 1.87E+09 | 0.834 | 46053906 | 0.834 | 1.92E+09 |
| 0.835 | 0 | 0.835 | 3.51E+09 | 0.835 | 88355123 | 0.835 | 3.59E+09 | 0.835 | 0 | 0.835 | 1.87E+09 | 0.835 | 47037938 | 0.835 | 1.91E+09 |
| 0.836 | 0 | 0.836 | 3.49E+09 | 0.836 | 90236759 | 0.836 | 3.58E+09 | 0.836 | 0 | 0.836 | 1.86E+09 | 0.836 | 48039671 | 0.836 | 1.91E+09 |
| 0.837 | 0 | 0.837 | 3.48E+09 | 0.837 | 92152089 | 0.837 | 3.57E+09 | 0.837 | 0 | 0.837 | 1.85E+09 | 0.837 | 49059343 | 0.837 | 1.9E+09 |
| 0.838 | 0 | 0.838 | 3.47E+09 | 0.838 | 94101562 | 0.838 | 3.56E+09 | 0.838 | 0 | 0.838 | 1.85E+09 | 0.838 | 50097190 | 0.838 | 1.9E+09 |
| 0.839 | 0 | 0.839 | 3.45E+09 | 0.839 | 96085626 | 0.839 | 3.55E+09 | 0.839 | 0 | 0.839 | 1.84E+09 | 0.839 | 51153454 | 0.839 | 1.89E+09 |
| 0.84 | 0 | 0.84 | 3.44E+09 | 0.84 | 98104733 | 0.84 | 3.54E+09 | 0.84 | 0 | 0.84 | 1.83E+09 | 0.84 | 52228373 | 0.84 | 1.88E+09 |
| 0.841 | 0 | 0.841 | 3.43E+09 | 0.841 | 1E+08 | 0.841 | 3.53E+09 | 0.841 | 0 | 0.841 | 1.82E+09 | 0.841 | 53322191 | 0.841 | 1.88E+09 |
| 0.842 | 0 | 0.842 | 3.41E+09 | 0.842 | 1.02E+08 | 0.842 | 3.52E+09 | 0.842 | 0 | 0.842 | 1.82E+09 | 0.842 | 54435150 | 0.842 | 1.87E+09 |
| 0.843 | 0 | 0.843 | 3.4E+09 | 0.843 | 1.04E+08 | 0.843 | 3.5E+09 | 0.843 | 0 | 0.843 | 1.81E+09 | 0.843 | 55567494 | 0.843 | 1.87E+09 |
| 0.844 | 0 | 0.844 | 3.39E+09 | 0.844 | 1.07E+08 | 0.844 | 3.49E+09 | 0.844 | 0 | 0.844 | 1.8E+09 | 0.844 | 56719468 | 0.844 | 1.86E+09 |
| 0.845 | 0 | 0.845 | 3.37E+09 | 0.845 | 1.09E+08 | 0.845 | 3.48E+09 | 0.845 | 0 | 0.845 | 1.8E+09 | 0.845 | 57891318 | 0.845 | 1.85E+09 |
| 0.846 | 0 | 0.846 | 3.36E+09 | 0.846 | 1.11E+08 | 0.846 | 3.47E+09 | 0.846 | 0 | 0.846 | 1.79E+09 | 0.846 | 59083291 | 0.846 | 1.85E+09 |
| 0.847 | 0 | 0.847 | 3.35E+09 | 0.847 | 1.13E+08 | 0.847 | 3.46E+09 | 0.847 | 0 | 0.847 | 1.78E+09 | 0.847 | 60295635 | 0.847 | 1.84E+09 |
| 0.848 | 0 | 0.848 | 3.33E+09 | 0.848 | 1.16E+08 | 0.848 | 3.45E+09 | 0.848 | 0 | 0.848 | 1.77E+09 | 0.848 | 61528598 | 0.848 | 1.84E+09 |
| 0.849 | 0 | 0.849 | 3.32E+09 | 0.849 | 1.18E+08 | 0.849 | 3.44E+09 | 0.849 | 0 | 0.849 | 1.77E+09 | 0.849 | 62782431 | 0.849 | 1.83E+09 |
| 0.85 | 0 | 0.85 | 3.31E+09 | 0.85 | 1.2E+08 | 0.85 | 3.43E+09 | 0.85 | 0 | 0.85 | 1.76E+09 | 0.85 | 64057383 | 0.85 | 1.82E+09 |
| 0.851 | 0 | 0.851 | 3.29E+09 | 0.851 | 1.23E+08 | 0.851 | 3.41E+09 | 0.851 | 0 | 0.851 | 1.75E+09 | 0.851 | 65353705 | 0.851 | 1.82E+09 |
| 0.852 | 0 | 0.852 | 3.28E+09 | 0.852 | 1.25E+08 | 0.852 | 3.4E+09 | 0.852 | 0 | 0.852 | 1.75E+09 | 0.852 | 66671650 | 0.852 | 1.81E+09 |
| 0.853 | 0 | 0.853 | 3.26E+09 | 0.853 | 1.28E+08 | 0.853 | 3.39E+09 | 0.853 | 0 | 0.853 | 1.74E+09 | 0.853 | 68011470 | 0.853 | 1.81E+09 |
| 0.854 | 0 | 0.854 | 3.25E+09 | 0.854 | 1.3E+08 | 0.854 | 3.38E+09 | 0.854 | 0 | 0.854 | 1.73E+09 | 0.854 | 69373417 | 0.854 | 1.8E+09 |
| 0.855 | 0 | 0.855 | 3.24E+09 | 0.855 | 1.33E+08 | 0.855 | 3.37E+09 | 0.855 | 0 | 0.855 | 1.72E+09 | 0.855 | 70757745 | 0.855 | 1.79E+09 |
| 0.856 | 0 | 0.856 | 3.22E+09 | 0.856 | 1.36E+08 | 0.856 | 3.36E+09 | 0.856 | 0 | 0.856 | 1.72E+09 | 0.856 | 72164709 | 0.856 | 1.79E+09 |
| 0.857 | 0 | 0.857 | 3.21E+09 | 0.857 | 1.38E+08 | 0.857 | 3.35E+09 | 0.857 | 0 | 0.857 | 1.71E+09 | 0.857 | 73594562 | 0.857 | 1.78E+09 |
| 0.858 | 0 | 0.858 | 3.2E+09 | 0.858 | 1.41E+08 | 0.858 | 3.34E+09 | 0.858 | 0 | 0.858 | 1.7E+09 | 0.858 | 75047560 | 0.858 | 1.78E+09 |
| 0.859 | 0 | 0.859 | 3.18E+09 | 0.859 | 1.44E+08 | 0.859 | 3.32E+09 | 0.859 | 0 | 0.859 | 1.69E+09 | 0.859 | 76523959 | 0.859 | 1.77E+09 |
| 0.86 | 0 | 0.86 | 3.17E+09 | 0.86 | 1.47E+08 | 0.86 | 3.31E+09 | 0.86 | 0 | 0.86 | 1.69E+09 | 0.86 | 78024013 | 0.86 | 1.76E+09 |
| 0.861 | 0 | 0.861 | 3.15E+09 | 0.861 | 1.49E+08 | 0.861 | 3.3E+09 | 0.861 | 0 | 0.861 | 1.68E+09 | 0.861 | 79547979 | 0.861 | 1.76E+09 |
| 0.862 | 0 | 0.862 | 3.14E+09 | 0.862 | 1.52E+08 | 0.862 | 3.29E+09 | 0.862 | 0 | 0.862 | 1.67E+09 | 0.862 | 81096114 | 0.862 | 1.75E+09 |
| 0.863 | 0 | 0.863 | 3.12E+09 | 0.863 | 1.55E+08 | 0.863 | 3.28E+09 | 0.863 | 0 | 0.863 | 1.66E+09 | 0.863 | 82668674 | 0.863 | 1.75E+09 |
| 0.864 | 0 | 0.864 | 3.11E+09 | 0.864 | 1.58E+08 | 0.864 | 3.27E+09 | 0.864 | 0 | 0.864 | 1.66E+09 | 0.864 | 84265916 | 0.864 | 1.74E+09 |
| 0.865 | 0 | 0.865 | 3.1E+09 | 0.865 | 1.61E+08 | 0.865 | 3.26E+09 | 0.865 | 0 | 0.865 | 1.65E+09 | 0.865 | 85888099 | 0.865 | 1.73E+09 |
| 0.866 | 0 | 0.866 | 3.08E+09 | 0.866 | 1.64E+08 | 0.866 | 3.25E+09 | 0.866 | 0 | 0.866 | 1.64E+09 | 0.866 | 87535479 | 0.866 | 1.73E+09 |
| 0.867 | 0 | 0.867 | 3.07E+09 | 0.867 | 1.68E+08 | 0.867 | 3.24E+09 | 0.867 | 0 | 0.867 | 1.63E+09 | 0.867 | 89208314 | 0.867 | 1.72E+09 |
| 0.868 | 0 | 0.868 | 3.05E+09 | 0.868 | 1.71E+08 | 0.868 | 3.22E+09 | 0.868 | 0 | 0.868 | 1.63E+09 | 0.868 | 90906864 | 0.868 | 1.72E+09 |
| 0.869 | 0 | 0.869 | 3.04E+09 | 0.869 | 1.74E+08 | 0.869 | 3.21E+09 | 0.869 | 0 | 0.869 | 1.62E+09 | 0.869 | 92631388 | 0.869 | 1.71E+09 |
| 0.87 | 0 | 0.87 | 3.03E+09 | 0.87 | 1.77E+08 | 0.87 | 3.2E+09 | 0.87 | 0 | 0.87 | 1.61E+09 | 0.87 | 94382143 | 0.87 | 1.7E+09 |
| 0.871 | 0 | 0.871 | 3.01E+09 | 0.871 | 1.81E+08 | 0.871 | 3.19E+09 | 0.871 | 0 | 0.871 | 1.6E+09 | 0.871 | 96159391 | 0.871 | 1.7E+09 |
| 0.872 | 0 | 0.872 | 3E+09 | 0.872 | 1.84E+08 | 0.872 | 3.18E+09 | 0.872 | 0 | 0.872 | 1.6E+09 | 0.872 | 97963393 | 0.872 | 1.69E+09 |
| 0.873 | 0 | 0.873 | 2.98E+09 | 0.873 | 1.87E+08 | 0.873 | 3.17E+09 | 0.873 | 0 | 0.873 | 1.59E+09 | 0.873 | 99794408 | 0.873 | 1.69E+09 |
| 0.874 | 0 | 0.874 | 2.97E+09 | 0.874 | 1.91E+08 | 0.874 | 3.16E+09 | 0.874 | 0 | 0.874 | 1.58E+09 | 0.874 | 1.02E+08 | 0.874 | 1.68E+09 |
| 0.875 | 0 | 0.875 | 2.95E+09 | 0.875 | 1.94E+08 | 0.875 | 3.15E+09 | 0.875 | 0 | 0.875 | 1.57E+09 | 0.875 | 1.04E+08 | 0.875 | 1.68E+09 |
| 0.876 | 0 | 0.876 | 2.94E+09 | 0.876 | 1.98E+08 | 0.876 | 3.14E+09 | 0.876 | 0 | 0.876 | 1.56E+09 | 0.876 | 1.05E+08 | 0.876 | 1.67E+09 |
| 0.877 | 0 | 0.877 | 2.92E+09 | 0.877 | 2.02E+08 | 0.877 | 3.13E+09 | 0.877 | 0 | 0.877 | 1.56E+09 | 0.877 | 1.07E+08 | 0.877 | 1.66E+09 |
| 0.878 | 0 | 0.878 | 2.91E+09 | 0.878 | 2.05E+08 | 0.878 | 3.12E+09 | 0.878 | 0 | 0.878 | 1.55E+09 | 0.878 | 1.09E+08 | 0.878 | 1.66E+09 |
| 0.879 | 0 | 0.879 | 2.9E+09 | 0.879 | 2.09E+08 | 0.879 | 3.1E+09 | 0.879 | 0 | 0.879 | 1.54E+09 | 0.879 | 1.11E+08 | 0.879 | 1.65E+09 |
| 0.88 | 0 | 0.88 | 2.88E+09 | 0.88 | 2.13E+08 | 0.88 | 3.09E+09 | 0.88 | 0 | 0.88 | 1.53E+09 | 0.88 | 1.13E+08 | 0.88 | 1.65E+09 |
| 0.881 | 0 | 0.881 | 2.87E+09 | 0.881 | 2.17E+08 | 0.881 | 3.08E+09 | 0.881 | 0 | 0.881 | 1.53E+09 | 0.881 | 1.15E+08 | 0.881 | 1.64E+09 |
| 0.882 | 0 | 0.882 | 2.85E+09 | 0.882 | 2.21E+08 | 0.882 | 3.07E+09 | 0.882 | 0 | 0.882 | 1.52E+09 | 0.882 | 1.18E+08 | 0.882 | 1.64E+09 |
| 0.883 | 0 | 0.883 | 2.84E+09 | 0.883 | 2.25E+08 | 0.883 | 3.06E+09 | 0.883 | 0 | 0.883 | 1.51E+09 | 0.883 | 1.2E+08 | 0.883 | 1.63E+09 |
| 0.884 | 0 | 0.884 | 2.82E+09 | 0.884 | 2.29E+08 | 0.884 | 3.05E+09 | 0.884 | 0 | 0.884 | 1.5E+09 | 0.884 | 1.22E+08 | 0.884 | 1.62E+09 |
| 0.885 | 0 | 0.885 | 2.81E+09 | 0.885 | 2.33E+08 | 0.885 | 3.04E+09 | 0.885 | 0 | 0.885 | 1.5E+09 | 0.885 | 1.24E+08 | 0.885 | 1.62E+09 |
| 0.886 | 0 | 0.886 | 2.79E+09 | 0.886 | 2.37E+08 | 0.886 | 3.03E+09 | 0.886 | 0 | 0.886 | 1.49E+09 | 0.886 | 1.26E+08 | 0.886 | 1.61E+09 |
| 0.887 | 0 | 0.887 | 2.78E+09 | 0.887 | 2.41E+08 | 0.887 | 3.02E+09 | 0.887 | 0 | 0.887 | 1.48E+09 | 0.887 | 1.28E+08 | 0.887 | 1.61E+09 |
| 0.888 | 0 | 0.888 | 2.76E+09 | 0.888 | 2.45E+08 | 0.888 | 3.01E+09 | 0.888 | 0 | 0.888 | 1.47E+09 | 0.888 | 1.31E+08 | 0.888 | 1.6E+09 |
| 0.889 | 0 | 0.889 | 2.75E+09 | 0.889 | 2.5E+08 | 0.889 | 3E+09 | 0.889 | 0 | 0.889 | 1.46E+09 | 0.889 | 1.33E+08 | 0.889 | 1.6E+09 |
| 0.89 | 0 | 0.89 | 2.74E+09 | 0.89 | 2.54E+08 | 0.89 | 2.99E+09 | 0.89 | 0 | 0.89 | 1.46E+09 | 0.89 | 1.35E+08 | 0.89 | 1.59E+09 |
| 0.891 | 0 | 0.891 | 2.72E+09 | 0.891 | 2.59E+08 | 0.891 | 2.98E+09 | 0.891 | 0 | 0.891 | 1.45E+09 | 0.891 | 1.38E+08 | 0.891 | 1.59E+09 |
| 0.892 | 0 | 0.892 | 2.71E+09 | 0.892 | 2.63E+08 | 0.892 | 2.97E+09 | 0.892 | 0 | 0.892 | 1.44E+09 | 0.892 | 1.4E+08 | 0.892 | 1.58E+09 |
| 0.893 | 0 | 0.893 | 2.69E+09 | 0.893 | 2.68E+08 | 0.893 | 2.96E+09 | 0.893 | 0 | 0.893 | 1.43E+09 | 0.893 | 1.43E+08 | 0.893 | 1.58E+09 |
| 0.894 | 0 | 0.894 | 2.68E+09 | 0.894 | 2.72E+08 | 0.894 | 2.95E+09 | 0.894 | 0 | 0.894 | 1.43E+09 | 0.894 | 1.45E+08 | 0.894 | 1.57E+09 |
| 0.895 | 0 | 0.895 | 2.66E+09 | 0.895 | 2.77E+08 | 0.895 | 2.94E+09 | 0.895 | 0 | 0.895 | 1.42E+09 | 0.895 | 1.47E+08 | 0.895 | 1.56E+09 |
| 0.896 | 0 | 0.896 | 2.65E+09 | 0.896 | 2.82E+08 | 0.896 | 2.93E+09 | 0.896 | 0 | 0.896 | 1.41E+09 | 0.896 | 1.5E+08 | 0.896 | 1.56E+09 |
| 0.897 | 0 | 0.897 | 2.63E+09 | 0.897 | 2.87E+08 | 0.897 | 2.92E+09 | 0.897 | 0 | 0.897 | 1.4E+09 | 0.897 | 1.53E+08 | 0.897 | 1.55E+09 |
| 0.898 | 0 | 0.898 | 2.62E+09 | 0.898 | 2.91E+08 | 0.898 | 2.91E+09 | 0.898 | 0 | 0.898 | 1.39E+09 | 0.898 | 1.55E+08 | 0.898 | 1.55E+09 |
| 0.899 | 0 | 0.899 | 2.6E+09 | 0.899 | 2.96E+08 | 0.899 | 2.9E+09 | 0.899 | 0 | 0.899 | 1.39E+09 | 0.899 | 1.58E+08 | 0.899 | 1.54E+09 |
| 0.9 | 0 | 0.9 | 2.59E+09 | 0.9 | 3.01E+08 | 0.9 | 2.89E+09 | 0.9 | 0 | 0.9 | 1.38E+09 | 0.9 | 1.6E+08 | 0.9 | 1.54E+09 |
| 0.901 | 0 | 0.901 | 2.58E+09 | 0.901 | 3.07E+08 | 0.901 | 2.88E+09 | 0.901 | 0 | 0.901 | 1.37E+09 | 0.901 | 1.63E+08 | 0.901 | 1.53E+09 |
| 0.902 | 0 | 0.902 | 2.56E+09 | 0.902 | 3.12E+08 | 0.902 | 2.87E+09 | 0.902 | 0 | 0.902 | 1.36E+09 | 0.902 | 1.66E+08 | 0.902 | 1.53E+09 |
| 0.903 | 0 | 0.903 | 2.55E+09 | 0.903 | 3.17E+08 | 0.903 | 2.86E+09 | 0.903 | 0 | 0.903 | 1.36E+09 | 0.903 | 1.69E+08 | 0.903 | 1.52E+09 |
| 0.904 | 0 | 0.904 | 2.53E+09 | 0.904 | 3.22E+08 | 0.904 | 2.85E+09 | 0.904 | 0 | 0.904 | 1.35E+09 | 0.904 | 1.72E+08 | 0.904 | 1.52E+09 |
| 0.905 | 0 | 0.905 | 2.52E+09 | 0.905 | 3.28E+08 | 0.905 | 2.84E+09 | 0.905 | 0 | 0.905 | 1.34E+09 | 0.905 | 1.74E+08 | 0.905 | 1.51E+09 |
| 0.906 | 0 | 0.906 | 2.5E+09 | 0.906 | 3.33E+08 | 0.906 | 2.84E+09 | 0.906 | 0 | 0.906 | 1.33E+09 | 0.906 | 1.77E+08 | 0.906 | 1.51E+09 |
| 0.907 | 0 | 0.907 | 2.49E+09 | 0.907 | 3.38E+08 | 0.907 | 2.83E+09 | 0.907 | 0 | 0.907 | 1.32E+09 | 0.907 | 1.8E+08 | 0.907 | 1.5E+09 |
| 0.908 | 0 | 0.908 | 2.47E+09 | 0.908 | 3.44E+08 | 0.908 | 2.82E+09 | 0.908 | 0 | 0.908 | 1.32E+09 | 0.908 | 1.83E+08 | 0.908 | 1.5E+09 |
| 0.909 | 0 | 0.909 | 2.46E+09 | 0.909 | 3.5E+08 | 0.909 | 2.81E+09 | 0.909 | 0 | 0.909 | 1.31E+09 | 0.909 | 1.86E+08 | 0.909 | 1.5E+09 |
| 0.91 | 0 | 0.91 | 2.44E+09 | 0.91 | 3.55E+08 | 0.91 | 2.8E+09 | 0.91 | 0 | 0.91 | 1.3E+09 | 0.91 | 1.89E+08 | 0.91 | 1.49E+09 |
| 0.911 | 0 | 0.911 | 2.43E+09 | 0.911 | 3.61E+08 | 0.911 | 2.79E+09 | 0.911 | 0 | 0.911 | 1.29E+09 | 0.911 | 1.92E+08 | 0.911 | 1.49E+09 |
| 0.912 | 0 | 0.912 | 2.42E+09 | 0.912 | 3.67E+08 | 0.912 | 2.78E+09 | 0.912 | 0 | 0.912 | 1.29E+09 | 0.912 | 1.96E+08 | 0.912 | 1.48E+09 |
| 0.913 | 0 | 0.913 | 2.4E+09 | 0.913 | 3.73E+08 | 0.913 | 2.77E+09 | 0.913 | 0 | 0.913 | 1.28E+09 | 0.913 | 1.99E+08 | 0.913 | 1.48E+09 |
| 0.914 | 0 | 0.914 | 2.39E+09 | 0.914 | 3.79E+08 | 0.914 | 2.77E+09 | 0.914 | 0 | 0.914 | 1.27E+09 | 0.914 | 2.02E+08 | 0.914 | 1.47E+09 |
| 0.915 | 0 | 0.915 | 2.37E+09 | 0.915 | 3.85E+08 | 0.915 | 2.76E+09 | 0.915 | 0 | 0.915 | 1.26E+09 | 0.915 | 2.05E+08 | 0.915 | 1.47E+09 |
| 0.916 | 0 | 0.916 | 2.36E+09 | 0.916 | 3.92E+08 | 0.916 | 2.75E+09 | 0.916 | 0 | 0.916 | 1.26E+09 | 0.916 | 2.09E+08 | 0.916 | 1.46E+09 |
| 0.917 | 0 | 0.917 | 2.34E+09 | 0.917 | 3.98E+08 | 0.917 | 2.74E+09 | 0.917 | 0 | 0.917 | 1.25E+09 | 0.917 | 2.12E+08 | 0.917 | 1.46E+09 |
| 0.918 | 0 | 0.918 | 2.33E+09 | 0.918 | 4.05E+08 | 0.918 | 2.73E+09 | 0.918 | 0 | 0.918 | 1.24E+09 | 0.918 | 2.15E+08 | 0.918 | 1.46E+09 |
| 0.919 | 0 | 0.919 | 2.32E+09 | 0.919 | 4.11E+08 | 0.919 | 2.73E+09 | 0.919 | 0 | 0.919 | 1.23E+09 | 0.919 | 2.19E+08 | 0.919 | 1.45E+09 |
| 0.92 | 0 | 0.92 | 2.3E+09 | 0.92 | 4.18E+08 | 0.92 | 2.72E+09 | 0.92 | 0 | 0.92 | 1.22E+09 | 0.92 | 2.22E+08 | 0.92 | 1.45E+09 |
| 0.921 | 0 | 0.921 | 2.29E+09 | 0.921 | 4.24E+08 | 0.921 | 2.71E+09 | 0.921 | 0 | 0.921 | 1.22E+09 | 0.921 | 2.26E+08 | 0.921 | 1.44E+09 |
| 0.922 | 0 | 0.922 | 2.27E+09 | 0.922 | 4.31E+08 | 0.922 | 2.7E+09 | 0.922 | 0 | 0.922 | 1.21E+09 | 0.922 | 2.3E+08 | 0.922 | 1.44E+09 |
| 0.923 | 0 | 0.923 | 2.26E+09 | 0.923 | 4.38E+08 | 0.923 | 2.7E+09 | 0.923 | 0 | 0.923 | 1.2E+09 | 0.923 | 2.33E+08 | 0.923 | 1.44E+09 |
| 0.924 | 0 | 0.924 | 2.24E+09 | 0.924 | 4.45E+08 | 0.924 | 2.69E+09 | 0.924 | 0 | 0.924 | 1.19E+09 | 0.924 | 2.37E+08 | 0.924 | 1.43E+09 |
| 0.925 | 0 | 0.925 | 2.23E+09 | 0.925 | 4.52E+08 | 0.925 | 2.68E+09 | 0.925 | 0 | 0.925 | 1.19E+09 | 0.925 | 2.41E+08 | 0.925 | 1.43E+09 |
| 0.926 | 0 | 0.926 | 2.22E+09 | 0.926 | 4.6E+08 | 0.926 | 2.68E+09 | 0.926 | 0 | 0.926 | 1.18E+09 | 0.926 | 2.45E+08 | 0.926 | 1.42E+09 |
| 0.927 | 0 | 0.927 | 2.2E+09 | 0.927 | 4.67E+08 | 0.927 | 2.67E+09 | 0.927 | 0 | 0.927 | 1.17E+09 | 0.927 | 2.49E+08 | 0.927 | 1.42E+09 |
| 0.928 | 0 | 0.928 | 2.19E+09 | 0.928 | 4.75E+08 | 0.928 | 2.66E+09 | 0.928 | 0 | 0.928 | 1.16E+09 | 0.928 | 2.53E+08 | 0.928 | 1.42E+09 |
| 0.929 | 0 | 0.929 | 2.17E+09 | 0.929 | 4.82E+08 | 0.929 | 2.66E+09 | 0.929 | 0 | 0.929 | 1.16E+09 | 0.929 | 2.57E+08 | 0.929 | 1.41E+09 |
| 0.93 | 0 | 0.93 | 2.16E+09 | 0.93 | 4.9E+08 | 0.93 | 2.65E+09 | 0.93 | 0 | 0.93 | 1.15E+09 | 0.93 | 2.61E+08 | 0.93 | 1.41E+09 |
| 0.931 | 0 | 0.931 | 2.14E+09 | 0.931 | 4.98E+08 | 0.931 | 2.64E+09 | 0.931 | 0 | 0.931 | 1.14E+09 | 0.931 | 2.65E+08 | 0.931 | 1.41E+09 |
| 0.932 | 0 | 0.932 | 2.13E+09 | 0.932 | 5.06E+08 | 0.932 | 2.64E+09 | 0.932 | 0 | 0.932 | 1.13E+09 | 0.932 | 2.69E+08 | 0.932 | 1.4E+09 |
| 0.933 | 0 | 0.933 | 2.12E+09 | 0.933 | 5.14E+08 | 0.933 | 2.63E+09 | 0.933 | 0 | 0.933 | 1.13E+09 | 0.933 | 2.74E+08 | 0.933 | 1.4E+09 |
| 0.934 | 0 | 0.934 | 2.1E+09 | 0.934 | 5.23E+08 | 0.934 | 2.63E+09 | 0.934 | 0 | 0.934 | 1.12E+09 | 0.934 | 2.78E+08 | 0.934 | 1.4E+09 |
| 0.935 | 0 | 0.935 | 2.09E+09 | 0.935 | 5.31E+08 | 0.935 | 2.62E+09 | 0.935 | 0 | 0.935 | 1.11E+09 | 0.935 | 2.83E+08 | 0.935 | 1.39E+09 |
| 0.936 | 0 | 0.936 | 2.08E+09 | 0.936 | 5.4E+08 | 0.936 | 2.61E+09 | 0.936 | 0 | 0.936 | 1.1E+09 | 0.936 | 2.87E+08 | 0.936 | 1.39E+09 |
| 0.937 | 0 | 0.937 | 2.06E+09 | 0.937 | 5.48E+08 | 0.937 | 2.61E+09 | 0.937 | 0 | 0.937 | 1.1E+09 | 0.937 | 2.92E+08 | 0.937 | 1.39E+09 |
| 0.938 | 0 | 0.938 | 2.05E+09 | 0.938 | 5.57E+08 | 0.938 | 2.6E+09 | 0.938 | 0 | 0.938 | 1.09E+09 | 0.938 | 2.97E+08 | 0.938 | 1.39E+09 |
| 0.939 | 0 | 0.939 | 2.03E+09 | 0.939 | 5.66E+08 | 0.939 | 2.6E+09 | 0.939 | 0 | 0.939 | 1.08E+09 | 0.939 | 3.02E+08 | 0.939 | 1.38E+09 |
| 0.94 | 0 | 0.94 | 2.02E+09 | 0.94 | 5.76E+08 | 0.94 | 2.6E+09 | 0.94 | 0 | 0.94 | 1.08E+09 | 0.94 | 3.06E+08 | 0.94 | 1.38E+09 |
| 0.941 | 0 | 0.941 | 2.01E+09 | 0.941 | 5.85E+08 | 0.941 | 2.59E+09 | 0.941 | 0 | 0.941 | 1.07E+09 | 0.941 | 3.11E+08 | 0.941 | 1.38E+09 |
| 0.942 | 0 | 0.942 | 1.99E+09 | 0.942 | 5.95E+08 | 0.942 | 2.59E+09 | 0.942 | 0 | 0.942 | 1.06E+09 | 0.942 | 3.17E+08 | 0.942 | 1.38E+09 |
| 0.943 | 0 | 0.943 | 1.98E+09 | 0.943 | 6.04E+08 | 0.943 | 2.58E+09 | 0.943 | 0 | 0.943 | 1.05E+09 | 0.943 | 3.22E+08 | 0.943 | 1.38E+09 |
| 0.944 | 0 | 0.944 | 1.97E+09 | 0.944 | 6.14E+08 | 0.944 | 2.58E+09 | 0.944 | 0 | 0.944 | 1.05E+09 | 0.944 | 3.27E+08 | 0.944 | 1.37E+09 |
| 0.945 | 0 | 0.945 | 1.95E+09 | 0.945 | 6.24E+08 | 0.945 | 2.58E+09 | 0.945 | 0 | 0.945 | 1.04E+09 | 0.945 | 3.32E+08 | 0.945 | 1.37E+09 |
| 0.946 | 0 | 0.946 | 1.94E+09 | 0.946 | 6.35E+08 | 0.946 | 2.57E+09 | 0.946 | 0 | 0.946 | 1.03E+09 | 0.946 | 3.38E+08 | 0.946 | 1.37E+09 |
| 0.947 | 0 | 0.947 | 1.92E+09 | 0.947 | 6.45E+08 | 0.947 | 2.57E+09 | 0.947 | 0 | 0.947 | 1.02E+09 | 0.947 | 3.43E+08 | 0.947 | 1.37E+09 |
| 0.948 | 0 | 0.948 | 1.91E+09 | 0.948 | 6.56E+08 | 0.948 | 2.57E+09 | 0.948 | 0 | 0.948 | 1.02E+09 | 0.948 | 3.49E+08 | 0.948 | 1.37E+09 |
| 0.949 | 0 | 0.949 | 1.9E+09 | 0.949 | 6.66E+08 | 0.949 | 2.56E+09 | 0.949 | 0 | 0.949 | 1.01E+09 | 0.949 | 3.55E+08 | 0.949 | 1.37E+09 |
| 0.95 | 0 | 0.95 | 1.88E+09 | 0.95 | 6.77E+08 | 0.95 | 2.56E+09 | 0.95 | 0 | 0.95 | 1E+09 | 0.95 | 3.61E+08 | 0.95 | 1.36E+09 |
| 0.951 | 0 | 0.951 | 1.87E+09 | 0.951 | 6.89E+08 | 0.951 | 2.56E+09 | 0.951 | 0 | 0.951 | 9.96E+08 | 0.951 | 3.67E+08 | 0.951 | 1.36E+09 |
| 0.952 | 0 | 0.952 | 1.86E+09 | 0.952 | 7E+08 | 0.952 | 2.56E+09 | 0.952 | 0 | 0.952 | 9.89E+08 | 0.952 | 3.73E+08 | 0.952 | 1.36E+09 |
| 0.953 | 0 | 0.953 | 1.84E+09 | 0.953 | 7.12E+08 | 0.953 | 2.56E+09 | 0.953 | 0 | 0.953 | 9.82E+08 | 0.953 | 3.79E+08 | 0.953 | 1.36E+09 |
| 0.954 | 0 | 0.954 | 1.83E+09 | 0.954 | 7.23E+08 | 0.954 | 2.55E+09 | 0.954 | 0 | 0.954 | 9.75E+08 | 0.954 | 3.85E+08 | 0.954 | 1.36E+09 |
| 0.955 | 0 | 0.955 | 1.82E+09 | 0.955 | 7.35E+08 | 0.955 | 2.55E+09 | 0.955 | 0 | 0.955 | 9.68E+08 | 0.955 | 3.91E+08 | 0.955 | 1.36E+09 |
| 0.956 | 0 | 0.956 | 1.8E+09 | 0.956 | 7.47E+08 | 0.956 | 2.55E+09 | 0.956 | 0 | 0.956 | 9.61E+08 | 0.956 | 3.98E+08 | 0.956 | 1.36E+09 |
| 0.957 | 0 | 0.957 | 1.79E+09 | 0.957 | 7.6E+08 | 0.957 | 2.55E+09 | 0.957 | 0 | 0.957 | 9.54E+08 | 0.957 | 4.04E+08 | 0.957 | 1.36E+09 |
| 0.958 | 0 | 0.958 | 1.78E+09 | 0.958 | 7.72E+08 | 0.958 | 2.55E+09 | 0.958 | 0 | 0.958 | 9.47E+08 | 0.958 | 4.11E+08 | 0.958 | 1.36E+09 |
| 0.959 | 0 | 0.959 | 1.77E+09 | 0.959 | 7.85E+08 | 0.959 | 2.55E+09 | 0.959 | 0 | 0.959 | 9.4E+08 | 0.959 | 4.18E+08 | 0.959 | 1.36E+09 |
| 0.96 | 0 | 0.96 | 1.75E+09 | 0.96 | 7.97E+08 | 0.96 | 2.55E+09 | 0.96 | 0 | 0.96 | 9.33E+08 | 0.96 | 4.24E+08 | 0.96 | 1.36E+09 |
| 0.961 | 0 | 0.961 | 1.74E+09 | 0.961 | 8.1E+08 | 0.961 | 2.55E+09 | 0.961 | 0 | 0.961 | 9.26E+08 | 0.961 | 4.31E+08 | 0.961 | 1.36E+09 |
| 0.962 | 0 | 0.962 | 1.73E+09 | 0.962 | 8.23E+08 | 0.962 | 2.55E+09 | 0.962 | 0 | 0.962 | 9.19E+08 | 0.962 | 4.38E+08 | 0.962 | 1.36E+09 |
| 0.963 | 0 | 0.963 | 1.71E+09 | 0.963 | 8.36E+08 | 0.963 | 2.55E+09 | 0.963 | 0 | 0.963 | 9.12E+08 | 0.963 | 4.45E+08 | 0.963 | 1.36E+09 |
| 0.964 | 0 | 0.964 | 1.7E+09 | 0.964 | 8.5E+08 | 0.964 | 2.55E+09 | 0.964 | 0 | 0.964 | 9.06E+08 | 0.964 | 4.52E+08 | 0.964 | 1.36E+09 |
| 0.965 | 0 | 0.965 | 1.69E+09 | 0.965 | 8.63E+08 | 0.965 | 2.55E+09 | 0.965 | 0 | 0.965 | 8.99E+08 | 0.965 | 4.6E+08 | 0.965 | 1.36E+09 |
| 0.966 | 0 | 0.966 | 1.68E+09 | 0.966 | 8.77E+08 | 0.966 | 2.55E+09 | 0.966 | 0 | 0.966 | 8.92E+08 | 0.966 | 4.67E+08 | 0.966 | 1.36E+09 |
| 0.967 | 0 | 0.967 | 1.66E+09 | 0.967 | 8.91E+08 | 0.967 | 2.55E+09 | 0.967 | 0 | 0.967 | 8.85E+08 | 0.967 | 4.74E+08 | 0.967 | 1.36E+09 |
| 0.968 | 0 | 0.968 | 1.65E+09 | 0.968 | 9.05E+08 | 0.968 | 2.55E+09 | 0.968 | 0 | 0.968 | 8.79E+08 | 0.968 | 4.82E+08 | 0.968 | 1.36E+09 |
| 0.969 | 0 | 0.969 | 1.64E+09 | 0.969 | 9.18E+08 | 0.969 | 2.56E+09 | 0.969 | 0 | 0.969 | 8.72E+08 | 0.969 | 4.89E+08 | 0.969 | 1.36E+09 |
| 0.97 | 0 | 0.97 | 1.63E+09 | 0.97 | 9.32E+08 | 0.97 | 2.56E+09 | 0.97 | 0 | 0.97 | 8.65E+08 | 0.97 | 4.96E+08 | 0.97 | 1.36E+09 |
| 0.971 | 0 | 0.971 | 1.61E+09 | 0.971 | 9.47E+08 | 0.971 | 2.56E+09 | 0.971 | 0 | 0.971 | 8.59E+08 | 0.971 | 5.04E+08 | 0.971 | 1.36E+09 |
| 0.972 | 0 | 0.972 | 1.6E+09 | 0.972 | 9.61E+08 | 0.972 | 2.56E+09 | 0.972 | 0 | 0.972 | 8.52E+08 | 0.972 | 5.11E+08 | 0.972 | 1.36E+09 |
| 0.973 | 0 | 0.973 | 1.59E+09 | 0.973 | 9.75E+08 | 0.973 | 2.56E+09 | 0.973 | 0 | 0.973 | 8.45E+08 | 0.973 | 5.19E+08 | 0.973 | 1.36E+09 |
| 0.974 | 0 | 0.974 | 1.58E+09 | 0.974 | 9.89E+08 | 0.974 | 2.56E+09 | 0.974 | 0 | 0.974 | 8.39E+08 | 0.974 | 5.27E+08 | 0.974 | 1.37E+09 |
| 0.975 | 0 | 0.975 | 1.56E+09 | 0.975 | 1E+09 | 0.975 | 2.57E+09 | 0.975 | 0 | 0.975 | 8.32E+08 | 0.975 | 5.34E+08 | 0.975 | 1.37E+09 |
| 0.976 | 0 | 0.976 | 1.55E+09 | 0.976 | 1.02E+09 | 0.976 | 2.57E+09 | 0.976 | 0 | 0.976 | 8.26E+08 | 0.976 | 5.42E+08 | 0.976 | 1.37E+09 |
| 0.977 | 0 | 0.977 | 1.54E+09 | 0.977 | 1.03E+09 | 0.977 | 2.57E+09 | 0.977 | 0 | 0.977 | 8.19E+08 | 0.977 | 5.49E+08 | 0.977 | 1.37E+09 |
| 0.978 | 0 | 0.978 | 1.53E+09 | 0.978 | 1.05E+09 | 0.978 | 2.57E+09 | 0.978 | 0 | 0.978 | 8.13E+08 | 0.978 | 5.57E+08 | 0.978 | 1.37E+09 |
| 0.979 | 0 | 0.979 | 1.51E+09 | 0.979 | 1.06E+09 | 0.979 | 2.58E+09 | 0.979 | 0 | 0.979 | 8.06E+08 | 0.979 | 5.65E+08 | 0.979 | 1.37E+09 |
| 0.98 | 0 | 0.98 | 1.5E+09 | 0.98 | 1.07E+09 | 0.98 | 2.58E+09 | 0.98 | 0 | 0.98 | 8E+08 | 0.98 | 5.72E+08 | 0.98 | 1.37E+09 |
| 0.981 | 0 | 0.981 | 1.49E+09 | 0.981 | 1.09E+09 | 0.981 | 2.58E+09 | 0.981 | 0 | 0.981 | 7.94E+08 | 0.981 | 5.8E+08 | 0.981 | 1.37E+09 |
| 0.982 | 0 | 0.982 | 1.48E+09 | 0.982 | 1.1E+09 | 0.982 | 2.58E+09 | 0.982 | 0 | 0.982 | 7.87E+08 | 0.982 | 5.87E+08 | 0.982 | 1.37E+09 |
| 0.983 | 0 | 0.983 | 1.47E+09 | 0.983 | 1.12E+09 | 0.983 | 2.58E+09 | 0.983 | 0 | 0.983 | 7.81E+08 | 0.983 | 5.95E+08 | 0.983 | 1.38E+09 |
| 0.984 | 0 | 0.984 | 1.46E+09 | 0.984 | 1.13E+09 | 0.984 | 2.59E+09 | 0.984 | 0 | 0.984 | 7.75E+08 | 0.984 | 6.02E+08 | 0.984 | 1.38E+09 |
| 0.985 | 0 | 0.985 | 1.44E+09 | 0.985 | 1.14E+09 | 0.985 | 2.59E+09 | 0.985 | 0 | 0.985 | 7.68E+08 | 0.985 | 6.09E+08 | 0.985 | 1.38E+09 |
| 0.986 | 0 | 0.986 | 1.43E+09 | 0.986 | 1.16E+09 | 0.986 | 2.59E+09 | 0.986 | 0 | 0.986 | 7.62E+08 | 0.986 | 6.17E+08 | 0.986 | 1.38E+09 |
| 0.987 | 0 | 0.987 | 1.42E+09 | 0.987 | 1.17E+09 | 0.987 | 2.59E+09 | 0.987 | 0 | 0.987 | 7.56E+08 | 0.987 | 6.24E+08 | 0.987 | 1.38E+09 |
| 0.988 | 0 | 0.988 | 1.41E+09 | 0.988 | 1.19E+09 | 0.988 | 2.59E+09 | 0.988 | 0 | 0.988 | 7.5E+08 | 0.988 | 6.31E+08 | 0.988 | 1.38E+09 |
| 0.989 | 0 | 0.989 | 1.4E+09 | 0.989 | 1.2E+09 | 0.989 | 2.6E+09 | 0.989 | 0 | 0.989 | 7.44E+08 | 0.989 | 6.38E+08 | 0.989 | 1.38E+09 |
| 0.99 | 0 | 0.99 | 1.39E+09 | 0.99 | 1.21E+09 | 0.99 | 2.6E+09 | 0.99 | 0 | 0.99 | 7.37E+08 | 0.99 | 6.45E+08 | 0.99 | 1.38E+09 |
| 0.991 | 0 | 0.991 | 1.37E+09 | 0.991 | 1.22E+09 | 0.991 | 2.6E+09 | 0.991 | 0 | 0.991 | 7.31E+08 | 0.991 | 6.52E+08 | 0.991 | 1.38E+09 |
| 0.992 | 0 | 0.992 | 1.36E+09 | 0.992 | 1.24E+09 | 0.992 | 2.6E+09 | 0.992 | 0 | 0.992 | 7.25E+08 | 0.992 | 6.59E+08 | 0.992 | 1.38E+09 |
| 0.993 | 0 | 0.993 | 1.35E+09 | 0.993 | 1.25E+09 | 0.993 | 2.6E+09 | 0.993 | 0 | 0.993 | 7.19E+08 | 0.993 | 6.65E+08 | 0.993 | 1.38E+09 |
| 0.994 | 0 | 0.994 | 1.34E+09 | 0.994 | 1.26E+09 | 0.994 | 2.6E+09 | 0.994 | 0 | 0.994 | 7.13E+08 | 0.994 | 6.72E+08 | 0.994 | 1.39E+09 |
| 0.995 | 0 | 0.995 | 1.33E+09 | 0.995 | 1.27E+09 | 0.995 | 2.6E+09 | 0.995 | 0 | 0.995 | 7.07E+08 | 0.995 | 6.78E+08 | 0.995 | 1.39E+09 |
| 0.996 | 0 | 0.996 | 1.32E+09 | 0.996 | 1.29E+09 | 0.996 | 2.6E+09 | 0.996 | 0 | 0.996 | 7.01E+08 | 0.996 | 6.85E+08 | 0.996 | 1.39E+09 |
| 0.997 | 0 | 0.997 | 1.31E+09 | 0.997 | 1.3E+09 | 0.997 | 2.6E+09 | 0.997 | 0 | 0.997 | 6.95E+08 | 0.997 | 6.91E+08 | 0.997 | 1.39E+09 |
| 0.998 | 0 | 0.998 | 1.3E+09 | 0.998 | 1.31E+09 | 0.998 | 2.6E+09 | 0.998 | 0 | 0.998 | 6.9E+08 | 0.998 | 6.97E+08 | 0.998 | 1.39E+09 |
| 0.999 | 0 | 0.999 | 1.28E+09 | 0.999 | 1.32E+09 | 0.999 | 2.6E+09 | 0.999 | 0 | 0.999 | 6.84E+08 | 0.999 | 7.03E+08 | 0.999 | 1.39E+09 |
| 1 | 0 | 1 | 1.27E+09 | 1 | 1.33E+09 | 1 | 2.6E+09 | 1 | 0 | 1 | 6.78E+08 | 1 | 7.09E+08 | 1 | 1.39E+09 |
| 1.001 | 0 | 1.001 | 1.26E+09 | 1.001 | 1.34E+09 | 1.001 | 2.6E+09 | 1.001 | 0 | 1.001 | 6.72E+08 | 1.001 | 7.15E+08 | 1.001 | 1.39E+09 |
| 1.002 | 0 | 1.002 | 1.25E+09 | 1.002 | 1.35E+09 | 1.002 | 2.6E+09 | 1.002 | 0 | 1.002 | 6.66E+08 | 1.002 | 7.2E+08 | 1.002 | 1.39E+09 |
| 1.003 | 0 | 1.003 | 1.24E+09 | 1.003 | 1.36E+09 | 1.003 | 2.6E+09 | 1.003 | 0 | 1.003 | 6.61E+08 | 1.003 | 7.26E+08 | 1.003 | 1.39E+09 |
| 1.004 | 0 | 1.004 | 1.23E+09 | 1.004 | 1.37E+09 | 1.004 | 2.6E+09 | 1.004 | 0 | 1.004 | 6.55E+08 | 1.004 | 7.32E+08 | 1.004 | 1.39E+09 |
| 1.005 | 0 | 1.005 | 1.22E+09 | 1.005 | 1.39E+09 | 1.005 | 2.6E+09 | 1.005 | 0 | 1.005 | 6.49E+08 | 1.005 | 7.38E+08 | 1.005 | 1.39E+09 |
| 1.006 | 0 | 1.006 | 1.21E+09 | 1.006 | 1.4E+09 | 1.006 | 2.6E+09 | 1.006 | 0 | 1.006 | 6.43E+08 | 1.006 | 7.43E+08 | 1.006 | 1.39E+09 |
| 1.007 | 0 | 1.007 | 1.2E+09 | 1.007 | 1.41E+09 | 1.007 | 2.61E+09 | 1.007 | 0 | 1.007 | 6.38E+08 | 1.007 | 7.49E+08 | 1.007 | 1.39E+09 |
| 1.008 | 0 | 1.008 | 1.19E+09 | 1.008 | 1.42E+09 | 1.008 | 2.61E+09 | 1.008 | 0 | 1.008 | 6.32E+08 | 1.008 | 7.55E+08 | 1.008 | 1.39E+09 |
| 1.009 | 0 | 1.009 | 1.18E+09 | 1.009 | 1.43E+09 | 1.009 | 2.61E+09 | 1.009 | 0 | 1.009 | 6.27E+08 | 1.009 | 7.6E+08 | 1.009 | 1.39E+09 |
| 1.01 | 0 | 1.01 | 1.17E+09 | 1.01 | 1.44E+09 | 1.01 | 2.61E+09 | 1.01 | 0 | 1.01 | 6.21E+08 | 1.01 | 7.66E+08 | 1.01 | 1.39E+09 |
| 1.011 | 0 | 1.011 | 1.16E+09 | 1.011 | 1.45E+09 | 1.011 | 2.61E+09 | 1.011 | 0 | 1.011 | 6.16E+08 | 1.011 | 7.72E+08 | 1.011 | 1.39E+09 |
| 1.012 | 0 | 1.012 | 1.15E+09 | 1.012 | 1.46E+09 | 1.012 | 2.61E+09 | 1.012 | 0 | 1.012 | 6.1E+08 | 1.012 | 7.77E+08 | 1.012 | 1.39E+09 |
| 1.013 | 0 | 1.013 | 1.14E+09 | 1.013 | 1.47E+09 | 1.013 | 2.61E+09 | 1.013 | 0 | 1.013 | 6.05E+08 | 1.013 | 7.83E+08 | 1.013 | 1.39E+09 |
| 1.014 | 0 | 1.014 | 1.13E+09 | 1.014 | 1.48E+09 | 1.014 | 2.61E+09 | 1.014 | 0 | 1.014 | 5.99E+08 | 1.014 | 7.89E+08 | 1.014 | 1.39E+09 |
| 1.015 | 0 | 1.015 | 1.12E+09 | 1.015 | 1.49E+09 | 1.015 | 2.61E+09 | 1.015 | 0 | 1.015 | 5.94E+08 | 1.015 | 7.94E+08 | 1.015 | 1.39E+09 |
| 1.016 | 0 | 1.016 | 1.11E+09 | 1.016 | 1.5E+09 | 1.016 | 2.61E+09 | 1.016 | 0 | 1.016 | 5.89E+08 | 1.016 | 8E+08 | 1.016 | 1.39E+09 |
| 1.017 | 0 | 1.017 | 1.1E+09 | 1.017 | 1.51E+09 | 1.017 | 2.61E+09 | 1.017 | 0 | 1.017 | 5.83E+08 | 1.017 | 8.05E+08 | 1.017 | 1.39E+09 |
| 1.018 | 0 | 1.018 | 1.09E+09 | 1.018 | 1.52E+09 | 1.018 | 2.61E+09 | 1.018 | 0 | 1.018 | 5.78E+08 | 1.018 | 8.11E+08 | 1.018 | 1.39E+09 |
| 1.019 | 0 | 1.019 | 1.08E+09 | 1.019 | 1.53E+09 | 1.019 | 2.61E+09 | 1.019 | 0 | 1.019 | 5.73E+08 | 1.019 | 8.17E+08 | 1.019 | 1.39E+09 |
| 1.02 | 0 | 1.02 | 1.07E+09 | 1.02 | 1.54E+09 | 1.02 | 2.61E+09 | 1.02 | 0 | 1.02 | 5.68E+08 | 1.02 | 8.22E+08 | 1.02 | 1.39E+09 |
| 1.021 | 0 | 1.021 | 1.06E+09 | 1.021 | 1.55E+09 | 1.021 | 2.61E+09 | 1.021 | 0 | 1.021 | 5.62E+08 | 1.021 | 8.28E+08 | 1.021 | 1.39E+09 |
| 1.022 | 0 | 1.022 | 1.05E+09 | 1.022 | 1.56E+09 | 1.022 | 2.61E+09 | 1.022 | 0 | 1.022 | 5.57E+08 | 1.022 | 8.33E+08 | 1.022 | 1.39E+09 |
| 1.023 | 0 | 1.023 | 1.04E+09 | 1.023 | 1.58E+09 | 1.023 | 2.61E+09 | 1.023 | 0 | 1.023 | 5.52E+08 | 1.023 | 8.38E+08 | 1.023 | 1.39E+09 |
| 1.024 | 0 | 1.024 | 1.03E+09 | 1.024 | 1.59E+09 | 1.024 | 2.61E+09 | 1.024 | 0 | 1.024 | 5.47E+08 | 1.024 | 8.44E+08 | 1.024 | 1.39E+09 |
| 1.025 | 0 | 1.025 | 1.02E+09 | 1.025 | 1.6E+09 | 1.025 | 2.61E+09 | 1.025 | 0 | 1.025 | 5.42E+08 | 1.025 | 8.49E+08 | 1.025 | 1.39E+09 |
| 1.026 | 0 | 1.026 | 1.01E+09 | 1.026 | 1.61E+09 | 1.026 | 2.61E+09 | 1.026 | 0 | 1.026 | 5.37E+08 | 1.026 | 8.55E+08 | 1.026 | 1.39E+09 |
| 1.027 | 0 | 1.027 | 9.99E+08 | 1.027 | 1.62E+09 | 1.027 | 2.61E+09 | 1.027 | 0 | 1.027 | 5.32E+08 | 1.027 | 8.6E+08 | 1.027 | 1.39E+09 |
| 1.028 | 0 | 1.028 | 9.9E+08 | 1.028 | 1.63E+09 | 1.028 | 2.62E+09 | 1.028 | 0 | 1.028 | 5.27E+08 | 1.028 | 8.65E+08 | 1.028 | 1.39E+09 |
| 1.029 | 0 | 1.029 | 9.8E+08 | 1.029 | 1.64E+09 | 1.029 | 2.62E+09 | 1.029 | 0 | 1.029 | 5.22E+08 | 1.029 | 8.71E+08 | 1.029 | 1.39E+09 |
| 1.03 | 0 | 1.03 | 9.71E+08 | 1.03 | 1.65E+09 | 1.03 | 2.62E+09 | 1.03 | 0 | 1.03 | 5.17E+08 | 1.03 | 8.76E+08 | 1.03 | 1.39E+09 |
| 1.031 | 0 | 1.031 | 9.62E+08 | 1.031 | 1.66E+09 | 1.031 | 2.62E+09 | 1.031 | 0 | 1.031 | 5.12E+08 | 1.031 | 8.81E+08 | 1.031 | 1.39E+09 |
| 1.032 | 0 | 1.032 | 9.53E+08 | 1.032 | 1.67E+09 | 1.032 | 2.62E+09 | 1.032 | 0 | 1.032 | 5.07E+08 | 1.032 | 8.87E+08 | 1.032 | 1.39E+09 |
| 1.033 | 0 | 1.033 | 9.44E+08 | 1.033 | 1.67E+09 | 1.033 | 2.62E+09 | 1.033 | 0 | 1.033 | 5.03E+08 | 1.033 | 8.92E+08 | 1.033 | 1.39E+09 |
| 1.034 | 0 | 1.034 | 9.35E+08 | 1.034 | 1.68E+09 | 1.034 | 2.62E+09 | 1.034 | 0 | 1.034 | 4.98E+08 | 1.034 | 8.97E+08 | 1.034 | 1.39E+09 |
| 1.035 | 0 | 1.035 | 9.26E+08 | 1.035 | 1.69E+09 | 1.035 | 2.62E+09 | 1.035 | 0 | 1.035 | 4.93E+08 | 1.035 | 9.02E+08 | 1.035 | 1.4E+09 |
| 1.036 | 0 | 1.036 | 9.17E+08 | 1.036 | 1.7E+09 | 1.036 | 2.62E+09 | 1.036 | 0 | 1.036 | 4.88E+08 | 1.036 | 9.07E+08 | 1.036 | 1.4E+09 |
| 1.037 | 0 | 1.037 | 9.08E+08 | 1.037 | 1.71E+09 | 1.037 | 2.62E+09 | 1.037 | 0 | 1.037 | 4.84E+08 | 1.037 | 9.12E+08 | 1.037 | 1.4E+09 |
| 1.038 | 0 | 1.038 | 9E+08 | 1.038 | 1.72E+09 | 1.038 | 2.62E+09 | 1.038 | 0 | 1.038 | 4.79E+08 | 1.038 | 9.17E+08 | 1.038 | 1.4E+09 |
| 1.039 | 0 | 1.039 | 8.91E+08 | 1.039 | 1.73E+09 | 1.039 | 2.62E+09 | 1.039 | 0 | 1.039 | 4.74E+08 | 1.039 | 9.22E+08 | 1.039 | 1.4E+09 |
| 1.04 | 0 | 1.04 | 8.82E+08 | 1.04 | 1.74E+09 | 1.04 | 2.62E+09 | 1.04 | 0 | 1.04 | 4.7E+08 | 1.04 | 9.27E+08 | 1.04 | 1.4E+09 |
| 1.041 | 0 | 1.041 | 8.74E+08 | 1.041 | 1.75E+09 | 1.041 | 2.62E+09 | 1.041 | 0 | 1.041 | 4.65E+08 | 1.041 | 9.32E+08 | 1.041 | 1.4E+09 |
| 1.042 | 0 | 1.042 | 8.65E+08 | 1.042 | 1.76E+09 | 1.042 | 2.63E+09 | 1.042 | 0 | 1.042 | 4.61E+08 | 1.042 | 9.37E+08 | 1.042 | 1.4E+09 |
| 1.043 | 0 | 1.043 | 8.57E+08 | 1.043 | 1.77E+09 | 1.043 | 2.63E+09 | 1.043 | 0 | 1.043 | 4.56E+08 | 1.043 | 9.42E+08 | 1.043 | 1.4E+09 |
| 1.044 | 0 | 1.044 | 8.48E+08 | 1.044 | 1.78E+09 | 1.044 | 2.63E+09 | 1.044 | 0 | 1.044 | 4.52E+08 | 1.044 | 9.47E+08 | 1.044 | 1.4E+09 |
| 1.045 | 0 | 1.045 | 8.4E+08 | 1.045 | 1.79E+09 | 1.045 | 2.63E+09 | 1.045 | 0 | 1.045 | 4.47E+08 | 1.045 | 9.52E+08 | 1.045 | 1.4E+09 |
| 1.046 | 0 | 1.046 | 8.32E+08 | 1.046 | 1.8E+09 | 1.046 | 2.63E+09 | 1.046 | 0 | 1.046 | 4.43E+08 | 1.046 | 9.57E+08 | 1.046 | 1.4E+09 |
| 1.047 | 0 | 1.047 | 8.23E+08 | 1.047 | 1.81E+09 | 1.047 | 2.63E+09 | 1.047 | 0 | 1.047 | 4.38E+08 | 1.047 | 9.62E+08 | 1.047 | 1.4E+09 |
| 1.048 | 0 | 1.048 | 8.15E+08 | 1.048 | 1.82E+09 | 1.048 | 2.63E+09 | 1.048 | 0 | 1.048 | 4.34E+08 | 1.048 | 9.66E+08 | 1.048 | 1.4E+09 |
| 1.049 | 0 | 1.049 | 8.07E+08 | 1.049 | 1.82E+09 | 1.049 | 2.63E+09 | 1.049 | 0 | 1.049 | 4.3E+08 | 1.049 | 9.71E+08 | 1.049 | 1.4E+09 |
| 1.05 | 0 | 1.05 | 7.99E+08 | 1.05 | 1.83E+09 | 1.05 | 2.63E+09 | 1.05 | 0 | 1.05 | 4.25E+08 | 1.05 | 9.76E+08 | 1.05 | 1.4E+09 |
| 1.051 | 0 | 1.051 | 7.91E+08 | 1.051 | 1.84E+09 | 1.051 | 2.63E+09 | 1.051 | 0 | 1.051 | 4.21E+08 | 1.051 | 9.8E+08 | 1.051 | 1.4E+09 |
| 1.052 | 0 | 1.052 | 7.83E+08 | 1.052 | 1.85E+09 | 1.052 | 2.63E+09 | 1.052 | 0 | 1.052 | 4.17E+08 | 1.052 | 9.85E+08 | 1.052 | 1.4E+09 |
| 1.053 | 0 | 1.053 | 7.75E+08 | 1.053 | 1.86E+09 | 1.053 | 2.63E+09 | 1.053 | 0 | 1.053 | 4.13E+08 | 1.053 | 9.9E+08 | 1.053 | 1.4E+09 |
| 1.054 | 0 | 1.054 | 7.67E+08 | 1.054 | 1.87E+09 | 1.054 | 2.63E+09 | 1.054 | 0 | 1.054 | 4.08E+08 | 1.054 | 9.94E+08 | 1.054 | 1.4E+09 |
| 1.055 | 0 | 1.055 | 7.59E+08 | 1.055 | 1.88E+09 | 1.055 | 2.63E+09 | 1.055 | 0 | 1.055 | 4.04E+08 | 1.055 | 9.99E+08 | 1.055 | 1.4E+09 |
| 1.056 | 0 | 1.056 | 7.52E+08 | 1.056 | 1.88E+09 | 1.056 | 2.64E+09 | 1.056 | 0 | 1.056 | 4E+08 | 1.056 | 1E+09 | 1.056 | 1.4E+09 |
| 1.057 | 0 | 1.057 | 7.44E+08 | 1.057 | 1.89E+09 | 1.057 | 2.64E+09 | 1.057 | 0 | 1.057 | 3.96E+08 | 1.057 | 1.01E+09 | 1.057 | 1.4E+09 |
| 1.058 | 0 | 1.058 | 7.36E+08 | 1.058 | 1.9E+09 | 1.058 | 2.64E+09 | 1.058 | 0 | 1.058 | 3.92E+08 | 1.058 | 1.01E+09 | 1.058 | 1.4E+09 |
| 1.059 | 0 | 1.059 | 7.29E+08 | 1.059 | 1.91E+09 | 1.059 | 2.64E+09 | 1.059 | 0 | 1.059 | 3.88E+08 | 1.059 | 1.02E+09 | 1.059 | 1.4E+09 |
| 1.06 | 0 | 1.06 | 7.21E+08 | 1.06 | 1.92E+09 | 1.06 | 2.64E+09 | 1.06 | 0 | 1.06 | 3.84E+08 | 1.06 | 1.02E+09 | 1.06 | 1.4E+09 |
| 1.061 | 0 | 1.061 | 7.14E+08 | 1.061 | 1.92E+09 | 1.061 | 2.64E+09 | 1.061 | 0 | 1.061 | 3.8E+08 | 1.061 | 1.02E+09 | 1.061 | 1.4E+09 |
| 1.062 | 0 | 1.062 | 7.07E+08 | 1.062 | 1.93E+09 | 1.062 | 2.64E+09 | 1.062 | 0 | 1.062 | 3.76E+08 | 1.062 | 1.03E+09 | 1.062 | 1.41E+09 |
| 1.063 | 0 | 1.063 | 6.99E+08 | 1.063 | 1.94E+09 | 1.063 | 2.64E+09 | 1.063 | 0 | 1.063 | 3.72E+08 | 1.063 | 1.03E+09 | 1.063 | 1.41E+09 |
| 1.064 | 0 | 1.064 | 6.92E+08 | 1.064 | 1.95E+09 | 1.064 | 2.64E+09 | 1.064 | 0 | 1.064 | 3.68E+08 | 1.064 | 1.04E+09 | 1.064 | 1.41E+09 |
| 1.065 | 0 | 1.065 | 6.85E+08 | 1.065 | 1.96E+09 | 1.065 | 2.64E+09 | 1.065 | 0 | 1.065 | 3.65E+08 | 1.065 | 1.04E+09 | 1.065 | 1.41E+09 |
| 1.066 | 0 | 1.066 | 6.78E+08 | 1.066 | 1.96E+09 | 1.066 | 2.64E+09 | 1.066 | 0 | 1.066 | 3.61E+08 | 1.066 | 1.05E+09 | 1.066 | 1.41E+09 |
| 1.067 | 0 | 1.067 | 6.71E+08 | 1.067 | 1.97E+09 | 1.067 | 2.64E+09 | 1.067 | 0 | 1.067 | 3.57E+08 | 1.067 | 1.05E+09 | 1.067 | 1.41E+09 |
| 1.068 | 0 | 1.068 | 6.63E+08 | 1.068 | 1.98E+09 | 1.068 | 2.64E+09 | 1.068 | 0 | 1.068 | 3.53E+08 | 1.068 | 1.05E+09 | 1.068 | 1.41E+09 |
| 1.069 | 0 | 1.069 | 6.56E+08 | 1.069 | 1.99E+09 | 1.069 | 2.64E+09 | 1.069 | 0 | 1.069 | 3.49E+08 | 1.069 | 1.06E+09 | 1.069 | 1.41E+09 |
| 1.07 | 0 | 1.07 | 6.5E+08 | 1.07 | 1.99E+09 | 1.07 | 2.64E+09 | 1.07 | 0 | 1.07 | 3.46E+08 | 1.07 | 1.06E+09 | 1.07 | 1.41E+09 |
| 1.071 | 0 | 1.071 | 6.43E+08 | 1.071 | 2E+09 | 1.071 | 2.64E+09 | 1.071 | 0 | 1.071 | 3.42E+08 | 1.071 | 1.06E+09 | 1.071 | 1.41E+09 |
| 1.072 | 0 | 1.072 | 6.36E+08 | 1.072 | 2.01E+09 | 1.072 | 2.64E+09 | 1.072 | 0 | 1.072 | 3.38E+08 | 1.072 | 1.07E+09 | 1.072 | 1.41E+09 |
| 1.073 | 0 | 1.073 | 6.29E+08 | 1.073 | 2.01E+09 | 1.073 | 2.64E+09 | 1.073 | 0 | 1.073 | 3.35E+08 | 1.073 | 1.07E+09 | 1.073 | 1.41E+09 |
| 1.074 | 0 | 1.074 | 6.22E+08 | 1.074 | 2.02E+09 | 1.074 | 2.64E+09 | 1.074 | 0 | 1.074 | 3.31E+08 | 1.074 | 1.08E+09 | 1.074 | 1.41E+09 |
| 1.075 | 0 | 1.075 | 6.16E+08 | 1.075 | 2.03E+09 | 1.075 | 2.64E+09 | 1.075 | 0 | 1.075 | 3.28E+08 | 1.075 | 1.08E+09 | 1.075 | 1.41E+09 |
| 1.076 | 0 | 1.076 | 6.09E+08 | 1.076 | 2.04E+09 | 1.076 | 2.64E+09 | 1.076 | 0 | 1.076 | 3.24E+08 | 1.076 | 1.08E+09 | 1.076 | 1.41E+09 |
| 1.077 | 0 | 1.077 | 6.02E+08 | 1.077 | 2.04E+09 | 1.077 | 2.64E+09 | 1.077 | 0 | 1.077 | 3.21E+08 | 1.077 | 1.09E+09 | 1.077 | 1.41E+09 |
| 1.078 | 0 | 1.078 | 5.96E+08 | 1.078 | 2.05E+09 | 1.078 | 2.64E+09 | 1.078 | 0 | 1.078 | 3.17E+08 | 1.078 | 1.09E+09 | 1.078 | 1.41E+09 |
| 1.079 | 0 | 1.079 | 5.89E+08 | 1.079 | 2.05E+09 | 1.079 | 2.64E+09 | 1.079 | 0 | 1.079 | 3.14E+08 | 1.079 | 1.09E+09 | 1.079 | 1.41E+09 |
| 1.08 | 0 | 1.08 | 5.83E+08 | 1.08 | 2.06E+09 | 1.08 | 2.64E+09 | 1.08 | 0 | 1.08 | 3.1E+08 | 1.08 | 1.1E+09 | 1.08 | 1.41E+09 |
| 1.081 | 0 | 1.081 | 5.77E+08 | 1.081 | 2.07E+09 | 1.081 | 2.64E+09 | 1.081 | 0 | 1.081 | 3.07E+08 | 1.081 | 1.1E+09 | 1.081 | 1.41E+09 |
| 1.082 | 0 | 1.082 | 5.7E+08 | 1.082 | 2.07E+09 | 1.082 | 2.64E+09 | 1.082 | 0 | 1.082 | 3.04E+08 | 1.082 | 1.1E+09 | 1.082 | 1.41E+09 |
| 1.083 | 0 | 1.083 | 5.64E+08 | 1.083 | 2.08E+09 | 1.083 | 2.64E+09 | 1.083 | 0 | 1.083 | 3E+08 | 1.083 | 1.11E+09 | 1.083 | 1.41E+09 |
| 1.084 | 0 | 1.084 | 5.58E+08 | 1.084 | 2.09E+09 | 1.084 | 2.64E+09 | 1.084 | 0 | 1.084 | 2.97E+08 | 1.084 | 1.11E+09 | 1.084 | 1.41E+09 |
| 1.085 | 0 | 1.085 | 5.52E+08 | 1.085 | 2.09E+09 | 1.085 | 2.64E+09 | 1.085 | 0 | 1.085 | 2.94E+08 | 1.085 | 1.11E+09 | 1.085 | 1.41E+09 |
| 1.086 | 0 | 1.086 | 5.46E+08 | 1.086 | 2.1E+09 | 1.086 | 2.64E+09 | 1.086 | 0 | 1.086 | 2.9E+08 | 1.086 | 1.12E+09 | 1.086 | 1.41E+09 |
| 1.087 | 0 | 1.087 | 5.4E+08 | 1.087 | 2.1E+09 | 1.087 | 2.64E+09 | 1.087 | 0 | 1.087 | 2.87E+08 | 1.087 | 1.12E+09 | 1.087 | 1.41E+09 |
| 1.088 | 0 | 1.088 | 5.34E+08 | 1.088 | 2.11E+09 | 1.088 | 2.64E+09 | 1.088 | 0 | 1.088 | 2.84E+08 | 1.088 | 1.12E+09 | 1.088 | 1.41E+09 |
| 1.089 | 0 | 1.089 | 5.28E+08 | 1.089 | 2.11E+09 | 1.089 | 2.64E+09 | 1.089 | 0 | 1.089 | 2.81E+08 | 1.089 | 1.13E+09 | 1.089 | 1.41E+09 |
| 1.09 | 0 | 1.09 | 5.22E+08 | 1.09 | 2.12E+09 | 1.09 | 2.64E+09 | 1.09 | 0 | 1.09 | 2.78E+08 | 1.09 | 1.13E+09 | 1.09 | 1.41E+09 |
| 1.091 | 0 | 1.091 | 5.16E+08 | 1.091 | 2.12E+09 | 1.091 | 2.64E+09 | 1.091 | 0 | 1.091 | 2.75E+08 | 1.091 | 1.13E+09 | 1.091 | 1.41E+09 |
| 1.092 | 0 | 1.092 | 5.1E+08 | 1.092 | 2.13E+09 | 1.092 | 2.64E+09 | 1.092 | 0 | 1.092 | 2.72E+08 | 1.092 | 1.13E+09 | 1.092 | 1.41E+09 |
| 1.093 | 0 | 1.093 | 5.04E+08 | 1.093 | 2.13E+09 | 1.093 | 2.64E+09 | 1.093 | 0 | 1.093 | 2.69E+08 | 1.093 | 1.14E+09 | 1.093 | 1.41E+09 |
| 1.094 | 0 | 1.094 | 4.99E+08 | 1.094 | 2.14E+09 | 1.094 | 2.64E+09 | 1.094 | 0 | 1.094 | 2.66E+08 | 1.094 | 1.14E+09 | 1.094 | 1.4E+09 |
| 1.095 | 0 | 1.095 | 4.93E+08 | 1.095 | 2.14E+09 | 1.095 | 2.64E+09 | 1.095 | 0 | 1.095 | 2.63E+08 | 1.095 | 1.14E+09 | 1.095 | 1.4E+09 |
| 1.096 | 0 | 1.096 | 4.87E+08 | 1.096 | 2.15E+09 | 1.096 | 2.64E+09 | 1.096 | 0 | 1.096 | 2.6E+08 | 1.096 | 1.14E+09 | 1.096 | 1.4E+09 |
| 1.097 | 0 | 1.097 | 4.82E+08 | 1.097 | 2.15E+09 | 1.097 | 2.64E+09 | 1.097 | 0 | 1.097 | 2.57E+08 | 1.097 | 1.15E+09 | 1.097 | 1.4E+09 |
| 1.098 | 0 | 1.098 | 4.76E+08 | 1.098 | 2.16E+09 | 1.098 | 2.64E+09 | 1.098 | 0 | 1.098 | 2.54E+08 | 1.098 | 1.15E+09 | 1.098 | 1.4E+09 |
| 1.099 | 0 | 1.099 | 4.71E+08 | 1.099 | 2.16E+09 | 1.099 | 2.63E+09 | 1.099 | 0 | 1.099 | 2.51E+08 | 1.099 | 1.15E+09 | 1.099 | 1.4E+09 |
| 1.1 | 0 | 1.1 | 4.66E+08 | 1.1 | 2.17E+09 | 1.1 | 2.63E+09 | 1.1 | 0 | 1.1 | 2.48E+08 | 1.1 | 1.15E+09 | 1.1 | 1.4E+09 |
| 1.101 | 0 | 1.101 | 4.6E+08 | 1.101 | 2.17E+09 | 1.101 | 2.63E+09 | 1.101 | 0 | 1.101 | 2.45E+08 | 1.101 | 1.16E+09 | 1.101 | 1.4E+09 |
| 1.102 | 0 | 1.102 | 4.55E+08 | 1.102 | 2.18E+09 | 1.102 | 2.63E+09 | 1.102 | 0 | 1.102 | 2.42E+08 | 1.102 | 1.16E+09 | 1.102 | 1.4E+09 |
| 1.103 | 0 | 1.103 | 4.5E+08 | 1.103 | 2.18E+09 | 1.103 | 2.63E+09 | 1.103 | 0 | 1.103 | 2.39E+08 | 1.103 | 1.16E+09 | 1.103 | 1.4E+09 |
| 1.104 | 0 | 1.104 | 4.45E+08 | 1.104 | 2.18E+09 | 1.104 | 2.63E+09 | 1.104 | 0 | 1.104 | 2.37E+08 | 1.104 | 1.16E+09 | 1.104 | 1.4E+09 |
| 1.105 | 0 | 1.105 | 4.39E+08 | 1.105 | 2.19E+09 | 1.105 | 2.63E+09 | 1.105 | 0 | 1.105 | 2.34E+08 | 1.105 | 1.16E+09 | 1.105 | 1.4E+09 |
| 1.106 | 0 | 1.106 | 4.34E+08 | 1.106 | 2.19E+09 | 1.106 | 2.62E+09 | 1.106 | 0 | 1.106 | 2.31E+08 | 1.106 | 1.17E+09 | 1.106 | 1.4E+09 |
| 1.107 | 0 | 1.107 | 4.29E+08 | 1.107 | 2.19E+09 | 1.107 | 2.62E+09 | 1.107 | 0 | 1.107 | 2.28E+08 | 1.107 | 1.17E+09 | 1.107 | 1.4E+09 |
| 1.108 | 0 | 1.108 | 4.24E+08 | 1.108 | 2.2E+09 | 1.108 | 2.62E+09 | 1.108 | 0 | 1.108 | 2.26E+08 | 1.108 | 1.17E+09 | 1.108 | 1.4E+09 |
| 1.109 | 0 | 1.109 | 4.19E+08 | 1.109 | 2.2E+09 | 1.109 | 2.62E+09 | 1.109 | 0 | 1.109 | 2.23E+08 | 1.109 | 1.17E+09 | 1.109 | 1.39E+09 |
| 1.11 | 0 | 1.11 | 4.14E+08 | 1.11 | 2.2E+09 | 1.11 | 2.62E+09 | 1.11 | 0 | 1.11 | 2.21E+08 | 1.11 | 1.17E+09 | 1.11 | 1.39E+09 |
| 1.111 | 0 | 1.111 | 4.09E+08 | 1.111 | 2.21E+09 | 1.111 | 2.62E+09 | 1.111 | 0 | 1.111 | 2.18E+08 | 1.111 | 1.17E+09 | 1.111 | 1.39E+09 |
| 1.112 | 0 | 1.112 | 4.05E+08 | 1.112 | 2.21E+09 | 1.112 | 2.61E+09 | 1.112 | 0 | 1.112 | 2.15E+08 | 1.112 | 1.18E+09 | 1.112 | 1.39E+09 |
| 1.113 | 0 | 1.113 | 4E+08 | 1.113 | 2.21E+09 | 1.113 | 2.61E+09 | 1.113 | 0 | 1.113 | 2.13E+08 | 1.113 | 1.18E+09 | 1.113 | 1.39E+09 |
| 1.114 | 0 | 1.114 | 3.95E+08 | 1.114 | 2.22E+09 | 1.114 | 2.61E+09 | 1.114 | 0 | 1.114 | 2.1E+08 | 1.114 | 1.18E+09 | 1.114 | 1.39E+09 |
| 1.115 | 0 | 1.115 | 3.9E+08 | 1.115 | 2.22E+09 | 1.115 | 2.61E+09 | 1.115 | 0 | 1.115 | 2.08E+08 | 1.115 | 1.18E+09 | 1.115 | 1.39E+09 |
| 1.116 | 0 | 1.116 | 3.86E+08 | 1.116 | 2.22E+09 | 1.116 | 2.61E+09 | 1.116 | 0 | 1.116 | 2.05E+08 | 1.116 | 1.18E+09 | 1.116 | 1.39E+09 |
| 1.117 | 0 | 1.117 | 3.81E+08 | 1.117 | 2.22E+09 | 1.117 | 2.6E+09 | 1.117 | 0 | 1.117 | 2.03E+08 | 1.117 | 1.18E+09 | 1.117 | 1.39E+09 |
| 1.118 | 0 | 1.118 | 3.76E+08 | 1.118 | 2.22E+09 | 1.118 | 2.6E+09 | 1.118 | 0 | 1.118 | 2E+08 | 1.118 | 1.18E+09 | 1.118 | 1.38E+09 |
| 1.119 | 0 | 1.119 | 3.72E+08 | 1.119 | 2.23E+09 | 1.119 | 2.6E+09 | 1.119 | 0 | 1.119 | 1.98E+08 | 1.119 | 1.19E+09 | 1.119 | 1.38E+09 |
| 1.12 | 0 | 1.12 | 3.67E+08 | 1.12 | 2.23E+09 | 1.12 | 2.6E+09 | 1.12 | 0 | 1.12 | 1.96E+08 | 1.12 | 1.19E+09 | 1.12 | 1.38E+09 |
| 1.121 | 0 | 1.121 | 3.63E+08 | 1.121 | 2.23E+09 | 1.121 | 2.59E+09 | 1.121 | 0 | 1.121 | 1.93E+08 | 1.121 | 1.19E+09 | 1.121 | 1.38E+09 |
| 1.122 | 0 | 1.122 | 3.59E+08 | 1.122 | 2.23E+09 | 1.122 | 2.59E+09 | 1.122 | 0 | 1.122 | 1.91E+08 | 1.122 | 1.19E+09 | 1.122 | 1.38E+09 |
| 1.123 | 0 | 1.123 | 3.54E+08 | 1.123 | 2.23E+09 | 1.123 | 2.59E+09 | 1.123 | 0 | 1.123 | 1.89E+08 | 1.123 | 1.19E+09 | 1.123 | 1.38E+09 |
| 1.124 | 0 | 1.124 | 3.5E+08 | 1.124 | 2.23E+09 | 1.124 | 2.58E+09 | 1.124 | 0 | 1.124 | 1.86E+08 | 1.124 | 1.19E+09 | 1.124 | 1.38E+09 |
| 1.125 | 0 | 1.125 | 3.46E+08 | 1.125 | 2.24E+09 | 1.125 | 2.58E+09 | 1.125 | 0 | 1.125 | 1.84E+08 | 1.125 | 1.19E+09 | 1.125 | 1.37E+09 |
| 1.126 | 0 | 1.126 | 3.42E+08 | 1.126 | 2.24E+09 | 1.126 | 2.58E+09 | 1.126 | 0 | 1.126 | 1.82E+08 | 1.126 | 1.19E+09 | 1.126 | 1.37E+09 |
| 1.127 | 0 | 1.127 | 3.37E+08 | 1.127 | 2.24E+09 | 1.127 | 2.58E+09 | 1.127 | 0 | 1.127 | 1.8E+08 | 1.127 | 1.19E+09 | 1.127 | 1.37E+09 |
| 1.128 | 0 | 1.128 | 3.33E+08 | 1.128 | 2.24E+09 | 1.128 | 2.57E+09 | 1.128 | 0 | 1.128 | 1.77E+08 | 1.128 | 1.19E+09 | 1.128 | 1.37E+09 |
| 1.129 | 0 | 1.129 | 3.29E+08 | 1.129 | 2.24E+09 | 1.129 | 2.57E+09 | 1.129 | 0 | 1.129 | 1.75E+08 | 1.129 | 1.19E+09 | 1.129 | 1.37E+09 |
| 1.13 | 0 | 1.13 | 3.25E+08 | 1.13 | 2.24E+09 | 1.13 | 2.57E+09 | 1.13 | 0 | 1.13 | 1.73E+08 | 1.13 | 1.19E+09 | 1.13 | 1.37E+09 |
| 1.131 | 0 | 1.131 | 3.21E+08 | 1.131 | 2.24E+09 | 1.131 | 2.56E+09 | 1.131 | 0 | 1.131 | 1.71E+08 | 1.131 | 1.19E+09 | 1.131 | 1.36E+09 |
| 1.132 | 0 | 1.132 | 3.17E+08 | 1.132 | 2.24E+09 | 1.132 | 2.56E+09 | 1.132 | 0 | 1.132 | 1.69E+08 | 1.132 | 1.19E+09 | 1.132 | 1.36E+09 |
| 1.133 | 0 | 1.133 | 3.13E+08 | 1.133 | 2.24E+09 | 1.133 | 2.55E+09 | 1.133 | 0 | 1.133 | 1.67E+08 | 1.133 | 1.19E+09 | 1.133 | 1.36E+09 |
| 1.134 | 0 | 1.134 | 3.09E+08 | 1.134 | 2.24E+09 | 1.134 | 2.55E+09 | 1.134 | 0 | 1.134 | 1.65E+08 | 1.134 | 1.19E+09 | 1.134 | 1.36E+09 |
| 1.135 | 0 | 1.135 | 3.05E+08 | 1.135 | 2.24E+09 | 1.135 | 2.55E+09 | 1.135 | 0 | 1.135 | 1.63E+08 | 1.135 | 1.19E+09 | 1.135 | 1.36E+09 |
| 1.136 | 0 | 1.136 | 3.02E+08 | 1.136 | 2.24E+09 | 1.136 | 2.54E+09 | 1.136 | 0 | 1.136 | 1.61E+08 | 1.136 | 1.19E+09 | 1.136 | 1.35E+09 |
| 1.137 | 0 | 1.137 | 2.98E+08 | 1.137 | 2.24E+09 | 1.137 | 2.54E+09 | 1.137 | 0 | 1.137 | 1.59E+08 | 1.137 | 1.19E+09 | 1.137 | 1.35E+09 |
| 1.138 | 0 | 1.138 | 2.94E+08 | 1.138 | 2.24E+09 | 1.138 | 2.53E+09 | 1.138 | 0 | 1.138 | 1.57E+08 | 1.138 | 1.19E+09 | 1.138 | 1.35E+09 |
| 1.139 | 0 | 1.139 | 2.9E+08 | 1.139 | 2.24E+09 | 1.139 | 2.53E+09 | 1.139 | 0 | 1.139 | 1.55E+08 | 1.139 | 1.19E+09 | 1.139 | 1.35E+09 |
| 1.14 | 0 | 1.14 | 2.87E+08 | 1.14 | 2.24E+09 | 1.14 | 2.53E+09 | 1.14 | 0 | 1.14 | 1.53E+08 | 1.14 | 1.19E+09 | 1.14 | 1.34E+09 |
| 1.141 | 0 | 1.141 | 2.83E+08 | 1.141 | 2.24E+09 | 1.141 | 2.52E+09 | 1.141 | 0 | 1.141 | 1.51E+08 | 1.141 | 1.19E+09 | 1.141 | 1.34E+09 |
| 1.142 | 0 | 1.142 | 2.79E+08 | 1.142 | 2.24E+09 | 1.142 | 2.52E+09 | 1.142 | 0 | 1.142 | 1.49E+08 | 1.142 | 1.19E+09 | 1.142 | 1.34E+09 |
| 1.143 | 0 | 1.143 | 2.76E+08 | 1.143 | 2.24E+09 | 1.143 | 2.51E+09 | 1.143 | 0 | 1.143 | 1.47E+08 | 1.143 | 1.19E+09 | 1.143 | 1.34E+09 |
| 1.144 | 0 | 1.144 | 2.72E+08 | 1.144 | 2.23E+09 | 1.144 | 2.51E+09 | 1.144 | 0 | 1.144 | 1.45E+08 | 1.144 | 1.19E+09 | 1.144 | 1.33E+09 |
| 1.145 | 0 | 1.145 | 2.69E+08 | 1.145 | 2.23E+09 | 1.145 | 2.5E+09 | 1.145 | 0 | 1.145 | 1.43E+08 | 1.145 | 1.19E+09 | 1.145 | 1.33E+09 |
| 1.146 | 0 | 1.146 | 2.65E+08 | 1.146 | 2.23E+09 | 1.146 | 2.5E+09 | 1.146 | 0 | 1.146 | 1.41E+08 | 1.146 | 1.19E+09 | 1.146 | 1.33E+09 |
| 1.147 | 0 | 1.147 | 2.62E+08 | 1.147 | 2.23E+09 | 1.147 | 2.49E+09 | 1.147 | 0 | 1.147 | 1.39E+08 | 1.147 | 1.19E+09 | 1.147 | 1.33E+09 |
| 1.148 | 0 | 1.148 | 2.59E+08 | 1.148 | 2.23E+09 | 1.148 | 2.49E+09 | 1.148 | 0 | 1.148 | 1.38E+08 | 1.148 | 1.19E+09 | 1.148 | 1.32E+09 |
| 1.149 | 0 | 1.149 | 2.55E+08 | 1.149 | 2.23E+09 | 1.149 | 2.48E+09 | 1.149 | 0 | 1.149 | 1.36E+08 | 1.149 | 1.19E+09 | 1.149 | 1.32E+09 |
| 1.15 | 0 | 1.15 | 2.52E+08 | 1.15 | 2.23E+09 | 1.15 | 2.48E+09 | 1.15 | 0 | 1.15 | 1.34E+08 | 1.15 | 1.19E+09 | 1.15 | 1.32E+09 |
| 1.151 | 0 | 1.151 | 2.49E+08 | 1.151 | 2.22E+09 | 1.151 | 2.47E+09 | 1.151 | 0 | 1.151 | 1.32E+08 | 1.151 | 1.18E+09 | 1.151 | 1.32E+09 |
| 1.152 | 0 | 1.152 | 2.46E+08 | 1.152 | 2.22E+09 | 1.152 | 2.47E+09 | 1.152 | 0 | 1.152 | 1.31E+08 | 1.152 | 1.18E+09 | 1.152 | 1.31E+09 |
| 1.153 | 0 | 1.153 | 2.42E+08 | 1.153 | 2.22E+09 | 1.153 | 2.46E+09 | 1.153 | 0 | 1.153 | 1.29E+08 | 1.153 | 1.18E+09 | 1.153 | 1.31E+09 |
| 1.154 | 0 | 1.154 | 2.39E+08 | 1.154 | 2.22E+09 | 1.154 | 2.46E+09 | 1.154 | 0 | 1.154 | 1.27E+08 | 1.154 | 1.18E+09 | 1.154 | 1.31E+09 |
| 1.155 | 0 | 1.155 | 2.36E+08 | 1.155 | 2.22E+09 | 1.155 | 2.45E+09 | 1.155 | 0 | 1.155 | 1.26E+08 | 1.155 | 1.18E+09 | 1.155 | 1.31E+09 |
| 1.156 | 0 | 1.156 | 2.33E+08 | 1.156 | 2.22E+09 | 1.156 | 2.45E+09 | 1.156 | 0 | 1.156 | 1.24E+08 | 1.156 | 1.18E+09 | 1.156 | 1.3E+09 |
| 1.157 | 0 | 1.157 | 2.3E+08 | 1.157 | 2.21E+09 | 1.157 | 2.44E+09 | 1.157 | 0 | 1.157 | 1.22E+08 | 1.157 | 1.18E+09 | 1.157 | 1.3E+09 |
| 1.158 | 0 | 1.158 | 2.27E+08 | 1.158 | 2.21E+09 | 1.158 | 2.44E+09 | 1.158 | 0 | 1.158 | 1.21E+08 | 1.158 | 1.18E+09 | 1.158 | 1.3E+09 |
| 1.159 | 0 | 1.159 | 2.24E+08 | 1.159 | 2.21E+09 | 1.159 | 2.43E+09 | 1.159 | 0 | 1.159 | 1.19E+08 | 1.159 | 1.18E+09 | 1.159 | 1.3E+09 |
| 1.16 | 0 | 1.16 | 2.21E+08 | 1.16 | 2.21E+09 | 1.16 | 2.43E+09 | 1.16 | 0 | 1.16 | 1.18E+08 | 1.16 | 1.17E+09 | 1.16 | 1.29E+09 |
| 1.161 | 0 | 1.161 | 2.18E+08 | 1.161 | 2.2E+09 | 1.161 | 2.42E+09 | 1.161 | 0 | 1.161 | 1.16E+08 | 1.161 | 1.17E+09 | 1.161 | 1.29E+09 |
| 1.162 | 0 | 1.162 | 2.15E+08 | 1.162 | 2.2E+09 | 1.162 | 2.42E+09 | 1.162 | 0 | 1.162 | 1.15E+08 | 1.162 | 1.17E+09 | 1.162 | 1.29E+09 |
| 1.163 | 0 | 1.163 | 2.12E+08 | 1.163 | 2.2E+09 | 1.163 | 2.41E+09 | 1.163 | 0 | 1.163 | 1.13E+08 | 1.163 | 1.17E+09 | 1.163 | 1.28E+09 |
| 1.164 | 0 | 1.164 | 2.09E+08 | 1.164 | 2.2E+09 | 1.164 | 2.41E+09 | 1.164 | 0 | 1.164 | 1.11E+08 | 1.164 | 1.17E+09 | 1.164 | 1.28E+09 |
| 1.165 | 0 | 1.165 | 2.07E+08 | 1.165 | 2.19E+09 | 1.165 | 2.4E+09 | 1.165 | 0 | 1.165 | 1.1E+08 | 1.165 | 1.17E+09 | 1.165 | 1.28E+09 |
| 1.166 | 0 | 1.166 | 2.04E+08 | 1.166 | 2.19E+09 | 1.166 | 2.4E+09 | 1.166 | 0 | 1.166 | 1.09E+08 | 1.166 | 1.17E+09 | 1.166 | 1.28E+09 |
| 1.167 | 0 | 1.167 | 2.01E+08 | 1.167 | 2.19E+09 | 1.167 | 2.39E+09 | 1.167 | 0 | 1.167 | 1.07E+08 | 1.167 | 1.17E+09 | 1.167 | 1.27E+09 |
| 1.168 | 0 | 1.168 | 1.98E+08 | 1.168 | 2.19E+09 | 1.168 | 2.38E+09 | 1.168 | 0 | 1.168 | 1.06E+08 | 1.168 | 1.16E+09 | 1.168 | 1.27E+09 |
| 1.169 | 0 | 1.169 | 1.96E+08 | 1.169 | 2.18E+09 | 1.169 | 2.38E+09 | 1.169 | 0 | 1.169 | 1.04E+08 | 1.169 | 1.16E+09 | 1.169 | 1.27E+09 |
| 1.17 | 0 | 1.17 | 1.93E+08 | 1.17 | 2.18E+09 | 1.17 | 2.37E+09 | 1.17 | 0 | 1.17 | 1.03E+08 | 1.17 | 1.16E+09 | 1.17 | 1.26E+09 |
| 1.171 | 0 | 1.171 | 1.9E+08 | 1.171 | 2.18E+09 | 1.171 | 2.37E+09 | 1.171 | 0 | 1.171 | 1.01E+08 | 1.171 | 1.16E+09 | 1.171 | 1.26E+09 |
| 1.172 | 0 | 1.172 | 1.88E+08 | 1.172 | 2.17E+09 | 1.172 | 2.36E+09 | 1.172 | 0 | 1.172 | 1E+08 | 1.172 | 1.16E+09 | 1.172 | 1.26E+09 |
| 1.173 | 0 | 1.173 | 1.85E+08 | 1.173 | 2.17E+09 | 1.173 | 2.36E+09 | 1.173 | 0 | 1.173 | 98673916 | 1.173 | 1.16E+09 | 1.173 | 1.25E+09 |
| 1.174 | 0 | 1.174 | 1.83E+08 | 1.174 | 2.17E+09 | 1.174 | 2.35E+09 | 1.174 | 0 | 1.174 | 97330034 | 1.174 | 1.15E+09 | 1.174 | 1.25E+09 |
| 1.175 | 0 | 1.175 | 1.8E+08 | 1.175 | 2.16E+09 | 1.175 | 2.34E+09 | 1.175 | 0 | 1.175 | 96001615 | 1.175 | 1.15E+09 | 1.175 | 1.25E+09 |
| 1.176 | 0 | 1.176 | 1.78E+08 | 1.176 | 2.16E+09 | 1.176 | 2.34E+09 | 1.176 | 0 | 1.176 | 94688525 | 1.176 | 1.15E+09 | 1.176 | 1.24E+09 |
| 1.177 | 0 | 1.177 | 1.75E+08 | 1.177 | 2.16E+09 | 1.177 | 2.33E+09 | 1.177 | 0 | 1.177 | 93390632 | 1.177 | 1.15E+09 | 1.177 | 1.24E+09 |
| 1.178 | 0 | 1.178 | 1.73E+08 | 1.178 | 2.15E+09 | 1.178 | 2.33E+09 | 1.178 | 0 | 1.178 | 92107805 | 1.178 | 1.15E+09 | 1.178 | 1.24E+09 |
| 1.179 | 0 | 1.179 | 1.71E+08 | 1.179 | 2.15E+09 | 1.179 | 2.32E+09 | 1.179 | 0 | 1.179 | 90839911 | 1.179 | 1.14E+09 | 1.179 | 1.23E+09 |
| 1.18 | 0 | 1.18 | 1.68E+08 | 1.18 | 2.15E+09 | 1.18 | 2.31E+09 | 1.18 | 0 | 1.18 | 89586819 | 1.18 | 1.14E+09 | 1.18 | 1.23E+09 |
| 1.181 | 0 | 1.181 | 1.66E+08 | 1.181 | 2.14E+09 | 1.181 | 2.31E+09 | 1.181 | 0 | 1.181 | 88348399 | 1.181 | 1.14E+09 | 1.181 | 1.23E+09 |
| 1.182 | 0 | 1.182 | 1.64E+08 | 1.182 | 2.14E+09 | 1.182 | 2.3E+09 | 1.182 | 0 | 1.182 | 87124522 | 1.182 | 1.14E+09 | 1.182 | 1.23E+09 |
| 1.183 | 0 | 1.183 | 1.61E+08 | 1.183 | 2.13E+09 | 1.183 | 2.29E+09 | 1.183 | 0 | 1.183 | 85915056 | 1.183 | 1.14E+09 | 1.183 | 1.22E+09 |
| 1.184 | 0 | 1.184 | 1.59E+08 | 1.184 | 2.13E+09 | 1.184 | 2.29E+09 | 1.184 | 0 | 1.184 | 84719874 | 1.184 | 1.13E+09 | 1.184 | 1.22E+09 |
| 1.185 | 0 | 1.185 | 1.57E+08 | 1.185 | 2.13E+09 | 1.185 | 2.28E+09 | 1.185 | 0 | 1.185 | 83538846 | 1.185 | 1.13E+09 | 1.185 | 1.21E+09 |
| 1.186 | 0 | 1.186 | 1.55E+08 | 1.186 | 2.12E+09 | 1.186 | 2.28E+09 | 1.186 | 0 | 1.186 | 82371846 | 1.186 | 1.13E+09 | 1.186 | 1.21E+09 |
| 1.187 | 0 | 1.187 | 1.53E+08 | 1.187 | 2.12E+09 | 1.187 | 2.27E+09 | 1.187 | 0 | 1.187 | 81218745 | 1.187 | 1.13E+09 | 1.187 | 1.21E+09 |
| 1.188 | 0 | 1.188 | 1.5E+08 | 1.188 | 2.11E+09 | 1.188 | 2.26E+09 | 1.188 | 0 | 1.188 | 80079416 | 1.188 | 1.12E+09 | 1.188 | 1.2E+09 |
| 1.189 | 0 | 1.189 | 1.48E+08 | 1.189 | 2.11E+09 | 1.189 | 2.26E+09 | 1.189 | 0 | 1.189 | 78953734 | 1.189 | 1.12E+09 | 1.189 | 1.2E+09 |
| 1.19 | 0 | 1.19 | 1.46E+08 | 1.19 | 2.1E+09 | 1.19 | 2.25E+09 | 1.19 | 0 | 1.19 | 77841573 | 1.19 | 1.12E+09 | 1.19 | 1.2E+09 |
| 1.191 | 0 | 1.191 | 1.44E+08 | 1.191 | 2.1E+09 | 1.191 | 2.24E+09 | 1.191 | 0 | 1.191 | 76742808 | 1.191 | 1.12E+09 | 1.191 | 1.19E+09 |
| 1.192 | 0 | 1.192 | 1.42E+08 | 1.192 | 2.09E+09 | 1.192 | 2.24E+09 | 1.192 | 0 | 1.192 | 75657313 | 1.192 | 1.11E+09 | 1.192 | 1.19E+09 |
| 1.193 | 0 | 1.193 | 1.4E+08 | 1.193 | 2.09E+09 | 1.193 | 2.23E+09 | 1.193 | 0 | 1.193 | 74584966 | 1.193 | 1.11E+09 | 1.193 | 1.19E+09 |
| 1.194 | 0 | 1.194 | 1.38E+08 | 1.194 | 2.08E+09 | 1.194 | 2.22E+09 | 1.194 | 0 | 1.194 | 73525643 | 1.194 | 1.11E+09 | 1.194 | 1.18E+09 |
| 1.195 | 0 | 1.195 | 1.36E+08 | 1.195 | 2.08E+09 | 1.195 | 2.22E+09 | 1.195 | 0 | 1.195 | 72479220 | 1.195 | 1.11E+09 | 1.195 | 1.18E+09 |
| 1.196 | 0 | 1.196 | 1.34E+08 | 1.196 | 2.07E+09 | 1.196 | 2.21E+09 | 1.196 | 0 | 1.196 | 71445577 | 1.196 | 1.1E+09 | 1.196 | 1.18E+09 |
| 1.197 | 0 | 1.197 | 1.32E+08 | 1.197 | 2.07E+09 | 1.197 | 2.2E+09 | 1.197 | 0 | 1.197 | 70424591 | 1.197 | 1.1E+09 | 1.197 | 1.17E+09 |
| 1.198 | 0 | 1.198 | 1.3E+08 | 1.198 | 2.06E+09 | 1.198 | 2.2E+09 | 1.198 | 0 | 1.198 | 69416142 | 1.198 | 1.1E+09 | 1.198 | 1.17E+09 |
| 1.199 | 0 | 1.199 | 1.29E+08 | 1.199 | 2.06E+09 | 1.199 | 2.19E+09 | 1.199 | 0 | 1.199 | 68420108 | 1.199 | 1.1E+09 | 1.199 | 1.16E+09 |
| 1.2 | 0 | 1.2 | 1.27E+08 | 1.2 | 2.05E+09 | 1.2 | 2.18E+09 | 1.2 | 0 | 1.2 | 67436372 | 1.2 | 1.09E+09 | 1.2 | 1.16E+09 |
| 1.201 | 0 | 1.201 | 1.25E+08 | 1.201 | 2.05E+09 | 1.201 | 2.17E+09 | 1.201 | 0 | 1.201 | 66464813 | 1.201 | 1.09E+09 | 1.201 | 1.16E+09 |
| 1.202 | 0 | 1.202 | 1.23E+08 | 1.202 | 2.04E+09 | 1.202 | 2.17E+09 | 1.202 | 0 | 1.202 | 65505313 | 1.202 | 1.09E+09 | 1.202 | 1.15E+09 |
| 1.203 | 0 | 1.203 | 1.21E+08 | 1.203 | 2.04E+09 | 1.203 | 2.16E+09 | 1.203 | 0 | 1.203 | 64557755 | 1.203 | 1.09E+09 | 1.203 | 1.15E+09 |
| 1.204 | 0 | 1.204 | 1.2E+08 | 1.204 | 2.03E+09 | 1.204 | 2.15E+09 | 1.204 | 0 | 1.204 | 63622021 | 1.204 | 1.08E+09 | 1.204 | 1.15E+09 |
| 1.205 | 0 | 1.205 | 1.18E+08 | 1.205 | 2.03E+09 | 1.205 | 2.15E+09 | 1.205 | 0 | 1.205 | 62697996 | 1.205 | 1.08E+09 | 1.205 | 1.14E+09 |
| 1.206 | 0 | 1.206 | 1.16E+08 | 1.206 | 2.02E+09 | 1.206 | 2.14E+09 | 1.206 | 0 | 1.206 | 61785562 | 1.206 | 1.08E+09 | 1.206 | 1.14E+09 |
| 1.207 | 0 | 1.207 | 1.14E+08 | 1.207 | 2.02E+09 | 1.207 | 2.13E+09 | 1.207 | 0 | 1.207 | 60884606 | 1.207 | 1.07E+09 | 1.207 | 1.13E+09 |
| 1.208 | 0 | 1.208 | 1.13E+08 | 1.208 | 2.01E+09 | 1.208 | 2.12E+09 | 1.208 | 0 | 1.208 | 59995012 | 1.208 | 1.07E+09 | 1.208 | 1.13E+09 |
| 1.209 | 0 | 1.209 | 1.11E+08 | 1.209 | 2E+09 | 1.209 | 2.12E+09 | 1.209 | 0 | 1.209 | 59116668 | 1.209 | 1.07E+09 | 1.209 | 1.13E+09 |
| 1.21 | 0 | 1.21 | 1.09E+08 | 1.21 | 2E+09 | 1.21 | 2.11E+09 | 1.21 | 0 | 1.21 | 58249459 | 1.21 | 1.06E+09 | 1.21 | 1.12E+09 |
| 1.211 | 0 | 1.211 | 1.08E+08 | 1.211 | 1.99E+09 | 1.211 | 2.1E+09 | 1.211 | 0 | 1.211 | 57393273 | 1.211 | 1.06E+09 | 1.211 | 1.12E+09 |
| 1.212 | 0 | 1.212 | 1.06E+08 | 1.212 | 1.99E+09 | 1.212 | 2.09E+09 | 1.212 | 0 | 1.212 | 56548000 | 1.212 | 1.06E+09 | 1.212 | 1.11E+09 |
| 1.213 | 0 | 1.213 | 1.05E+08 | 1.213 | 1.98E+09 | 1.213 | 2.09E+09 | 1.213 | 0 | 1.213 | 55713527 | 1.213 | 1.05E+09 | 1.213 | 1.11E+09 |
| 1.214 | 0 | 1.214 | 1.03E+08 | 1.214 | 1.97E+09 | 1.214 | 2.08E+09 | 1.214 | 0 | 1.214 | 54889744 | 1.214 | 1.05E+09 | 1.214 | 1.11E+09 |
| 1.215 | 0 | 1.215 | 1.02E+08 | 1.215 | 1.97E+09 | 1.215 | 2.07E+09 | 1.215 | 0 | 1.215 | 54076541 | 1.215 | 1.05E+09 | 1.215 | 1.1E+09 |
| 1.216 | 0 | 1.216 | 1E+08 | 1.216 | 1.96E+09 | 1.216 | 2.06E+09 | 1.216 | 0 | 1.216 | 53273811 | 1.216 | 1.04E+09 | 1.216 | 1.1E+09 |
| 1.217 | 0 | 1.217 | 98580095 | 1.217 | 1.96E+09 | 1.217 | 2.05E+09 | 1.217 | 0 | 1.217 | 52481443 | 1.217 | 1.04E+09 | 1.217 | 1.09E+09 |
| 1.218 | 0 | 1.218 | 97110992 | 1.218 | 1.95E+09 | 1.218 | 2.05E+09 | 1.218 | 0 | 1.218 | 51699332 | 1.218 | 1.04E+09 | 1.218 | 1.09E+09 |
| 1.219 | 0 | 1.219 | 95660952 | 1.219 | 1.94E+09 | 1.219 | 2.04E+09 | 1.219 | 0 | 1.219 | 50927369 | 1.219 | 1.03E+09 | 1.219 | 1.09E+09 |
| 1.22 | 0 | 1.22 | 94229776 | 1.22 | 1.94E+09 | 1.22 | 2.03E+09 | 1.22 | 0 | 1.22 | 50165448 | 1.22 | 1.03E+09 | 1.22 | 1.08E+09 |
| 1.221 | 0 | 1.221 | 92817266 | 1.221 | 1.93E+09 | 1.221 | 2.02E+09 | 1.221 | 0 | 1.221 | 49413465 | 1.221 | 1.03E+09 | 1.221 | 1.08E+09 |
| 1.222 | 0 | 1.222 | 91423224 | 1.222 | 1.92E+09 | 1.222 | 2.02E+09 | 1.222 | 0 | 1.222 | 48671314 | 1.222 | 1.02E+09 | 1.222 | 1.07E+09 |
| 1.223 | 0 | 1.223 | 90047455 | 1.223 | 1.92E+09 | 1.223 | 2.01E+09 | 1.223 | 0 | 1.223 | 47938891 | 1.223 | 1.02E+09 | 1.223 | 1.07E+09 |
| 1.224 | 0 | 1.224 | 88689766 | 1.224 | 1.91E+09 | 1.224 | 2E+09 | 1.224 | 0 | 1.224 | 47216093 | 1.224 | 1.02E+09 | 1.224 | 1.06E+09 |
| 1.225 | 0 | 1.225 | 87349963 | 1.225 | 1.9E+09 | 1.225 | 1.99E+09 | 1.225 | 0 | 1.225 | 46502817 | 1.225 | 1.01E+09 | 1.225 | 1.06E+09 |
| 1.226 | 0 | 1.226 | 86027854 | 1.226 | 1.9E+09 | 1.226 | 1.98E+09 | 1.226 | 0 | 1.226 | 45798961 | 1.226 | 1.01E+09 | 1.226 | 1.06E+09 |
| 1.227 | 0 | 1.227 | 84723250 | 1.227 | 1.89E+09 | 1.227 | 1.98E+09 | 1.227 | 0 | 1.227 | 45104425 | 1.227 | 1.01E+09 | 1.227 | 1.05E+09 |
| 1.228 | 0 | 1.228 | 83435962 | 1.228 | 1.88E+09 | 1.228 | 1.97E+09 | 1.228 | 0 | 1.228 | 44419106 | 1.228 | 1E+09 | 1.228 | 1.05E+09 |
| 1.229 | 0 | 1.229 | 82165801 | 1.229 | 1.88E+09 | 1.229 | 1.96E+09 | 1.229 | 0 | 1.229 | 43742906 | 1.229 | 9.99E+08 | 1.229 | 1.04E+09 |
| 1.23 | 0 | 1.23 | 80912583 | 1.23 | 1.87E+09 | 1.23 | 1.95E+09 | 1.23 | 0 | 1.23 | 43075726 | 1.23 | 9.95E+08 | 1.23 | 1.04E+09 |
| 1.231 | 0 | 1.231 | 79676122 | 1.231 | 1.86E+09 | 1.231 | 1.94E+09 | 1.231 | 0 | 1.231 | 42417467 | 1.231 | 9.92E+08 | 1.231 | 1.03E+09 |
| 1.232 | 0 | 1.232 | 78456235 | 1.232 | 1.86E+09 | 1.232 | 1.93E+09 | 1.232 | 0 | 1.232 | 41768031 | 1.232 | 9.88E+08 | 1.232 | 1.03E+09 |
| 1.233 | 0 | 1.233 | 77252739 | 1.233 | 1.85E+09 | 1.233 | 1.93E+09 | 1.233 | 0 | 1.233 | 41127321 | 1.233 | 9.84E+08 | 1.233 | 1.03E+09 |
| 1.234 | 0 | 1.234 | 76065453 | 1.234 | 1.84E+09 | 1.234 | 1.92E+09 | 1.234 | 0 | 1.234 | 40495242 | 1.234 | 9.8E+08 | 1.234 | 1.02E+09 |
| 1.235 | 0 | 1.235 | 74894200 | 1.235 | 1.83E+09 | 1.235 | 1.91E+09 | 1.235 | 0 | 1.235 | 39871697 | 1.235 | 9.76E+08 | 1.235 | 1.02E+09 |
| 1.236 | 0 | 1.236 | 73738799 | 1.236 | 1.83E+09 | 1.236 | 1.9E+09 | 1.236 | 0 | 1.236 | 39256593 | 1.236 | 9.72E+08 | 1.236 | 1.01E+09 |
| 1.237 | 0 | 1.237 | 72599075 | 1.237 | 1.82E+09 | 1.237 | 1.89E+09 | 1.237 | 0 | 1.237 | 38649834 | 1.237 | 9.69E+08 | 1.237 | 1.01E+09 |
| 1.238 | 0 | 1.238 | 71474853 | 1.238 | 1.81E+09 | 1.238 | 1.88E+09 | 1.238 | 0 | 1.238 | 38051327 | 1.238 | 9.65E+08 | 1.238 | 1E+09 |
| 1.239 | 0 | 1.239 | 70365957 | 1.239 | 1.8E+09 | 1.239 | 1.87E+09 | 1.239 | 0 | 1.239 | 37460980 | 1.239 | 9.61E+08 | 1.239 | 9.98E+08 |
| 1.24 | 0 | 1.24 | 69272216 | 1.24 | 1.8E+09 | 1.24 | 1.87E+09 | 1.24 | 0 | 1.24 | 36878701 | 1.24 | 9.57E+08 | 1.24 | 9.94E+08 |
| 1.241 | 0 | 1.241 | 68193458 | 1.241 | 1.79E+09 | 1.241 | 1.86E+09 | 1.241 | 0 | 1.241 | 36304399 | 1.241 | 9.53E+08 | 1.241 | 9.89E+08 |
| 1.242 | 0 | 1.242 | 67129513 | 1.242 | 1.78E+09 | 1.242 | 1.85E+09 | 1.242 | 0 | 1.242 | 35737983 | 1.242 | 9.49E+08 | 1.242 | 9.85E+08 |
| 1.243 | 0 | 1.243 | 66080212 | 1.243 | 1.77E+09 | 1.243 | 1.84E+09 | 1.243 | 0 | 1.243 | 35179363 | 1.243 | 9.45E+08 | 1.243 | 9.8E+08 |
| 1.244 | 0 | 1.244 | 65045389 | 1.244 | 1.77E+09 | 1.244 | 1.83E+09 | 1.244 | 0 | 1.244 | 34628450 | 1.244 | 9.41E+08 | 1.244 | 9.75E+08 |
| 1.245 | 0 | 1.245 | 64024877 | 1.245 | 1.76E+09 | 1.245 | 1.82E+09 | 1.245 | 0 | 1.245 | 34085156 | 1.245 | 9.37E+08 | 1.245 | 9.71E+08 |
| 1.246 | 0 | 1.246 | 63018511 | 1.246 | 1.75E+09 | 1.246 | 1.81E+09 | 1.246 | 0 | 1.246 | 33549394 | 1.246 | 9.33E+08 | 1.246 | 9.66E+08 |
| 1.247 | 0 | 1.247 | 62026129 | 1.247 | 1.74E+09 | 1.247 | 1.81E+09 | 1.247 | 0 | 1.247 | 33021075 | 1.247 | 9.28E+08 | 1.247 | 9.61E+08 |
| 1.248 | 0 | 1.248 | 61047568 | 1.248 | 1.74E+09 | 1.248 | 1.8E+09 | 1.248 | 0 | 1.248 | 32500115 | 1.248 | 9.24E+08 | 1.248 | 9.57E+08 |
| 1.249 | 0 | 1.249 | 60082667 | 1.249 | 1.73E+09 | 1.249 | 1.79E+09 | 1.249 | 0 | 1.249 | 31986428 | 1.249 | 9.2E+08 | 1.249 | 9.52E+08 |
| 1.25 | 0 | 1.25 | 59131269 | 1.25 | 1.72E+09 | 1.25 | 1.78E+09 | 1.25 | 0 | 1.25 | 31479928 | 1.25 | 9.16E+08 | 1.25 | 9.48E+08 |
| 1.251 | 0 | 1.251 | 58193213 | 1.251 | 1.71E+09 | 1.251 | 1.77E+09 | 1.251 | 0 | 1.251 | 30980532 | 1.251 | 9.12E+08 | 1.251 | 9.43E+08 |
| 1.252 | 0 | 1.252 | 57268345 | 1.252 | 1.7E+09 | 1.252 | 1.76E+09 | 1.252 | 0 | 1.252 | 30488157 | 1.252 | 9.08E+08 | 1.252 | 9.38E+08 |
| 1.253 | 0 | 1.253 | 56356508 | 1.253 | 1.7E+09 | 1.253 | 1.75E+09 | 1.253 | 0 | 1.253 | 30002719 | 1.253 | 9.03E+08 | 1.253 | 9.33E+08 |
| 1.254 | 0 | 1.254 | 55457549 | 1.254 | 1.69E+09 | 1.254 | 1.74E+09 | 1.254 | 0 | 1.254 | 29524137 | 1.254 | 8.99E+08 | 1.254 | 9.29E+08 |
| 1.255 | 0 | 1.255 | 54571315 | 1.255 | 1.68E+09 | 1.255 | 1.74E+09 | 1.255 | 0 | 1.255 | 29052329 | 1.255 | 8.95E+08 | 1.255 | 9.24E+08 |
| 1.256 | 0 | 1.256 | 53697655 | 1.256 | 1.67E+09 | 1.256 | 1.73E+09 | 1.256 | 0 | 1.256 | 28587216 | 1.256 | 8.91E+08 | 1.256 | 9.19E+08 |
| 1.257 | 0 | 1.257 | 52836418 | 1.257 | 1.66E+09 | 1.257 | 1.72E+09 | 1.257 | 0 | 1.257 | 28128716 | 1.257 | 8.86E+08 | 1.257 | 9.15E+08 |
| 1.258 | 0 | 1.258 | 51987456 | 1.258 | 1.66E+09 | 1.258 | 1.71E+09 | 1.258 | 0 | 1.258 | 27676751 | 1.258 | 8.82E+08 | 1.258 | 9.1E+08 |
| 1.259 | 0 | 1.259 | 51150622 | 1.259 | 1.65E+09 | 1.259 | 1.7E+09 | 1.259 | 0 | 1.259 | 27231242 | 1.259 | 8.78E+08 | 1.259 | 9.05E+08 |
| 1.26 | 0 | 1.26 | 50325769 | 1.26 | 1.64E+09 | 1.26 | 1.69E+09 | 1.26 | 0 | 1.26 | 26792112 | 1.26 | 8.73E+08 | 1.26 | 9E+08 |
| 1.261 | 0 | 1.261 | 49512753 | 1.261 | 1.63E+09 | 1.261 | 1.68E+09 | 1.261 | 0 | 1.261 | 26359284 | 1.261 | 8.69E+08 | 1.261 | 8.95E+08 |
| 1.262 | 0 | 1.262 | 48711430 | 1.262 | 1.62E+09 | 1.262 | 1.67E+09 | 1.262 | 0 | 1.262 | 25932681 | 1.262 | 8.65E+08 | 1.262 | 8.91E+08 |
| 1.263 | 0 | 1.263 | 47921658 | 1.263 | 1.62E+09 | 1.263 | 1.66E+09 | 1.263 | 0 | 1.263 | 25512227 | 1.263 | 8.6E+08 | 1.263 | 8.86E+08 |
| 1.264 | 0 | 1.264 | 47143295 | 1.264 | 1.61E+09 | 1.264 | 1.66E+09 | 1.264 | 0 | 1.264 | 25097848 | 1.264 | 8.56E+08 | 1.264 | 8.81E+08 |
| 1.265 | 0 | 1.265 | 46376204 | 1.265 | 1.6E+09 | 1.265 | 1.65E+09 | 1.265 | 0 | 1.265 | 24689468 | 1.265 | 8.52E+08 | 1.265 | 8.76E+08 |
| 1.266 | 0 | 1.266 | 45620244 | 1.266 | 1.59E+09 | 1.266 | 1.64E+09 | 1.266 | 0 | 1.266 | 24287015 | 1.266 | 8.47E+08 | 1.266 | 8.71E+08 |
| 1.267 | 0 | 1.267 | 44875279 | 1.267 | 1.58E+09 | 1.267 | 1.63E+09 | 1.267 | 0 | 1.267 | 23890415 | 1.267 | 8.43E+08 | 1.267 | 8.67E+08 |
| 1.268 | 0 | 1.268 | 44141173 | 1.268 | 1.57E+09 | 1.268 | 1.62E+09 | 1.268 | 0 | 1.268 | 23499597 | 1.268 | 8.38E+08 | 1.268 | 8.62E+08 |
| 1.269 | 0 | 1.269 | 43417792 | 1.269 | 1.57E+09 | 1.269 | 1.61E+09 | 1.269 | 0 | 1.269 | 23114488 | 1.269 | 8.34E+08 | 1.269 | 8.57E+08 |
| 1.27 | 0 | 1.27 | 42705002 | 1.27 | 1.56E+09 | 1.27 | 1.6E+09 | 1.27 | 0 | 1.27 | 22735017 | 1.27 | 8.29E+08 | 1.27 | 8.52E+08 |
| 1.271 | 0 | 1.271 | 42002671 | 1.271 | 1.55E+09 | 1.271 | 1.59E+09 | 1.271 | 0 | 1.271 | 22361115 | 1.271 | 8.25E+08 | 1.271 | 8.47E+08 |
| 1.272 | 0 | 1.272 | 41310669 | 1.272 | 1.54E+09 | 1.272 | 1.58E+09 | 1.272 | 0 | 1.272 | 21992711 | 1.272 | 8.2E+08 | 1.272 | 8.42E+08 |
| 1.273 | 0 | 1.273 | 40628865 | 1.273 | 1.53E+09 | 1.273 | 1.57E+09 | 1.273 | 0 | 1.273 | 21629737 | 1.273 | 8.16E+08 | 1.273 | 8.38E+08 |
| 1.274 | 0 | 1.274 | 39957132 | 1.274 | 1.52E+09 | 1.274 | 1.56E+09 | 1.274 | 0 | 1.274 | 21272124 | 1.274 | 8.12E+08 | 1.274 | 8.33E+08 |
| 1.275 | 0 | 1.275 | 39295342 | 1.275 | 1.52E+09 | 1.275 | 1.56E+09 | 1.275 | 0 | 1.275 | 20919804 | 1.275 | 8.07E+08 | 1.275 | 8.28E+08 |
| 1.276 | 0 | 1.276 | 38643370 | 1.276 | 1.51E+09 | 1.276 | 1.55E+09 | 1.276 | 0 | 1.276 | 20572711 | 1.276 | 8.03E+08 | 1.276 | 8.23E+08 |
| 1.277 | 0 | 1.277 | 38001091 | 1.277 | 1.5E+09 | 1.277 | 1.54E+09 | 1.277 | 0 | 1.277 | 20230779 | 1.277 | 7.98E+08 | 1.277 | 8.18E+08 |
| 1.278 | 0 | 1.278 | 37368381 | 1.278 | 1.49E+09 | 1.278 | 1.53E+09 | 1.278 | 0 | 1.278 | 19893941 | 1.278 | 7.93E+08 | 1.278 | 8.13E+08 |
| 1.279 | 0 | 1.279 | 36745119 | 1.279 | 1.48E+09 | 1.279 | 1.52E+09 | 1.279 | 0 | 1.279 | 19562132 | 1.279 | 7.89E+08 | 1.279 | 8.08E+08 |
| 1.28 | 0 | 1.28 | 36131183 | 1.28 | 1.47E+09 | 1.28 | 1.51E+09 | 1.28 | 0 | 1.28 | 19235289 | 1.28 | 7.84E+08 | 1.28 | 8.04E+08 |
| 1.281 | 0 | 1.281 | 35526453 | 1.281 | 1.46E+09 | 1.281 | 1.5E+09 | 1.281 | 0 | 1.281 | 18913347 | 1.281 | 7.8E+08 | 1.281 | 7.99E+08 |
| 1.282 | 0 | 1.282 | 34930812 | 1.282 | 1.46E+09 | 1.282 | 1.49E+09 | 1.282 | 0 | 1.282 | 18596243 | 1.282 | 7.75E+08 | 1.282 | 7.94E+08 |
| 1.283 | 0 | 1.283 | 34344141 | 1.283 | 1.45E+09 | 1.283 | 1.48E+09 | 1.283 | 0 | 1.283 | 18283915 | 1.283 | 7.71E+08 | 1.283 | 7.89E+08 |
| 1.284 | 0 | 1.284 | 33766324 | 1.284 | 1.44E+09 | 1.284 | 1.47E+09 | 1.284 | 0 | 1.284 | 17976301 | 1.284 | 7.66E+08 | 1.284 | 7.84E+08 |
| 1.285 | 0 | 1.285 | 33197246 | 1.285 | 1.43E+09 | 1.285 | 1.46E+09 | 1.285 | 0 | 1.285 | 17673339 | 1.285 | 7.62E+08 | 1.285 | 7.79E+08 |
| 1.286 | 0 | 1.286 | 32636794 | 1.286 | 1.42E+09 | 1.286 | 1.45E+09 | 1.286 | 0 | 1.286 | 17374969 | 1.286 | 7.57E+08 | 1.286 | 7.74E+08 |
| 1.287 | 0 | 1.287 | 32084854 | 1.287 | 1.41E+09 | 1.287 | 1.45E+09 | 1.287 | 0 | 1.287 | 17081130 | 1.287 | 7.52E+08 | 1.287 | 7.7E+08 |
| 1.288 | 0 | 1.288 | 31541315 | 1.288 | 1.4E+09 | 1.288 | 1.44E+09 | 1.288 | 0 | 1.288 | 16791765 | 1.288 | 7.48E+08 | 1.288 | 7.65E+08 |
| 1.289 | 0 | 1.289 | 31006067 | 1.289 | 1.4E+09 | 1.289 | 1.43E+09 | 1.289 | 0 | 1.289 | 16506813 | 1.289 | 7.43E+08 | 1.289 | 7.6E+08 |
| 1.29 | 0 | 1.29 | 30479000 | 1.29 | 1.39E+09 | 1.29 | 1.42E+09 | 1.29 | 0 | 1.29 | 16226216 | 1.29 | 7.39E+08 | 1.29 | 7.55E+08 |
| 1.291 | 0 | 1.291 | 29960006 | 1.291 | 1.38E+09 | 1.291 | 1.41E+09 | 1.291 | 0 | 1.291 | 15949917 | 1.291 | 7.34E+08 | 1.291 | 7.5E+08 |
| 1.292 | 0 | 1.292 | 29448979 | 1.292 | 1.37E+09 | 1.292 | 1.4E+09 | 1.292 | 0 | 1.292 | 15677860 | 1.292 | 7.3E+08 | 1.292 | 7.45E+08 |
| 1.293 | 0 | 1.293 | 28945811 | 1.293 | 1.36E+09 | 1.293 | 1.39E+09 | 1.293 | 0 | 1.293 | 15409987 | 1.293 | 7.25E+08 | 1.293 | 7.4E+08 |
| 1.294 | 0 | 1.294 | 28450399 | 1.294 | 1.35E+09 | 1.294 | 1.38E+09 | 1.294 | 0 | 1.294 | 15146242 | 1.294 | 7.2E+08 | 1.294 | 7.36E+08 |
| 1.295 | 0 | 1.295 | 27962639 | 1.295 | 1.34E+09 | 1.295 | 1.37E+09 | 1.295 | 0 | 1.295 | 14886572 | 1.295 | 7.16E+08 | 1.295 | 7.31E+08 |
| 1.296 | 0 | 1.296 | 27482427 | 1.296 | 1.34E+09 | 1.296 | 1.36E+09 | 1.296 | 0 | 1.296 | 14630920 | 1.296 | 7.11E+08 | 1.296 | 7.26E+08 |
| 1.297 | 0 | 1.297 | 27009664 | 1.297 | 1.33E+09 | 1.297 | 1.35E+09 | 1.297 | 0 | 1.297 | 14379233 | 1.297 | 7.07E+08 | 1.297 | 7.21E+08 |
| 1.298 | 0 | 1.298 | 26544248 | 1.298 | 1.32E+09 | 1.298 | 1.35E+09 | 1.298 | 0 | 1.298 | 14131458 | 1.298 | 7.02E+08 | 1.298 | 7.16E+08 |
| 1.299 | 0 | 1.299 | 26086080 | 1.299 | 1.31E+09 | 1.299 | 1.34E+09 | 1.299 | 0 | 1.299 | 13887541 | 1.299 | 6.97E+08 | 1.299 | 7.11E+08 |
| 1.3 | 0 | 1.3 | 25635061 | 1.3 | 1.3E+09 | 1.3 | 1.33E+09 | 1.3 | 0 | 1.3 | 13647431 | 1.3 | 6.93E+08 | 1.3 | 7.07E+08 |
| 1.301 | 0 | 1.301 | 25191096 | 1.301 | 1.29E+09 | 1.301 | 1.32E+09 | 1.301 | 0 | 1.301 | 13411075 | 1.301 | 6.88E+08 | 1.301 | 7.02E+08 |
| 1.302 | 0 | 1.302 | 24754086 | 1.302 | 1.28E+09 | 1.302 | 1.31E+09 | 1.302 | 0 | 1.302 | 13178423 | 1.302 | 6.84E+08 | 1.302 | 6.97E+08 |
| 1.303 | 0 | 1.303 | 24323939 | 1.303 | 1.28E+09 | 1.303 | 1.3E+09 | 1.303 | 0 | 1.303 | 12949424 | 1.303 | 6.79E+08 | 1.303 | 6.92E+08 |
| 1.304 | 0 | 1.304 | 23900558 | 1.304 | 1.27E+09 | 1.304 | 1.29E+09 | 1.304 | 0 | 1.304 | 12724027 | 1.304 | 6.75E+08 | 1.304 | 6.87E+08 |
| 1.305 | 0 | 1.305 | 23483853 | 1.305 | 1.26E+09 | 1.305 | 1.28E+09 | 1.305 | 0 | 1.305 | 12502184 | 1.305 | 6.7E+08 | 1.305 | 6.83E+08 |
| 1.306 | 0 | 1.306 | 23073729 | 1.306 | 1.25E+09 | 1.306 | 1.27E+09 | 1.306 | 0 | 1.306 | 12283845 | 1.306 | 6.65E+08 | 1.306 | 6.78E+08 |
| 1.307 | 0 | 1.307 | 22670098 | 1.307 | 1.24E+09 | 1.307 | 1.26E+09 | 1.307 | 0 | 1.307 | 12068962 | 1.307 | 6.61E+08 | 1.307 | 6.73E+08 |
| 1.308 | 0 | 1.308 | 22272868 | 1.308 | 1.23E+09 | 1.308 | 1.26E+09 | 1.308 | 0 | 1.308 | 11857488 | 1.308 | 6.56E+08 | 1.308 | 6.68E+08 |
| 1.309 | 0 | 1.309 | 21881952 | 1.309 | 1.22E+09 | 1.309 | 1.25E+09 | 1.309 | 0 | 1.309 | 11649374 | 1.309 | 6.52E+08 | 1.309 | 6.63E+08 |
| 1.31 | 0 | 1.31 | 21497260 | 1.31 | 1.22E+09 | 1.31 | 1.24E+09 | 1.31 | 0 | 1.31 | 11444574 | 1.31 | 6.47E+08 | 1.31 | 6.59E+08 |
| 1.311 | 0 | 1.311 | 21118706 | 1.311 | 1.21E+09 | 1.311 | 1.23E+09 | 1.311 | 0 | 1.311 | 11243042 | 1.311 | 6.43E+08 | 1.311 | 6.54E+08 |
| 1.312 | 0 | 1.312 | 20746205 | 1.312 | 1.2E+09 | 1.312 | 1.22E+09 | 1.312 | 0 | 1.312 | 11044733 | 1.312 | 6.38E+08 | 1.312 | 6.49E+08 |
| 1.313 | 0 | 1.313 | 20379671 | 1.313 | 1.19E+09 | 1.313 | 1.21E+09 | 1.313 | 0 | 1.313 | 10849600 | 1.313 | 6.34E+08 | 1.313 | 6.44E+08 |
| 1.314 | 0 | 1.314 | 20019021 | 1.314 | 1.18E+09 | 1.314 | 1.2E+09 | 1.314 | 0 | 1.314 | 10657599 | 1.314 | 6.29E+08 | 1.314 | 6.4E+08 |
| 1.315 | 0 | 1.315 | 19664171 | 1.315 | 1.17E+09 | 1.315 | 1.19E+09 | 1.315 | 0 | 1.315 | 10468686 | 1.315 | 6.25E+08 | 1.315 | 6.35E+08 |
| 1.316 | 0 | 1.316 | 19315040 | 1.316 | 1.16E+09 | 1.316 | 1.18E+09 | 1.316 | 0 | 1.316 | 10282818 | 1.316 | 6.2E+08 | 1.316 | 6.3E+08 |
| 1.317 | 0 | 1.317 | 18971546 | 1.317 | 1.16E+09 | 1.317 | 1.18E+09 | 1.317 | 0 | 1.317 | 10099951 | 1.317 | 6.16E+08 | 1.317 | 6.26E+08 |
| 1.318 | 0 | 1.318 | 18633609 | 1.318 | 1.15E+09 | 1.318 | 1.17E+09 | 1.318 | 0 | 1.318 | 9920042 | 1.318 | 6.11E+08 | 1.318 | 6.21E+08 |
| 1.319 | 0 | 1.319 | 18301151 | 1.319 | 1.14E+09 | 1.319 | 1.16E+09 | 1.319 | 0 | 1.319 | 9743050 | 1.319 | 6.07E+08 | 1.319 | 6.16E+08 |
| 1.32 | 0 | 1.32 | 17974092 | 1.32 | 1.13E+09 | 1.32 | 1.15E+09 | 1.32 | 0 | 1.32 | 9568933 | 1.32 | 6.02E+08 | 1.32 | 6.12E+08 |
| 1.321 | 0 | 1.321 | 17652356 | 1.321 | 1.12E+09 | 1.321 | 1.14E+09 | 1.321 | 0 | 1.321 | 9397649 | 1.321 | 5.98E+08 | 1.321 | 6.07E+08 |
| 1.322 | 0 | 1.322 | 17335866 | 1.322 | 1.11E+09 | 1.322 | 1.13E+09 | 1.322 | 0 | 1.322 | 9229158 | 1.322 | 5.93E+08 | 1.322 | 6.02E+08 |
| 1.323 | 0 | 1.323 | 17024547 | 1.323 | 1.11E+09 | 1.323 | 1.12E+09 | 1.323 | 0 | 1.323 | 9063420 | 1.323 | 5.89E+08 | 1.323 | 5.98E+08 |
| 1.324 | 0 | 1.324 | 16718324 | 1.324 | 1.1E+09 | 1.324 | 1.11E+09 | 1.324 | 0 | 1.324 | 8900395 | 1.324 | 5.84E+08 | 1.324 | 5.93E+08 |
| 1.325 | 0 | 1.325 | 16417123 | 1.325 | 1.09E+09 | 1.325 | 1.11E+09 | 1.325 | 0 | 1.325 | 8740043 | 1.325 | 5.8E+08 | 1.325 | 5.89E+08 |
| 1.326 | 0 | 1.326 | 16120872 | 1.326 | 1.08E+09 | 1.326 | 1.1E+09 | 1.326 | 0 | 1.326 | 8582327 | 1.326 | 5.75E+08 | 1.326 | 5.84E+08 |
| 1.327 | 0 | 1.327 | 15829498 | 1.327 | 1.07E+09 | 1.327 | 1.09E+09 | 1.327 | 0 | 1.327 | 8427207 | 1.327 | 5.71E+08 | 1.327 | 5.79E+08 |
| 1.328 | 0 | 1.328 | 15542931 | 1.328 | 1.06E+09 | 1.328 | 1.08E+09 | 1.328 | 0 | 1.328 | 8274647 | 1.328 | 5.67E+08 | 1.328 | 5.75E+08 |
| 1.329 | 0 | 1.329 | 15261100 | 1.329 | 1.06E+09 | 1.329 | 1.07E+09 | 1.329 | 0 | 1.329 | 8124607 | 1.329 | 5.62E+08 | 1.329 | 5.7E+08 |
| 1.33 | 0 | 1.33 | 14983936 | 1.33 | 1.05E+09 | 1.33 | 1.06E+09 | 1.33 | 0 | 1.33 | 7977053 | 1.33 | 5.58E+08 | 1.33 | 5.66E+08 |
| 1.331 | 0 | 1.331 | 14711371 | 1.331 | 1.04E+09 | 1.331 | 1.05E+09 | 1.331 | 0 | 1.331 | 7831946 | 1.331 | 5.53E+08 | 1.331 | 5.61E+08 |
| 1.332 | 0 | 1.332 | 14443336 | 1.332 | 1.03E+09 | 1.332 | 1.05E+09 | 1.332 | 0 | 1.332 | 7689251 | 1.332 | 5.49E+08 | 1.332 | 5.57E+08 |
| 1.333 | 0 | 1.333 | 14179766 | 1.333 | 1.02E+09 | 1.333 | 1.04E+09 | 1.333 | 0 | 1.333 | 7548933 | 1.333 | 5.45E+08 | 1.333 | 5.52E+08 |
| 1.334 | 0 | 1.334 | 13920593 | 1.334 | 1.01E+09 | 1.334 | 1.03E+09 | 1.334 | 0 | 1.334 | 7410957 | 1.334 | 5.4E+08 | 1.334 | 5.48E+08 |
| 1.335 | 0 | 1.335 | 13665753 | 1.335 | 1.01E+09 | 1.335 | 1.02E+09 | 1.335 | 0 | 1.335 | 7275286 | 1.335 | 5.36E+08 | 1.335 | 5.43E+08 |
| 1.336 | 0 | 1.336 | 13415181 | 1.336 | 9.99E+08 | 1.336 | 1.01E+09 | 1.336 | 0 | 1.336 | 7141889 | 1.336 | 5.32E+08 | 1.336 | 5.39E+08 |
| 1.337 | 0 | 1.337 | 13168814 | 1.337 | 9.91E+08 | 1.337 | 1E+09 | 1.337 | 0 | 1.337 | 7010730 | 1.337 | 5.27E+08 | 1.337 | 5.34E+08 |
| 1.338 | 0 | 1.338 | 12926590 | 1.338 | 9.83E+08 | 1.338 | 9.96E+08 | 1.338 | 0 | 1.338 | 6881775 | 1.338 | 5.23E+08 | 1.338 | 5.3E+08 |
| 1.339 | 0 | 1.339 | 12688445 | 1.339 | 9.75E+08 | 1.339 | 9.87E+08 | 1.339 | 0 | 1.339 | 6754993 | 1.339 | 5.19E+08 | 1.339 | 5.26E+08 |
| 1.34 | 0 | 1.34 | 12454319 | 1.34 | 9.67E+08 | 1.34 | 9.79E+08 | 1.34 | 0 | 1.34 | 6630351 | 1.34 | 5.15E+08 | 1.34 | 5.21E+08 |
| 1.341 | 0 | 1.341 | 12224151 | 1.341 | 9.59E+08 | 1.341 | 9.71E+08 | 1.341 | 0 | 1.341 | 6507816 | 1.341 | 5.1E+08 | 1.341 | 5.17E+08 |
| 1.342 | 0 | 1.342 | 11997883 | 1.342 | 9.51E+08 | 1.342 | 9.63E+08 | 1.342 | 0 | 1.342 | 6387356 | 1.342 | 5.06E+08 | 1.342 | 5.12E+08 |
| 1.343 | 0 | 1.343 | 11775454 | 1.343 | 9.43E+08 | 1.343 | 9.55E+08 | 1.343 | 0 | 1.343 | 6268941 | 1.343 | 5.02E+08 | 1.343 | 5.08E+08 |
| 1.344 | 0 | 1.344 | 11556806 | 1.344 | 9.35E+08 | 1.344 | 9.46E+08 | 1.344 | 0 | 1.344 | 6152539 | 1.344 | 4.98E+08 | 1.344 | 5.04E+08 |
| 1.345 | 0 | 1.345 | 11341883 | 1.345 | 9.27E+08 | 1.345 | 9.38E+08 | 1.345 | 0 | 1.345 | 6038120 | 1.345 | 4.94E+08 | 1.345 | 5E+08 |
| 1.346 | 0 | 1.346 | 11130628 | 1.346 | 9.19E+08 | 1.346 | 9.3E+08 | 1.346 | 0 | 1.346 | 5925653 | 1.346 | 4.89E+08 | 1.346 | 4.95E+08 |
| 1.347 | 0 | 1.347 | 10922984 | 1.347 | 9.11E+08 | 1.347 | 9.22E+08 | 1.347 | 0 | 1.347 | 5815109 | 1.347 | 4.85E+08 | 1.347 | 4.91E+08 |
| 1.348 | 0 | 1.348 | 10718897 | 1.348 | 9.04E+08 | 1.348 | 9.14E+08 | 1.348 | 0 | 1.348 | 5706458 | 1.348 | 4.81E+08 | 1.348 | 4.87E+08 |
| 1.349 | 0 | 1.349 | 10518312 | 1.349 | 8.96E+08 | 1.349 | 9.06E+08 | 1.349 | 0 | 1.349 | 5599672 | 1.349 | 4.77E+08 | 1.349 | 4.83E+08 |
| 1.35 | 0 | 1.35 | 10321175 | 1.35 | 8.88E+08 | 1.35 | 8.98E+08 | 1.35 | 0 | 1.35 | 5494722 | 1.35 | 4.73E+08 | 1.35 | 4.78E+08 |
| 1.351 | 0 | 1.351 | 10127433 | 1.351 | 8.8E+08 | 1.351 | 8.91E+08 | 1.351 | 0 | 1.351 | 5391579 | 1.351 | 4.69E+08 | 1.351 | 4.74E+08 |
| 1.352 | 0 | 1.352 | 9937034 | 1.352 | 8.73E+08 | 1.352 | 8.83E+08 | 1.352 | 0 | 1.352 | 5290215 | 1.352 | 4.65E+08 | 1.352 | 4.7E+08 |
| 1.353 | 0 | 1.353 | 9749927 | 1.353 | 8.65E+08 | 1.353 | 8.75E+08 | 1.353 | 0 | 1.353 | 5190604 | 1.353 | 4.61E+08 | 1.353 | 4.66E+08 |
| 1.354 | 0 | 1.354 | 9566059 | 1.354 | 8.58E+08 | 1.354 | 8.67E+08 | 1.354 | 0 | 1.354 | 5092717 | 1.354 | 4.57E+08 | 1.354 | 4.62E+08 |
| 1.355 | 0 | 1.355 | 9385381 | 1.355 | 8.5E+08 | 1.355 | 8.59E+08 | 1.355 | 0 | 1.355 | 4996529 | 1.355 | 4.52E+08 | 1.355 | 4.57E+08 |
| 1.356 | 0 | 1.356 | 9207843 | 1.356 | 8.42E+08 | 1.356 | 8.52E+08 | 1.356 | 0 | 1.356 | 4902013 | 1.356 | 4.48E+08 | 1.356 | 4.53E+08 |
| 1.357 | 0 | 1.357 | 9033396 | 1.357 | 8.35E+08 | 1.357 | 8.44E+08 | 1.357 | 0 | 1.357 | 4809142 | 1.357 | 4.44E+08 | 1.357 | 4.49E+08 |
| 1.358 | 0 | 1.358 | 8861992 | 1.358 | 8.27E+08 | 1.358 | 8.36E+08 | 1.358 | 0 | 1.358 | 4717891 | 1.358 | 4.41E+08 | 1.358 | 4.45E+08 |
| 1.359 | 0 | 1.359 | 8693583 | 1.359 | 8.2E+08 | 1.359 | 8.29E+08 | 1.359 | 0 | 1.359 | 4628234 | 1.359 | 4.37E+08 | 1.359 | 4.41E+08 |
| 1.36 | 0 | 1.36 | 8528122 | 1.36 | 8.13E+08 | 1.36 | 8.21E+08 | 1.36 | 0 | 1.36 | 4540147 | 1.36 | 4.33E+08 | 1.36 | 4.37E+08 |
| 1.361 | 0 | 1.361 | 8365563 | 1.361 | 8.05E+08 | 1.361 | 8.14E+08 | 1.361 | 0 | 1.361 | 4453605 | 1.361 | 4.29E+08 | 1.361 | 4.33E+08 |
| 1.362 | 0 | 1.362 | 8205860 | 1.362 | 7.98E+08 | 1.362 | 8.06E+08 | 1.362 | 0 | 1.362 | 4368583 | 1.362 | 4.25E+08 | 1.362 | 4.29E+08 |
| 1.363 | 0 | 1.363 | 8048967 | 1.363 | 7.91E+08 | 1.363 | 7.99E+08 | 1.363 | 0 | 1.363 | 4285058 | 1.363 | 4.21E+08 | 1.363 | 4.25E+08 |
| 1.364 | 0 | 1.364 | 7894840 | 1.364 | 7.83E+08 | 1.364 | 7.91E+08 | 1.364 | 0 | 1.364 | 4203005 | 1.364 | 4.17E+08 | 1.364 | 4.21E+08 |
| 1.365 | 0 | 1.365 | 7743436 | 1.365 | 7.76E+08 | 1.365 | 7.84E+08 | 1.365 | 0 | 1.365 | 4122401 | 1.365 | 4.13E+08 | 1.365 | 4.17E+08 |
| 1.366 | 0 | 1.366 | 7594711 | 1.366 | 7.69E+08 | 1.366 | 7.76E+08 | 1.366 | 0 | 1.366 | 4043224 | 1.366 | 4.09E+08 | 1.366 | 4.13E+08 |
| 1.367 | 0 | 1.367 | 7448621 | 1.367 | 7.62E+08 | 1.367 | 7.69E+08 | 1.367 | 0 | 1.367 | 3965450 | 1.367 | 4.06E+08 | 1.367 | 4.09E+08 |
| 1.368 | 0 | 1.368 | 7305126 | 1.368 | 7.55E+08 | 1.368 | 7.62E+08 | 1.368 | 0 | 1.368 | 3889057 | 1.368 | 4.02E+08 | 1.368 | 4.06E+08 |
| 1.369 | 0 | 1.369 | 7164183 | 1.369 | 7.47E+08 | 1.369 | 7.55E+08 | 1.369 | 0 | 1.369 | 3814022 | 1.369 | 3.98E+08 | 1.369 | 4.02E+08 |
| 1.37 | 0 | 1.37 | 7025752 | 1.37 | 7.4E+08 | 1.37 | 7.47E+08 | 1.37 | 0 | 1.37 | 3740325 | 1.37 | 3.94E+08 | 1.37 | 3.98E+08 |
| 1.371 | 0 | 1.371 | 6889791 | 1.371 | 7.33E+08 | 1.371 | 7.4E+08 | 1.371 | 0 | 1.371 | 3667943 | 1.371 | 3.9E+08 | 1.371 | 3.94E+08 |
| 1.372 | 0 | 1.372 | 6756262 | 1.372 | 7.26E+08 | 1.372 | 7.33E+08 | 1.372 | 0 | 1.372 | 3596856 | 1.372 | 3.87E+08 | 1.372 | 3.9E+08 |
| 1.373 | 0 | 1.373 | 6625125 | 1.373 | 7.19E+08 | 1.373 | 7.26E+08 | 1.373 | 0 | 1.373 | 3527042 | 1.373 | 3.83E+08 | 1.373 | 3.87E+08 |
| 1.374 | 0 | 1.374 | 6496340 | 1.374 | 7.13E+08 | 1.374 | 7.19E+08 | 1.374 | 0 | 1.374 | 3458480 | 1.374 | 3.79E+08 | 1.374 | 3.83E+08 |
| 1.375 | 0 | 1.375 | 6369871 | 1.375 | 7.06E+08 | 1.375 | 7.12E+08 | 1.375 | 0 | 1.375 | 3391151 | 1.375 | 3.76E+08 | 1.375 | 3.79E+08 |
| 1.376 | 0 | 1.376 | 6245679 | 1.376 | 6.99E+08 | 1.376 | 7.05E+08 | 1.376 | 0 | 1.376 | 3325035 | 1.376 | 3.72E+08 | 1.376 | 3.75E+08 |
| 1.377 | 0 | 1.377 | 6123727 | 1.377 | 6.92E+08 | 1.377 | 6.98E+08 | 1.377 | 0 | 1.377 | 3260111 | 1.377 | 3.68E+08 | 1.377 | 3.72E+08 |
| 1.378 | 0 | 1.378 | 6003979 | 1.378 | 6.85E+08 | 1.378 | 6.91E+08 | 1.378 | 0 | 1.378 | 3196360 | 1.378 | 3.65E+08 | 1.378 | 3.68E+08 |
| 1.379 | 0 | 1.379 | 5886398 | 1.379 | 6.79E+08 | 1.379 | 6.84E+08 | 1.379 | 0 | 1.379 | 3133763 | 1.379 | 3.61E+08 | 1.379 | 3.64E+08 |
| 1.38 | 0 | 1.38 | 5770950 | 1.38 | 6.72E+08 | 1.38 | 6.78E+08 | 1.38 | 0 | 1.38 | 3072301 | 1.38 | 3.58E+08 | 1.38 | 3.61E+08 |
| 1.381 | 0 | 1.381 | 5657598 | 1.381 | 6.65E+08 | 1.381 | 6.71E+08 | 1.381 | 0 | 1.381 | 3011956 | 1.381 | 3.54E+08 | 1.381 | 3.57E+08 |
| 1.382 | 0 | 1.382 | 5546308 | 1.382 | 6.59E+08 | 1.382 | 6.64E+08 | 1.382 | 0 | 1.382 | 2952708 | 1.382 | 3.51E+08 | 1.382 | 3.54E+08 |
| 1.383 | 0 | 1.383 | 5437047 | 1.383 | 6.52E+08 | 1.383 | 6.58E+08 | 1.383 | 0 | 1.383 | 2894540 | 1.383 | 3.47E+08 | 1.383 | 3.5E+08 |
| 1.384 | 0 | 1.384 | 5329780 | 1.384 | 6.46E+08 | 1.384 | 6.51E+08 | 1.384 | 0 | 1.384 | 2837434 | 1.384 | 3.44E+08 | 1.384 | 3.47E+08 |
| 1.385 | 0 | 1.385 | 5224475 | 1.385 | 6.39E+08 | 1.385 | 6.44E+08 | 1.385 | 0 | 1.385 | 2781373 | 1.385 | 3.4E+08 | 1.385 | 3.43E+08 |
| 1.386 | 0 | 1.386 | 5121100 | 1.386 | 6.33E+08 | 1.386 | 6.38E+08 | 1.386 | 0 | 1.386 | 2726338 | 1.386 | 3.37E+08 | 1.386 | 3.4E+08 |
| 1.387 | 0 | 1.387 | 5019621 | 1.387 | 6.26E+08 | 1.387 | 6.31E+08 | 1.387 | 0 | 1.387 | 2672314 | 1.387 | 3.33E+08 | 1.387 | 3.36E+08 |
| 1.388 | 0 | 1.388 | 4920007 | 1.388 | 6.2E+08 | 1.388 | 6.25E+08 | 1.388 | 0 | 1.388 | 2619282 | 1.388 | 3.3E+08 | 1.388 | 3.33E+08 |
| 1.389 | 0 | 1.389 | 4822228 | 1.389 | 6.14E+08 | 1.389 | 6.18E+08 | 1.389 | 0 | 1.389 | 2567227 | 1.389 | 3.27E+08 | 1.389 | 3.29E+08 |
| 1.39 | 0 | 1.39 | 4726252 | 1.39 | 6.07E+08 | 1.39 | 6.12E+08 | 1.39 | 0 | 1.39 | 2516132 | 1.39 | 3.23E+08 | 1.39 | 3.26E+08 |
| 1.391 | 0 | 1.391 | 4632049 | 1.391 | 6.01E+08 | 1.391 | 6.06E+08 | 1.391 | 0 | 1.391 | 2465981 | 1.391 | 3.2E+08 | 1.391 | 3.22E+08 |
| 1.392 | 0 | 1.392 | 4539590 | 1.392 | 5.95E+08 | 1.392 | 5.99E+08 | 1.392 | 0 | 1.392 | 2416758 | 1.392 | 3.17E+08 | 1.392 | 3.19E+08 |
| 1.393 | 0 | 1.393 | 4448844 | 1.393 | 5.89E+08 | 1.393 | 5.93E+08 | 1.393 | 0 | 1.393 | 2368447 | 1.393 | 3.13E+08 | 1.393 | 3.16E+08 |
| 1.394 | 0 | 1.394 | 4359784 | 1.394 | 5.82E+08 | 1.394 | 5.87E+08 | 1.394 | 0 | 1.394 | 2321034 | 1.394 | 3.1E+08 | 1.394 | 3.12E+08 |
| 1.395 | 0 | 1.395 | 4272380 | 1.395 | 5.76E+08 | 1.395 | 5.81E+08 | 1.395 | 0 | 1.395 | 2274502 | 1.395 | 3.07E+08 | 1.395 | 3.09E+08 |
| 1.396 | 0 | 1.396 | 4186604 | 1.396 | 5.7E+08 | 1.396 | 5.74E+08 | 1.396 | 0 | 1.396 | 2228838 | 1.396 | 3.04E+08 | 1.396 | 3.06E+08 |
| 1.397 | 0 | 1.397 | 4102429 | 1.397 | 5.64E+08 | 1.397 | 5.68E+08 | 1.397 | 0 | 1.397 | 2184025 | 1.397 | 3E+08 | 1.397 | 3.03E+08 |
| 1.398 | 0 | 1.398 | 4019827 | 1.398 | 5.58E+08 | 1.398 | 5.62E+08 | 1.398 | 0 | 1.398 | 2140050 | 1.398 | 2.97E+08 | 1.398 | 2.99E+08 |
| 1.399 | 0 | 1.399 | 3938773 | 1.399 | 5.52E+08 | 1.399 | 5.56E+08 | 1.399 | 0 | 1.399 | 2096899 | 1.399 | 2.94E+08 | 1.399 | 2.96E+08 |
| 1.4 | 0 | 1.4 | 3859238 | 1.4 | 5.46E+08 | 1.4 | 5.5E+08 | 1.4 | 0 | 1.4 | 2054557 | 1.4 | 2.91E+08 | 1.4 | 2.93E+08 |
| 1.401 | 0 | 1.401 | 3781197 | 1.401 | 5.41E+08 | 1.401 | 5.44E+08 | 1.401 | 0 | 1.401 | 2013010 | 1.401 | 2.88E+08 | 1.401 | 2.9E+08 |
| 1.402 | 0 | 1.402 | 3704626 | 1.402 | 5.35E+08 | 1.402 | 5.39E+08 | 1.402 | 0 | 1.402 | 1972245 | 1.402 | 2.85E+08 | 1.402 | 2.87E+08 |
| 1.403 | 0 | 1.403 | 3629497 | 1.403 | 5.29E+08 | 1.403 | 5.33E+08 | 1.403 | 0 | 1.403 | 1932248 | 1.403 | 2.82E+08 | 1.403 | 2.84E+08 |
| 1.404 | 0 | 1.404 | 3555786 | 1.404 | 5.23E+08 | 1.404 | 5.27E+08 | 1.404 | 0 | 1.404 | 1893007 | 1.404 | 2.79E+08 | 1.404 | 2.81E+08 |
| 1.405 | 0 | 1.405 | 3483470 | 1.405 | 5.18E+08 | 1.405 | 5.21E+08 | 1.405 | 0 | 1.405 | 1854508 | 1.405 | 2.76E+08 | 1.405 | 2.77E+08 |
| 1.406 | 0 | 1.406 | 3412523 | 1.406 | 5.12E+08 | 1.406 | 5.15E+08 | 1.406 | 0 | 1.406 | 1816737 | 1.406 | 2.73E+08 | 1.406 | 2.74E+08 |
| 1.407 | 0 | 1.407 | 3342923 | 1.407 | 5.06E+08 | 1.407 | 5.1E+08 | 1.407 | 0 | 1.407 | 1779684 | 1.407 | 2.7E+08 | 1.407 | 2.71E+08 |
| 1.408 | 0 | 1.408 | 3274645 | 1.408 | 5.01E+08 | 1.408 | 5.04E+08 | 1.408 | 0 | 1.408 | 1743335 | 1.408 | 2.67E+08 | 1.408 | 2.68E+08 |
| 1.409 | 0 | 1.409 | 3207666 | 1.409 | 4.95E+08 | 1.409 | 4.99E+08 | 1.409 | 0 | 1.409 | 1707677 | 1.409 | 2.64E+08 | 1.409 | 2.65E+08 |
| 1.41 | 0 | 1.41 | 3141965 | 1.41 | 4.9E+08 | 1.41 | 4.93E+08 | 1.41 | 0 | 1.41 | 1672699 | 1.41 | 2.61E+08 | 1.41 | 2.62E+08 |
| 1.411 | 0 | 1.411 | 3077518 | 1.411 | 4.84E+08 | 1.411 | 4.87E+08 | 1.411 | 0 | 1.411 | 1638390 | 1.411 | 2.58E+08 | 1.411 | 2.6E+08 |
| 1.412 | 0 | 1.412 | 3014305 | 1.412 | 4.79E+08 | 1.412 | 4.82E+08 | 1.412 | 0 | 1.412 | 1604736 | 1.412 | 2.55E+08 | 1.412 | 2.57E+08 |
| 1.413 | 0 | 1.413 | 2952302 | 1.413 | 4.74E+08 | 1.413 | 4.77E+08 | 1.413 | 0 | 1.413 | 1571728 | 1.413 | 2.52E+08 | 1.413 | 2.54E+08 |
| 1.414 | 0 | 1.414 | 2891489 | 1.414 | 4.68E+08 | 1.414 | 4.71E+08 | 1.414 | 0 | 1.414 | 1539352 | 1.414 | 2.49E+08 | 1.414 | 2.51E+08 |
| 1.415 | 0 | 1.415 | 2831845 | 1.415 | 4.63E+08 | 1.415 | 4.66E+08 | 1.415 | 0 | 1.415 | 1507599 | 1.415 | 2.46E+08 | 1.415 | 2.48E+08 |
| 1.416 | 0 | 1.416 | 2773349 | 1.416 | 4.58E+08 | 1.416 | 4.61E+08 | 1.416 | 0 | 1.416 | 1476458 | 1.416 | 2.44E+08 | 1.416 | 2.45E+08 |
| 1.417 | 0 | 1.417 | 2715981 | 1.417 | 4.53E+08 | 1.417 | 4.55E+08 | 1.417 | 0 | 1.417 | 1445917 | 1.417 | 2.41E+08 | 1.417 | 2.42E+08 |
| 1.418 | 0 | 1.418 | 2659721 | 1.418 | 4.47E+08 | 1.418 | 4.5E+08 | 1.418 | 0 | 1.418 | 1415965 | 1.418 | 2.38E+08 | 1.418 | 2.4E+08 |
| 1.419 | 0 | 1.419 | 2604549 | 1.419 | 4.42E+08 | 1.419 | 4.45E+08 | 1.419 | 0 | 1.419 | 1386593 | 1.419 | 2.35E+08 | 1.419 | 2.37E+08 |
| 1.42 | 0 | 1.42 | 2550447 | 1.42 | 4.37E+08 | 1.42 | 4.4E+08 | 1.42 | 0 | 1.42 | 1357791 | 1.42 | 2.33E+08 | 1.42 | 2.34E+08 |
| 1.421 | 0 | 1.421 | 2497394 | 1.421 | 4.32E+08 | 1.421 | 4.35E+08 | 1.421 | 0 | 1.421 | 1329547 | 1.421 | 2.3E+08 | 1.421 | 2.31E+08 |
| 1.422 | 0 | 1.422 | 2445373 | 1.422 | 4.27E+08 | 1.422 | 4.3E+08 | 1.422 | 0 | 1.422 | 1301852 | 1.422 | 2.27E+08 | 1.422 | 2.29E+08 |
| 1.423 | 0 | 1.423 | 2394364 | 1.423 | 4.22E+08 | 1.423 | 4.25E+08 | 1.423 | 0 | 1.423 | 1274696 | 1.423 | 2.25E+08 | 1.423 | 2.26E+08 |
| 1.424 | 0 | 1.424 | 2344350 | 1.424 | 4.17E+08 | 1.424 | 4.2E+08 | 1.424 | 0 | 1.424 | 1248070 | 1.424 | 2.22E+08 | 1.424 | 2.23E+08 |
| 1.425 | 0 | 1.425 | 2295313 | 1.425 | 4.12E+08 | 1.425 | 4.15E+08 | 1.425 | 0 | 1.425 | 1221964 | 1.425 | 2.2E+08 | 1.425 | 2.21E+08 |
| 1.426 | 0 | 1.426 | 2247235 | 1.426 | 4.08E+08 | 1.426 | 4.1E+08 | 1.426 | 0 | 1.426 | 1196369 | 1.426 | 2.17E+08 | 1.426 | 2.18E+08 |
| 1.427 | 0 | 1.427 | 2200099 | 1.427 | 4.03E+08 | 1.427 | 4.05E+08 | 1.427 | 0 | 1.427 | 1171275 | 1.427 | 2.14E+08 | 1.427 | 2.16E+08 |
| 1.428 | 0 | 1.428 | 2153888 | 1.428 | 3.98E+08 | 1.428 | 4E+08 | 1.428 | 0 | 1.428 | 1146673 | 1.428 | 2.12E+08 | 1.428 | 2.13E+08 |
| 1.429 | 0 | 1.429 | 2108585 | 1.429 | 3.93E+08 | 1.429 | 3.95E+08 | 1.429 | 0 | 1.429 | 1122555 | 1.429 | 2.09E+08 | 1.429 | 2.11E+08 |
| 1.43 | 0 | 1.43 | 2064174 | 1.43 | 3.89E+08 | 1.43 | 3.91E+08 | 1.43 | 0 | 1.43 | 1098912 | 1.43 | 2.07E+08 | 1.43 | 2.08E+08 |
| 1.431 | 0 | 1.431 | 2020638 | 1.431 | 3.84E+08 | 1.431 | 3.86E+08 | 1.431 | 0 | 1.431 | 1075735 | 1.431 | 2.04E+08 | 1.431 | 2.06E+08 |
| 1.432 | 0 | 1.432 | 1977963 | 1.432 | 3.79E+08 | 1.432 | 3.81E+08 | 1.432 | 0 | 1.432 | 1053015 | 1.432 | 2.02E+08 | 1.432 | 2.03E+08 |
| 1.433 | 0 | 1.433 | 1936131 | 1.433 | 3.75E+08 | 1.433 | 3.77E+08 | 1.433 | 0 | 1.433 | 1030745 | 1.433 | 2E+08 | 1.433 | 2.01E+08 |
| 1.434 | 0 | 1.434 | 1895128 | 1.434 | 3.7E+08 | 1.434 | 3.72E+08 | 1.434 | 0 | 1.434 | 1008916 | 1.434 | 1.97E+08 | 1.434 | 1.98E+08 |
| 1.435 | 0 | 1.435 | 1854938 | 1.435 | 3.66E+08 | 1.435 | 3.68E+08 | 1.435 | 0 | 1.435 | 987520.2 | 1.435 | 1.95E+08 | 1.435 | 1.96E+08 |
| 1.436 | 0 | 1.436 | 1815547 | 1.436 | 3.61E+08 | 1.436 | 3.63E+08 | 1.436 | 0 | 1.436 | 966549.4 | 1.436 | 1.92E+08 | 1.436 | 1.93E+08 |
| 1.437 | 0 | 1.437 | 1776940 | 1.437 | 3.57E+08 | 1.437 | 3.59E+08 | 1.437 | 0 | 1.437 | 945996 | 1.437 | 1.9E+08 | 1.437 | 1.91E+08 |
| 1.438 | 0 | 1.438 | 1739102 | 1.438 | 3.53E+08 | 1.438 | 3.54E+08 | 1.438 | 0 | 1.438 | 925852.3 | 1.438 | 1.88E+08 | 1.438 | 1.89E+08 |
| 1.439 | 0 | 1.439 | 1702020 | 1.439 | 3.48E+08 | 1.439 | 3.5E+08 | 1.439 | 0 | 1.439 | 906110.6 | 1.439 | 1.85E+08 | 1.439 | 1.86E+08 |
| 1.44 | 0 | 1.44 | 1665679 | 1.44 | 3.44E+08 | 1.44 | 3.46E+08 | 1.44 | 0 | 1.44 | 886763.7 | 1.44 | 1.83E+08 | 1.44 | 1.84E+08 |
| 1.441 | 0 | 1.441 | 1630066 | 1.441 | 3.4E+08 | 1.441 | 3.41E+08 | 1.441 | 0 | 1.441 | 867804.2 | 1.441 | 1.81E+08 | 1.441 | 1.82E+08 |
| 1.442 | 0 | 1.442 | 1595167 | 1.442 | 3.36E+08 | 1.442 | 3.37E+08 | 1.442 | 0 | 1.442 | 849225 | 1.442 | 1.79E+08 | 1.442 | 1.79E+08 |
| 1.443 | 0 | 1.443 | 1560969 | 1.443 | 3.31E+08 | 1.443 | 3.33E+08 | 1.443 | 0 | 1.443 | 831018.9 | 1.443 | 1.76E+08 | 1.443 | 1.77E+08 |
| 1.444 | 0 | 1.444 | 1527459 | 1.444 | 3.27E+08 | 1.444 | 3.29E+08 | 1.444 | 0 | 1.444 | 813179.1 | 1.444 | 1.74E+08 | 1.444 | 1.75E+08 |
| 1.445 | 0 | 1.445 | 1494625 | 1.445 | 3.23E+08 | 1.445 | 3.25E+08 | 1.445 | 0 | 1.445 | 795698.7 | 1.445 | 1.72E+08 | 1.445 | 1.73E+08 |
| 1.446 | 0 | 1.446 | 1462452 | 1.446 | 3.19E+08 | 1.446 | 3.21E+08 | 1.446 | 0 | 1.446 | 778571 | 1.446 | 1.7E+08 | 1.446 | 1.71E+08 |
| 1.447 | 0 | 1.447 | 1430930 | 1.447 | 3.15E+08 | 1.447 | 3.17E+08 | 1.447 | 0 | 1.447 | 761789.5 | 1.447 | 1.68E+08 | 1.447 | 1.69E+08 |
| 1.448 | 0 | 1.448 | 1400046 | 1.448 | 3.11E+08 | 1.448 | 3.13E+08 | 1.448 | 0 | 1.448 | 745347.6 | 1.448 | 1.66E+08 | 1.448 | 1.66E+08 |
| 1.449 | 0 | 1.449 | 1369788 | 1.449 | 3.07E+08 | 1.449 | 3.09E+08 | 1.449 | 0 | 1.449 | 729239.1 | 1.449 | 1.64E+08 | 1.449 | 1.64E+08 |
| 1.45 | 0 | 1.45 | 1340144 | 1.45 | 3.03E+08 | 1.45 | 3.05E+08 | 1.45 | 0 | 1.45 | 713457.6 | 1.45 | 1.61E+08 | 1.45 | 1.62E+08 |
| 1.451 | 0 | 1.451 | 1311103 | 1.451 | 2.99E+08 | 1.451 | 3.01E+08 | 1.451 | 0 | 1.451 | 697996.9 | 1.451 | 1.59E+08 | 1.451 | 1.6E+08 |
| 1.452 | 0 | 1.452 | 1282654 | 1.452 | 2.96E+08 | 1.452 | 2.97E+08 | 1.452 | 0 | 1.452 | 682851.1 | 1.452 | 1.57E+08 | 1.452 | 1.58E+08 |
| 1.453 | 0 | 1.453 | 1254784 | 1.453 | 2.92E+08 | 1.453 | 2.93E+08 | 1.453 | 0 | 1.453 | 668014.1 | 1.453 | 1.55E+08 | 1.453 | 1.56E+08 |
| 1.454 | 0 | 1.454 | 1227484 | 1.454 | 2.88E+08 | 1.454 | 2.89E+08 | 1.454 | 0 | 1.454 | 653480.2 | 1.454 | 1.53E+08 | 1.454 | 1.54E+08 |
| 1.455 | 0 | 1.455 | 1200742 | 1.455 | 2.84E+08 | 1.455 | 2.86E+08 | 1.455 | 0 | 1.455 | 639243.6 | 1.455 | 1.51E+08 | 1.455 | 1.52E+08 |
| 1.456 | 0 | 1.456 | 1174549 | 1.456 | 2.81E+08 | 1.456 | 2.82E+08 | 1.456 | 0 | 1.456 | 625298.7 | 1.456 | 1.49E+08 | 1.456 | 1.5E+08 |
| 1.457 | 0 | 1.457 | 1148892 | 1.457 | 2.77E+08 | 1.457 | 2.78E+08 | 1.457 | 0 | 1.457 | 611639.9 | 1.457 | 1.47E+08 | 1.457 | 1.48E+08 |
| 1.458 | 0 | 1.458 | 1123763 | 1.458 | 2.73E+08 | 1.458 | 2.75E+08 | 1.458 | 0 | 1.458 | 598261.7 | 1.458 | 1.46E+08 | 1.458 | 1.46E+08 |
| 1.459 | 0 | 1.459 | 1099151 | 1.459 | 2.7E+08 | 1.459 | 2.71E+08 | 1.459 | 0 | 1.459 | 585158.8 | 1.459 | 1.44E+08 | 1.459 | 1.44E+08 |
| 1.46 | 0 | 1.46 | 1075046 | 1.46 | 2.66E+08 | 1.46 | 2.67E+08 | 1.46 | 0 | 1.46 | 572326 | 1.46 | 1.42E+08 | 1.46 | 1.42E+08 |
| 1.461 | 0 | 1.461 | 1051438 | 1.461 | 2.63E+08 | 1.461 | 2.64E+08 | 1.461 | 0 | 1.461 | 559758 | 1.461 | 1.4E+08 | 1.461 | 1.4E+08 |
| 1.462 | 0 | 1.462 | 1028319 | 1.462 | 2.59E+08 | 1.462 | 2.6E+08 | 1.462 | 0 | 1.462 | 547449.8 | 1.462 | 1.38E+08 | 1.462 | 1.39E+08 |
| 1.463 | 0 | 1.463 | 1005678 | 1.463 | 2.56E+08 | 1.463 | 2.57E+08 | 1.463 | 0 | 1.463 | 535396.5 | 1.463 | 1.36E+08 | 1.463 | 1.37E+08 |
| 1.464 | 0 | 1.464 | 983506.6 | 1.464 | 2.52E+08 | 1.464 | 2.53E+08 | 1.464 | 0 | 1.464 | 523593 | 1.464 | 1.34E+08 | 1.464 | 1.35E+08 |
| 1.465 | 0 | 1.465 | 961795.5 | 1.465 | 2.49E+08 | 1.465 | 2.5E+08 | 1.465 | 0 | 1.465 | 512034.6 | 1.465 | 1.33E+08 | 1.465 | 1.33E+08 |
| 1.466 | 0 | 1.466 | 940535.9 | 1.466 | 2.46E+08 | 1.466 | 2.47E+08 | 1.466 | 0 | 1.466 | 500716.5 | 1.466 | 1.31E+08 | 1.466 | 1.31E+08 |
| 1.467 | 0 | 1.467 | 919719 | 1.467 | 2.42E+08 | 1.467 | 2.43E+08 | 1.467 | 0 | 1.467 | 489634.1 | 1.467 | 1.29E+08 | 1.467 | 1.3E+08 |
| 1.468 | 0 | 1.468 | 899336.2 | 1.468 | 2.39E+08 | 1.468 | 2.4E+08 | 1.468 | 0 | 1.468 | 478782.9 | 1.468 | 1.27E+08 | 1.468 | 1.28E+08 |
| 1.469 | 0 | 1.469 | 879379.1 | 1.469 | 2.36E+08 | 1.469 | 2.37E+08 | 1.469 | 0 | 1.469 | 468158.2 | 1.469 | 1.26E+08 | 1.469 | 1.26E+08 |
| 1.47 | 0 | 1.47 | 859839.4 | 1.47 | 2.33E+08 | 1.47 | 2.34E+08 | 1.47 | 0 | 1.47 | 457755.8 | 1.47 | 1.24E+08 | 1.47 | 1.24E+08 |
| 1.471 | 0 | 1.471 | 840709.1 | 1.471 | 2.3E+08 | 1.471 | 2.3E+08 | 1.471 | 0 | 1.471 | 447571.3 | 1.471 | 1.22E+08 | 1.471 | 1.23E+08 |
| 1.472 | 0 | 1.472 | 821980 | 1.472 | 2.27E+08 | 1.472 | 2.27E+08 | 1.472 | 0 | 1.472 | 437600.5 | 1.472 | 1.21E+08 | 1.472 | 1.21E+08 |
| 1.473 | 0 | 1.473 | 803644.4 | 1.473 | 2.23E+08 | 1.473 | 2.24E+08 | 1.473 | 0 | 1.473 | 427839.1 | 1.473 | 1.19E+08 | 1.473 | 1.19E+08 |
| 1.474 | 0 | 1.474 | 785694.6 | 1.474 | 2.2E+08 | 1.474 | 2.21E+08 | 1.474 | 0 | 1.474 | 418283.1 | 1.474 | 1.17E+08 | 1.474 | 1.18E+08 |
| 1.475 | 0 | 1.475 | 768122.9 | 1.475 | 2.17E+08 | 1.475 | 2.18E+08 | 1.475 | 0 | 1.475 | 408928.4 | 1.475 | 1.16E+08 | 1.475 | 1.16E+08 |
| 1.476 | 0 | 1.476 | 750922 | 1.476 | 2.14E+08 | 1.476 | 2.15E+08 | 1.476 | 0 | 1.476 | 399771.1 | 1.476 | 1.14E+08 | 1.476 | 1.15E+08 |
| 1.477 | 0 | 1.477 | 734084.6 | 1.477 | 2.11E+08 | 1.477 | 2.12E+08 | 1.477 | 0 | 1.477 | 390807.3 | 1.477 | 1.13E+08 | 1.477 | 1.13E+08 |
| 1.478 | 0 | 1.478 | 717603.5 | 1.478 | 2.08E+08 | 1.478 | 2.09E+08 | 1.478 | 0 | 1.478 | 382033.2 | 1.478 | 1.11E+08 | 1.478 | 1.11E+08 |
| 1.479 | 0 | 1.479 | 701471.7 | 1.479 | 2.06E+08 | 1.479 | 2.06E+08 | 1.479 | 0 | 1.479 | 373445 | 1.479 | 1.09E+08 | 1.479 | 1.1E+08 |
| 1.48 | 0 | 1.48 | 685682.2 | 1.48 | 2.03E+08 | 1.48 | 2.03E+08 | 1.48 | 0 | 1.48 | 365039.1 | 1.48 | 1.08E+08 | 1.48 | 1.08E+08 |
| 1.481 | 0 | 1.481 | 670228.3 | 1.481 | 2E+08 | 1.481 | 2.01E+08 | 1.481 | 0 | 1.481 | 356811.9 | 1.481 | 1.06E+08 | 1.481 | 1.07E+08 |
| 1.482 | 0 | 1.482 | 655103.3 | 1.482 | 1.97E+08 | 1.482 | 1.98E+08 | 1.482 | 0 | 1.482 | 348759.7 | 1.482 | 1.05E+08 | 1.482 | 1.05E+08 |
| 1.483 | 0 | 1.483 | 640300.7 | 1.483 | 1.94E+08 | 1.483 | 1.95E+08 | 1.483 | 0 | 1.483 | 340879.2 | 1.483 | 1.03E+08 | 1.483 | 1.04E+08 |
| 1.484 | 0 | 1.484 | 625814.1 | 1.484 | 1.92E+08 | 1.484 | 1.92E+08 | 1.484 | 0 | 1.484 | 333166.9 | 1.484 | 1.02E+08 | 1.484 | 1.02E+08 |
| 1.485 | 0 | 1.485 | 611637.1 | 1.485 | 1.89E+08 | 1.485 | 1.89E+08 | 1.485 | 0 | 1.485 | 325619.5 | 1.485 | 1.01E+08 | 1.485 | 1.01E+08 |
| 1.486 | 0 | 1.486 | 597763.6 | 1.486 | 1.86E+08 | 1.486 | 1.87E+08 | 1.486 | 0 | 1.486 | 318233.6 | 1.486 | 99121015 | 1.486 | 99439249 |
| 1.487 | 0 | 1.487 | 584187.5 | 1.487 | 1.84E+08 | 1.487 | 1.84E+08 | 1.487 | 0 | 1.487 | 311006 | 1.487 | 97711894 | 1.487 | 98022900 |
| 1.488 | 0 | 1.488 | 570902.8 | 1.488 | 1.81E+08 | 1.488 | 1.81E+08 | 1.488 | 0 | 1.488 | 303933.6 | 1.488 | 96318826 | 1.488 | 96622759 |
| 1.489 | 0 | 1.489 | 557903.7 | 1.489 | 1.78E+08 | 1.489 | 1.79E+08 | 1.489 | 0 | 1.489 | 297013.2 | 1.489 | 94941695 | 1.489 | 95238708 |
| 1.49 | 0 | 1.49 | 545184.5 | 1.49 | 1.76E+08 | 1.49 | 1.76E+08 | 1.49 | 0 | 1.49 | 290241.9 | 1.49 | 93580386 | 1.49 | 93870628 |
| 1.491 | 0 | 1.491 | 532739.5 | 1.491 | 1.73E+08 | 1.491 | 1.74E+08 | 1.491 | 0 | 1.491 | 283616.5 | 1.491 | 92234785 | 1.491 | 92518402 |
| 1.492 | 0 | 1.492 | 520563.2 | 1.492 | 1.71E+08 | 1.492 | 1.71E+08 | 1.492 | 0 | 1.492 | 277134.1 | 1.492 | 90904776 | 1.492 | 91181911 |
| 1.493 | 0 | 1.493 | 508650.1 | 1.493 | 1.68E+08 | 1.493 | 1.69E+08 | 1.493 | 0 | 1.493 | 270791.9 | 1.493 | 89590244 | 1.493 | 89861036 |
| 1.494 | 0 | 1.494 | 496995 | 1.494 | 1.66E+08 | 1.494 | 1.66E+08 | 1.494 | 0 | 1.494 | 264587 | 1.494 | 88291072 | 1.494 | 88555659 |
| 1.495 | 0 | 1.495 | 485592.5 | 1.495 | 1.63E+08 | 1.495 | 1.64E+08 | 1.495 | 0 | 1.495 | 258516.7 | 1.495 | 87007144 | 1.495 | 87265661 |
| 1.496 | 0 | 1.496 | 474437.7 | 1.496 | 1.61E+08 | 1.496 | 1.62E+08 | 1.496 | 0 | 1.496 | 252578.1 | 1.496 | 85738344 | 1.496 | 85990922 |
| 1.497 | 0 | 1.497 | 463525.3 | 1.497 | 1.59E+08 | 1.497 | 1.59E+08 | 1.497 | 0 | 1.497 | 246768.7 | 1.497 | 84484555 | 1.497 | 84731324 |
| 1.498 | 0 | 1.498 | 452850.6 | 1.498 | 1.56E+08 | 1.498 | 1.57E+08 | 1.498 | 0 | 1.498 | 241085.7 | 1.498 | 83245661 | 1.498 | 83486747 |
| 1.499 | 0 | 1.499 | 442408.5 | 1.499 | 1.54E+08 | 1.499 | 1.55E+08 | 1.499 | 0 | 1.499 | 235526.6 | 1.499 | 82021545 | 1.499 | 82257072 |
| 1.5 | 0 | 1.5 | 432194.5 | 1.5 | 1.52E+08 | 1.5 | 1.52E+08 | 1.5 | 0 | 1.5 | 230089 | 1.5 | 80812090 | 1.5 | 81042179 |
| 1.501 | 0 | 1.501 | 422203.8 | 1.501 | 1.5E+08 | 1.501 | 1.5E+08 | 1.501 | 0 | 1.501 | 224770.2 | 1.501 | 79617180 | 1.501 | 79841950 |
| 1.502 | 0 | 1.502 | 412431.9 | 1.502 | 1.47E+08 | 1.502 | 1.48E+08 | 1.502 | 0 | 1.502 | 219567.9 | 1.502 | 78436696 | 1.502 | 78656264 |
| 1.503 | 0 | 1.503 | 402874.2 | 1.503 | 1.45E+08 | 1.503 | 1.46E+08 | 1.503 | 0 | 1.503 | 214479.6 | 1.503 | 77270522 | 1.503 | 77485001 |
| 1.504 | 0 | 1.504 | 393526.3 | 1.504 | 1.43E+08 | 1.504 | 1.43E+08 | 1.504 | 0 | 1.504 | 209503 | 1.504 | 76118541 | 1.504 | 76328044 |
| 1.505 | 0 | 1.505 | 384384 | 1.505 | 1.41E+08 | 1.505 | 1.41E+08 | 1.505 | 0 | 1.505 | 204635.9 | 1.505 | 74980636 | 1.505 | 75185272 |
| 1.506 | 0 | 1.506 | 375443 | 1.506 | 1.39E+08 | 1.506 | 1.39E+08 | 1.506 | 0 | 1.506 | 199875.9 | 1.506 | 73856689 | 1.506 | 74056565 |
| 1.507 | 0 | 1.507 | 366699 | 1.507 | 1.37E+08 | 1.507 | 1.37E+08 | 1.507 | 0 | 1.507 | 195220.9 | 1.507 | 72746584 | 1.507 | 72941805 |
| 1.508 | 0 | 1.508 | 358148.2 | 1.508 | 1.35E+08 | 1.508 | 1.35E+08 | 1.508 | 0 | 1.508 | 190668.6 | 1.508 | 71650204 | 1.508 | 71840873 |
| 1.509 | 0 | 1.509 | 349786.3 | 1.509 | 1.33E+08 | 1.509 | 1.33E+08 | 1.509 | 0 | 1.509 | 186217 | 1.509 | 70567432 | 1.509 | 70753649 |
| 1.51 | 0 | 1.51 | 341609.6 | 1.51 | 1.31E+08 | 1.51 | 1.31E+08 | 1.51 | 0 | 1.51 | 181864 | 1.51 | 69498150 | 1.51 | 69680014 |
| 1.511 | 0 | 1.511 | 333614.2 | 1.511 | 1.29E+08 | 1.511 | 1.29E+08 | 1.511 | 0 | 1.511 | 177607.4 | 1.511 | 68442242 | 1.511 | 68619850 |
| 1.512 | 0 | 1.512 | 325796.3 | 1.512 | 1.27E+08 | 1.512 | 1.27E+08 | 1.512 | 0 | 1.512 | 173445.3 | 1.512 | 67399593 | 1.512 | 67573038 |
| 1.513 | 0 | 1.513 | 318152.1 | 1.513 | 1.25E+08 | 1.513 | 1.25E+08 | 1.513 | 0 | 1.513 | 169375.8 | 1.513 | 66370084 | 1.513 | 66539460 |
| 1.514 | 0 | 1.514 | 310678.1 | 1.514 | 1.23E+08 | 1.514 | 1.23E+08 | 1.514 | 0 | 1.514 | 165396.8 | 1.514 | 65353600 | 1.514 | 65518997 |
| 1.515 | 0 | 1.515 | 303370.7 | 1.515 | 1.21E+08 | 1.515 | 1.21E+08 | 1.515 | 0 | 1.515 | 161506.6 | 1.515 | 64350025 | 1.515 | 64511531 |
| 1.516 | 0 | 1.516 | 296226.5 | 1.516 | 1.19E+08 | 1.516 | 1.19E+08 | 1.516 | 0 | 1.516 | 157703.2 | 1.516 | 63359242 | 1.516 | 63516945 |
| 1.517 | 0 | 1.517 | 289241.9 | 1.517 | 1.17E+08 | 1.517 | 1.17E+08 | 1.517 | 0 | 1.517 | 153984.8 | 1.517 | 62381137 | 1.517 | 62535122 |
| 1.518 | 0 | 1.518 | 282413.6 | 1.518 | 1.15E+08 | 1.518 | 1.16E+08 | 1.518 | 0 | 1.518 | 150349.6 | 1.518 | 61415593 | 1.518 | 61565943 |
| 1.519 | 0 | 1.519 | 275738.4 | 1.519 | 1.14E+08 | 1.519 | 1.14E+08 | 1.519 | 0 | 1.519 | 146795.9 | 1.519 | 60462495 | 1.519 | 60609291 |
| 1.52 | 0 | 1.52 | 269213 | 1.52 | 1.12E+08 | 1.52 | 1.12E+08 | 1.52 | 0 | 1.52 | 143321.9 | 1.52 | 59521729 | 1.52 | 59665051 |
| 1.521 | 0 | 1.521 | 262834.3 | 1.521 | 1.1E+08 | 1.521 | 1.1E+08 | 1.521 | 0 | 1.521 | 139926 | 1.521 | 58593180 | 1.521 | 58733106 |
| 1.522 | 0 | 1.522 | 256599.1 | 1.522 | 1.08E+08 | 1.522 | 1.09E+08 | 1.522 | 0 | 1.522 | 136606.6 | 1.522 | 57676732 | 1.522 | 57813339 |
| 1.523 | 0 | 1.523 | 250504.4 | 1.523 | 1.07E+08 | 1.523 | 1.07E+08 | 1.523 | 0 | 1.523 | 133361.9 | 1.523 | 56772273 | 1.523 | 56905635 |
| 1.524 | 0 | 1.524 | 244547.2 | 1.524 | 1.05E+08 | 1.524 | 1.05E+08 | 1.524 | 0 | 1.524 | 130190.5 | 1.524 | 55879688 | 1.524 | 56009878 |
| 1.525 | 0 | 1.525 | 238724.6 | 1.525 | 1.03E+08 | 1.525 | 1.04E+08 | 1.525 | 0 | 1.525 | 127090.7 | 1.525 | 54998863 | 1.525 | 55125954 |
| 1.526 | 0 | 1.526 | 233033.8 | 1.526 | 1.02E+08 | 1.526 | 1.02E+08 | 1.526 | 0 | 1.526 | 124061 | 1.526 | 54129686 | 1.526 | 54253747 |
| 1.527 | 0 | 1.527 | 227471.9 | 1.527 | 1E+08 | 1.527 | 1E+08 | 1.527 | 0 | 1.527 | 121100 | 1.527 | 53272044 | 1.527 | 53393144 |
| 1.528 | 0 | 1.528 | 222036.2 | 1.528 | 98475620 | 1.528 | 98697656 | 1.528 | 0 | 1.528 | 118206.2 | 1.528 | 52425824 | 1.528 | 52544030 |
| 1.529 | 0 | 1.529 | 216723.9 | 1.529 | 96907342 | 1.529 | 97124066 | 1.529 | 0 | 1.529 | 115378.1 | 1.529 | 51590914 | 1.529 | 51706292 |
| 1.53 | 0 | 1.53 | 211532.5 | 1.53 | 95360099 | 1.53 | 95571632 | 1.53 | 0 | 1.53 | 112614.3 | 1.53 | 50767202 | 1.53 | 50879817 |
| 1.531 | 0 | 1.531 | 206459.4 | 1.531 | 93833682 | 1.531 | 94040142 | 1.531 | 0 | 1.531 | 109913.5 | 1.531 | 49954578 | 1.531 | 50064492 |
| 1.532 | 0 | 1.532 | 201501.9 | 1.532 | 92327883 | 1.532 | 92529385 | 1.532 | 0 | 1.532 | 107274.3 | 1.532 | 49152931 | 1.532 | 49260205 |
| 1.533 | 0 | 1.533 | 196657.7 | 1.533 | 90842495 | 1.533 | 91039152 | 1.533 | 0 | 1.533 | 104695.4 | 1.533 | 48362149 | 1.533 | 48466844 |
| 1.534 | 0 | 1.534 | 191924.2 | 1.534 | 89377310 | 1.534 | 89569234 | 1.534 | 0 | 1.534 | 102175.4 | 1.534 | 47582123 | 1.534 | 47684299 |
| 1.535 | 0 | 1.535 | 187299.2 | 1.535 | 87932124 | 1.535 | 88119423 | 1.535 | 0 | 1.535 | 99713.13 | 1.535 | 46812744 | 1.535 | 46912457 |
| 1.536 | 0 | 1.536 | 182780.1 | 1.536 | 86506730 | 1.536 | 86689511 | 1.536 | 0 | 1.536 | 97307.33 | 1.536 | 46053902 | 1.536 | 46151210 |
| 1.537 | 0 | 1.537 | 178364.9 | 1.537 | 85100926 | 1.537 | 85279291 | 1.537 | 0 | 1.537 | 94956.76 | 1.537 | 45305489 | 1.537 | 45400446 |
| 1.538 | 0 | 1.538 | 174051.1 | 1.538 | 83714508 | 1.538 | 83888559 | 1.538 | 0 | 1.538 | 92660.23 | 1.538 | 44567397 | 1.538 | 44660057 |
| 1.539 | 0 | 1.539 | 169836.7 | 1.539 | 82347274 | 1.539 | 82517111 | 1.539 | 0 | 1.539 | 90416.57 | 1.539 | 43839518 | 1.539 | 43929934 |
| 1.54 | 0 | 1.54 | 165719.4 | 1.54 | 80999023 | 1.54 | 81164742 | 1.54 | 0 | 1.54 | 88224.63 | 1.54 | 43121744 | 1.54 | 43209969 |
| 1.541 | 0 | 1.541 | 161697.1 | 1.541 | 79669554 | 1.541 | 79831251 | 1.541 | 0 | 1.541 | 86083.27 | 1.541 | 42413970 | 1.541 | 42500053 |
| 1.542 | 0 | 1.542 | 157767.8 | 1.542 | 78358668 | 1.542 | 78516436 | 1.542 | 0 | 1.542 | 83991.41 | 1.542 | 41716089 | 1.542 | 41800080 |
| 1.543 | 0 | 1.543 | 153929.4 | 1.543 | 77066167 | 1.543 | 77220096 | 1.543 | 0 | 1.543 | 81947.95 | 1.543 | 41027995 | 1.543 | 41109943 |
| 1.544 | 0 | 1.544 | 150180 | 1.544 | 75791853 | 1.544 | 75942033 | 1.544 | 0 | 1.544 | 79951.85 | 1.544 | 40349584 | 1.544 | 40429536 |
| 1.545 | 0 | 1.545 | 146517.5 | 1.545 | 74535530 | 1.545 | 74682048 | 1.545 | 0 | 1.545 | 78002.06 | 1.545 | 39680751 | 1.545 | 39758753 |
| 1.546 | 0 | 1.546 | 142940.1 | 1.546 | 73297004 | 1.546 | 73439944 | 1.546 | 0 | 1.546 | 76097.57 | 1.546 | 39021392 | 1.546 | 39097490 |
| 1.547 | 0 | 1.547 | 139446 | 1.547 | 72076079 | 1.547 | 72215525 | 1.547 | 0 | 1.547 | 74237.38 | 1.547 | 38371404 | 1.547 | 38445641 |
| 1.548 | 0 | 1.548 | 136033.3 | 1.548 | 70872562 | 1.548 | 71008595 | 1.548 | 0 | 1.548 | 72420.52 | 1.548 | 37730683 | 1.548 | 37803104 |
| 1.549 | 0 | 1.549 | 132700.1 | 1.549 | 69686262 | 1.549 | 69818962 | 1.549 | 0 | 1.549 | 70646.03 | 1.549 | 37099129 | 1.549 | 37169775 |
| 1.55 | 0 | 1.55 | 129444.8 | 1.55 | 68516988 | 1.55 | 68646432 | 1.55 | 0 | 1.55 | 68912.99 | 1.55 | 36476638 | 1.55 | 36545551 |
| 1.551 | 0 | 1.551 | 126265.6 | 1.551 | 67364549 | 1.551 | 67490814 | 1.551 | 0 | 1.551 | 67220.47 | 1.551 | 35863110 | 1.551 | 35930330 |
| 1.552 | 0 | 1.552 | 123160.8 | 1.552 | 66228757 | 1.552 | 66351918 | 1.552 | 0 | 1.552 | 65567.58 | 1.552 | 35258444 | 1.552 | 35324012 |
| 1.553 | 0 | 1.553 | 120128.9 | 1.553 | 65109425 | 1.553 | 65229553 | 1.553 | 0 | 1.553 | 63953.44 | 1.553 | 34662541 | 1.553 | 34726494 |
| 1.554 | 0 | 1.554 | 117168.1 | 1.554 | 64006365 | 1.554 | 64123533 | 1.554 | 0 | 1.554 | 62377.19 | 1.554 | 34075301 | 1.554 | 34137678 |
| 1.555 | 0 | 1.555 | 114276.9 | 1.555 | 62919393 | 1.555 | 63033670 | 1.555 | 0 | 1.555 | 60837.99 | 1.555 | 33496626 | 1.555 | 33557464 |
| 1.556 | 0 | 1.556 | 111453.7 | 1.556 | 61848324 | 1.556 | 61959778 | 1.556 | 0 | 1.556 | 59335.02 | 1.556 | 32926417 | 1.556 | 32985752 |
| 1.557 | 0 | 1.557 | 108697.1 | 1.557 | 60792976 | 1.557 | 60901674 | 1.557 | 0 | 1.557 | 57867.47 | 1.557 | 32364578 | 1.557 | 32422445 |
| 1.558 | 0 | 1.558 | 106005.5 | 1.558 | 59753167 | 1.558 | 59859173 | 1.558 | 0 | 1.558 | 56434.54 | 1.558 | 31811011 | 1.558 | 31867445 |
| 1.559 | 0 | 1.559 | 103377.5 | 1.559 | 58728716 | 1.559 | 58832093 | 1.559 | 0 | 1.559 | 55035.47 | 1.559 | 31265620 | 1.559 | 31320655 |
| 1.56 | 0 | 1.56 | 100811.7 | 1.56 | 57719444 | 1.56 | 57820255 | 1.56 | 0 | 1.56 | 53669.49 | 1.56 | 30728310 | 1.56 | 30781979 |
| 1.561 | 0 | 1.561 | 98306.65 | 1.561 | 56725172 | 1.561 | 56823478 | 1.561 | 0 | 1.561 | 52335.87 | 1.561 | 30198986 | 1.561 | 30251322 |
| 1.562 | 0 | 1.562 | 95861.01 | 1.562 | 55745724 | 1.562 | 55841585 | 1.562 | 0 | 1.562 | 51033.88 | 1.562 | 29677554 | 1.562 | 29728587 |
| 1.563 | 0 | 1.563 | 93473.45 | 1.563 | 54780924 | 1.563 | 54874397 | 1.563 | 0 | 1.563 | 49762.8 | 1.563 | 29163919 | 1.563 | 29213682 |
| 1.564 | 0 | 1.564 | 91142.66 | 1.564 | 53830597 | 1.564 | 53921740 | 1.564 | 0 | 1.564 | 48521.95 | 1.564 | 28657991 | 1.564 | 28706513 |
| 1.565 | 0 | 1.565 | 88867.36 | 1.565 | 52894571 | 1.565 | 52983438 | 1.565 | 0 | 1.565 | 47310.64 | 1.565 | 28159675 | 1.565 | 28206985 |
| 1.566 | 0 | 1.566 | 86646.3 | 1.566 | 51972673 | 1.566 | 52059319 | 1.566 | 0 | 1.566 | 46128.2 | 1.566 | 27668881 | 1.566 | 27715009 |
| 1.567 | 0 | 1.567 | 84478.24 | 1.567 | 51064732 | 1.567 | 51149210 | 1.567 | 0 | 1.567 | 44973.99 | 1.567 | 27185517 | 1.567 | 27230491 |
| 1.568 | 0 | 1.568 | 82362 | 1.568 | 50170580 | 1.568 | 50252942 | 1.568 | 0 | 1.568 | 43847.36 | 1.568 | 26709494 | 1.568 | 26753341 |
| 1.569 | 0 | 1.569 | 80296.4 | 1.569 | 49290047 | 1.569 | 49370344 | 1.569 | 0 | 1.569 | 42747.69 | 1.569 | 26240722 | 1.569 | 26283469 |
| 1.57 | 0 | 1.57 | 78280.29 | 1.57 | 48422968 | 1.57 | 48501248 | 1.57 | 0 | 1.57 | 41674.36 | 1.57 | 25779111 | 1.57 | 25820786 |
| 1.571 | 0 | 1.571 | 76312.54 | 1.571 | 47569176 | 1.571 | 47645489 | 1.571 | 0 | 1.571 | 40626.78 | 1.571 | 25324575 | 1.571 | 25365202 |
| 1.572 | 0 | 1.572 | 74392.05 | 1.572 | 46728507 | 1.572 | 46802899 | 1.572 | 0 | 1.572 | 39604.36 | 1.572 | 24877025 | 1.572 | 24916630 |
| 1.573 | 0 | 1.573 | 72517.75 | 1.573 | 45900798 | 1.573 | 45973316 | 1.573 | 0 | 1.573 | 38606.54 | 1.573 | 24436375 | 1.573 | 24474981 |
| 1.574 | 0 | 1.574 | 70688.57 | 1.574 | 45085888 | 1.574 | 45156576 | 1.574 | 0 | 1.574 | 37632.73 | 1.574 | 24002538 | 1.574 | 24040171 |
| 1.575 | 0 | 1.575 | 68903.5 | 1.575 | 44283615 | 1.575 | 44352518 | 1.575 | 0 | 1.575 | 36682.41 | 1.575 | 23575429 | 1.575 | 23612111 |
| 1.576 | 0 | 1.576 | 67161.52 | 1.576 | 43493820 | 1.576 | 43560982 | 1.576 | 0 | 1.576 | 35755.02 | 1.576 | 23154963 | 1.576 | 23190718 |
| 1.577 | 0 | 1.577 | 65461.65 | 1.577 | 42716347 | 1.577 | 42781809 | 1.577 | 0 | 1.577 | 34850.06 | 1.577 | 22741057 | 1.577 | 22775907 |
| 1.578 | 0 | 1.578 | 63802.91 | 1.578 | 41951038 | 1.578 | 42014840 | 1.578 | 0 | 1.578 | 33966.99 | 1.578 | 22333626 | 1.578 | 22367593 |
| 1.579 | 0 | 1.579 | 62184.36 | 1.579 | 41197737 | 1.579 | 41259921 | 1.579 | 0 | 1.579 | 33105.31 | 1.579 | 21932589 | 1.579 | 21965694 |
| 1.58 | 0 | 1.58 | 60605.07 | 1.58 | 40456292 | 1.58 | 40516897 | 1.58 | 0 | 1.58 | 32264.54 | 1.58 | 21537863 | 1.58 | 21570128 |
| 1.581 | 0 | 1.581 | 59064.15 | 1.581 | 39726548 | 1.581 | 39785613 | 1.581 | 0 | 1.581 | 31444.2 | 1.581 | 21149367 | 1.581 | 21180811 |
| 1.582 | 0 | 1.582 | 57560.71 | 1.582 | 39008356 | 1.582 | 39065917 | 1.582 | 0 | 1.582 | 30643.8 | 1.582 | 20767020 | 1.582 | 20797664 |
| 1.583 | 0 | 1.583 | 56093.87 | 1.583 | 38301565 | 1.583 | 38357659 | 1.583 | 0 | 1.583 | 29862.9 | 1.583 | 20390743 | 1.583 | 20420606 |
| 1.584 | 0 | 1.584 | 54662.8 | 1.584 | 37606026 | 1.584 | 37660689 | 1.584 | 0 | 1.584 | 29101.03 | 1.584 | 20020457 | 1.584 | 20049558 |
| 1.585 | 0 | 1.585 | 53266.66 | 1.585 | 36921593 | 1.585 | 36974859 | 1.585 | 0 | 1.585 | 28357.76 | 1.585 | 19656082 | 1.585 | 19684440 |
| 1.586 | 0 | 1.586 | 51904.64 | 1.586 | 36248118 | 1.586 | 36300022 | 1.586 | 0 | 1.586 | 27632.66 | 1.586 | 19297542 | 1.586 | 19325175 |
| 1.587 | 0 | 1.587 | 50575.95 | 1.587 | 35585457 | 1.587 | 35636033 | 1.587 | 0 | 1.587 | 26925.3 | 1.587 | 18944759 | 1.587 | 18971684 |
| 1.588 | 0 | 1.588 | 49279.82 | 1.588 | 34933467 | 1.588 | 34982747 | 1.588 | 0 | 1.588 | 26235.28 | 1.588 | 18597657 | 1.588 | 18623892 |
| 1.589 | 0 | 1.589 | 48015.49 | 1.589 | 34292005 | 1.589 | 34340021 | 1.589 | 0 | 1.589 | 25562.18 | 1.589 | 18256159 | 1.589 | 18281722 |
| 1.59 | 0 | 1.59 | 46782.2 | 1.59 | 33660931 | 1.59 | 33707714 | 1.59 | 0 | 1.59 | 24905.61 | 1.59 | 17920192 | 1.59 | 17945098 |
| 1.591 | 0 | 1.591 | 45579.25 | 1.591 | 33040106 | 1.591 | 33085685 | 1.591 | 0 | 1.591 | 24265.19 | 1.591 | 17589681 | 1.591 | 17613947 |
| 1.592 | 0 | 1.592 | 44405.92 | 1.592 | 32429391 | 1.592 | 32473797 | 1.592 | 0 | 1.592 | 23640.54 | 1.592 | 17264553 | 1.592 | 17288193 |
| 1.593 | 0 | 1.593 | 43261.51 | 1.593 | 31828649 | 1.593 | 31871910 | 1.593 | 0 | 1.593 | 23031.29 | 1.593 | 16944733 | 1.593 | 16967765 |
| 1.594 | 0 | 1.594 | 42145.34 | 1.594 | 31237744 | 1.594 | 31279890 | 1.594 | 0 | 1.594 | 22437.07 | 1.594 | 16630151 | 1.594 | 16652588 |
| 1.595 | 0 | 1.595 | 41056.76 | 1.595 | 30656543 | 1.595 | 30697600 | 1.595 | 0 | 1.595 | 21857.54 | 1.595 | 16320735 | 1.595 | 16342593 |
| 1.596 | 0 | 1.596 | 39995.11 | 1.596 | 30084913 | 1.596 | 30124908 | 1.596 | 0 | 1.596 | 21292.34 | 1.596 | 16016414 | 1.596 | 16037707 |
| 1.597 | 0 | 1.597 | 38959.77 | 1.597 | 29522721 | 1.597 | 29561681 | 1.597 | 0 | 1.597 | 20741.15 | 1.597 | 15717118 | 1.597 | 15737859 |
| 1.598 | 0 | 1.598 | 37950.1 | 1.598 | 28969838 | 1.598 | 29007788 | 1.598 | 0 | 1.598 | 20203.63 | 1.598 | 15422778 | 1.598 | 15442981 |
| 1.599 | 0 | 1.599 | 36965.5 | 1.599 | 28426134 | 1.599 | 28463099 | 1.599 | 0 | 1.599 | 19679.46 | 1.599 | 15133324 | 1.599 | 15153004 |
| 1.6 | 0 | 1.6 | 36005.39 | 1.6 | 27891482 | 1.6 | 27927487 | 1.6 | 0 | 1.6 | 19168.32 | 1.6 | 14848689 | 1.6 | 14867858 |
| 1.601 | 0 | 1.601 | 35069.17 | 1.601 | 27365754 | 1.601 | 27400824 | 1.601 | 0 | 1.601 | 18669.9 | 1.601 | 14568806 | 1.601 | 14587476 |
| 1.602 | 0 | 1.602 | 34156.29 | 1.602 | 26848827 | 1.602 | 26882984 | 1.602 | 0 | 1.602 | 18183.91 | 1.602 | 14293608 | 1.602 | 14311792 |
| 1.603 | 0 | 1.603 | 33266.18 | 1.603 | 26340576 | 1.603 | 26373842 | 1.603 | 0 | 1.603 | 17710.04 | 1.603 | 14023028 | 1.603 | 14040738 |
| 1.604 | 0 | 1.604 | 32398.31 | 1.604 | 25840879 | 1.604 | 25873277 | 1.604 | 0 | 1.604 | 17248.01 | 1.604 | 13757002 | 1.604 | 13774250 |
| 1.605 | 0 | 1.605 | 31552.16 | 1.605 | 25349613 | 1.605 | 25381165 | 1.605 | 0 | 1.605 | 16797.54 | 1.605 | 13495466 | 1.605 | 13512263 |
| 1.606 | 0 | 1.606 | 30727.19 | 1.606 | 24866659 | 1.606 | 24897387 | 1.606 | 0 | 1.606 | 16358.34 | 1.606 | 13238354 | 1.606 | 13254712 |
| 1.607 | 0 | 1.607 | 29922.9 | 1.607 | 24391899 | 1.607 | 24421822 | 1.607 | 0 | 1.607 | 15930.16 | 1.607 | 12985604 | 1.607 | 13001534 |
| 1.608 | 0 | 1.608 | 29138.81 | 1.608 | 23925214 | 1.608 | 23954353 | 1.608 | 0 | 1.608 | 15512.73 | 1.608 | 12737153 | 1.608 | 12752666 |
| 1.609 | 0 | 1.609 | 28374.42 | 1.609 | 23466488 | 1.609 | 23494863 | 1.609 | 0 | 1.609 | 15105.79 | 1.609 | 12492940 | 1.609 | 12508045 |
| 1.61 | 0 | 1.61 | 27629.27 | 1.61 | 23015607 | 1.61 | 23043236 | 1.61 | 0 | 1.61 | 14709.09 | 1.61 | 12252902 | 1.61 | 12267611 |
| 1.611 | 0 | 1.611 | 26902.89 | 1.611 | 22572456 | 1.611 | 22599359 | 1.611 | 0 | 1.611 | 14322.39 | 1.611 | 12016980 | 1.611 | 12031303 |
| 1.612 | 0 | 1.612 | 26194.83 | 1.612 | 22136923 | 1.612 | 22163118 | 1.612 | 0 | 1.612 | 13945.44 | 1.612 | 11785114 | 1.612 | 11799059 |
| 1.613 | 0 | 1.613 | 25504.65 | 1.613 | 21708896 | 1.613 | 21734401 | 1.613 | 0 | 1.613 | 13578 | 1.613 | 11557244 | 1.613 | 11570822 |
| 1.614 | 0 | 1.614 | 24831.92 | 1.614 | 21288266 | 1.614 | 21313098 | 1.614 | 0 | 1.614 | 13219.86 | 1.614 | 11333311 | 1.614 | 11346531 |
| 1.615 | 0 | 1.615 | 24176.22 | 1.615 | 20874923 | 1.615 | 20899099 | 1.615 | 0 | 1.615 | 12870.78 | 1.615 | 11113259 | 1.615 | 11126129 |
| 1.616 | 0 | 1.616 | 23537.14 | 1.616 | 20468760 | 1.616 | 20492297 | 1.616 | 0 | 1.616 | 12530.56 | 1.616 | 10897028 | 1.616 | 10909559 |
| 1.617 | 0 | 1.617 | 22914.28 | 1.617 | 20069670 | 1.617 | 20092584 | 1.617 | 0 | 1.617 | 12198.96 | 1.617 | 10684563 | 1.617 | 10696762 |
| 1.618 | 0 | 1.618 | 22307.24 | 1.618 | 19677549 | 1.618 | 19699856 | 1.618 | 0 | 1.618 | 11875.78 | 1.618 | 10475808 | 1.618 | 10487684 |
| 1.619 | 0 | 1.619 | 21715.63 | 1.619 | 19292291 | 1.619 | 19314007 | 1.619 | 0 | 1.619 | 11560.83 | 1.619 | 10270707 | 1.619 | 10282268 |
| 1.62 | 0 | 1.62 | 21139.09 | 1.62 | 18913795 | 1.62 | 18934934 | 1.62 | 0 | 1.62 | 11253.9 | 1.62 | 10069206 | 1.62 | 10080460 |
| 1.621 | 0 | 1.621 | 20577.25 | 1.621 | 18541958 | 1.621 | 18562535 | 1.621 | 0 | 1.621 | 10954.79 | 1.621 | 9871250 | 1.621 | 9882204 |
| 1.622 | 0 | 1.622 | 20029.75 | 1.622 | 18176680 | 1.622 | 18196710 | 1.622 | 0 | 1.622 | 10663.31 | 1.622 | 9676785 | 1.622 | 9687449 |
| 1.623 | 0 | 1.623 | 19496.24 | 1.623 | 17817862 | 1.623 | 17837359 | 1.623 | 0 | 1.623 | 10379.29 | 1.623 | 9485760 | 1.623 | 9496139 |
| 1.624 | 0 | 1.624 | 18976.38 | 1.624 | 17465406 | 1.624 | 17484382 | 1.624 | 0 | 1.624 | 10102.53 | 1.624 | 9298122 | 1.624 | 9308224 |
| 1.625 | 0 | 1.625 | 18469.84 | 1.625 | 17119214 | 1.625 | 17137684 | 1.625 | 0 | 1.625 | 9832.854 | 1.625 | 9113818 | 1.625 | 9123651 |
| 1.626 | 0 | 1.626 | 17976.28 | 1.626 | 16779191 | 1.626 | 16797167 | 1.626 | 0 | 1.626 | 9570.097 | 1.626 | 8932799 | 1.626 | 8942369 |
| 1.627 | 0 | 1.627 | 17495.4 | 1.627 | 16445241 | 1.627 | 16462737 | 1.627 | 0 | 1.627 | 9314.087 | 1.627 | 8755013 | 1.627 | 8764327 |
| 1.628 | 0 | 1.628 | 17026.87 | 1.628 | 16117273 | 1.628 | 16134300 | 1.628 | 0 | 1.628 | 9064.657 | 1.628 | 8580411 | 1.628 | 8589476 |
| 1.629 | 0 | 1.629 | 16570.4 | 1.629 | 15795192 | 1.629 | 15811762 | 1.629 | 0 | 1.629 | 8821.646 | 1.629 | 8408944 | 1.629 | 8417765 |
| 1.63 | 0 | 1.63 | 16125.7 | 1.63 | 15478908 | 1.63 | 15495033 | 1.63 | 0 | 1.63 | 8584.895 | 1.63 | 8240562 | 1.63 | 8249147 |
| 1.631 | 0 | 1.631 | 15692.46 | 1.631 | 15168330 | 1.631 | 15184023 | 1.631 | 0 | 1.631 | 8354.252 | 1.631 | 8075219 | 1.631 | 8083573 |
| 1.632 | 0 | 1.632 | 15270.41 | 1.632 | 14863370 | 1.632 | 14878640 | 1.632 | 0 | 1.632 | 8129.564 | 1.632 | 7912866 | 1.632 | 7920996 |
| 1.633 | 0 | 1.633 | 14859.27 | 1.633 | 14563939 | 1.633 | 14578798 | 1.633 | 0 | 1.633 | 7910.685 | 1.633 | 7753457 | 1.633 | 7761368 |
| 1.634 | 0 | 1.634 | 14458.78 | 1.634 | 14269951 | 1.634 | 14284410 | 1.634 | 0 | 1.634 | 7697.471 | 1.634 | 7596946 | 1.634 | 7604643 |
| 1.635 | 0 | 1.635 | 14068.66 | 1.635 | 13981319 | 1.635 | 13995388 | 1.635 | 0 | 1.635 | 7489.783 | 1.635 | 7443286 | 1.635 | 7450776 |
| 1.636 | 0 | 1.636 | 13688.66 | 1.636 | 13697960 | 1.636 | 13711648 | 1.636 | 0 | 1.636 | 7287.483 | 1.636 | 7292433 | 1.636 | 7299720 |
| 1.637 | 0 | 1.637 | 13318.53 | 1.637 | 13419788 | 1.637 | 13433107 | 1.637 | 0 | 1.637 | 7090.437 | 1.637 | 7144342 | 1.637 | 7151432 |
| 1.638 | 0 | 1.638 | 12958.03 | 1.638 | 13146723 | 1.638 | 13159681 | 1.638 | 0 | 1.638 | 6898.514 | 1.638 | 6998969 | 1.638 | 7005867 |
| 1.639 | 0 | 1.639 | 12606.91 | 1.639 | 12878681 | 1.639 | 12891288 | 1.639 | 0 | 1.639 | 6711.589 | 1.639 | 6856270 | 1.639 | 6862982 |
| 1.64 | 0 | 1.64 | 12264.95 | 1.64 | 12615584 | 1.64 | 12627849 | 1.64 | 0 | 1.64 | 6529.535 | 1.64 | 6716204 | 1.64 | 6722734 |
| 1.641 | 0 | 1.641 | 11931.9 | 1.641 | 12357350 | 1.641 | 12369282 | 1.641 | 0 | 1.641 | 6352.231 | 1.641 | 6578727 | 1.641 | 6585080 |
| 1.642 | 0 | 1.642 | 11607.56 | 1.642 | 12103902 | 1.642 | 12115510 | 1.642 | 0 | 1.642 | 6179.559 | 1.642 | 6443798 | 1.642 | 6449978 |
| 1.643 | 0 | 1.643 | 11291.7 | 1.643 | 11855162 | 1.643 | 11866454 | 1.643 | 0 | 1.643 | 6011.403 | 1.643 | 6311376 | 1.643 | 6317387 |
| 1.644 | 0 | 1.644 | 10984.11 | 1.644 | 11611055 | 1.644 | 11622039 | 1.644 | 0 | 1.644 | 5847.65 | 1.644 | 6181419 | 1.644 | 6187267 |

[illegible]

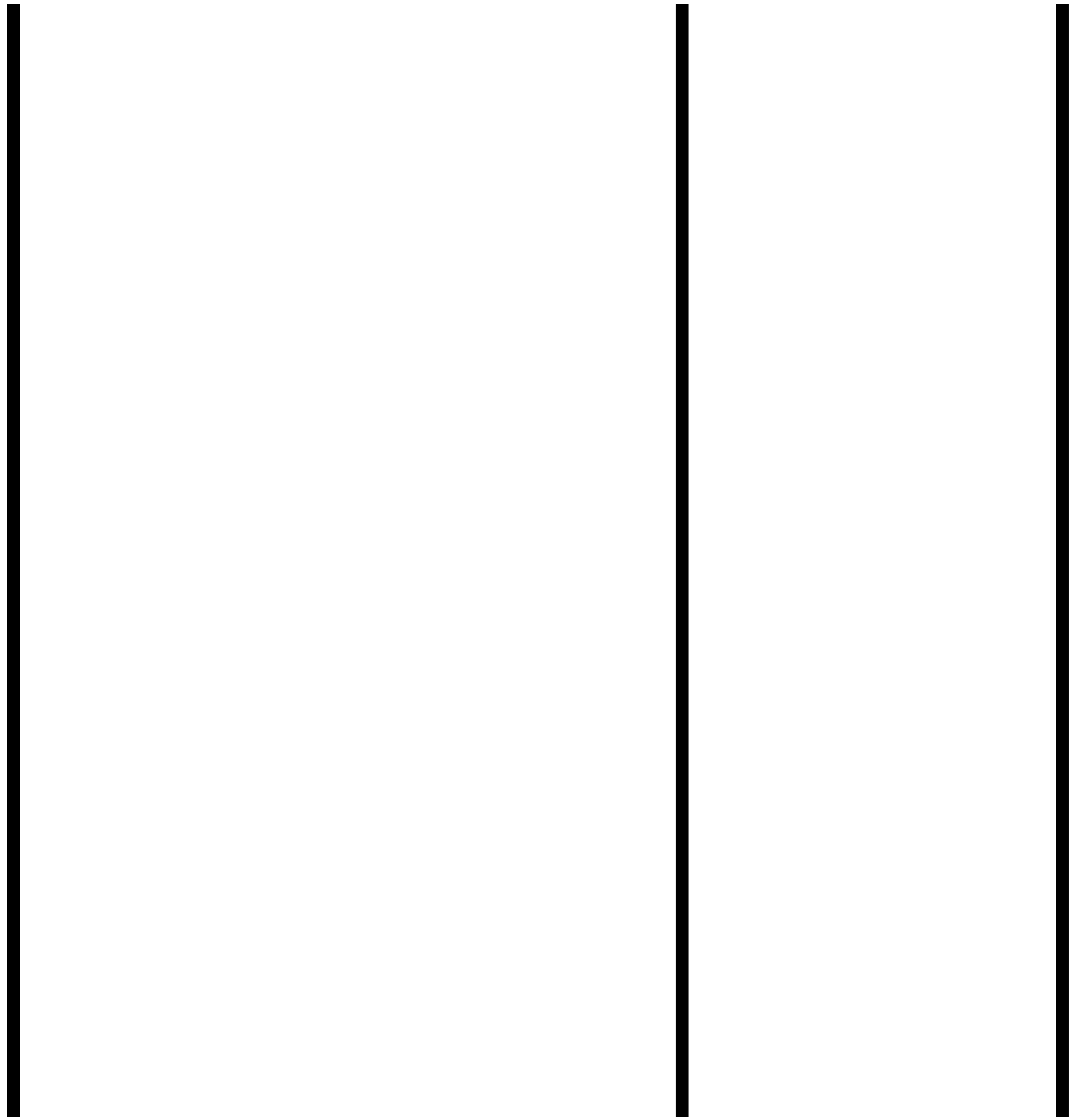

| Pt$_1$-post CO TPD | | Pt$_1$-post $C_2D_4$ TPD | |
|---|---|---|---|
| 0.3497 | 2553.33 | 0.3497 | 2300 |
| 0.3517 | 1786.67 | 0.3517 | 2255.55 |
| 0.3537 | 2233.33 | 0.3537 | 1666.667 |
| 0.3557 | 1793.33 | 0.3557 | 1788.883 |
| 0.3577 | 1820 | 0.3577 | 1888.883 |
| 0.3597 | 1693.33 | 0.3597 | 1611.112 |
| 0.3617 | 1913.33 | 0.3617 | 2044.45 |
| 0.3637 | 2306.67 | 0.3637 | 2188.883 |
| 0.3657 | 2240 | 0.3657 | 2588.883 |
| 0.3677 | 2033.33 | 0.3677 | 1688.883 |
| 0.3697 | 2166.67 | 0.3697 | 2200 |
| 0.3717 | 2500 | 0.3717 | 1033.333 |
| 0.3737 | 1780 | 0.3737 | 1800 |
| 0.3757 | 2346.67 | 0.3757 | 1722.217 |
| 0.3777 | 2320 | 0.3777 | 1755.55 |
| 0.3797 | 2266.67 | 0.3797 | 1655.555 |
| 0.3817 | 2380 | 0.3817 | 2422.217 |
| 0.3837 | 2853.33 | 0.3837 | 2355.55 |
| 0.3857 | 2480 | 0.3857 | 2388.883 |
| 0.3877 | 2366.67 | 0.3877 | 2511.117 |
| 0.3897 | 2546.67 | 0.3897 | 2344.45 |
| 0.3917 | 2906.67 | 0.3917 | 3211.117 |
| 0.3937 | 3553.33 | 0.3937 | 3744.45 |
| 0.3957 | 4760 | 0.3957 | 5100 |
| 0.3977 | 3906.67 | 0.3977 | 4733.333 |
| 0.3997 | 4466.67 | 0.3997 | 5722.217 |
| 0.4017 | 5266.67 | 0.4017 | 4755.55 |
| 0.4037 | 6193.33 | 0.4037 | 6733.333 |
| 0.4057 | 5413.33 | 0.4057 | 7100 |
| 0.4077 | 5500 | 0.4077 | 7077.783 |
| 0.4097 | 7493.33 | 0.4097 | 8600 |
| 0.4117 | 8093.33 | 0.4117 | 8066.667 |

Fig. 5

| Pt$_1$Sn$_{1.1}$-post CO TPD | | Pt$_1$Sn$_{1.1}$-post $C_2D_4$ TPD | | Pt$_1$Sn$_{0.1}$-post CO TPD | | Pt$_1$Sn$_{0.1}$-post $C_2D_4$ TPD | |
|---|---|---|---|---|---|---|---|

Fig. 6

| 4th CO (Pt$_1$) | | after 4 $C_2D_4$ (Pt$_1$) | | 4th CO (Pt$_2$) | | after 4 $C_2D_4$ (Pt$_2$) | | $SiO_2$ | |
|---|---|---|---|---|---|---|---|---|---|

Fig. 7

| 4th CO (Pt$_1$Sn$_{1.1}$) | | after 4 $C_2D_4$ (Pt$_1$Sn$_{1.1}$) | | 4th CO (Pt$_1$Sn$_{0.1}$) | | after 4 $C_2D_4$ (Pt$_1$Sn$_{0.1}$) | | $SiO_2$ | |
|---|---|---|---|---|---|---|---|---|---|

Fig. S1

| 1st TPD(Pt$_1$) | $C_2D_4$ | 1st TPD(Pt$_1$) | $D_2$ | 2nd TPD(Pt$_1$) | $C_2D_4$ | 2nd TPD(Pt$_1$) | $D_2$ | 3rd TPD(Pt$_1$) | $C_2D_4$ | 3rd TPD(Pt$_1$) | $D_2$ | 4th TPD(Pt$_1$) | $C_2D_4$ | 4th TPD (Pt$_1$) | $D_2$ | 1st TPD(Pt$_2$) | $C_2D_4$ | 1st TPD(Pt$_2$) | $D_2$ | 2nd TPD(Pt$_2$) | $C_2D_4$ | 2nd TPD(Pt$_2$) | $D_2$ | 3rd TPD(Pt$_2$) | $C_2D_4$ | 3rd TPD(Pt$_2$) | $D_2$ | 4th TPD(Pt$_2$) |
|---|---|---|---|---|---|---|---|---|---|---|---|---|---|---|---|---|---|---|---|---|---|---|---|---|---|---|---|---|

| | |
|---|---|
| 0.6653 | 114.2855 |
| 0.6663 | 114.2855 |
| 0.6674 | 371.4286 |
| 0.6684 | 285.7145 |
| 0.6694 | 133.3331 |
| 0.6704 | 247.6186 |
| 0.6715 | 285.7143 |
| 0.6725 | 247.6193 |
| 0.6735 | 142.8571 |
| 0.6745 | 85.7143 |
| 0.6756 | 47.6191 |
| 0.6766 | 19.0476 |
| 0.6776 | 123.8093 |
| 0.6786 | 495.2379 |
| 0.6797 | 485.7143 |
| 0.6807 | 133.3334 |
| 0.6817 | 228.5712 |
| 0.6828 | 304.7614 |
| 0.6838 | 247.6186 |
| 0.6848 | 209.5236 |
| 0.6858 | 161.905 |
| 0.6869 | 276.1907 |
| 0.6879 | 409.5236 |
| 0.6889 | 304.7616 |
| 0.6899 | 438.0952 |
| 0.691 | 390.4762 |
| 0.692 | 190.4762 |
| 0.693 | 180.9524 |
| 0.694 | 142.8574 |
| 0.6951 | 133.3336 |
| 0.6961 | 114.2857 |
| 0.6971 | 133.3334 |
| 0.6982 | 19.0476 |
| 0.6992 | 171.4286 |
| 0.7002 | 304.7621 |
| 0.7012 | 133.3336 |
| 0.7023 | 0 |
| 0.7033 | 161.905 |
| 0.7043 | 161.905 |
| 0.7053 | 85.7143 |
| 0.7064 | 200 |
| 0.7074 | 219.0479 |
| 0.7084 | 104.7621 |
| 0.7094 | 466.6664 |
| 0.7105 | 552.3807 |
| 0.7115 | 238.095 |
| 0.7125 | 228.5714 |
| 0.7136 | 257.1429 |
| 0.7146 | 295.2379 |
| 0.7156 | 304.7621 |
| 0.7166 | 304.7621 |
| 0.7177 | 142.8571 |
| 0.7187 | 28.5714 |
| 0.7197 | 66.6666 |
| 0.7207 | 66.6666 |
| 0.7218 | 28.5714 |
| 0.7228 | 28.5714 |
| 0.7238 | 0 |
| 0.7248 | 0 |
| 0.7259 | 57.1429 |
| 0.7269 | 171.4286 |
| 0.7279 | 228.5714 |
| 0.729 | 114.2857 |
| 0.73 | 161.905 |
| 0.731 | 180.9526 |
| 0.732 | 19.0476 |
| 0.7331 | 0 |
| 0.7341 | 38.0952 |
| 0.7351 | 38.0952 |
| 0.7361 | 9.5238 |
| 0.7372 | 9.5238 |
| 0.7382 | 0 |
| 0.7392 | 0 |
| 0.7402 | 0 |
| 0.7413 | 0 |
| 0.7423 | 0 |
| 0.7433 | 85.7143 |
| 0.7444 | 85.7143 |
| 0.7454 | 95.2379 |
| 0.7464 | 200 |
| 0.7474 | 104.7621 |
| 0.7485 | 9.5238 |
| 0.7495 | 95.2381 |
| 0.7505 | 85.7143 |
| 0.7515 | 0 |
| 0.7526 | 95.2379 |
| 0.7536 | 104.7616 |
| 0.7546 | 9.5238 |
| 0.7556 | 0 |
| 0.7567 | 0 |
| 0.7577 | 0 |
| 0.7587 | 0 |
| 0.7598 | 0 |
| 0.7608 | 0 |
| 0.7618 | 104.7621 |
| 0.7628 | 104.7621 |
| 0.7639 | 19.0476 |
| 0.7649 | 104.7619 |
| 0.7659 | 95.2381 |
| 0.7669 | 19.0476 |
| 0.768 | 9.5238 |
| 0.769 | 0 |
| 0.77 | 0 |
| 0.771 | 0 |
| 0.7721 | 85.7143 |
| 0.7731 | 171.4286 |
| 0.7741 | 85.7143 |
| 0.7752 | 0 |
| 0.7762 | 0 |
| 0.7772 | 0 |
| 0.7782 | 0 |
| 0.7793 | 0 |
| 0.7803 | 47.6191 |
| 0.7813 | 47.6191 |
| 0.7823 | 28.5714 |
| 0.7834 | 28.5714 |
| 0.7844 | 0 |
| 0.7854 | 0 |
| 0.7864 | 0 |
| 0.7875 | 0 |
| 0.7885 | 0 |
| 0.7895 | 0 |
| 0.7906 | 0 |
| 0.7916 | 85.7143 |
| 0.7926 | 85.7143 |
| 0.7936 | 28.5714 |
| 0.7947 | 104.7621 |
| 0.7957 | 85.7145 |
| 0.7967 | 9.5238 |
| 0.7977 | 0 |
| 0.7988 | 9.5238 |
| 0.7998 | 9.5238 |
| 0.8008 | 0 |
| 0.8018 | 219.0479 |
| 0.8029 | 219.0479 |
| 0.8039 | 0 |
| 0.8049 | 0 |
| 0.806 | 0 |
| 0.807 | 0 |
| 0.808 | 9.5238 |
| 0.809 | 9.5238 |
| 0.8101 | 0 |
| 0.8111 | 0 |
| 0.8121 | 0 |
| 0.8131 | 47.6191 |
| 0.8142 | 114.2857 |
| 0.8152 | 76.1904 |
| 0.8162 | 123.8095 |
| 0.8172 | 114.2857 |
| 0.8183 | 47.6191 |
| 0.8193 | 47.6191 |
| 0.8203 | 9.5238 |
| 0.8214 | 19.0476 |
| 0.8224 | 9.5238 |
| 0.8234 | 0 |
| 0.8244 | 0 |
| 0.8255 | 0 |
| 0.8265 | 0 |
| 0.8275 | 0 |
| 0.8285 | 0 |
| 0.8296 | 0 |
| 0.8306 | 47.6191 |
| 0.8316 | 85.7143 |
| 0.8326 | 47.619 |
| 0.8337 | 9.5238 |
| 0.8347 | 0 |
| 0.8357 | 9.5238 |
| 0.8368 | 76.1904 |
| 0.8378 | 66.6666 |
| 0.8388 | 76.1907 |
| 0.8398 | 190.4764 |
| 0.8409 | 114.2857 |
| 0.8419 | 19.0476 |
| 0.8429 | 19.0476 |
| 0.8439 | 0 |
| 0.845 | 0 |
| 0.846 | 0 |
| 0.847 | 9.5238 |
| 0.848 | 9.5238 |
| 0.8491 | 180.9521 |
| 0.8501 | 190.4759 |
| 0.8511 | 9.5238 |
| 0.8522 | 19.0476 |
| 0.8532 | 19.0476 |
| 0.8542 | 0 |
| 0.8552 | 0 |
| 0.8563 | 9.5238 |
| 0.8573 | 9.5238 |
| 0.8583 | 38.0952 |
| 0.8593 | 38.0952 |
| 0.8604 | 0 |
| 0.8614 | 0 |
| 0.8624 | 95.2379 |
| 0.8634 | 95.2379 |
| 0.8645 | 180.9521 |
| 0.8655 | 209.5236 |
| 0.8665 | 76.1905 |
| 0.8676 | 57.1429 |
| 0.8686 | 9.5238 |
| 0.8696 | 76.1907 |
| 0.8706 | 76.1907 |
| 0.8717 | 9.5238 |
| 0.8727 | 123.8095 |
| 0.8737 | 400 |
| 0.8747 | 400 |
| 0.8758 | 114.2857 |
| 0.8768 | 180.9521 |
| 0.8778 | 190.4759 |
| 0.8789 | 190.4759 |
| 0.8799 | 276.19 |
| 0.8809 | 333.3329 |
| 0.8819 | 257.1426 |
| 0.883 | 85.7143 |
| 0.884 | 171.4288 |
| 0.885 | 152.3812 |
| 0.886 | 152.3812 |
| 0.8871 | 123.8098 |
| 0.8881 | 371.4284 |
| 0.8891 | 571.4286 |
| 0.8901 | 219.0479 |
| 0.8912 | 209.5236 |
| 0.8922 | 514.2857 |
| 0.8932 | 400 |
| 0.8943 | 104.7616 |
| 0.8953 | 114.2859 |
| 0.8963 | 428.5714 |
| 0.8973 | 571.4286 |
| 0.8984 | 447.6193 |
| 0.8994 | 361.905 |
| 0.9004 | 533.3336 |
| 0.9014 | 809.5236 |
| 0.9025 | 561.9043 |
| 0.9035 | 438.095 |
| 0.9045 | 676.1907 |
| 0.9055 | 466.6671 |
| 0.9066 | 314.2857 |
| 0.9076 | 580.9521 |
| 0.9086 | 590.4764 |
| 0.9097 | 371.4286 |
| 0.9107 | 257.1429 |
| 0.9117 | 400 |
| 0.9127 | 304.7616 |
| 0.9138 | 142.8574 |
| 0.9148 | 276.1907 |
| 0.9158 | 161.9048 |
| 0.9168 | 104.7619 |
| 0.9179 | 123.8095 |
| 0.9189 | 285.7145 |
| 0.9199 | 247.6193 |
| 0.9209 | 161.905 |
| 0.922 | 247.6193 |
| 0.923 | 104.7619 |
| 0.924 | 28.5714 |
| 0.9251 | 104.7616 |
| 0.9261 | 104.7616 |
| 0.9271 | 114.2859 |
| 0.9281 | 228.5714 |
| 0.9292 | 295.2379 |
| 0.9302 | 371.4286 |
| 0.9312 | 209.5238 |
| 0.9322 | 257.1431 |
| 0.9333 | 561.905 |
| 0.9343 | 695.2379 |
| 0.9353 | 633.3329 |
| 0.9363 | 776.19 |
| 0.9374 | 876.19 |
| 0.9384 | 519.0471 |
| 0.9394 | 619.0471 |
| 0.9405 | 861.9043 |
| 0.9415 | 952.3807 |
| 0.9425 | 1028.571 |
| 0.9435 | 1128.571 |
| 0.9446 | 1290.476 |
| 0.9456 | 1071.426 |
| 0.9466 | 1133.329 |
| 0.9476 | 1400 |
| 0.9487 | 1400 |
| 0.9497 | 1357.141 |
| 0.9507 | 1476.191 |
| 0.9517 | 1100 |
| 0.9528 | 609.5236 |
| 0.9538 | 966.6664 |
| 0.9548 | 1057.143 |
| 0.9559 | 609.5238 |
| 0.9569 | 638.0952 |
| 0.9579 | 823.8095 |
| 0.9589 | 571.4288 |
| 0.96 | 223.81 |
| 0.961 | 90.4764 |
| 0.962 | 222.1429 |
| 0.963 | 340 |
| 0.9641 | 189.2857 |
| 0.9651 | 0 |
| 0.9661 | 9.5238 |
| 0.9671 | 95.2381 |
| 0.9682 | 85.7143 |
| 0.9692 | 0 |
| 0.9702 | 0 |
| 0.9713 | 0 |
| 0.9723 | 28.5714 |
| 0.9733 | 28.5714 |
| 0.9743 | 0 |
| 0.9754 | 76.1907 |
| 0.9764 | 76.1907 |
| 0.9774 | 19.0476 |
| 0.9784 | 19.0476 |
| 0.9795 | 0 |
| 0.9805 | 0 |
| 0.9815 | 9.5238 |
| 0.9825 | 9.5238 |
| 0.9836 | 0 |
| 0.9846 | 0 |
| 0.9856 | 133.3336 |
| 0.9867 | 228.5714 |
| 0.9877 | 219.0471 |
| 0.9887 | 123.8093 |
| 0.9897 | 19.0476 |
| 0.9908 | 19.0476 |
| 0.9918 | 0 |
| 0.9928 | 0 |
| 0.9938 | 95.2379 |
| 0.9949 | 95.2379 |
| 0.9959 | 9.5238 |
| 0.9969 | 9.5238 |
| 0.9979 | 0 |
| 0.999 | 0 |
| 1 | 0 |
| 1.001 | 0 |
| 1.0021 | 0 |
| 1.0031 | 152.3807 |
| 1.0041 | 333.3329 |
| 1.0051 | 180.9521 |
| 1.0062 | 0 |
| 1.0072 | 0 |
| 1.0082 | 114.2857 |
| 1.0092 | 133.3334 |
| 1.0103 | 95.2384 |
| 1.0113 | 161.905 |
| 1.0123 | 85.7143 |
| 1.0133 | 0 |
| 1.0144 | 0 |
| 1.0154 | 0 |
| 1.0164 | 0 |
| 1.0175 | 28.5714 |
| 1.0185 | 28.5714 |
| 1.0195 | 0 |
| 1.0205 | 9.5238 |
| 1.0216 | 9.5238 |
| 1.0226 | 0 |
| 1.0236 | 0 |
| 1.0246 | 76.1907 |
| 1.0257 | 76.1907 |
| 1.0267 | 0 |

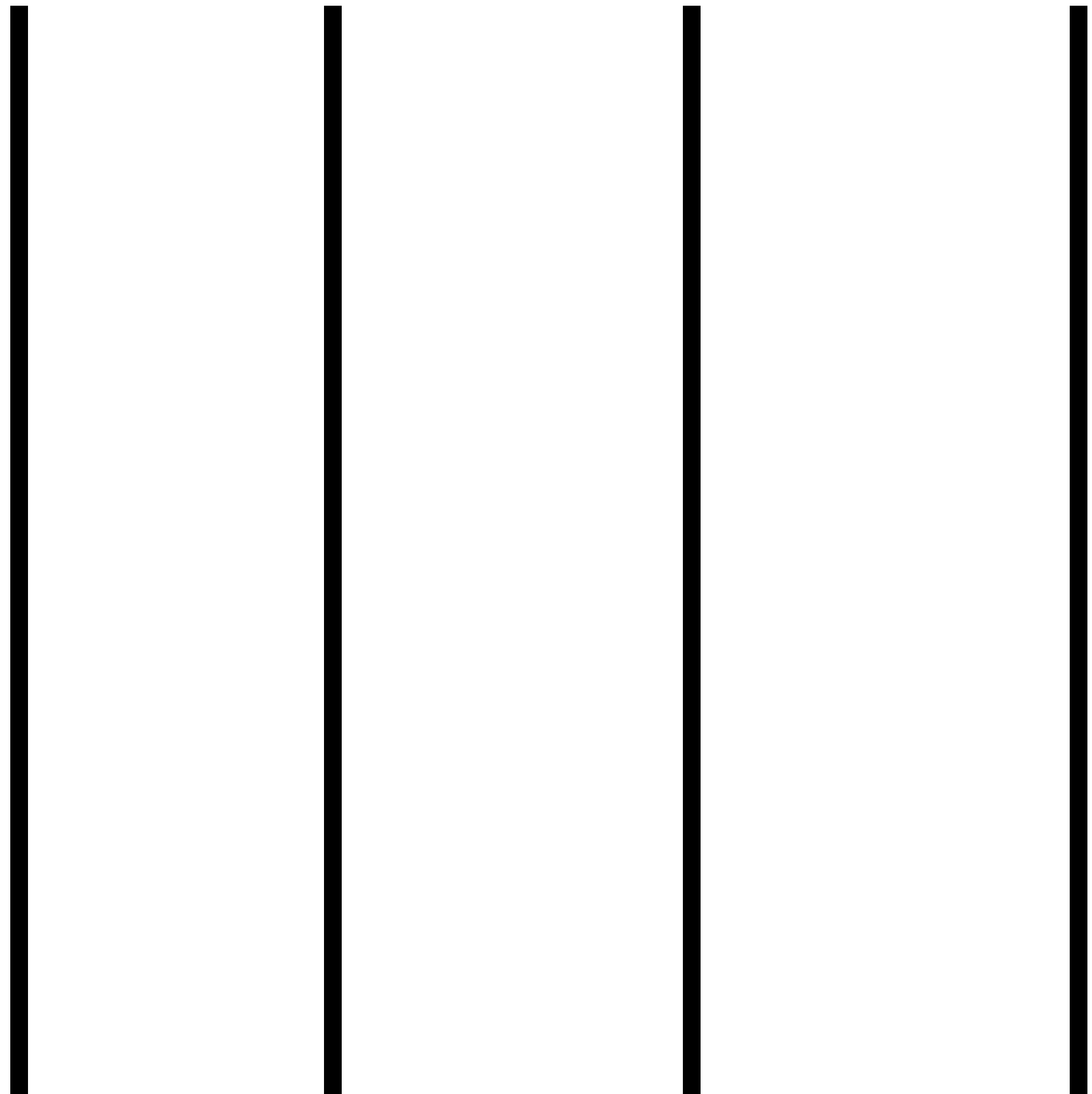

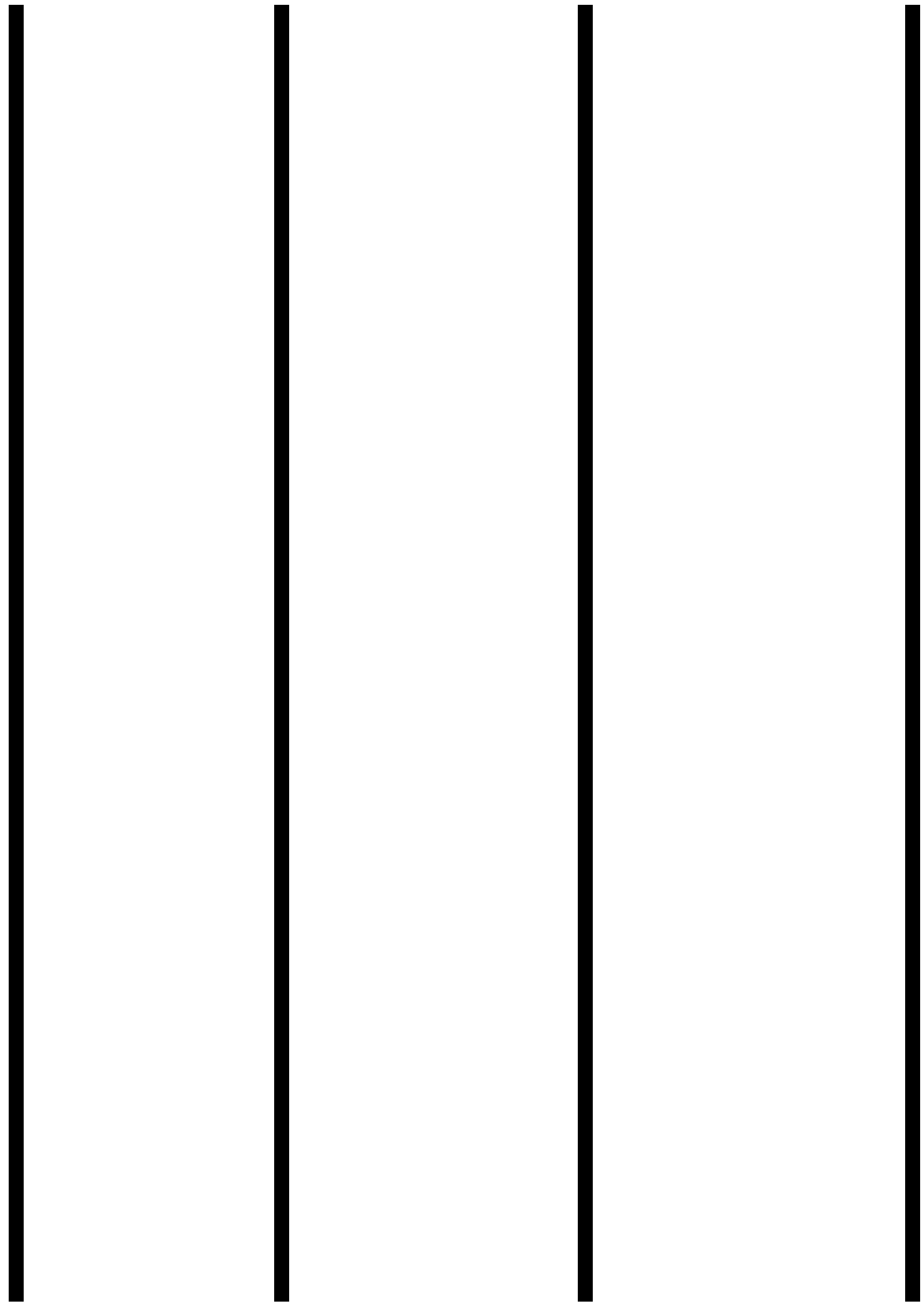

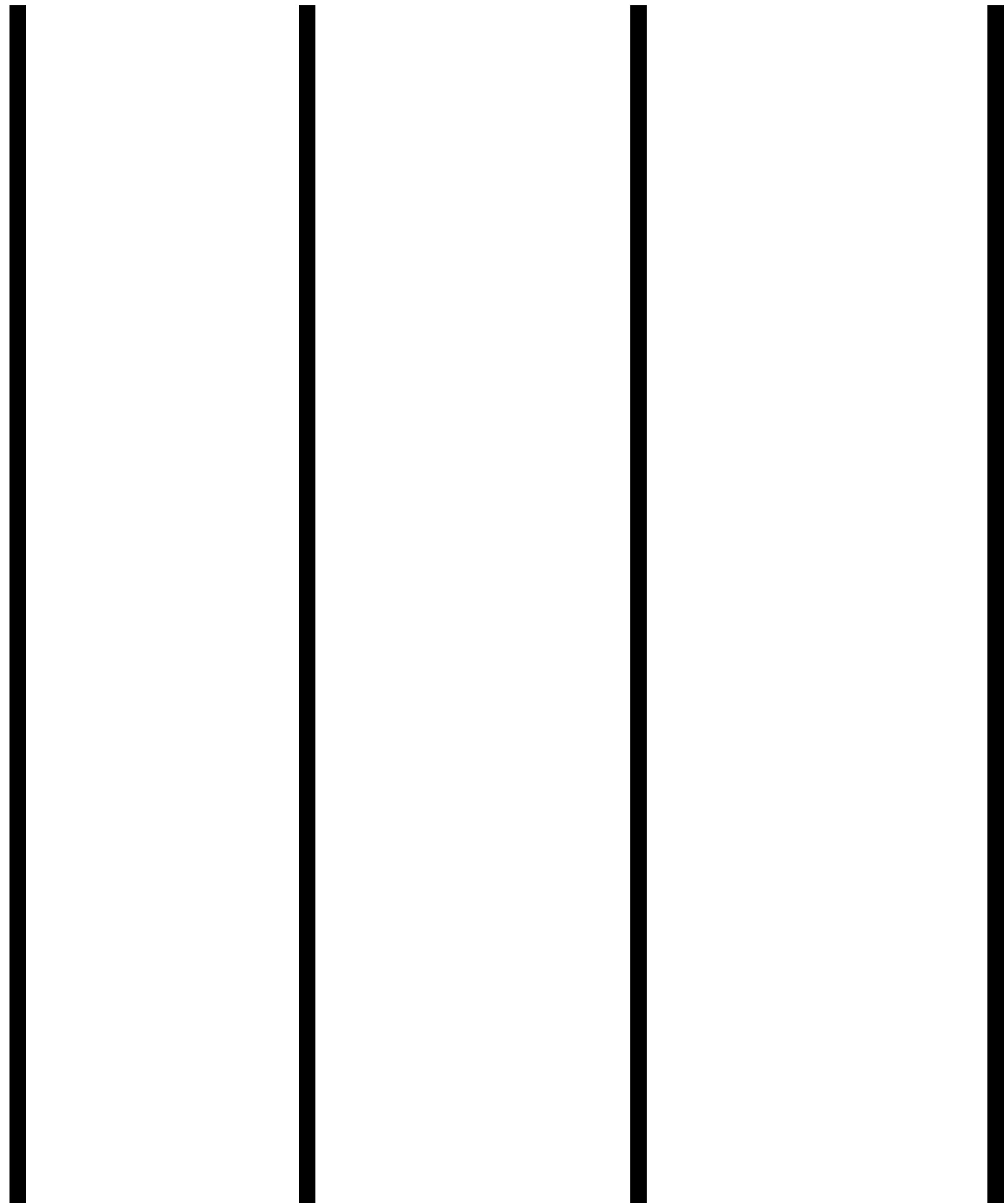

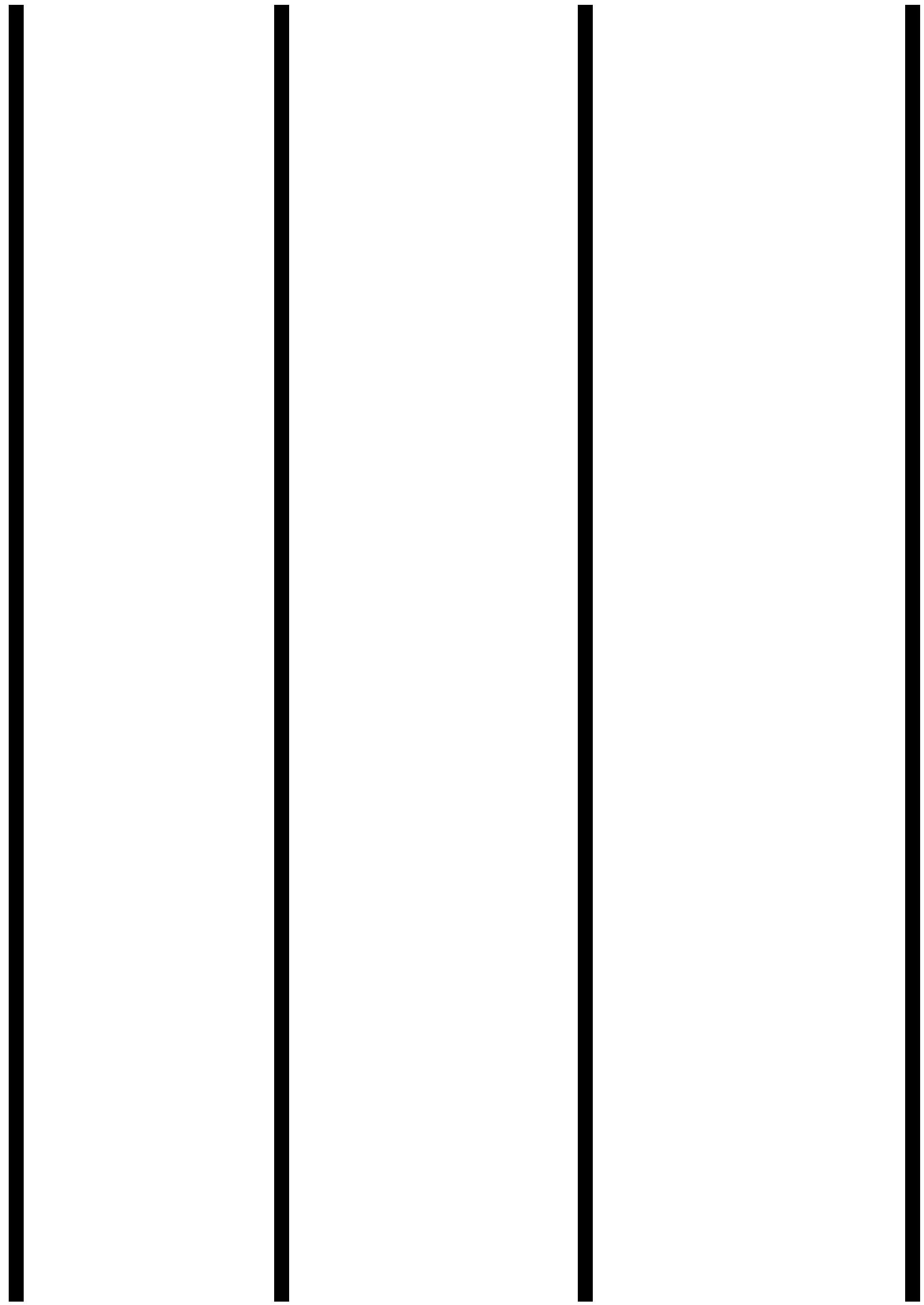

| C2D4 | 4th TPD (Pt1) | D2 | SiO2 | |
|---|---|---|---|---|
| 3652129429 | 179.2471 | -350405511.4 | 178.3067 | 619865914 |
| 1323731925 | 179.6078 | -194536650.1 | 181.8359 | 2256348902 |
| -912139556.9 | 179.532 | -168558705.5 | 185.7283 | 951455621.5 |
| [illegible] | [illegible] | [illegible] | [illegible] | [illegible] |

Fig. S2

Top

| 1st TPD(Pt1Sn1.5) | C2D4 | 1st TPD(Pt1Sn1.5) | (Inset)D2 | 2nd TPD(Pt1Sn1.5) | C2D4 | 2nd TPD(Pt1Sn1.5) | (Inset)D2 | 3rd TPD(Pt1Sn1.5) | C2D4 | 3rd TPD(Pt1Sn1.5) | (Inset)D2 | 4th TPD(Pt1Sn1.5) | C2D4 | 4th TPD(Pt1Sn1.5) | (Inset)D2 |
|---|---|---|---|---|---|---|---|---|---|---|---|---|---|---|---|
| 177.8219 | 448944229.7 | 178.2503 | -379831873.7 | 177.5267 | -835118040.3 | 178.1409 | -597690472.4 | 177.359 | -699368205.2 | 177.8769 | -425244552.5 | 177.3277 | -234651362.8 | 177.0432 | -852616219.2 |
| 177.4456 | 1462340234 | 177.4951 | -202574811.8 | 176.9107 | 2369747952 | 177.0819 | 382653964.3 | 176.8734 | 848900213.8 | 177.0213 | -426307817.3 | 176.9008 | 20885413.56 | 177.2622 | 81572793.24 |
| 177.8347 | 2212794896 | 177.5625 | -336049181 | 176.8116 | 494236947 | 176.9429 | -509448451.1 | 176.9212 | 2778803034 | 176.8915 | 287701194.7 | 176.8958 | 181183306.5 | 177.0517 | 170440810.2 |
| [illegible] | [illegible] | [illegible] | [illegible] | [illegible] | [illegible] | [illegible] | [illegible] | [illegible] | [illegible] | [illegible] | [illegible] | [illegible] | [illegible] | [illegible] | [illegible] |

Bottom

| 1st TPD(Pt1Sn1.5) | C2D4 | 1st TPD(Pt1Sn1.5) | (Inset)D2 | 2nd TPD(Pt1Sn1.5) | C2D4 | 2nd TPD(Pt1Sn1.5) | (Inset)D2 | 3rd TPD(Pt1Sn1.5) | C2D4 | 3rd TPD(Pt1Sn1.5) | (Inset)D2 | 4th TPD(Pt1Sn1.5) | C2D4 | 4th TPD(Pt1Sn1.5) | (Inset)D2 | SiO2 | |
|---|---|---|---|---|---|---|---|---|---|---|---|---|---|---|---|---|---|
| 179.8303 | 3992384597 | 179.817 | 180181617.6 | 180.1414 | 1726243903 | 180.1938 | -5E+08 | 179.9429 | -614022153.3 | 180.0082 | -146140172.8 | 179.6874 | 828892677.5 | 179.707 | -257797483.3 | 180.2807 | 800100021.9 |
| 179.5969 | -111972203.1 | 179.685 | -138212348.3 | 179.8079 | 651883636.6 | 180.0636 | 3.1E+08 | 179.8618 | 388231722.4 | 179.8629 | 221312600.9 | 179.9035 | -947684835.6 | 180.0442 | 305650847.8 | 179.7449 | 140238600.2 |
| 179.5895 | 890810217.8 | 179.6447 | 449203968.1 | 179.5725 | -1861305769 | 179.6689 | 1.5E+07 | 179.7286 | 1215172953 | 179.9091 | -268538083.9 | 179.6694 | -2347973040 | 179.8881 | -37321277.21 | 179.2837 | -219822470.7 |
| [illegible] | [illegible] | [illegible] | [illegible] | [illegible] | [illegible] | [illegible] | [illegible] | [illegible] | [illegible] | [illegible] | [illegible] | [illegible] | [illegible] | [illegible] | [illegible] | [illegible] | [illegible] |

Fig. S3

| HCl TPD(Pt1Sn1.5) | | HCl TPD(Pt1Sn1.5) | |
|---|---|---|---|
| 179.8388 | 564742386.5 | 180.0332 | 114605686.4 |
| 179.6082 | 124461452.3 | 179.6723 | 211939492.3 |
| 179.5959 | 251181422.9 | 179.4135 | -92111343.94 |
| [illegible] | [illegible] | [illegible] | [illegible] |

Fig. S4

| Top: Pt1Sn1.5/SiO2-post TPD | | Fit | | Bottom: Pt1Sn1.5/SiO2-post TPD | | Fit | | Both: SiO2 raw | | SiO2 Fit | |
|---|---|---|---|---|---|---|---|---|---|---|---|
| 300.3 | 0 | 300.3 | 0 | 300.6 | 0 | 300.6 | 0 | 300 | 29.31 | 300 | 0 |
| 300.2 | 17.46 | 300.2 | 0 | 300.5 | 0 | 300.5 | 0 | 299.9 | -21.67 | 299.9 | 0 |
| 300.1 | 37.5 | 300.1 | 0 | 300.4 | 0 | 300.4 | 0 | 299.8 | -41.67 | 299.8 | 0 |
| [illegible] | [illegible] | [illegible] | [illegible] | [illegible] | [illegible] | [illegible] | [illegible] | [illegible] | [illegible] | [illegible] | [illegible] |

Fig. S5

| 1st C2D4 TPD Raw Data | | SiO2 Fit | | Low Temp Fit | |
|---|---|---|---|---|---|
| 177.928 | -988995891.7 | 131 | 14526.715 | 131 | 20607.88 |
| 177.29 | 1830902237 | 132 | 20529.85 | 132 | 26853.46 |
| 176.896 | -1143244080 | 133 | 28869.068 | 133 | 35058.08 |
| [illegible] | [illegible] | [illegible] | [illegible] | [illegible] | [illegible] |

| | | | |
|---|---|---|---|
| 479 | 0 | 479 | 33680033 |
| 480 | 0 | 480 | 31673991 |
| 481 | 0 | 481 | 29847834 |
| 482 | 0 | 482 | 28119141 |
| 483 | 0 | 483 | 26483321 |
| 484 | 0 | 484 | 24935935 |
| 485 | 0 | 485 | 23472701 |
| 486 | 0 | 486 | 22089496 |
| 487 | 0 | 487 | 20782361 |
| 488 | 0 | 488 | 19547495 |
| 489 | 0 | 489 | 18381258 |
| 490 | 0 | 490 | 17280164 |
| 491 | 0 | 491 | 16240880 |
| 492 | 0 | 492 | 15260219 |
| 493 | 0 | 493 | 14335139 |
| 494 | 0 | 494 | 13462738 |
| 495 | 0 | 495 | 12640245 |
| 496 | 0 | 496 | 11865022 |
| 497 | 0 | 497 | 11134553 |
| 498 | 0 | 498 | 10446443 |
| 499 | 0 | 499 | 9798414 |
| 500 | 0 | 500 | 9188294 |
| 501 | 0 | 501 | 8614023 |
| 502 | 0 | 502 | 8073639 |
| 503 | 0 | 503 | 7565279 |
| 504 | 0 | 504 | 7087172 |
| 505 | 0 | 505 | 6637638 |
| 506 | 0 | 506 | 6215081 |
| 507 | 0 | 507 | 5817988 |
| 508 | 0 | 508 | 5444922 |
| 509 | 0 | 509 | 5094523 |
| 510 | 0 | 510 | 4765499 |
| 511 | 0 | 511 | 4456628 |
| 512 | 0 | 512 | 4166751 |
| 513 | 0 | 513 | 3894771 |
| 514 | 0 | 514 | 3639650 |
| 515 | 0 | 515 | 3400406 |
| 516 | 0 | 516 | 3176108 |
| 517 | 0 | 517 | 2965878 |
| 518 | 0 | 518 | 2768884 |
| 519 | 0 | 519 | 2584341 |
| 520 | 0 | 520 | 2411507 |
| 521 | 0 | 521 | 2249681 |
| 522 | 0 | 522 | 2098200 |
| 523 | 0 | 523 | 1956441 |
| 524 | 0 | 524 | 1823813 |
| 525 | 0 | 525 | 1699760 |
| 526 | 0 | 526 | 1583758 |
| 527 | 0 | 527 | 1475311 |
| 528 | 0 | 528 | 1373955 |
| 529 | 0 | 529 | 1279249 |
| 530 | 0 | 530 | 1190781 |
| 531 | 0 | 531 | 1108159 |
| 532 | 0 | 532 | 1031019 |
| 533 | 0 | 533 | 959013.7 |
| 534 | 0 | 534 | 891819.5 |
| 535 | 0 | 535 | 829130.7 |
| 536 | 0 | 536 | 770660.2 |
| 537 | 0 | 537 | 716137.9 |
| 538 | 0 | 538 | 665310.2 |
| 539 | 0 | 539 | 617938.9 |
| 540 | 0 | 540 | 573800 |
| 541 | 0 | 541 | 532683.5 |
| 542 | 0 | 542 | 494392 |
| 543 | 0 | 543 | 458740.6 |
| 544 | 0 | 544 | 425555.5 |
| 545 | 0 | 545 | 394674 |
| 546 | 0 | 546 | 365943.3 |
| 547 | 0 | 547 | 339220.5 |
| 548 | 0 | 548 | 314371.3 |
| 549 | 0 | 549 | 291270.2 |
| 550 | 0 | 550 | 269799.7 |
| 551 | 0 | 551 | 249849.5 |
| 552 | 0 | 552 | 231316.7 |
| 553 | 0 | 553 | 214104.7 |
| 554 | 0 | 554 | 198123.4 |
| 555 | 0 | 555 | 183288.5 |
| 556 | 0 | 556 | 169521.1 |
| 557 | 0 | 557 | 156747.4 |
| 558 | 0 | 558 | 144898.6 |
| 559 | 0 | 559 | 133910.4 |
| 560 | 0 | 560 | 123722.7 |
| 561 | 0 | 561 | 114279.4 |
| 562 | 0 | 562 | 105528.2 |
| 563 | 0 | 563 | 97420.3 |
| 564 | 0 | 564 | 89910.07 |
| 565 | 0 | 565 | 82955.05 |
| 566 | 0 | 566 | 76515.65 |
| 567 | 0 | 567 | 70554.96 |
| 568 | 0 | 568 | 65038.61 |
| 569 | 0 | 569 | 59934.55 |
| 570 | 0 | 570 | 55212.97 |
| 571 | 0 | 571 | 50846.11 |
| 572 | 0 | 572 | 46808.13 |
| 573 | 0 | 573 | 43075 |
| 574 | 0 | 574 | 39624.36 |
| 575 | 0 | 575 | 36435.45 |
| 576 | 0 | 576 | 33488.93 |
| 577 | 0 | 577 | 30766.87 |
| 578 | 0 | 578 | 28252.61 |
| 579 | 0 | 579 | 25930.68 |
| 580 | 0 | 580 | 23786.71 |
| 581 | 0 | 581 | 21807.39 |
| 582 | 0 | 582 | 19980.37 |
| 583 | 0 | 583 | 18294.21 |
| 584 | 0 | 584 | 16738.3 |
| 585 | 0 | 585 | 15302.81 |
| 586 | 0 | 586 | 13978.68 |
| 587 | 0 | 587 | 12757.49 |
| 588 | 0 | 588 | 11631.48 |
| 589 | 0 | 589 | 10593.47 |
| 590 | 0 | 590 | 9636.834 |
| 591 | 0 | 591 | 8755.457 |
| 592 | 0 | 592 | 7943.701 |
| 593 | 0 | 593 | 7196.371 |
| 594 | 0 | 594 | 6508.681 |
| 595 | 0 | 595 | 5876.224 |
| 596 | 0 | 596 | 5294.945 |
| 597 | 0 | 597 | 4761.114 |
| 598 | 0 | 598 | 4271.299 |
| 599 | 0 | 599 | 3822.342 |
| 600 | 0 | 600 | 3411.336 |
| 601 | 0 | 601 | 3035.605 |
| 602 | 0 | 602 | 2692.68 |
| 603 | 0 | 603 | 2380.279 |
| 604 | 0 | 604 | 2096.291 |
| 605 | 0 | 605 | 1838.757 |
| 606 | 0 | 606 | 1605.851 |
| 607 | 0 | 607 | 1395.866 |
| 608 | 0 | 608 | 1207.201 |
| 609 | 0 | 609 | 1038.343 |
| 610 | 0 | 610 | 887.861 |
| 611 | 0 | 611 | 754.393 |
| 612 | 0 | 612 | 636.635 |
| 613 | 0 | 613 | 533.338 |
| 614 | 0 | 614 | 443.299 |
| 615 | 0 | 615 | 365.361 |
| 616 | 0 | 616 | 298.407 |
| 617 | 0 | 617 | 241.363 |
| 618 | 0 | 618 | 193.197 |
| 619 | 0 | 619 | 152.921 |
| 620 | 0 | 620 | 119.597 |
| 621 | 0 | 621 | 92.339 |
| 622 | 0 | 622 | 70.315 |
| 623 | 0 | 623 | 52.757 |
| 624 | 0 | 624 | 38.959 |
| 625 | 0 | 625 | 28.283 |
| 626 | 0 | 626 | 20.159 |
| 627 | 0 | 627 | 14.088 |
| 628 | 0 | 628 | 9.639 |
| 629 | 0 | 629 | 6.446 |
| 630 | 0 | 630 | 4.206 |
| 631 | 0 | 631 | 2.672 |
| 632 | 0 | 632 | 1.649 |
| 633 | 0 | 633 | 0.987 |
| 634 | 0 | 634 | 0.571 |
| 635 | 0 | 635 | 0.318 |
| 636 | 0 | 636 | 0.17 |
| 637 | 0 | 637 | 0.087 |
| 638 | 0 | 638 | 0.043 |
| 639 | 0 | 639 | 0.02 |
| 640 | 0 | 640 | 9.00E-03 |
| 641 | 0 | 641 | 4.00E-03 |
| 642 | 0 | 642 | 1.00E-03 |
| 643 | 0 | 643 | 0 |
| 644 | 0 | 644 | 0 |
| 645 | 0 | 645 | 0 |
| 646 | 0 | 646 | 0 |
| 647 | 0 | 647 | 0 |
| 648 | 0 | 648 | 0 |
| 649 | 0 | 649 | 0 |
| 650 | 0 | 650 | 0 |
| 651 | 0 | 651 | 0 |
| 652 | 0 | 652 | 0 |
| 653 | 0 | 653 | 0 |
| 654 | 0 | 654 | 0 |
| 655 | 0 | 655 | 0 |
| 656 | 0 | 656 | 0 |
| 657 | 0 | 657 | 0 |
| 658 | 0 | 658 | 0 |
| 659 | 0 | 659 | 0 |
| 660 | 0 | 660 | 0 |
| 661 | 0 | 661 | 0 |
| 662 | 0 | 662 | 0 |
| 663 | 0 | 663 | 0 |
| 664 | 0 | 664 | 0 |
| 665 | 0 | 665 | 0 |
| 666 | 0 | 666 | 0 |
| 667 | 0 | 667 | 0 |
| 668 | 0 | 668 | 0 |
| 669 | 0 | 669 | 0 |
| 670 | 0 | 670 | 0 |
| 671 | 0 | 671 | 0 |
| 672 | 0 | 672 | 0 |
| 673 | 0 | 673 | 0 |
| 674 | 0 | 674 | 0 |
| 675 | 0 | 675 | 0 |
| 676 | 0 | 676 | 0 |
| 677 | 0 | 677 | 0 |
| 678 | 0 | 678 | 0 |
| 679 | 0 | 679 | 0 |
| 680 | 0 | 680 | 0 |
| 681 | 0 | 681 | 0 |
| 682 | 0 | 682 | 0 |
| 683 | 0 | 683 | 0 |
| 684 | 0 | 684 | 0 |
| 685 | 0 | 685 | 0 |
| 686 | 0 | 686 | 0 |
| 687 | 0 | 687 | 0 |
| 688 | 0 | 688 | 0 |
| 689 | 0 | 689 | 0 |
| 690 | 0 | 690 | 0 |
| 691 | 0 | 691 | 0 |
| 692 | 0 | 692 | 0 |
| 693 | 0 | 693 | 0 |
| 694 | 0 | 694 | 0 |
| 695 | 0 | 695 | 0 |
| 696 | 0 | 696 | 0 |
| 697 | 0 | 697 | 0 |
| 698 | 0 | 698 | 0 |
| 699 | 0 | 699 | 0 |
| 700 | 0 | 700 | 0 |
| 701 | 0 | 701 | 0 |
| 702 | 0 | 702 | 0 |
| 703 | 0 | 703 | 0 |
| 704 | 0 | 704 | 0 |
| 705 | 0 | 705 | 0 |
| 706 | 0 | 706 | 0 |
| 707 | 0 | 707 | 0 |
| 708 | 0 | 708 | 0 |
| 709 | 0 | 709 | 0 |
| 710 | 0 | 710 | 0 |
| 711 | 0 | 711 | 0 |
| 712 | 0 | 712 | 0 |
| 713 | 0 | 713 | 0 |
| 714 | 0 | 714 | 0 |
| 715 | 0 | 715 | 0 |
| 716 | 0 | 716 | 0 |
| 717 | 0 | 717 | 0 |
| 718 | 0 | 718 | 0 |
| 719 | 0 | 719 | 0 |
| 720 | 0 | 720 | 0 |
| 721 | 0 | 721 | 0 |
| 722 | 0 | 722 | 0 |
| 723 | 0 | 723 | 0 |
| 724 | 0 | 724 | 0 |
| 725 | 0 | 725 | 0 |
| 726 | 0 | 726 | 0 |
| 727 | 0 | 727 | 0 |
| 728 | 0 | 728 | 0 |
| 729 | 0 | 729 | 0 |
| 730 | 0 | 730 | 0 |
| 731 | 0 | 731 | 0 |
| 732 | 0 | 732 | 0 |
| 733 | 0 | 733 | 0 |
| 734 | 0 | 734 | 0 |
| 735 | 0 | 735 | 0 |
| 736 | 0 | 736 | 0 |
| 737 | 0 | 737 | 0 |
| 738 | 0 | 738 | 0 |
| 739 | 0 | 739 | 0 |
| 740 | 0 | 740 | 0 |
| 741 | 0 | 741 | 0 |
| 742 | 0 | 742 | 0 |
| 743 | 0 | 743 | 0 |
| 744 | 0 | 744 | 0 |
| 745 | 0 | 745 | 0 |
| 746 | 0 | 746 | 0 |
| 747 | 0 | 747 | 0 |
| 748 | 0 | 748 | 0 |
| | 0 | | 0 |
| | 0 | | 0 |
| | 0 | | 0 |
| | 0 | | 0 |
| | 0 | | 0 |
| | 0 | | 0 |
| | 0 | | 0 |
| | 0 | | 0 |
| | 0 | | 0 |
| | 0 | | 0 |
| | 0 | | 0 |
| | 0 | | 0 |
| | 0 | | 0 |
| | 0 | | 0 |
| | 0 | | 0 |
| | 0 | | 0 |
| | 0 | | 0 |
| | 0 | | 0 |
| | 0 | | 0 |
| | 0 | | 0 |
| | 0 | | 0 |
| | 0 | | 0 |
| | 0 | | 0 |
| | 0 | | 0 |
| | 0 | | 0 |
| | 0 | | 0 |
| | 0 | | 0 |
| | 0 | | 0 |
| | 0 | | 0 |
| | 0 | | 0 |
| | 0 | | 0 |
| | 0 | | 0 |
| | 0 | | 0 |
| | 0 | | 0 |
| | 0 | | 0 |
| | 0 | | 0 |
| | 0 | | 0 |
| | 0 | | 0 |
| | 0 | | 0 |
| | 0 | | 0 |
| | 0 | | 0 |
| | 0 | | 0 |
| | 0 | | 0 |
| | 0 | | 0 |
| | 0 | | 0 |
| | 0 | | 0 |
| | 0 | | 0 |
| | 0 | | 0 |
| | 0 | | 0 |
| | 0 | | 0 |
| | 0 | | 0 |
| | 0 | | 0 |
| | 0 | | 0 |
| | 0 | | 0 |
| | 0 | | 0 |
| | 0 | | 0 |
| | 0 | | 0 |
| | 0 | | 0 |
| | 0 | | 0 |
| | 0 | | 0 |
| | 0 | | 0 |
| | 0 | | 0 |
| | 0 | | 0 |
| | 0 | | 0 |
| | 0 | | 0 |
| | 0 | | 0 |
| | 0 | | 0 |
| | 0 | | 0 |
| | 0 | | 0 |
| | 0 | | 0 |
| | 0 | | 0 |
| | 0 | | 0 |
| | 0 | | 0 |
| | 0 | | 0 |
| | 0 | | 0 |
| | 0 | | 0 |
| | 0 | | 0 |
| | 0 | | 0 |
| | 0 | | 0 |

| High Temp Fit | | Envelope | | Fig. S6 | SiO₂ (Pt₄) | | Low Temp(Pt₄) | | High Temp(Pt₄) | |
|---|---|---|---|---|---|---|---|---|---|---|
| 131 | 15.733 | 131 | 35150.32 | Top | 131 | 35150.323 | 0.25 | 159608.508 | 0.25 | 0.021 |
| 132 | 19.336 | 132 | 47402.64 | | 132 | 47402.643 | 0.251 | 165440.185 | 0.251 | 0.023 |
| 133 | 23.944 | 133 | 63951.09 | | 133 | 63951.091 | 0.252 | 171473.738 | 0.252 | 0.024 |
| 134 | 29.694 | 134 | 86157.19 | | 134 | 86157.185 | 0.253 | 177715.726 | 0.253 | 0.025 |
| 135 | 36.781 | 135 | 115822.2 | | 135 | 115822.169 | 0.254 | 184172.906 | 0.254 | 0.027 |
| 136 | 45.442 | 136 | 155302.1 | | 136 | 155302.132 | 0.255 | 190852.24 | 0.255 | 0.029 |
| 137 | 55.959 | 137 | 207662.8 | | 137 | 207662.784 | 0.256 | 197760.895 | 0.256 | 0.03 |
| 138 | 68.662 | 138 | 276878.2 | | 138 | 276878.186 | 0.257 | 204906.255 | 0.257 | 0.032 |
| 139 | 83.937 | 139 | 368082.7 | | 139 | 368082.701 | 0.258 | 212295.921 | 0.258 | 0.034 |
| 140 | 102.232 | 140 | 487888.4 | | 140 | 487888.431 | 0.259 | 219937.722 | 0.259 | 0.036 |
| 141 | 124.069 | 141 | 644783.5 | | 141 | 644783.509 | 0.26 | 227839.717 | 0.26 | 0.038 |
| 142 | 150.05 | 142 | 849630.2 | | 142 | 849630.221 | 0.261 | 236010.204 | 0.261 | 0.041 |
| 143 | 180.87 | 143 | 1116286 | | 143 | 1116286.267 | 0.262 | 244457.725 | 0.262 | 0.043 |
| 144 | 217.328 | 144 | 1462377 | | 144 | 1462377.284 | 0.263 | 253191.073 | 0.263 | 0.045 |
| 145 | 260.343 | 145 | 1910254 | | 145 | 1910254.008 | 0.264 | 262219.298 | 0.264 | 0.048 |
| 146 | 310.975 | 146 | 2488174 | | 146 | 2488173.72 | 0.265 | 271551.715 | 0.265 | 0.051 |
| 147 | 370.442 | 147 | 3231753 | | 147 | 3231753.044 | 0.266 | 281197.909 | 0.266 | 0.054 |
| 148 | 440.145 | 148 | 4185748 | | 148 | 4185747.952 | 0.267 | 291167.745 | 0.267 | 0.057 |
| 149 | 521.694 | 149 | 5406226 | | 149 | 5406225.785 | 0.268 | 301471.371 | 0.268 | 0.061 |
| 150 | 616.938 | 150 | 6963205 | | 150 | 6963204.757 | 0.269 | 312119.23 | 0.269 | 0.064 |
| 151 | 728.001 | 151 | 8943847 | | 151 | 8943847.024 | 0.27 | 323122.065 | 0.27 | 0.068 |
| 152 | 857.317 | 152 | 11456303 | | 152 | 11456302.62 | 0.271 | 334490.927 | 0.271 | 0.072 |
| 153 | 1007.677 | 153 | 14634314 | | 153 | 14634314.5 | 0.272 | 346237.184 | 0.272 | 0.076 |
| 154 | 1182.275 | 154 | 18642702 | | 154 | 18642701.7 | 0.273 | 358372.528 | 0.273 | 0.081 |
| 155 | 1384.764 | 155 | 23683845 | | 155 | 23683844.57 | 0.274 | 370908.982 | 0.274 | 0.086 |
| 156 | 1619.322 | 156 | 30005309 | | 156 | 30005308.55 | 0.275 | 383858.913 | 0.275 | 0.091 |
| 157 | 1890.714 | 157 | 37908721 | | 157 | 37908720.73 | 0.276 | 397235.035 | 0.276 | 0.096 |
| 158 | 2204.38 | 158 | 47760016 | | 158 | 47760016.27 | 0.277 | 411050.423 | 0.277 | 0.102 |
| 159 | 2566.519 | 159 | 60001142 | | 159 | 60001141.57 | 0.278 | 425318.518 | 0.278 | 0.108 |
| 160 | 2984.185 | 160 | 75163224 | | 160 | 75163223.57 | 0.279 | 440053.138 | 0.279 | 0.114 |
| 161 | 3465.408 | 161 | 93881206 | | 161 | 93881205.76 | 0.28 | 455268.487 | 0.28 | 0.121 |
| 162 | 4019.308 | 162 | 1.17E+08 | | 162 | 116909748.3 | 0.281 | 470979.166 | 0.281 | 0.128 |
| 163 | 4656.247 | 163 | 1.45E+08 | | 163 | 145140152.8 | 0.282 | 487200.18 | 0.282 | 0.135 |
| 164 | 5387.975 | 164 | 1.8E+08 | | 164 | 179617771.5 | 0.283 | 503946.951 | 0.283 | 0.143 |
| 165 | 6227.814 | 165 | 2.22E+08 | | 165 | 221559212.7 | 0.284 | 521235.326 | 0.284 | 0.151 |
| 166 | 7190.849 | 166 | 2.72E+08 | | 166 | 272368279.4 | 0.285 | 539081.588 | 0.285 | 0.16 |
| 167 | 8294.149 | 167 | 3.34E+08 | | 167 | 333649290.9 | 0.286 | 557502.468 | 0.286 | 0.169 |
| 168 | 9557.008 | 168 | 4.07E+08 | | 168 | 407216013.5 | 0.287 | 576515.155 | 0.287 | 0.179 |
| 169 | 11001.22 | 169 | 4.95E+08 | | 169 | 495094064.5 | 0.288 | 596137.304 | 0.288 | 0.189 |
| 170 | 12651.39 | 170 | 6E+08 | | 170 | 599514221.5 | 0.289 | 616387.054 | 0.289 | 0.2 |
| 171 | 14535.25 | 171 | 7.23E+08 | | 171 | 722893859 | 0.29 | 637283.033 | 0.29 | 0.212 |
| 172 | 16684.03 | 172 | 8.68E+08 | | 172 | 867803369.5 | 0.291 | 658844.372 | 0.291 | 0.224 |
| 173 | 19132.92 | 173 | 1.04E+09 | | 173 | 1036914786 | 0.292 | 681090.721 | 0.292 | 0.237 |
| 174 | 21921.47 | 174 | 1.23E+09 | | 174 | 1232929620 | 0.293 | 704042.254 | 0.293 | 0.251 |
| 175 | 25094.15 | 175 | 1.46E+09 | | 175 | 1458484472 | 0.294 | 727719.686 | 0.294 | 0.265 |
| 176 | 28700.85 | 176 | 1.72E+09 | | 176 | 1716032928 | 0.295 | 752144.285 | 0.295 | 0.28 |
| 177 | 32797.58 | 177 | 2.01E+09 | | 177 | 2007705559 | 0.296 | 777337.886 | 0.296 | 0.296 |
| [illegible] | [illegible] | [illegible] | [illegible] | | [illegible] | [illegible] | [illegible] | [illegible] | [illegible] | [illegible] |

| Envelope(Pt₄) | | | SiO₂ (Pt₄Sn₃) | | Low Temp(Pt₄Sn₃) | | High Temp(Pt₄Sn₃) | | Envelope(Pt₄Sn₃) | |
|---|---|---|---|---|---|---|---|---|---|---|
| 0.25 | 159608.53 | Bottom | 0.25 | 0 | 0.25 | 1.00E-03 | 0.25 | 0 | 0.25 | 1.00E-03 |
| 0.251 | 165440.208 | | 0.251 | 0 | 0.251 | 1.00E-03 | 0.251 | 0 | 0.251 | 1.00E-03 |
| 0.252 | 171473.762 | | 0.252 | 0 | 0.252 | 1.00E-03 | 0.252 | 0 | 0.252 | 2.00E-03 |
| 0.253 | 177715.751 | | 0.253 | 0 | 0.253 | 2.00E-03 | 0.253 | 0 | 0.253 | 2.00E-03 |
| 0.254 | 184172.933 | | 0.254 | 0 | 0.254 | 2.00E-03 | 0.254 | 0 | 0.254 | 2.00E-03 |
| 0.255 | 190852.269 | | 0.255 | 0 | 0.255 | 2.00E-03 | 0.255 | 0 | 0.255 | 2.00E-03 |
| 0.256 | 197760.926 | | 0.256 | 0 | 0.256 | 2.00E-03 | 0.256 | 0 | 0.256 | 3.00E-03 |
| 0.257 | 204906.287 | | 0.257 | 0 | 0.257 | 3.00E-03 | 0.257 | 0 | 0.257 | 3.00E-03 |
| 0.258 | 212295.955 | | 0.258 | 0 | 0.258 | 3.00E-03 | 0.258 | 0 | 0.258 | 3.00E-03 |
| 0.259 | 219937.758 | | 0.259 | 0 | 0.259 | 4.00E-03 | 0.259 | 0 | 0.259 | 4.00E-03 |
| 0.26 | 227839.755 | | 0.26 | 0 | 0.26 | 4.00E-03 | 0.26 | 0 | 0.26 | 4.00E-03 |
| 0.261 | 236010.244 | | 0.261 | 0 | 0.261 | 4.00E-03 | 0.261 | 0 | 0.261 | 5.00E-03 |
| 0.262 | 244457.768 | | 0.262 | 0 | 0.262 | 5.00E-03 | 0.262 | 0 | 0.262 | 5.00E-03 |
| 0.263 | 253191.118 | | 0.263 | 0 | 0.263 | 6.00E-03 | 0.263 | 0 | 0.263 | 6.00E-03 |
| 0.264 | 262219.346 | | 0.264 | 0 | 0.264 | 6.00E-03 | 0.264 | 0 | 0.264 | 7.00E-03 |
| 0.265 | 271551.766 | | 0.265 | 0 | 0.265 | 7.00E-03 | 0.265 | 0 | 0.265 | 7.00E-03 |
| 0.266 | 281197.963 | | 0.266 | 0 | 0.266 | 8.00E-03 | 0.266 | 0 | 0.266 | 8.00E-03 |
| 0.267 | 291167.802 | | 0.267 | 0 | 0.267 | 9.00E-03 | 0.267 | 0 | 0.267 | 9.00E-03 |
| 0.268 | 301471.431 | | 0.268 | 0 | 0.268 | 0.01 | 0.268 | 0 | 0.268 | 0.011 |
| 0.269 | 312119.294 | | 0.269 | 0 | 0.269 | 0.012 | 0.269 | 0 | 0.269 | 0.012 |
| 0.27 | 323122.133 | | 0.27 | 0 | 0.27 | 0.013 | 0.27 | 0 | 0.27 | 0.014 |
| 0.271 | 334490.999 | | 0.271 | 0 | 0.271 | 0.015 | 0.271 | 0 | 0.271 | 0.015 |
| 0.272 | 346237.261 | | 0.272 | 0 | 0.272 | 0.017 | 0.272 | 0 | 0.272 | 0.017 |
| 0.273 | 358372.609 | | 0.273 | 0 | 0.273 | 0.019 | 0.273 | 0 | 0.273 | 0.019 |
| 0.274 | 370909.068 | | 0.274 | 0 | 0.274 | 0.021 | 0.274 | 0 | 0.274 | 0.022 |
| 0.275 | 383859.004 | | 0.275 | 0 | 0.275 | 0.024 | 0.275 | 0 | 0.275 | 0.024 |
| 0.276 | 397235.131 | | 0.276 | 0 | 0.276 | 0.027 | 0.276 | 0 | 0.276 | 0.027 |
| 0.277 | 411050.525 | | 0.277 | 0 | 0.277 | 0.03 | 0.277 | 0 | 0.277 | 0.031 |
| 0.278 | 425318.626 | | 0.278 | 0 | 0.278 | 0.034 | 0.278 | 1.00E-03 | 0.278 | 0.035 |
| 0.279 | 440053.252 | | 0.279 | 0 | 0.279 | 0.038 | 0.279 | 1.00E-03 | 0.279 | 0.039 |
| 0.28 | 455268.608 | | 0.28 | 0 | 0.28 | 0.043 | 0.28 | 1.00E-03 | 0.28 | 0.044 |
| 0.281 | 470979.293 | | 0.281 | 0 | 0.281 | 0.048 | 0.281 | 1.00E-03 | 0.281 | 0.049 |
| [illegible] | [illegible] | | [illegible] | [illegible] | [illegible] | [illegible] | [illegible] | [illegible] | [illegible] | [illegible] |

**Theoretical Data**

C2H4/Pt4/SiO2 Reactants, Transition states, and Products xyz coordinates
Format:
# of atoms
Name (Isomer-R/TS/P-pathway_number)
Element symbol X Y Z

| Pt4-Main-Pathways | | | | Pt4Sn3-Main-Pathways | | | | Pt4-All-Pathways | | | | Pt4Sn3-All-Pathways | | | |
|---|---|---|---|---|---|---|---|---|---|---|---|---|---|---|---|
| 193 | | | | 196 | | | | 193 | | | | 196 | | | |
| I-R | | | | I-R | | | | I-R | | | | I-R | | | |
| H | 6.58656000 | 12.90919000 | 0.00000000 | H | 6.58656000 | 12.90919000 | 0.00000000 | H | 6.58656000 | 12.90919000 | 0.00000000 | H | 6.58656000 | 12.90919000 | 0.00000000 |
| O | 12.34643000 | 6.23675000 | 0.32400000 | O | 12.34643000 | 6.23675000 | 0.32400000 | O | 12.34643000 | 6.23675000 | 0.32400000 | O | 12.34643000 | 6.23675000 | 0.32400000 |
| O | 6.32035000 | 0.68143000 | 0.34265000 | O | 6.32035000 | 0.68143000 | 0.34265000 | O | 6.32035000 | 0.68143000 | 0.34265000 | O | 6.32035000 | 0.68143000 | 0.34265000 |
| O | 6.64696000 | 10.90101000 | 0.43734000 | O | 6.64696000 | 10.90101000 | 0.43734000 | O | 6.64696000 | 10.90101000 | 0.43734000 | O | 6.64696000 | 10.90101000 | 0.43734000 |
| O | 1.43306000 | 8.88837000 | 0.50018000 | O | 1.43306000 | 8.88837000 | 0.50018000 | O | 1.43306000 | 8.88837000 | 0.50018000 | O | 1.43306000 | 8.88837000 | 0.50018000 |
| H | 9.03195000 | 2.06583000 | 0.63119000 | H | 9.03195000 | 2.06583000 | 0.63119000 | H | 9.03195000 | 2.06583000 | 0.63119000 | H | 9.03195000 | 2.06583000 | 0.63119000 |
| O | 3.86303000 | 10.46476000 | 0.65412000 | O | 3.86303000 | 10.46476000 | 0.65412000 | O | 3.86303000 | 10.46476000 | 0.65412000 | O | 3.86303000 | 10.46476000 | 0.65412000 |
| O | 9.40856000 | 10.33665000 | 0.75816000 | O | 9.40856000 | 10.33665000 | 0.75816000 | O | 9.40856000 | 10.33665000 | 0.75816000 | O | 9.40856000 | 10.33665000 | 0.75816000 |
| O | 8.97135000 | 1.15729000 | 0.95693000 | O | 8.97135000 | 1.15729000 | 0.95693000 | O | 8.97135000 | 1.15729000 | 0.95693000 | O | 8.97135000 | 1.15729000 | 0.95693000 |
| Si | 0.33529000 | 7.71048000 | 1.04689000 | Si | 0.33529000 | 7.71048000 | 1.04689000 | Si | 0.33529000 | 7.71048000 | 1.04689000 | Si | 0.33529000 | 7.71048000 | 1.04689000 |
| O | 11.41461000 | 3.24415000 | 1.12214000 | O | 11.41461000 | 3.24415000 | 1.12214000 | O | 11.41461000 | 3.24415000 | 1.12214000 | O | 11.41461000 | 3.24415000 | 1.12214000 |
| O | 6.05809000 | 6.39254000 | 1.24787000 | O | 6.05809000 | 6.39254000 | 1.24787000 | O | 6.05809000 | 6.39254000 | 1.24787000 | O | 6.05809000 | 6.39254000 | 1.24787000 |
| O | 11.35216000 | 8.50616000 | 1.29237000 | O | 11.35216000 | 8.50616000 | 1.29237000 | O | 11.35216000 | 8.50616000 | 1.29237000 | O | 11.35216000 | 8.50616000 | 1.29237000 |
| Si | 11.95804000 | 4.88343000 | 1.29616000 | Si | 11.95804000 | 4.88343000 | 1.29616000 | Si | 11.95804000 | 4.88343000 | 1.29616000 | Si | 11.95804000 | 4.88343000 | 1.29616000 |
| O | 5.57676000 | 8.67061000 | 1.29731000 | O | 5.57676000 | 8.67061000 | 1.29731000 | O | 5.57676000 | 8.67061000 | 1.29731000 | O | 5.57676000 | 8.67061000 | 1.29731000 |
| Si | 5.36204000 | 10.27901000 | 1.31001000 | Si | 5.36204000 | 10.27901000 | 1.31001000 | Si | 5.36204000 | 10.27901000 | 1.31001000 | Si | 5.36204000 | 10.27901000 | 1.31001000 |
| Si | 2.45532000 | 9.93546000 | 1.35284000 | Si | 2.45532000 | 9.93546000 | 1.35284000 | Si | 2.45532000 | 9.93546000 | 1.35284000 | Si | 2.45532000 | 9.93546000 | 1.35284000 |
| Si | 7.97589000 | 10.75088000 | 1.45379000 | Si | 7.97589000 | 10.75088000 | 1.45379000 | Si | 7.97589000 | 10.75088000 | 1.45379000 | Si | 7.97589000 | 10.75088000 | 1.45379000 |
| Si | 7.46857000 | 0.73103000 | 1.50179000 | Si | 7.46857000 | 0.73103000 | 1.50179000 | Si | 7.46857000 | 0.73103000 | 1.50179000 | Si | 7.46857000 | 0.73103000 | 1.50179000 |
| O | 11.74039000 | 11.12450000 | 1.65148000 | O | 11.74039000 | 11.12450000 | 1.65148000 | O | 11.74039000 | 11.12450000 | 1.65148000 | O | 11.74039000 | 11.12450000 | 1.65148000 |
| Si | 10.67104000 | 9.93229000 | 1.72914000 | Si | 10.67104000 | 9.93229000 | 1.72914000 | Si | 10.67104000 | 9.93229000 | 1.72914000 | Si | 10.67104000 | 9.93229000 | 1.72914000 |
| O | 11.94920000 | 0.43492000 | 1.76809000 | O | 11.94920000 | 0.43492000 | 1.76809000 | O | 11.94920000 | 0.43492000 | 1.76809000 | O | 11.94920000 | 0.43492000 | 1.76809000 |
| O | 1.79839000 | 11.37620000 | 1.80136000 | O | 1.79839000 | 11.37620000 | 1.80136000 | O | 1.79839000 | 11.37620000 | 1.80136000 | O | 1.79839000 | 11.37620000 | 1.80136000 |
| O | 1.06378000 | 4.55670000 | 1.82871000 | O | 1.06378000 | 4.55670000 | 1.82871000 | O | 1.06378000 | 4.55670000 | 1.82871000 | O | 1.06378000 | 4.55670000 | 1.82871000 |
| O | 8.39548000 | 7.04101000 | 1.84762000 | O | 8.39548000 | 7.04101000 | 1.84762000 | O | 8.39548000 | 7.04101000 | 1.84762000 | O | 8.39548000 | 7.04101000 | 1.84762000 |
| Si | 6.89164000 | 7.82098000 | 1.90040000 | Si | 6.89164000 | 7.82098000 | 1.90040000 | Si | 6.89164000 | 7.82098000 | 1.90040000 | Si | 6.89164000 | 7.82098000 | 1.90040000 |
| O | 8.12958000 | 12.28788000 | 2.10634000 | O | 8.12958000 | 12.28788000 | 2.10634000 | O | 8.12958000 | 12.28788000 | 2.10634000 | O | 8.12958000 | 12.28788000 | 2.10634000 |
| O | 10.66672000 | 5.50028000 | 2.11341000 | O | 10.66672000 | 5.50028000 | 2.11341000 | O | 10.66672000 | 5.50028000 | 2.11341000 | O | 10.66672000 | 5.50028000 | 2.11341000 |
| Si | 11.59967000 | 1.97594000 | 2.23463000 | Si | 11.59967000 | 1.97594000 | 2.23463000 | Si | 11.59967000 | 1.97594000 | 2.23463000 | Si | 11.59967000 | 1.97594000 | 2.23463000 |
| O | 7.64251000 | 9.37060000 | 2.35451000 | O | 7.64251000 | 9.37060000 | 2.35451000 | O | 7.64251000 | 9.37060000 | 2.35451000 | O | 7.64251000 | 9.37060000 | 2.35451000 |
| [illegible] | [illegible] | [illegible] | [illegible] | [illegible] | [illegible] | [illegible] | [illegible] | [illegible] | [illegible] | [illegible] | [illegible] | [illegible] | [illegible] | [illegible] | [illegible] |

| | | | | | | | | | | |
|---|---|---|---|---|---|---|---|---|---|---|
| 479 | 4.31E+09 | 479 | 4.34E+09 | [redacted] | 479 | 4341069529 | 0.598 | 820689438.2 | 0.598 | 250405.519 |
| 480 | 4.17E+09 | 480 | 4.2E+09 | | 480 | 4196909322 | 0.599 | 831560411.6 | 0.599 | 259248.631 |
| 481 | 4.03E+09 | 481 | 4.06E+09 | | 481 | 4063392048 | 0.6 | 842520360 | 0.6 | 268385.464 |
| 482 | 3.9E+09 | 482 | 3.93E+09 | | 482 | 3932715454 | 0.601 | 853569015.2 | 0.601 | 277825.084 |
| 483 | 3.78E+09 | 483 | 3.8E+09 | | 483 | 3804890967 | 0.602 | 864706088 | 0.602 | 287576.811 |
| 484 | 3.65E+09 | 484 | 3.68E+09 | | 484 | 3679923801 | 0.603 | 875931267.9 | 0.603 | 297650.227 |
| 485 | 3.53E+09 | 485 | 3.56E+09 | | 485 | 3557813790 | 0.604 | 887244223 | 0.604 | 308055.181 |
| 486 | 3.42E+09 | 486 | 3.44E+09 | | 486 | 3438556037 | 0.605 | 898644600 | 0.605 | 318801.798 |
| 487 | 3.3E+09 | 487 | 3.32E+09 | | 487 | 3322141447 | 0.606 | 910132023.5 | 0.606 | 329900.483 |
| 488 | 3.19E+09 | 488 | 3.21E+09 | | 488 | 3208557166 | 0.607 | 921706096.4 | 0.607 | 341361.93 |
| 489 | 3.08E+09 | 489 | 3.1E+09 | | 489 | 3097786951 | 0.608 | 933366399.2 | 0.608 | 353197.129 |
| 490 | 2.97E+09 | 490 | 2.99E+09 | | 490 | 2989811481 | 0.609 | 945112490.4 | 0.609 | 365417.371 |
| 491 | 2.87E+09 | 491 | 2.88E+09 | | 491 | 2884608640 | 0.61 | 956943905.6 | 0.61 | 378034.258 |
| 492 | 2.77E+09 | 492 | 2.78E+09 | | 492 | 2782153754 | 0.611 | 968860158.2 | 0.611 | 391059.709 |
| 493 | 2.67E+09 | 493 | 2.68E+09 | | 493 | 2682419809 | 0.612 | 980860738.5 | 0.612 | 404505.966 |
| 494 | 2.57E+09 | 494 | 2.59E+09 | | 494 | 2585377649 | 0.613 | 992945113.9 | 0.613 | 418385.607 |
| 495 | 2.48E+09 | 495 | 2.49E+09 | | 495 | 2490996151 | 0.614 | 1005112729 | 0.614 | 432711.548 |
| 496 | 2.39E+09 | 496 | 2.4E+09 | | 496 | 2399242395 | 0.615 | 1017363004 | 0.615 | 447497.053 |
| 497 | 2.3E+09 | 497 | 2.31E+09 | | 497 | 2310081811 | 0.616 | 1029695338 | 0.616 | 462755.745 |
| 498 | 2.21E+09 | 498 | 2.22E+09 | | 498 | 2223478326 | 0.617 | 1042109104 | 0.617 | 478501.61 |
| 499 | 2.13E+09 | 499 | 2.14E+09 | | 499 | 2139394496 | 0.618 | 1054603653 | 0.618 | 494749.009 |
| 500 | 2.05E+09 | 500 | 2.06E+09 | | 500 | 2057791632 | 0.619 | 1067178313 | 0.619 | 511512.684 |
| 501 | 1.97E+09 | 501 | 1.98E+09 | | 501 | 1978629920 | 0.62 | 1079832385 | 0.62 | 528807.769 |
| 502 | 1.89E+09 | 502 | 1.9E+09 | | 502 | 1901868537 | 0.621 | 1092565150 | 0.621 | 546649.798 |
| 503 | 1.82E+09 | 503 | 1.83E+09 | | 503 | 1827465756 | 0.622 | 1105375862 | 0.622 | 565054.713 |
| 504 | 1.75E+09 | 504 | 1.76E+09 | | 504 | 1755379051 | 0.623 | 1118263752 | 0.623 | 584038.875 |
| 505 | 1.68E+09 | 505 | 1.69E+09 | | 505 | 1685565196 | 0.624 | 1131228027 | 0.624 | 603619.075 |
| 506 | 1.61E+09 | 506 | 1.62E+09 | | 506 | 1617980357 | 0.625 | 1144267870 | 0.625 | 623812.539 |
| 507 | 1.55E+09 | 507 | 1.55E+09 | | 507 | 1552580187 | 0.626 | 1157382438 | 0.626 | 644636.942 |
| 508 | 1.48E+09 | 508 | 1.49E+09 | | 508 | 1489319906 | 0.627 | 1170570866 | 0.627 | 666110.415 |
| 509 | 1.42E+09 | 509 | 1.43E+09 | | 509 | 1428154390 | 0.628 | 1183832263 | 0.628 | 688251.559 |
| 510 | 1.36E+09 | 510 | 1.37E+09 | | 510 | 1369038245 | 0.629 | 1197165713 | 0.629 | 711079.451 |
| 511 | 1.31E+09 | 511 | 1.31E+09 | | 511 | 1311925885 | 0.63 | 1210570277 | 0.63 | 734613.657 |
| 512 | 1.25E+09 | 512 | 1.26E+09 | | 512 | 1256771605 | 0.631 | 1224044990 | 0.631 | 758874.243 |
| 513 | 1.2E+09 | 513 | 1.2E+09 | | 513 | 1203529648 | 0.632 | 1237588864 | 0.632 | 783881.784 |
| 514 | 1.15E+09 | 514 | 1.15E+09 | | 514 | 1152154270 | 0.633 | 1251200886 | 0.633 | 809657.377 |
| 515 | 1.1E+09 | 515 | 1.1E+09 | | 515 | 1102599805 | 0.634 | 1264880016 | 0.634 | 836222.65 |
| 516 | 1.05E+09 | 516 | 1.05E+09 | | 516 | 1054820724 | 0.635 | 1278625193 | 0.635 | 863599.777 |
| 517 | 1.01E+09 | 517 | 1.01E+09 | | 517 | 1008771687 | 0.636 | 1292435329 | 0.636 | 891811.486 |
| 518 | 9.62E+08 | 518 | 9.64E+08 | | 518 | 964407601.6 | 0.637 | 1306309312 | 0.637 | 920881.071 |
| 519 | 9.19E+08 | 519 | 9.22E+08 | | 519 | 921683671.5 | 0.638 | 1320246006 | 0.638 | 950832.409 |
| 520 | 8.78E+08 | 520 | 8.81E+08 | | 520 | 880555443 | 0.639 | 1334244251 | 0.639 | 981689.963 |
| 521 | 8.39E+08 | 521 | 8.41E+08 | | 521 | 840978850.4 | 0.64 | 1348302859 | 0.64 | 1013478.805 |
| 522 | 8.01E+08 | 522 | 8.03E+08 | | 522 | 802910257.9 | 0.641 | 1362420623 | 0.641 | 1046224.619 |
| 523 | 7.64E+08 | 523 | 7.66E+08 | | 523 | 766306498.4 | 0.642 | 1376596308 | 0.642 | 1079953.721 |
| 524 | 7.29E+08 | 524 | 7.31E+08 | | 524 | 731124910.7 | 0.643 | 1390828654 | 0.643 | 1114693.068 |

| | | | | | | | | | |
|---|---|---|---|---|---|---|---|---|---|
| 0 | 0.947 | 2446516229 | 0.947 | 692384150.9 | 0.947 | 3138900380 | 0.947 | 0 | [illegible] |
| 0 | 0.948 | 2438510220 | 0.948 | 699442730.4 | 0.948 | 3137952951 | 0.948 | 0 | [illegible] |
| 0 | 0.949 | 2430476399 | 0.949 | 706530987.5 | 0.949 | 3137007387 | 0.949 | 0 | [illegible] |
| 0 | 0.95 | 2422415214 | 0.95 | 713648263.2 | 0.95 | 3136063477 | 0.95 | 0 | [illegible] |
| 0 | 0.951 | 2414327113 | 0.951 | 720793883.7 | 0.951 | 3135120997 | 0.951 | 0 | [illegible] |
| 0 | 0.952 | 2406212545 | 0.952 | 727967160.4 | 0.952 | 3134179705 | 0.952 | 0 | [illegible] |
| 0 | 0.953 | 2398071959 | 0.953 | 735167389.9 | 0.953 | 3133239349 | 0.953 | 0 | [illegible] |
| 0 | 0.954 | 2389905804 | 0.954 | 742393854.2 | 0.954 | 3132299659 | 0.954 | 0 | [illegible] |
| 0 | 0.955 | 2381714530 | 0.955 | 749645820.7 | 0.955 | 3131360351 | 0.955 | 0 | [illegible] |
| 0 | 0.956 | 2373498587 | 0.956 | 756922542.1 | 0.956 | 3130421129 | 0.956 | 0 | [illegible] |
| 0 | 0.957 | 2365258422 | 0.957 | 764223256.9 | 0.957 | 3129481679 | 0.957 | 0 | [illegible] |
| 0 | 0.958 | 2356994488 | 0.958 | 771547189.1 | 0.958 | 3128541677 | 0.958 | 0 | [illegible] |
| 0 | 0.959 | 2348707232 | 0.959 | 778893548.5 | 0.959 | 3127600781 | 0.959 | 0 | [illegible] |
| 0 | 0.96 | 2340397105 | 0.96 | 786261530.7 | 0.96 | 3126658636 | 0.96 | 0 | [illegible] |
| 0 | 0.961 | 2332064557 | 0.961 | 793650317.4 | 0.961 | 3125714874 | 0.961 | 0 | [illegible] |
| 0 | 0.962 | 2323710037 | 0.962 | 801059076.2 | 0.962 | 3124769113 | 0.962 | 0 | [illegible] |
| 0 | 0.963 | 2315333994 | 0.963 | 808486961.1 | 0.963 | 3123820955 | 0.963 | 0 | [illegible] |
| 0 | 0.964 | 2306936877 | 0.964 | 815933112.5 | 0.964 | 3122869989 | 0.964 | 0 | [illegible] |
| 0 | 0.965 | 2298519136 | 0.965 | 823396657.1 | 0.965 | 3121915793 | 0.965 | 0 | [illegible] |
| 0 | 0.966 | 2290081218 | 0.966 | 830876708.3 | 0.966 | 3120957927 | 0.966 | 0 | [illegible] |
| 0 | 0.967 | 2281623574 | 0.967 | 838372366.5 | 0.967 | 3119995940 | 0.967 | 0 | [illegible] |
| 0 | 0.968 | 2273146649 | 0.968 | 845882718.8 | 0.968 | 3119029368 | 0.968 | 0 | [illegible] |
| 0 | 0.969 | 2264650893 | 0.969 | 853406839.7 | 0.969 | 3118057733 | 0.969 | 0 | [illegible] |
| 0 | 0.97 | 2256136753 | 0.97 | 860943790.7 | 0.97 | 3117080543 | 0.97 | 0 | [illegible] |
| 0 | 0.971 | 2247604675 | 0.971 | 868492621 | 0.971 | 3116097296 | 0.971 | 0 | [illegible] |
| 0 | 0.972 | 2239055105 | 0.972 | 876052367.6 | 0.972 | 3115107472 | 0.972 | 0 | [illegible] |
| 0 | 0.973 | 2230488490 | 0.973 | 883622055 | 0.973 | 3114110545 | 0.973 | 0 | [illegible] |
| 0 | 0.974 | 2221905274 | 0.974 | 891200686.1 | 0.974 | 3113105970 | 0.974 | 0 | [illegible] |
| 0 | 0.975 | 2213305903 | 0.975 | 898787291.9 | 0.975 | 3112093194 | 0.975 | 0 | [illegible] |
| 0 | 0.976 | 2204690819 | 0.976 | 906380832.1 | 0.976 | 3111071651 | 0.976 | 0 | [illegible] |
| 0 | 0.977 | 2196060468 | 0.977 | 913980294.8 | 0.977 | 3110040762 | 0.977 | 0 | [illegible] |
| 0 | 0.978 | 2187415290 | 0.978 | 921584647.4 | 0.978 | 3108999937 | 0.978 | 0 | [illegible] |
| 0 | 0.979 | 2178755728 | 0.979 | 929192846.2 | 0.979 | 3107948574 | 0.979 | 0 | [illegible] |
| 0 | 0.98 | 2170082223 | 0.98 | 936803837 | 0.98 | 3106886060 | 0.98 | 0 | [illegible] |
| 0 | 0.981 | 2161395216 | 0.981 | 944416555.2 | 0.981 | 3105811771 | 0.981 | 0 | [illegible] |
| 0 | 0.982 | 2152695145 | 0.982 | 952029926.2 | 0.982 | 3104725071 | 0.982 | 0 | [illegible] |
| 0 | 0.983 | 2143982449 | 0.983 | 959642865.4 | 0.983 | 3103625314 | 0.983 | 0 | [illegible] |
| 0 | 0.984 | 2135257566 | 0.984 | 967254278.8 | 0.984 | 3102511844 | 0.984 | 0 | [illegible] |
| 0 | 0.985 | 2126520931 | 0.985 | 974863063 | 0.985 | 3101383994 | 0.985 | 0 | [illegible] |
| 0 | 0.986 | 2117772982 | 0.986 | 982468105.4 | 0.986 | 3100241087 | 0.986 | 0 | [illegible] |
| 0 | 0.987 | 2109014152 | 0.987 | 990068284.8 | 0.987 | 3099082436 | 0.987 | 0 | [illegible] |
| 0 | 0.988 | 2100244874 | 0.988 | 997662471.4 | 0.988 | 3097907346 | 0.988 | 0 | [illegible] |
| 0 | 0.989 | 2091465582 | 0.989 | 1005249527 | 0.989 | 3096715109 | 0.989 | 0 | [illegible] |
| 0 | 0.99 | 2082676706 | 0.99 | 1012828306 | 0.99 | 3095505012 | 0.99 | 0 | [illegible] |
| 0 | 0.991 | 2073878676 | 0.991 | 1020397655 | 0.991 | 3094276331 | 0.991 | 0 | [illegible] |
| 0 | 0.992 | 2065071922 | 0.992 | 1027956413 | 0.992 | 3093028335 | 0.992 | 0 | [illegible] |
| 0 | 0.993 | 2056256870 | 0.993 | 1035503412 | 0.993 | 3091760282 | 0.993 | 0 | [illegible] |
| 0 | 0.994 | 2047433948 | 0.994 | 1043037477 | 0.994 | 3090471425 | 0.994 | 0 | [illegible] |

[illegible]

| | | | | | | | | |
|---|---|---|---|---|---|---|---|---|
| [illegible] | 0 | 1.296 | 202149814 | 1.296 | 971070410.5 | 1.296 | 1173220224 | 1.296 [illegible] |
| [illegible] | 0 | 1.297 | 199931686.7 | 1.297 | 966543176.2 | 1.297 | 1166474863 | 1.297 [illegible] |
| [illegible] | 0 | 1.298 | 197733504.2 | 1.298 | 962014170.2 | 1.298 | 1159747674 | 1.298 [illegible] |
| [illegible] | 0 | 1.299 | 195555144.2 | 1.299 | 957483615.8 | 1.299 | 1153038760 | 1.299 [illegible] |
| [illegible] | 0 | 1.3 | 193396484.7 | 1.3 | 952951735.2 | 1.3 | 1146348220 | 1.3 [illegible] |
| [illegible] | 0 | 1.301 | 191257403.6 | 1.301 | 948418749.6 | 1.301 | 1139676153 | 1.301 [illegible] |
| [illegible] | 0 | 1.302 | 189137778.9 | 1.302 | 943884879.5 | 1.302 | 1133022658 | 1.302 [illegible] |
| [illegible] | 0 | 1.303 | 187037488.7 | 1.303 | 939350344.2 | 1.303 | 1126387833 | 1.303 [illegible] |
| [illegible] | 0 | 1.304 | 184956411.1 | 1.304 | 934815362 | 1.304 | 1119771773 | 1.304 [illegible] |
| [illegible] | 0 | 1.305 | 182894424.2 | 1.305 | 930280150.3 | 1.305 | 1113174575 | 1.305 [illegible] |
| [illegible] | 0 | 1.306 | 180851406.5 | 1.306 | 925744925.3 | 1.306 | 1106596332 | 1.306 [illegible] |
| [illegible] | 0 | 1.307 | 178827236.3 | 1.307 | 921209902.4 | 1.307 | 1100037139 | 1.307 [illegible] |
| [illegible] | 0 | 1.308 | 176821792 | 1.308 | 916675295.6 | 1.308 | 1093497088 | 1.308 [illegible] |
| [illegible] | 0 | 1.309 | 174834952.5 | 1.309 | 912141318.1 | 1.309 | 1086976271 | 1.309 [illegible] |
| [illegible] | 0 | 1.31 | 172866596.3 | 1.31 | 907608181.9 | 1.31 | 1080474778 | 1.31 [illegible] |
| [illegible] | 0 | 1.311 | 170916602.4 | 1.311 | 903076097.9 | 1.311 | 1073992700 | 1.311 [illegible] |
| [illegible] | 0 | 1.312 | 168984849.8 | 1.312 | 898545275.7 | 1.312 | 1067530126 | 1.312 [illegible] |
| [illegible] | 0 | 1.313 | 167071217.7 | 1.313 | 894015924.1 | 1.313 | 1061087142 | 1.313 [illegible] |
| [illegible] | 0 | 1.314 | 165175585.6 | 1.314 | 889488250.6 | 1.314 | 1054663836 | 1.314 [illegible] |
| [illegible] | 0 | 1.315 | 163297832.9 | 1.315 | 884962461.3 | 1.315 | 1048260294 | 1.315 [illegible] |
| [illegible] | 0 | 1.316 | 161437839.3 | 1.316 | 880438761.5 | 1.316 | 1041876601 | 1.316 [illegible] |
| [illegible] | 0 | 1.317 | 159595484.9 | 1.317 | 875917355 | 1.317 | 1035512840 | 1.317 [illegible] |
| [illegible] | 0 | 1.318 | 157770649.6 | 1.318 | 871398444.6 | 1.318 | 1029169094 | 1.318 [illegible] |
| [illegible] | 0 | 1.319 | 155963213.8 | 1.319 | 866882231.8 | 1.319 | 1022845446 | 1.319 [illegible] |
| [illegible] | 0 | 1.32 | 154173058.1 | 1.32 | 862368916.9 | 1.32 | 1016541975 | 1.32 [illegible] |
| [illegible] | 0 | 1.321 | 152400063.2 | 1.321 | 857858698.7 | 1.321 | 1010258762 | 1.321 [illegible] |
| [illegible] | 0 | 1.322 | 150644110.1 | 1.322 | 853351775.2 | 1.322 | 1003995885 | 1.322 [illegible] |
| [illegible] | 0 | 1.323 | 148905080.1 | 1.323 | 848848342.8 | 1.323 | 997753422.9 | 1.323 [illegible] |
| [illegible] | 0 | 1.324 | 147182854.6 | 1.324 | 844348596.7 | 1.324 | 991531451.3 | 1.324 [illegible] |
| [illegible] | 0 | 1.325 | 145477315.4 | 1.325 | 839852730.8 | 1.325 | 985330046.1 | 1.325 [illegible] |
| [illegible] | 0 | 1.326 | 143788344.4 | 1.326 | 835360937.6 | 1.326 | 979149282 | 1.326 [illegible] |
| [illegible] | 0 | 1.327 | 142115824 | 1.327 | 830873408.6 | 1.327 | 972989232.6 | 1.327 [illegible] |
| [illegible] | 0 | 1.328 | 140459636.7 | 1.328 | 826390333.6 | 1.328 | 966849970.3 | 1.328 [illegible] |
| [illegible] | 0 | 1.329 | 138819665.3 | 1.329 | 821911901.3 | 1.329 | 960731566.6 | 1.329 [illegible] |
| [illegible] | 0 | 1.33 | 137195793 | 1.33 | 817438298.9 | 1.33 | 954634091.9 | 1.33 [illegible] |
| [illegible] | 0 | 1.331 | 135587903.2 | 1.331 | 812969712.4 | 1.331 | 948557615.6 | 1.331 [illegible] |
| [illegible] | 0 | 1.332 | 133995879.6 | 1.332 | 808506326.2 | 1.332 | 942502205.8 | 1.332 [illegible] |
| [illegible] | 0 | 1.333 | 132419606.2 | 1.333 | 804048323.5 | 1.333 | 936467929.8 | 1.333 [illegible] |
| [illegible] | 0 | 1.334 | 130858967.5 | 1.334 | 799595886.1 | 1.334 | 930454853.7 | 1.334 [illegible] |
| [illegible] | 0 | 1.335 | 129313848.2 | 1.335 | 795149194.4 | 1.335 | 924463042.6 | 1.335 [illegible] |
| [illegible] | 0 | 1.336 | 127784133.1 | 1.336 | 790708427.2 | 1.336 | 918492560.4 | 1.336 [illegible] |
| [illegible] | 0 | 1.337 | 126269707.8 | 1.337 | 786273762.2 | 1.337 | 912543470 | 1.337 [illegible] |
| [illegible] | 0 | 1.338 | 124770457.9 | 1.338 | 781845375.4 | 1.338 | 906615833.3 | 1.338 [illegible] |
| [illegible] | 0 | 1.339 | 123286269.4 | 1.339 | 777423441.6 | 1.339 | 900709711 | 1.339 [illegible] |
| [illegible] | 0 | 1.34 | 121817028.8 | 1.34 | 773008133.8 | 1.34 | 894825162.6 | 1.34 [illegible] |
| [illegible] | 0 | 1.341 | 120362622.8 | 1.341 | 768599624 | 1.341 | 888962246.8 | 1.341 [illegible] |
| [illegible] | | | | | | | | |

[illegible]

| | | | | | | | | | |
|---|---|---|---|---|---|---|---|---|---|
| 0 | 1.645 | 1115060.594 | 1.645 | 44852572.69 | 1.645 | 45967633.28 | 1.645 | 0 | 1.645 |
| 0 | 1.646 | 1094305.437 | 1.646 | 44274473.7 | 1.646 | 45368779.14 | 1.646 | 0 | 1.646 |
| 0 | 1.647 | 1073912.74 | 1.647 | 43702786.43 | 1.647 | 44776699.17 | 1.647 | 0 | 1.647 |
| 0 | 1.648 | 1053876.648 | 1.648 | 43137455.12 | 1.648 | 44191331.77 | 1.648 | 0 | 1.648 |
| 0 | 1.649 | 1034191.388 | 1.649 | 42578424.25 | 1.649 | 43612615.64 | 1.649 | 0 | 1.649 |
| 0 | 1.65 | 1014851.275 | 1.65 | 42025638.59 | 1.65 | 43040489.87 | 1.65 | 0 | 1.65 |
| 0 | 1.651 | 995850.706 | 1.651 | 41479043.18 | 1.651 | 42474893.89 | 1.651 | 0 | 1.651 |
| 0 | 1.652 | 977184.16 | 1.652 | 40938583.35 | 1.652 | 41915767.51 | 1.652 | 0 | 1.652 |
| 0 | 1.653 | 958846.197 | 1.653 | 40404204.67 | 1.653 | 41363050.87 | 1.653 | 0 | 1.653 |
| 0 | 1.654 | 940831.46 | 1.654 | 39875853.03 | 1.654 | 40816684.49 | 1.654 | 0 | 1.654 |
| 0 | 1.655 | 923134.668 | 1.655 | 39353474.57 | 1.655 | 40276609.24 | 1.655 | 0 | 1.655 |
| 0 | 1.656 | 905750.619 | 1.656 | 38837015.73 | 1.656 | 39742766.35 | 1.656 | 0 | 1.656 |
| 0 | 1.657 | 888674.191 | 1.657 | 38326423.22 | 1.657 | 39215097.41 | 1.657 | 0 | 1.657 |
| 0 | 1.658 | 871900.335 | 1.658 | 37821644.05 | 1.658 | 38693544.39 | 1.658 | 0 | 1.658 |
| 0 | 1.659 | 855424.078 | 1.659 | 37322625.5 | 1.659 | 38178049.58 | 1.659 | 0 | 1.659 |
| 0 | 1.66 | 839240.522 | 1.66 | 36829315.14 | 1.66 | 37668555.66 | 1.66 | 0 | 1.66 |
| 0 | 1.661 | 823344.842 | 1.661 | 36341660.83 | 1.661 | 37165005.67 | 1.661 | 0 | 1.661 |
| 0 | 1.662 | 807732.285 | 1.662 | 35859610.72 | 1.662 | 36667343.01 | 1.662 | 0 | 1.662 |
| 0 | 1.663 | 792398.17 | 1.663 | 35383113.25 | 1.663 | 36175511.42 | 1.663 | 0 | 1.663 |
| 0 | 1.664 | 777337.886 | 1.664 | 34912117.15 | 1.664 | 35689455.04 | 1.664 | 0 | 1.664 |
| 0 | 1.665 | 762546.891 | 1.665 | 34446571.44 | 1.665 | 35209118.33 | 1.665 | 0 | 1.665 |
| 0 | 1.666 | 748020.713 | 1.666 | 33986425.44 | 1.666 | 34734446.15 | 1.666 | 0 | 1.666 |
| 0 | 1.667 | 733754.946 | 1.667 | 33531628.75 | 1.667 | 34265383.69 | 1.667 | 0 | 1.667 |
| 0 | 1.668 | 719745.252 | 1.668 | 33082131.27 | 1.668 | 33801876.53 | 1.668 | 0 | 1.668 |
| 0 | 1.669 | 705987.358 | 1.669 | 32637883.22 | 1.669 | 33343870.58 | 1.669 | 0 | 1.669 |
| 0 | 1.67 | 692477.058 | 1.67 | 32198835.07 | 1.67 | 32891312.13 | 1.67 | 0 | 1.67 |
| 0 | 1.671 | 679210.207 | 1.671 | 31764937.63 | 1.671 | 32444147.84 | 1.671 | 0 | 1.671 |
| 0 | 1.672 | 666182.725 | 1.672 | 31336141.98 | 1.672 | 32002324.71 | 1.672 | 0 | 1.672 |
| 0 | 1.673 | 653390.594 | 1.673 | 30912399.51 | 1.673 | 31565790.1 | 1.673 | 0 | 1.673 |
| 0 | 1.674 | 640829.859 | 1.674 | 30493661.9 | 1.674 | 31134491.76 | 1.674 | 0 | 1.674 |
| 0 | 1.675 | 628496.624 | 1.675 | 30079881.15 | 1.675 | 30708377.78 | 1.675 | 0 | 1.675 |
| 0 | 1.676 | 616387.054 | 1.676 | 29671009.53 | 1.676 | 30287396.59 | 1.676 | 0 | 1.676 |
| 0 | 1.677 | 604497.371 | 1.677 | 29266999.64 | 1.677 | 29871497.01 | 1.677 | 0 | 1.677 |
| 0 | 1.678 | 592823.858 | 1.678 | 28867804.35 | 1.678 | 29460628.21 | 1.678 | 0 | 1.678 |
| 0 | 1.679 | 581362.855 | 1.679 | 28473376.86 | 1.679 | 29054739.71 | 1.679 | 0 | 1.679 |
| 0 | 1.68 | 570110.756 | 1.68 | 28083670.64 | 1.68 | 28653781.4 | 1.68 | 0 | 1.68 |
| 0 | 1.681 | 559064.015 | 1.681 | 27698639.51 | 1.681 | 28257703.52 | 1.681 | 0 | 1.681 |
| 0 | 1.682 | 548219.138 | 1.682 | 27318237.53 | 1.682 | 27866456.67 | 1.682 | 0 | 1.682 |
| 0 | 1.683 | 537572.686 | 1.683 | 26942419.11 | 1.683 | 27479991.8 | 1.683 | 0 | 1.683 |
| 0 | 1.684 | 527121.275 | 1.684 | 26571138.94 | 1.684 | 27098260.22 | 1.684 | 0 | 1.684 |
| 0 | 1.685 | 516861.574 | 1.685 | 26204352.03 | 1.685 | 26721213.6 | 1.685 | 0 | 1.685 |
| 0 | 1.686 | 506790.301 | 1.686 | 25842013.66 | 1.686 | 26348803.96 | 1.686 | 0 | 1.686 |
| 0 | 1.687 | 496904.229 | 1.687 | 25484079.44 | 1.687 | 25980983.67 | 1.687 | 0 | 1.687 |
| 0 | 1.688 | 487200.18 | 1.688 | 25130505.28 | 1.688 | 25617705.46 | 1.688 | 0 | 1.688 |
| 0 | 1.689 | 477675.026 | 1.689 | 24781247.39 | 1.689 | 25258922.42 | 1.689 | 0 | 1.689 |
| 0 | 1.69 | 468325.69 | 1.69 | 24436262.27 | 1.69 | 24904587.96 | 1.69 | 0 | 1.69 |
| 0 | 1.691 | 459149.14 | 1.691 | 24095506.74 | 1.691 | 24554655.88 | 1.691 | 0 | 1.691 |
| 0 | 1.692 | 450142.397 | 1.692 | 23758937.9 | 1.692 | 24209080.3 | 1.692 | 0 | 1.692 |

[illegible]

| | | | |
|---|---|---|---|
| O | 6.64696000 | 10.90101000 | 0.43734000 |
| O | 1.43306000 | 8.88837000 | 0.50018000 |
| H | 9.03195000 | 2.06583000 | 0.63119000 |
| O | 3.86303000 | 10.46476000 | 0.65412000 |
| O | 9.40856000 | 10.33665000 | 0.75816000 |
| O | 8.97135000 | 1.15729000 | 0.95693000 |
| Si | 0.33529000 | 7.71048000 | 1.04689000 |
| O | 11.41461000 | 3.24415000 | 1.12214000 |
| O | 6.05809000 | 6.39254000 | 1.24787000 |
| O | 11.35216000 | 8.50616000 | 1.29237000 |
| Si | 11.95804000 | 4.88343000 | 1.29616000 |
| O | 5.57676000 | 8.67061000 | 1.29731000 |
| Si | 5.36204000 | 10.27901000 | 1.31001000 |
| Si | 2.45532000 | 9.93546000 | 1.35284000 |
| Si | 7.97589000 | 10.75088000 | 1.45379000 |
| Si | 7.46857000 | 0.73103000 | 1.50179000 |
| O | 11.74039000 | 11.12450000 | 1.65148000 |
| Si | 10.67104000 | 9.93229000 | 1.72914000 |
| O | 11.94920000 | 0.43492000 | 1.76809000 |
| O | 1.79839000 | 11.37620000 | 1.80136000 |
| O | 1.06378000 | 4.55670000 | 1.82871000 |
| O | 8.39548000 | 7.04101000 | 1.84762000 |
| Si | 6.89164000 | 7.82098000 | 1.90040000 |
| O | 8.12958000 | 12.28788000 | 2.10634000 |
| O | 10.66672000 | 5.50028000 | 2.11341000 |
| Si | 11.59967000 | 1.97594000 | 2.23463000 |
| O | 7.64251000 | 9.37060000 | 2.35451000 |
| Si | 0.41183000 | 12.06493000 | 2.35560000 |
| O | 8.23978000 | 4.36766000 | 2.37970000 |
| O | 1.05524000 | 7.20221000 | 2.42764000 |
| O | 6.96701000 | 1.98259000 | 2.49582000 |
| Si | 5.19710000 | 5.95740000 | 2.53516000 |
| Si | 9.11036000 | 5.76089000 | 2.59097000 |
| O | 2.94267000 | 9.03534000 | 2.65423000 |
| O | 5.57685000 | 11.01483000 | 2.79376000 |
| O | 5.55014000 | 4.38570000 | 2.83243000 |
| O | 3.63872000 | 6.39622000 | 2.88391000 |
| O | 0.55106000 | 2.29399000 | 3.02640000 |
| Si | 6.98768000 | 3.52843000 | 3.03292000 |
| Si | 1.42896000 | 3.64159000 | 3.13315000 |
| Si | 2.46711000 | 7.55467000 | 3.17378000 |
| O | 10.16690000 | 9.96555000 | 3.28935000 |
| O | 2.92356000 | 3.04287000 | 3.38900000 |
| O | 10.43410000 | 1.98437000 | 3.43561000 |
| O | 6.09824000 | 7.13813000 | 3.49108000 |
| O | 0.48071000 | 11.99165000 | 4.02327000 |
| O | 9.16071000 | 6.09578000 | 4.18112000 |
| Si | 6.19584000 | 10.88914000 | 4.31980000 |
| O | 0.97667000 | 4.54705000 | 4.38853000 |
| O | 3.08859000 | 0.48030000 | 4.59870000 |
| O | 7.12128000 | 3.54351000 | 4.68117000 |
| Si | 3.24937000 | 2.09393000 | 4.69687000 |
| O | 12.16214000 | 9.04504000 | 4.71645000 |
| O | 2.01498000 | 7.68646000 | 4.73946000 |
| Si | 10.60485000 | 9.57516000 | 4.79299000 |
| O | 6.00410000 | 9.36258000 | 4.82024000 |
| O | 7.69724000 | 11.39131000 | 4.91130000 |
| Si | 10.68433000 | 2.45872000 | 4.99262000 |
| O | 4.79955000 | 2.37986000 | 5.01078000 |
| O | 10.79898000 | 4.10323000 | 5.01979000 |
| Si | 6.04217000 | 7.79614000 | 5.06574000 |
| O | 11.97163000 | 6.52034000 | 5.16816000 |
| O | 5.27731000 | 11.86749000 | 5.19492000 |
| Si | 12.17232000 | 4.92029000 | 5.36653000 |
| O | 9.61395000 | 8.45242000 | 5.37118000 |
| O | 2.78230000 | 11.22876000 | 5.39252000 |
| Si | 0.52618000 | 7.80404000 | 5.42080000 |
| Si | 1.24491000 | 11.77372000 | 5.47459000 |
| Si | 8.97684000 | 6.98953000 | 5.51034000 |
| O | 12.12776000 | 2.03321000 | 5.54723000 |
| Si | 5.79442000 | 3.48931000 | 5.63946000 |
| Si | 3.88078000 | 12.41462000 | 5.72509000 |
| O | 7.36934000 | 7.19216000 | 5.73077000 |
| O | 5.07310000 | 4.93270000 | 5.77145000 |
| O | 4.69389000 | 7.40689000 | 5.89179000 |
| O | 10.51297000 | 10.77216000 | 5.91766000 |
| O | 2.24135000 | 2.44179000 | 5.95153000 |
| O | 9.68518000 | 1.77422000 | 6.10531000 |
| O | 0.99593000 | 0.04056000 | 6.30419000 |
| O | 6.62583000 | 0.44991000 | 6.32625000 |
| O | 2.60175000 | 5.71772000 | 6.42522000 |
| Si | 7.59978000 | 12.00862000 | 6.46258000 |
| Si | 0.91102000 | 1.63242000 | 6.49053000 |
| Si | 4.19754000 | 6.01659000 | 6.60349000 |
| O | 0.52427000 | 10.85444000 | 6.61729000 |
| O | 9.59361000 | 6.33156000 | 6.88213000 |
| O | 0.24089000 | 4.58204000 | 6.89383000 |
| O | 0.63894000 | 7.98388000 | 7.03464000 |
| O | 6.09524000 | 2.98598000 | 7.14604000 |
| O | 8.85396000 | 3.87869000 | 7.14825000 |
| O | 6.56019000 | 11.02720000 | 7.22128000 |
| Si | 11.46811000 | 10.87769000 | 7.25600000 |
| O | 4.04474000 | 12.85904000 | 7.33218000 |
| Si | 1.66166000 | 5.02240000 | 7.57693000 |
| Si | 9.28667000 | 2.38491000 | 7.58584000 |
| Si | 6.52179000 | 1.48441000 | 7.58891000 |
| O | 8.76445000 | 12.06825000 | 7.62044000 |
| Si | 8.81881000 | 5.29454000 | 7.88978000 |
| O | 7.29605000 | 5.87780000 | 8.06985000 |
| O | 0.72451000 | 1.93019000 | 8.08120000 |
| O | 6.89743000 | 8.56528000 | 8.09660000 |
| O | 11.16503000 | 12.17028000 | 8.21368000 |
| O | 4.36474000 | 6.02266000 | 8.21442000 |
| O | 8.00193000 | 1.62577000 | 8.21556000 |
| O | 2.52145000 | 3.83121000 | 8.25785000 |
| O | 4.60298000 | 9.87454000 | 8.29784000 |
| O | 11.23294000 | 9.65737000 | 8.33763000 |

[illegible]

| | | | |
|---|---|---|---|
| O | 2.76434234 | 8.06965657 | 12.96733092 |
| Si | 6.96161890 | 12.96055448 | 13.20024112 |
| O | 7.74882715 | 1.32796533 | 13.24080813 |
| O | 8.35006392 | 3.83536361 | 13.36992560 |
| O | 0.15772314 | 9.75990248 | 13.41390659 |
| O | 10.53983616 | 5.57246316 | 13.68544452 |
| O | 7.67352059 | 6.30296674 | 13.98524423 |
| Si | 8.94062930 | 5.18180534 | 14.11285287 |
| O | 9.10398333 | 4.87123714 | 15.71372111 |
| Pt | 7.11900450 | 11.96499779 | 15.24340785 |
| Sn | 8.90274723 | 0.51123385 | 15.85965329 |
| Sn | 10.56595719 | 3.39044239 | 16.54100449 |
| H | 5.40020416 | 10.90920509 | 16.98393245 |
| C | 6.37033295 | 10.56457308 | 17.35197577 |
| H | 6.51907168 | 9.47784507 | 17.44932368 |
| Sn | 8.06041252 | 5.19814128 | 17.67193686 |
| H | 6.91184867 | 1.99973823 | 17.83872494 |
| C | 7.31664460 | 11.41137805 | 17.89613170 |
| Pt | 7.37752022 | 0.29602060 | 17.91425414 |
| Pt | 8.49081849 | 2.65415782 | 18.19626201 |
| H | 8.19117052 | 10.92863444 | 18.36820094 |
| Pt | 10.24086013 | 4.45014003 | 18.89534856 |

196

II-R

| | | | |
|---|---|---|---|
| H | 6.58656000 | 12.90919000 | 0.00000000 |
| O | 12.34643000 | 6.23675000 | 0.32400000 |
| O | 6.32035000 | 0.68143000 | 0.34265000 |
| O | 6.64696000 | 10.90101000 | 0.43734000 |
| O | 1.43306000 | 8.88837000 | 0.50018000 |
| H | 9.03195000 | 2.06583000 | 0.63119000 |
| O | 3.86303000 | 10.46476000 | 0.65412000 |
| O | 9.40856000 | 10.33665000 | 0.75816000 |
| O | 8.97135000 | 1.15729000 | 0.95693000 |
| Si | 0.33529000 | 7.71048000 | 1.04689000 |
| O | 11.41461000 | 3.24415000 | 1.12214000 |
| O | 6.05809000 | 6.39254000 | 1.24787000 |
| O | 11.35216000 | 8.50616000 | 1.29237000 |
| Si | 11.95804000 | 4.88343000 | 1.29616000 |
| O | 5.57676000 | 8.67061000 | 1.29731000 |
| Si | 5.36204000 | 10.27901000 | 1.31001000 |
| Si | 2.45532000 | 9.93546000 | 1.35284000 |
| Si | 7.97589000 | 10.75088000 | 1.45379000 |
| Si | 7.46857000 | 0.73103000 | 1.50179000 |
| O | 11.74039000 | 11.12450000 | 1.65148000 |
| Si | 10.67104000 | 9.93229000 | 1.72914000 |
| O | 11.94920000 | 0.43492000 | 1.76809000 |
| O | 1.79839000 | 11.37620000 | 1.80136000 |
| O | 1.06378000 | 4.55670000 | 1.82871000 |
| O | 8.39548000 | 7.04101000 | 1.84762000 |
| Si | 6.89164000 | 7.82098000 | 1.90040000 |
| O | 8.12958000 | 12.28788000 | 2.10634000 |
| O | 10.66672000 | 5.50028000 | 2.11341000 |
| Si | 11.59967000 | 1.97594000 | 2.23463000 |
| O | 7.64251000 | 9.37060000 | 2.35451000 |
| Si | 0.41183000 | 12.06493000 | 2.35560000 |
| O | 8.23978000 | 4.36766000 | 2.37970000 |
| O | 1.05524000 | 7.20221000 | 2.42764000 |
| O | 6.96701000 | 1.98259000 | 2.49582000 |
| Si | 5.19710000 | 5.95740000 | 2.53516000 |
| Si | 9.11036000 | 5.76089000 | 2.59097000 |
| O | 2.94267000 | 9.03534000 | 2.65423000 |
| O | 5.57685000 | 11.01483000 | 2.79376000 |
| O | 5.55014000 | 4.38570000 | 2.83243000 |
| O | 3.63872000 | 6.39622000 | 2.88391000 |
| O | 0.55106000 | 2.29399000 | 3.02640000 |
| Si | 6.98768000 | 3.52843000 | 3.03292000 |
| Si | 1.42896000 | 3.64159000 | 3.13315000 |
| Si | 2.46711000 | 7.55467000 | 3.17378000 |
| O | 10.16690000 | 9.96555000 | 3.28935000 |
| O | 2.92356000 | 3.04287000 | 3.38900000 |
| O | 10.43410000 | 1.98437000 | 3.43561000 |
| O | 6.09824000 | 7.13813000 | 3.49108000 |
| O | 0.48071000 | 11.99165000 | 4.02327000 |
| O | 9.16071000 | 6.09578000 | 4.18112000 |
| Si | 6.19584000 | 10.88914000 | 4.31980000 |
| O | 0.97667000 | 4.54705000 | 4.38853000 |
| O | 3.08859000 | 0.48030000 | 4.59870000 |
| O | 7.12128000 | 3.54351000 | 4.68117000 |
| Si | 3.24937000 | 2.09393000 | 4.69687000 |
| O | 12.16214000 | 9.04504000 | 4.71645000 |
| O | 2.01498000 | 7.68646000 | 4.73946000 |
| Si | 10.60485000 | 9.57516000 | 4.79299000 |
| O | 6.00410000 | 9.36258000 | 4.82024000 |
| O | 7.69724000 | 11.39131000 | 4.91130000 |
| Si | 10.68433000 | 2.45872000 | 4.99262000 |
| O | 4.79955000 | 2.37986000 | 5.01078000 |
| O | 10.79898000 | 4.10323000 | 5.01979000 |
| Si | 6.04217000 | 7.79614000 | 5.06574000 |
| O | 11.97163000 | 6.52034000 | 5.16816000 |
| O | 5.27731000 | 11.86749000 | 5.19492000 |
| Si | 12.17232000 | 4.92029000 | 5.36653000 |
| O | 9.61395000 | 8.45242000 | 5.37118000 |
| O | 2.78230000 | 11.22876000 | 5.39252000 |
| Si | 0.52618000 | 7.80404000 | 5.42080000 |
| Si | 1.24491000 | 11.77372000 | 5.47459000 |
| Si | 8.97684000 | 6.98953000 | 5.51034000 |
| O | 12.12776000 | 2.03321000 | 5.54723000 |
| Si | 5.79442000 | 3.48931000 | 5.63946000 |
| Si | 3.88078000 | 12.41462000 | 5.72509000 |
| O | 7.36934000 | 7.19216000 | 5.73077000 |
| O | 5.07310000 | 4.93270000 | 5.77145000 |
| O | 4.69389000 | 7.40689000 | 5.89179000 |
| O | 10.51297000 | 10.77216000 | 5.91766000 |
| O | 2.24135000 | 2.44179000 | 5.95153000 |
| O | 9.68518000 | 1.77422000 | 6.10531000 |
| O | 0.99593000 | 0.04056000 | 6.30419000 |
| O | 6.62583000 | 0.44991000 | 6.32625000 |

[illegible]

O 10.79898000 4.10323000 5.01979000
Si 6.04217000 7.79614000 5.06574000
O 11.97163000 6.52034000 5.16816000
O 5.27731000 11.86749000 5.19492000
Si 12.17232000 4.92029000 5.36653000
O 9.61395000 8.45242000 5.37118000
O 2.78230000 11.22876000 5.39252000
Si 0.52618000 7.80404000 5.42080000
Si 1.24491000 11.77372000 5.47459000
Si 8.97684000 6.98953000 5.51034000
O 12.12776000 2.03321000 5.54723000
Si 5.79442000 3.48931000 5.63946000
Si 3.88078000 12.41462000 5.72509000
O 7.36934000 7.19216000 5.73077000
O 5.07310000 4.93270000 5.77145000
O 4.69389000 7.40689000 5.89179000
O 10.51297000 10.77216000 5.91766000
O 2.24135000 2.44179000 5.95153000
O 9.68518000 1.77422000 6.10531000
O 0.99593000 0.04056000 6.30419000
O 6.62583000 0.44991000 6.32625000
O 2.60175000 5.71772000 6.42522000
Si 7.59978000 12.00862000 6.46258000
Si 0.91102000 1.63242000 6.49053000
Si 4.19754000 6.01659000 6.60349000
O 0.52427000 10.85444000 6.61729000
O 9.59361000 6.33156000 6.88213000
O 0.24089000 4.58204000 6.89383000
O 0.63894000 7.98388000 7.03464000
O 6.09524000 2.98598000 7.14604000
O 8.85396000 3.87869000 7.14825000
O 6.56019000 11.02720000 7.22128000
Si 11.46811000 10.87769000 7.25600000
O 4.04474000 12.85904000 7.33218000
Si 1.66166000 5.02240000 7.57693000
Si 9.28667000 2.38491000 7.58584000
Si 6.52179000 1.48441000 7.58891000
O 8.76445000 12.06825000 7.62044000
Si 8.81881000 5.29454000 7.88978000
O 7.29605000 5.87780000 8.06985000
O 0.72451000 1.93019000 8.08120000
O 6.89743000 8.56528000 8.09660000
O 11.16503000 12.17028000 8.21368000
O 4.36474000 6.02266000 8.21442000
O 8.00193000 1.62577000 8.21556000
O 2.52145000 3.83121000 8.25785000
O 4.60298000 9.87454000 8.29784000
O 11.23294000 9.65737000 8.33763000
Si 6.20759000 10.01218000 8.39279000
Si 1.16603000 7.67143000 8.54078000
O 2.62367000 8.34195000 8.68505000
O 5.53447000 0.91970000 8.72556000
O 1.28774000 6.06899000 8.77017000
Si 4.03346000 0.30045000 8.83076000
O 10.31381000 2.32344000 8.86802000
Si 9.68524000 12.14075000 8.96686000
Si 7.16648000 7.23733000 8.97783000
Si 1.87646000 2.56144000 9.05035000
O 3.02209000 1.46592000 9.34606000
Si 3.45236000 9.54199000 9.37580000
O 9.41126000 5.06335000 9.39370000
O 2.55276000 10.91413000 9.53046000
O 8.59502000 7.45156000 9.76356000
O 0.25735000 8.20553000 9.78759000
Si 4.61912000 5.94270000 9.80685000
O 6.67757000 10.56640000 9.82734000
Si 11.57529000 9.37814000 9.91657000
O 9.35747000 10.86631000 9.95108000
O 5.93269000 6.91789000 9.99594000
O 9.20339000 0.22140000 10.04828000
O 4.65432000 12.45007000 10.15129000
O 4.77645000 4.38801000 10.27452000
Si 9.84758000 1.68549000 10.33173000
O 1.23122000 3.17126000 10.40270000
Si 9.17939000 6.14520000 10.61421000
O 1.22910000 12.81664000 10.76714000
O 4.13857000 9.06001000 10.78539000
O 12.00916000 10.72988000 10.81921000
O 10.42087000 8.80291000 10.93948000
O 3.59533000 6.56215000 10.94387000
Si 2.02201000 11.36570000 11.01221000
O 8.69596000 2.57416000 11.09874000
O 10.68623000 6.10892000 11.26061000
O 11.16631000 1.47556000 11.27849000
Si 9.45776000 10.06018000 11.38731000
Si 6.89245000 10.75082000 11.41329000
O 1.69607000 9.81736000 11.56143000
O 7.83751000 5.87395000 11.56351000
Si 4.62983000 12.22691000 11.78878000
O 7.58990000 12.19780000 11.79330000
Si 4.77658000 3.93278000 11.86347000
O 7.96866000 9.60059000 11.88240000
Si 0.00000000 0.64100000 11.89328000
Si 4.04675000 7.84950000 11.90235000
O 6.05743000 2.87874000 11.92654000
Si 1.46098962 3.35843631 12.00784337
O 3.09638979 11.86982305 12.17562770
O 5.46644717 10.82828352 12.17628586
O 3.15495761 3.56601593 12.26900976
Si 7.66414317 2.74901020 12.37221758
O 0.41615574 4.61550849 12.45234921
Si 11.78928508 10.90349972 12.51295766
Si 11.72177517 5.89392136 12.53831256
O 0.91668767 1.87064147 12.58468925
O 5.56165778 7.60310970 12.59715321
O 0.24691077 7.28007083 12.64874786
Si 1.21431364 8.66100654 12.65520626
O 10.17305443 10.76176257 12.67944562
O 5.36897259 0.24668561 12.69797789
Si 6.63365791 6.28945366 12.70186657
O 5.44695419 5.12874953 12.79150655
O 12.09931569 12.48648589 12.87786994
O 2.76473566 8.07597760 12.96400259
Si 6.95094709 13.00111514 13.12637108
O 7.72910894 1.35492847 13.28060953
O 8.35401970 3.88131861 13.31417247
O 0.15869836 9.76334951 13.41548651
O 10.52723285 5.61759714 13.68565970
O 7.69789029 6.33064561 13.97526400
Si 8.94700897 5.19234211 14.14400112
Pt 7.23364809 11.92887640 15.04061281
Pt 8.43475979 0.91856381 15.62490827
O 9.06007026 4.86581559 15.72823246
Pt 9.07522031 3.11188009 16.55184454
H 9.04740976 13.05202918 16.78354435
Pt 9.07106444 1.16219315 18.12968727
H 11.47248308 1.21378965 19.46859046
C 10.39635454 1.05487760 19.53998687
C 9.30336658 1.60537681 20.21097546
H 9.27489636 2.66931754 20.48992211
H 8.60347478 0.97765134 20.77927462
193
II-TS2
C 9.93993112500459 0.04678293106019 19.47530057484640
C 9.09969443935759 0.72642917023372 20.24919185121417
H 6.58656000000001 12.90918999999996 0.00000000000000
H 9.03195000000001 2.06583000000001 0.63119000000006
H 10.62084601787045 12.33888200285938 19.50248474137993
H 10.31503809274385 2.19393751141602 18.46386917799622
H 8.19966963727987 0.40414680086499 20.59110970309802
H 9.40357303271168 1.47313999942958 20.80666680196144
O 12.34642999999997 6.23675000000002 0.32400000000007
O 6.32035000000003 0.68143000000001 0.34265000000005
O 6.64696000000003 10.90101000000003 0.43733999999995
O 1.43306000000000 8.88837000000004 0.50018000000000
O 3.86302999999998 10.46476000000001 0.65411999999992
O 9.40855999999997 10.33665000000001 0.75816000000009
O 8.97135000000001 1.15729000000001 0.95693000000006
O 11.41461000000000 3.24415000000001 1.12213999999995
O 6.05808999999996 6.39254000000003 1.24786999999992
O 11.35215999999998 8.50616000000004 1.29236999999989
O 5.57675999999996 8.67061000000003 1.29730999999993
O 11.74038999999997 11.12449999999999 1.65147999999999
O 11.94920000000000 0.43491999999997 1.76809000000003
O 1.79838999999998 11.37620000000001 1.80135999999993
O 1.06378000000003 4.55670000000004 1.82871000000000
O 8.39548000000004 7.04101000000000 1.84762000000001
O 8.12957999999999 12.28788000000000 2.10634000000004
O 10.66671999999996 5.50028000000003 2.11340999999993
O 7.64250999999997 9.37059999999996 2.35450999999989
O 8.23978000000001 4.36766000000002 2.37969999999996
O 1.05523999999999 7.20221000000000 2.42764000000011
O 6.96701000000000 1.98259000000004 2.49582000000009
O 2.94267000000003 9.03534000000004 2.65423000000010
O 5.57684999999999 11.01483000000002 2.79376000000002
O 5.55013999999998 4.38570000000005 2.83242999999993
O 3.63872000000004 6.39621999999996 2.88391000000002
O 0.55106000000004 2.29398999999997 3.02639999999997
O 10.16689999999996 9.96554999999999 3.28935000000000
O 2.92355999999999 3.04286999999996 3.38899999999990
O 10.43409999999997 1.98437000000002 3.43561000000000
O 6.09823999999998 7.13813000000000 3.49108000000000
O 0.48070999999996 11.99164999999995 4.02326999999990
O 9.16071000000002 6.09578000000000 4.18111999999997
O 0.97666999999999 4.54704999999998 4.38852999999994
O 3.08859000000003 0.48030000000000 4.59870000000000
O 7.12127999999999 3.54350999999997 4.68117000000006
O 12.16213999999997 9.04504000000001 4.71644999999990
O 2.01498000000001 7.68645999999997 4.73946000000000
O 6.00409999999997 9.36258000000003 4.82024000000000
O 7.69723999999998 11.39130999999999 4.91129999999997
O 4.79955000000004 2.37986000000001 5.01078000000006
O 10.79898000000004 4.10323000000004 5.01979000000006
O 11.97162999999998 6.52033999999996 5.16815999999994
O 5.27730999999998 11.86749000000002 5.19491999999990
O 9.61395000000002 8.45241999999998 5.37118000000010
O 2.78230000000004 11.22875999999996 5.39252000000010
O 12.12775999999999 2.03320999999998 5.54722999999990
O 7.36934000000001 7.19216000000004 5.73076999999990
O 5.07310000000001 4.93268999999997 5.77144999999997
O 4.69389000000001 7.40688999999997 5.89178999999990
O 10.51296999999998 10.77216000000004 5.91766000000006
O 2.24134999999998 2.44178999999999 5.95153000000003
O 9.68518000000001 1.77421999999996 6.10530999999990
O 0.99593000000004 0.04055999999999 6.30419000000006
O 6.62583000000004 0.44990999999996 6.32625000000006
O 2.60175000000003 5.71772000000001 6.42522000000010
O 0.52427000000003 10.85443999999999 6.61728999999990
O 9.59361000000001 6.33155999999997 6.88212999999997
O 0.24088999999996 4.58203999999997 6.89382999999997
O 0.63894000000003 7.98388000000003 7.03464000000010
O 6.09523999999999 2.98598000000003 7.14604000000010
O 8.85395999999999 3.87869000000004 7.14824999999997
O 6.56018999999997 11.02720000000000 7.22127999999997
O 4.04473999999998 12.85904000000001 7.33218000000010
O 8.76444999999999 12.06824999999998 7.62043999999990
O 7.29605000000001 5.87779999999998 8.06985000000010
O 0.72450999999998 1.93019000000004 8.08120000000010
O 6.89742999999996 8.56527999999996 8.09660000000010
O 11.16503000000000 12.17027999999997 8.21368000000006
O 4.36473999999996 6.02265999999997 8.21442000000003
O 8.00193000000000 1.62576999999997 8.21556000000010
O 2.52144999999999 3.83120999999999 8.25784999999997
O 4.60298000000000 9.87453999999997 8.29783999999994
O 11.23293999999997 9.65737000000002 8.33763000000000
O 2.62367000000001 8.34195000000002 8.68505000000003
O 5.53447000000000 0.91970000000001 8.72556000000010
O 1.28774000000001 6.06899000000004 8.77017000000000
O 10.31381000000000 2.32343999999997 8.86801999999987
O 3.02209000000000 1.46592000000000 9.34606000000006
O 9.41126000000003 5.06335000000000 9.39370000000010
O 2.55275999999996 10.91413000000001 9.53045999999994
O 8.59502000000002 7.45155999999996 9.76356000000010
O 0.25734999999999 8.20553000000001 9.78758999999990
O 6.67756999999997 10.56639999999996 9.82734000000006
O 9.35747000000002 10.86630999999996 9.95108000000003
O 5.93269000000002 6.91789000000004 9.99594000000003
O 9.20339000000000 0.22140000000002 10.04827999999997
O 4.65431999999998 12.45006999999997 10.15129000000003
O 4.77644999999996 4.38800999999999 10.27451999999994
O 1.23121999999997 3.17125999999997 10.40270000000010
O 1.22909999999998 12.81664000000003 10.76713999999994
O 4.13856999999997 9.06001000000002 10.78539000000000

O 7.64251000 9.37060000 2.35451000
Si 0.41183000 12.06493000 2.35560000
O 8.23978000 4.36766000 2.37970000
O 1.05524000 7.20221000 2.42764000
O 6.96701000 1.98259000 2.49582000
Si 5.19710000 5.95740000 2.53516000
Si 9.11036000 5.76089000 2.59097000
O 2.94267000 9.03534000 2.65423000
O 5.57685000 11.01483000 2.79376000
O 5.55014000 4.38570000 2.83243000
O 3.63872000 6.39622000 2.88391000
O 0.55106000 2.29399000 3.02640000
Si 6.98768000 3.52843000 3.03292000
Si 1.42896000 3.64159000 3.13315000
Si 2.46711000 7.55467000 3.17378000
O 10.16690000 9.96555000 3.28935000
O 2.92356000 3.04287000 3.38900000
O 10.43410000 1.98437000 3.43561000
O 6.09824000 7.13813000 3.49108000
O 0.48071000 11.99165000 4.02327000
O 9.16071000 6.09578000 4.18112000
Si 6.19584000 10.88914000 4.31980000
O 0.97667000 4.54705000 4.38853000
O 3.08859000 0.48030000 4.59870000
O 7.12128000 3.54351000 4.68117000
Si 3.24937000 2.09393000 4.69687000
O 12.16214000 9.04504000 4.71645000
O 2.01498000 7.68646000 4.73946000
Si 10.60485000 9.57516000 4.79299000
O 6.00410000 9.36258000 4.82024000
O 7.69724000 11.39131000 4.91130000
Si 10.68433000 2.45872000 4.99262000
O 4.79955000 2.37986000 5.01078000
O 10.79898000 4.10323000 5.01979000
Si 6.04217000 7.79614000 5.06574000
O 11.97163000 6.52034000 5.16816000
O 5.27731000 11.86749000 5.19492000
Si 12.17232000 4.92029000 5.36653000
O 9.61395000 8.45242000 5.37118000
O 2.78230000 11.22876000 5.39252000
Si 0.52618000 7.80404000 5.42080000
Si 1.24491000 11.77372000 5.47459000
Si 8.97684000 6.98953000 5.51034000
O 12.12776000 2.03321000 5.54723000
Si 5.79442000 3.48931000 5.63946000
Si 3.88078000 12.41462000 5.72509000
O 7.36934000 7.19216000 5.73077000
O 5.07310000 4.93270000 5.77145000
O 4.69389000 7.40689000 5.89179000
O 10.51297000 10.77216000 5.91766000
O 2.24135000 2.44179000 5.95153000
O 9.68518000 1.77422000 6.10531000
O 0.99593000 0.04056000 6.30419000
O 6.62583000 0.44991000 6.32625000
O 2.60175000 5.71772000 6.42522000
Si 7.59978000 12.00862000 6.46258000
Si 0.91102000 1.63242000 6.49053000
Si 4.19754000 6.01659000 6.60349000
O 0.52427000 10.85444000 6.61729000
O 9.59361000 6.33156000 6.88213000
O 0.24089000 4.58204000 6.89383000
O 0.63894000 7.98388000 7.03464000
O 6.09524000 2.98598000 7.14604000
O 8.85396000 3.87869000 7.14825000
O 6.56019000 11.02720000 7.22128000
Si 11.46811000 10.87769000 7.25600000
O 4.04474000 12.85904000 7.33218000
Si 1.66166000 5.02240000 7.57693000
Si 9.28667000 2.38491000 7.58584000
Si 6.52179000 1.48441000 7.58891000
O 8.76445000 12.06825000 7.62044000
Si 8.81881000 5.29454000 7.88978000
O 7.29605000 5.87780000 8.06985000
O 0.72451000 1.93019000 8.08120000
O 6.89743000 8.56528000 8.09660000
O 11.16503000 12.17028000 8.21368000
O 4.36474000 6.02266000 8.21442000
O 8.00193000 1.62577000 8.21556000
O 2.52145000 3.83121000 8.25785000
O 4.60298000 9.87454000 8.29784000
O 11.23294000 9.65737000 8.33763000
Si 6.20759000 10.01218000 8.39279000
Si 1.16603000 7.67143000 8.54078000
O 2.62367000 8.34195000 8.68505000
O 5.53447000 0.91970000 8.72556000
O 1.28774000 6.06899000 8.77017000
Si 4.03346000 0.30045000 8.83076000
O 10.31381000 2.32344000 8.86802000
Si 9.68524000 12.14075000 8.96686000
Si 7.16648000 7.23733000 8.97783000
Si 1.87646000 2.56144000 9.05035000
O 3.02209000 1.46592000 9.34606000
Si 3.45236000 9.54199000 9.37580000
O 9.41126000 5.06335000 9.39370000
O 2.55276000 10.91413000 9.53046000
O 8.59502000 7.45156000 9.76356000
O 0.25735000 8.20553000 9.78759000
Si 4.61912000 5.94270000 9.80685000
O 6.67757000 10.56640000 9.82734000
Si 11.57529000 9.37814000 9.91657000
O 9.35747000 10.86631000 9.95108000
O 5.93269000 6.91789000 9.99594000
O 9.20339000 0.22140000 10.04828000
O 4.65432000 12.45007000 10.15129000
O 4.77645000 4.38801000 10.27452000
Si 9.84758000 1.68549000 10.33173000
O 1.23122000 3.17126000 10.40270000
Si 9.17939000 6.14520000 10.61421000
O 1.22910000 12.81664000 10.76714000
O 4.13857000 9.06001000 10.78539000
O 12.00916000 10.72988000 10.81921000
O 10.42087000 8.80291000 10.93948000
O 3.59533000 6.56215000 10.94387000
Si 2.02201000 11.36570000 11.01221000
O 8.69596000 2.57416000 11.09874000
O 10.68623000 6.10892000 11.26061000
O 11.16631000 1.47556000 11.27849000
Si 9.45776000 10.06018000 11.38731000
Si 6.89245000 10.75082000 11.41329000
O 1.69607000 9.81736000 11.56143000
O 7.83751000 5.87395000 11.56351000
Si 4.62983000 12.22691000 11.78878000
O 7.58990000 12.19780000 11.79330000
Si 4.77658000 3.93278000 11.86347000
O 7.96866000 9.60059000 11.88240000
Si 0.00000000 0.64100000 11.89328000
Si 4.04675000 7.84950000 11.90235000
O 6.05743000 2.87874000 11.92654000
Si 1.46800278 3.35445980 12.00681559
O 5.45526257 10.81771106 12.15895877
O 3.09361757 11.86737315 12.17532967
O 3.15331689 3.56484201 12.27103197
Si 7.67303953 2.72440118 12.40235014
O 0.42608277 4.61157019 12.44909505
Si 11.80011981 10.90057549 12.50488972
Si 11.72344810 5.87769966 12.53762373
O 0.91775416 1.86733823 12.58642250
O 5.55722582 7.61928204 12.58776976
Si 6.61113884 6.30286624 12.65006865
Si 1.21373735 8.65337492 12.65048034
O 0.24081142 7.27472983 12.65170021
O 10.18206825 10.75319314 12.68929039
O 5.35732437 0.23032431 12.72083098
O 5.43189510 5.14398129 12.77960284
O 12.09948171 12.48399320 12.88115374
O 2.76323492 8.07244912 12.96431660
Si 6.95652989 13.00219743 13.16251079
O 7.73201235 1.33255566 13.25062011
O 8.35754582 3.86511235 13.39405000
O 0.16186130 9.76034582 13.41792461
O 10.54355219 5.56176443 13.68885596
O 7.68451419 6.37057863 13.90502638
Si 8.94832842 5.21458983 14.17393598
O 8.94650295 5.21306013 15.79688652
Pt 7.25673069 11.60003444 14.99559739
H 6.54019124 7.84050868 16.02326267
Sn 9.51461748 11.58406676 16.17471630
Pt 7.82848492 6.83100167 16.37576527
Pt 8.03165792 9.47352122 16.49348483
Sn 5.95475878 10.90514676 17.28394586
Pt 5.59035792 8.23159709 17.48279668
H 3.30233227 6.36036052 17.83604433
C 3.01559768 7.27899219 18.36204696
C 3.82711292 8.34660680 18.42798446
Sn 7.89980963 7.89408629 18.72208826
H 2.01523025 7.27237933 18.81944986
H 3.47358343 9.23524828 18.96778573
196
II-TS2
H 6.58656000 12.90919000 0.00000000
O 12.34643000 6.23675000 0.32400000
O 6.32035000 0.68143000 0.34265000
O 6.64696000 10.90101000 0.43734000
O 1.43306000 8.88837000 0.50018000
H 9.03195000 2.06583000 0.63119000
O 3.86303000 10.46476000 0.65412000
O 9.40856000 10.33665000 0.75816000
O 8.97135000 1.15729000 0.95693000
Si 0.33529000 7.71048000 1.04689000
O 11.41461000 3.24415000 1.12214000
O 6.05809000 6.39254000 1.24787000
O 11.35216000 8.50616000 1.29237000
Si 11.95804000 4.88343000 1.29616000
O 5.57676000 8.67061000 1.29731000
Si 5.36204000 10.27901000 1.31001000
Si 2.45532000 9.93546000 1.35284000
Si 7.97589000 10.75088000 1.45379000
Si 7.46857000 0.73103000 1.50179000
O 11.74039000 11.12450000 1.65148000
Si 10.67104000 9.93229000 1.72914000
O 11.94920000 0.43492000 1.76809000
O 1.79839000 11.37620000 1.80136000
O 1.06378000 4.55670000 1.82871000
O 8.39548000 7.04101000 1.84762000
Si 6.89164000 7.82098000 1.90040000
O 8.12958000 12.28788000 2.10634000
O 10.66672000 5.50028000 2.11341000
Si 11.59967000 1.97594000 2.23463000
O 7.64251000 9.37060000 2.35451000
Si 0.41183000 12.06493000 2.35560000
O 8.23978000 4.36766000 2.37970000
O 1.05524000 7.20221000 2.42764000
O 6.96701000 1.98259000 2.49582000
Si 5.19710000 5.95740000 2.53516000
Si 9.11036000 5.76089000 2.59097000
O 2.94267000 9.03534000 2.65423000
O 5.57685000 11.01483000 2.79376000
O 5.55014000 4.38570000 2.83243000
O 3.63872000 6.39622000 2.88391000
O 0.55106000 2.29399000 3.02640000
Si 6.98768000 3.52843000 3.03292000
Si 1.42896000 3.64159000 3.13315000
Si 2.46711000 7.55467000 3.17378000
O 10.16690000 9.96555000 3.28935000
O 2.92356000 3.04287000 3.38900000
O 10.43410000 1.98437000 3.43561000
O 6.09824000 7.13813000 3.49108000
O 0.48071000 11.99165000 4.02327000
O 9.16071000 6.09578000 4.18112000
Si 6.19584000 10.88914000 4.31980000
O 0.97667000 4.54705000 4.38853000
O 3.08859000 0.48030000 4.59870000
O 7.12128000 3.54351000 4.68117000
Si 3.24937000 2.09393000 4.69687000
O 12.16214000 9.04504000 4.71645000
O 2.01498000 7.68646000 4.73946000
Si 10.60485000 9.57516000 4.79299000
O 6.00410000 9.36258000 4.82024000
O 7.69724000 11.39131000 4.91130000
Si 10.68433000 2.45872000 4.99262000
O 4.79955000 2.37986000 5.01078000
O 10.79898000 4.10323000 5.01979000
Si 6.04217000 7.79614000 5.06574000

| | | | |
|---|---|---|---|
| O | 12.00916000000002 | 10.72987999999997 | 10.81921000000010 |
| O | 10.42087000000003 | 8.80290999999998 | 10.93948000000000 |
| O | 3.59532999999998 | 6.56214999999997 | 10.94387000000006 |
| O | 8.69596000000002 | 2.57416000000001 | 11.09873999999990 |
| O | 10.68622999999996 | 6.10892000000003 | 11.26061000000003 |
| O | 11.16631000000004 | 1.47555999999997 | 11.27848999999990 |
| O | 1.69607000000000 | 9.81736000000004 | 11.56142999999997 |
| O | 7.83750999999998 | 5.87395000000004 | 11.56350999999994 |
| O | 7.58989999999996 | 12.19780000000004 | 11.79330000000003 |
| O | 7.96866000000004 | 9.60059000000000 | 11.88239999999997 |
| O | 6.05742999999996 | 2.87874000000000 | 11.92653999999994 |
| O | 5.46798287302002 | 10.82807405866915 | 12.17812075116490 |
| O | 3.09658044253444 | 11.86899606235746 | 12.17513947350038 |
| O | 3.15437183196883 | 3.56650294122463 | 12.26885027224374 |
| O | 0.41501858363516 | 4.61436723011201 | 12.45335015169373 |
| O | 0.91730588718479 | 1.87009631017530 | 12.58493508660605 |
| O | 5.56261220151274 | 7.60129398222955 | 12.59809679797309 |
| O | 5.36715264962089 | 0.23734702907422 | 12.71030393358298 |
| O | 0.24409512012476 | 7.27856930393390 | 12.65183367829674 |
| O | 10.17237104113143 | 10.76149040396494 | 12.68020847200979 |
| O | 5.44609082177247 | 5.12814458177882 | 12.79296818723856 |
| O | 12.10024787675254 | 12.48599144429568 | 12.87795466798957 |
| O | 2.76361054164142 | 8.07489992979829 | 12.96326351507424 |
| O | 7.73946301944007 | 1.34868462247914 | 13.21821803289072 |
| O | 8.35251232421204 | 3.87229740595521 | 13.32298923879558 |
| O | 0.15811171494411 | 9.76272953982306 | 13.41513312686128 |
| O | 10.52508142361034 | 5.60827462165196 | 13.69010050310726 |
| O | 7.69873603343230 | 6.32152759752830 | 13.98055225900246 |
| O | 9.05128516167592 | 4.85503304837562 | 15.74074193868477 |
| Pt | 8.28345010507660 | 1.06695323848867 | 15.73426460027258 |
| Pt | 7.26335103224831 | 12.04906304704380 | 15.11598284549744 |
| Pt | 9.11088089371830 | 3.14708242672396 | 16.67782380941229 |
| Pt | 9.06710048142512 | 0.87928912597777 | 18.10127385515328 |
| Si | 0.33529000000003 | 7.71048000000000 | 1.04689000000008 |
| Si | 11.95804000000004 | 4.88342999999995 | 1.29616000000010 |
| Si | 5.36203999999997 | 10.27900999999999 | 1.31000999999992 |
| Si | 2.45532000000003 | 9.93546000000002 | 1.35284000000002 |
| Si | 7.97589000000000 | 10.75088000000000 | 1.45379000000003 |
| Si | 7.46856999999998 | 0.73103000000003 | 1.50179000000003 |
| Si | 10.67104000000004 | 9.93228999999998 | 1.72913999999992 |
| Si | 6.89163999999996 | 7.82098000000001 | 1.90039999999999 |
| Si | 11.59966999999997 | 1.97594000000002 | 2.23462999999992 |
| Si | 0.41182999999997 | 12.06493000000001 | 2.35560000000010 |
| Si | 5.19710000000004 | 5.95740000000003 | 2.53515999999990 |
| Si | 9.11035999999999 | 5.76089000000002 | 2.59096999999997 |
| Si | 6.98767999999998 | 3.52843000000003 | 3.03292000000010 |
| [illegible] | | | |
| O | 11.97163000 | 6.52034000 | 5.16816000 |
| O | 5.27731000 | 11.86749000 | 5.19492000 |
| Si | 12.17232000 | 4.92029000 | 5.36653000 |
| O | 9.61395000 | 8.45242000 | 5.37118000 |
| O | 2.78230000 | 11.22876000 | 5.39252000 |
| Si | 0.52618000 | 7.80404000 | 5.42080000 |
| Si | 1.24491000 | 11.77372000 | 5.47459000 |
| Si | 8.97684000 | 6.98953000 | 5.51034000 |
| O | 12.12776000 | 2.03321000 | 5.54723000 |
| Si | 5.79442000 | 3.48931000 | 5.63946000 |
| Si | 3.88078000 | 12.41462000 | 5.72509000 |
| O | 7.36934000 | 7.19216000 | 5.73077000 |
| O | 5.07310000 | 4.93270000 | 5.77145000 |
| O | 4.69389000 | 7.40689000 | 5.89179000 |
| O | 10.51297000 | 10.77216000 | 5.91766000 |
| O | 2.24135000 | 2.44179000 | 5.95153000 |
| O | 9.68518000 | 1.77422000 | 6.10531000 |
| O | 0.99593000 | 0.04056000 | 6.30419000 |
| O | 6.62583000 | 0.44991000 | 6.32625000 |
| O | 2.60175000 | 5.71772000 | 6.42522000 |
| Si | 7.59978000 | 12.00862000 | 6.46258000 |
| Si | 0.91102000 | 1.63242000 | 6.49053000 |
| Si | 4.19754000 | 6.01659000 | 6.60349000 |
| O | 0.52427000 | 10.85444000 | 6.61729000 |
| O | 9.59361000 | 6.33156000 | 6.88213000 |
| O | 0.24089000 | 4.58204000 | 6.89383000 |
| O | 0.63894000 | 7.98388000 | 7.03464000 |
| O | 6.09524000 | 2.98598000 | 7.14604000 |
| O | 8.85396000 | 3.87869000 | 7.14825000 |
| O | 6.56019000 | 11.02720000 | 7.22128000 |
| Si | 11.46811000 | 10.87769000 | 7.25600000 |
| O | 4.04474000 | 12.85904000 | 7.33218000 |
| Si | 1.66166000 | 5.02240000 | 7.57693000 |
| Si | 9.28667000 | 2.38491000 | 7.58584000 |
| Si | 6.52179000 | 1.48441000 | 7.58891000 |
| O | 8.76445000 | 12.06825000 | 7.62044000 |
| Si | 8.81881000 | 5.29454000 | 7.88978000 |
| O | 7.29605000 | 5.87780000 | 8.06985000 |
| O | 0.72451000 | 1.93019000 | 8.08120000 |
| O | 6.89743000 | 8.56528000 | 8.09660000 |
| O | 11.16503000 | 12.17028000 | 8.21368000 |
| O | 4.36474000 | 6.02266000 | 8.21442000 |
| O | 8.00193000 | 1.62577000 | 8.21556000 |
| O | 2.52145000 | 3.83121000 | 8.25785000 |
| O | 4.60298000 | 9.87454000 | 8.29784000 |
| O | 11.22294000 | 9.65737000 | 8.33763000 |
| [illegible] | | | |
| O | 2.24134999999998 | 2.44178999999999 | 5.95153000000003 |
| O | 9.68518000000001 | 1.77421999999996 | 6.10530999999990 |
| O | 0.99593000000004 | 0.04055999999996 | 6.30419000000006 |
| O | 6.62583000000004 | 0.44990999999996 | 6.32625000000006 |
| O | 2.60175000000003 | 5.71772000000001 | 6.42522000000010 |
| O | 0.52427000000003 | 10.85443999999999 | 6.61728999999990 |
| O | 9.59361000000001 | 6.33155999999997 | 6.88212999999997 |
| O | 0.24088999999996 | 4.58203999999997 | 6.89382999999997 |
| O | 0.63894000000003 | 7.98388000000003 | 7.03464000000010 |
| O | 6.09523999999999 | 2.98598000000003 | 7.14604000000010 |
| O | 8.85395999999999 | 3.87869000000004 | 7.14824999999997 |
| O | 6.56018999999997 | 11.02720000000000 | 7.22127999999997 |
| O | 4.04473999999998 | 12.85904000000001 | 7.33218000000010 |
| O | 8.76444999999999 | 12.06824999999998 | 7.62043999999990 |
| O | 7.29605000000001 | 5.87779999999998 | 8.06985000000010 |
| O | 0.72450999999998 | 1.93019000000004 | 8.08120000000010 |
| O | 6.89742999999996 | 8.56527999999996 | 8.09660000000010 |
| O | 11.16503000000000 | 12.17027999999997 | 8.21368000000006 |
| O | 4.36473999999996 | 6.02265999999997 | 8.21442000000003 |
| O | 8.00193000000000 | 1.62576999999997 | 8.21556000000010 |
| O | 2.52144999999999 | 3.83120999999999 | 8.25784999999997 |
| O | 4.60298000000000 | 9.87453999999997 | 8.29783999999994 |
| O | 11.23293999999997 | 9.65737000000002 | 8.33763000000000 |
| O | 2.62367000000001 | 8.34195000000002 | 8.68505000000003 |
| O | 5.53447000000000 | 0.91970000000001 | 8.72556000000010 |
| O | 1.28774000000001 | 6.06899000000004 | 8.77017000000000 |
| O | 10.31381000000000 | 2.22343999999997 | 8.86801999999987 |
| O | 3.02209000000000 | 1.46592000000000 | 9.34606000000006 |
| O | 9.41126000000003 | 5.06335000000000 | 9.39370000000010 |
| O | 2.55275999999996 | 10.91413000000001 | 9.53045999999994 |
| O | 8.59502000000002 | 7.45155999999996 | 9.76356000000010 |
| O | 0.25734999999999 | 8.20553000000001 | 9.78758999999990 |
| O | 6.67756999999997 | 10.56639999999996 | 9.82734000000006 |
| O | 9.35747000000002 | 10.86630999999996 | 9.95108000000003 |
| O | 5.93269000000002 | 6.91789000000004 | 9.99594000000003 |
| O | 9.20339000000000 | 0.22140000000002 | 10.04827999999997 |
| O | 4.65431999999998 | 12.45006999999997 | 10.15129000000003 |
| O | 4.77644999999996 | 4.38800999999999 | 10.27451999999994 |
| O | 1.23121999999997 | 3.17125999999997 | 10.40270000000010 |
| O | 1.22909999999998 | 12.81664000000003 | 10.76713999999994 |
| O | 4.13856999999997 | 9.06001000000002 | 10.78539000000000 |
| O | 12.00916000000002 | 10.72987999999997 | 10.81921000000010 |
| O | 10.42087000000003 | 8.80290999999998 | 10.93948000000000 |
| O | 3.59532999999998 | 6.56214999999997 | 10.94387000000006 |
| O | 8.69596000000002 | 2.57416000000001 | 11.09873999999990 |
| O | 10.68622999999996 | 6.10892000000003 | 11.26061000000003 |
| O | 11.16631000000004 | 1.47555999999997 | 11.27848999999990 |
| O | 1.69607000000000 | 9.81736000000004 | 11.56142999999997 |
| O | 7.83750999999998 | 5.87395000000004 | 11.56350999999994 |
| O | 7.58989999999996 | 12.19780000000004 | 11.79330000000003 |
| O | 7.96866000000004 | 9.60059000000000 | 11.88239999999997 |
| O | 6.05742999999996 | 2.87874000000000 | 11.92653999999994 |
| O | 5.46520083800054 | 10.82687858562490 | 12.17345880347149 |
| O | 3.09563522418907 | 11.86927001705096 | 12.17520339768202 |
| O | 3.15477057383068 | 3.56663620361611 | 12.26919188652368 |
| O | 0.41634523477572 | 4.61512787429911 | 12.45271131115555 |
| O | 0.91622006659524 | 1.87110526542205 | 12.58553884044477 |
| O | 5.56141427317800 | 7.60320556134290 | 12.59695992659546 |
| O | 5.36965209042659 | 0.24873458840660 | 12.69050081328214 |
| O | 0.24560488527658 | 7.27963661619327 | 12.64997748830614 |
| O | 10.17341681591071 | 10.76085395838143 | 12.67984820521315 |
| O | 5.44698233835762 | 5.12877898896068 | 12.79168180061994 |
| O | 12.09969466638575 | 12.48658653787969 | 12.87865194635357 |
| O | 2.76403850760655 | 8.07457501671568 | 12.96364856512035 |
| O | 7.73118834977698 | 1.35185831587092 | 13.26242866455846 |
| O | 8.35758457149369 | 3.87345349922186 | 13.32574417895581 |
| O | 0.15905140493383 | 9.76338416403525 | 13.41580091379546 |
| O | 10.52732913303452 | 5.61062505231915 | 13.68584178509037 |
| O | 7.69818251247211 | 6.33083605305872 | 13.97181025105443 |
| O | 9.05975454496311 | 4.91729713946868 | 15.74082317116269 |
| Pt | 8.05762583430250 | 1.14509459435455 | 15.66938302891958 |
| Pt | 7.21238244730303 | 11.99371523034742 | 15.08299540037206 |
| Pt | 8.88256116576180 | 3.27326696654027 | 16.71477263420752 |
| Pt | 9.38809657283653 | 1.13348873258342 | 17.81558317107261 |
| Si | 0.33529000000003 | 7.71048000000000 | 1.04689000000008 |
| Si | 11.95804000000004 | 4.88342999999995 | 1.29616000000010 |
| Si | 5.36203999999997 | 10.27900999999999 | 1.31000999999992 |
| Si | 2.45532000000003 | 9.93546000000002 | 1.35284000000002 |
| Si | 7.97589000000000 | 10.75088000000000 | 1.45379000000003 |
| Si | 7.46856999999998 | 0.73103000000003 | 1.50179000000003 |
| Si | 10.67104000000004 | 9.93228999999998 | 1.72913999999992 |
| Si | 6.89163999999996 | 7.82098000000001 | 1.90039999999999 |
| Si | 11.59966999999997 | 1.97594000000002 | 2.23462999999992 |
| Si | 0.41182999999997 | 12.06493000000001 | 2.35560000000010 |
| Si | 5.19710000000004 | 5.95740000000003 | 2.53515999999990 |
| Si | 9.11035999999999 | 5.76089000000002 | 2.59096999999997 |
| Si | 6.98767999999998 | 3.52843000000003 | 3.03292000000010 |
| Si | 1.42896000000001 | 3.64158999999998 | 3.13314999999989 |
| Si | 2.46711000000000 | 7.55467000000003 | 3.17378000000008 |
| Si | 6.19584000000002 | 10.88913999999998 | 4.31980000000000 |
| Si | 3.24937000000000 | 2.09392999999998 | 4.69687000000000 |
| Si | 10.60485000000002 | 9.57516000000005 | 4.79298999999990 |
| Si | 10.68432999999996 | 2.45871999999998 | 4.99262000000000 |
| Si | 6.04216999999995 | 7.79614000000002 | 5.06574000000000 |
| Si | 12.17232000000003 | 4.92029000000004 | 5.36653000000000 |
| Si | 0.52618000000003 | 7.80403999999996 | 5.42080000000010 |
| Si | 1.24491000000002 | 11.77372000000004 | 5.47459000000003 |
| Si | 8.97683999999999 | 6.98952999999997 | 5.51034000000003 |
| Si | 5.79442000000005 | 3.48931000000001 | 5.63946000000010 |
| Si | 3.88077999999997 | 12.41462000000000 | 5.72508999999990 |
| Si | 7.59978000000002 | 12.00861999999996 | 6.46257999999990 |
| Si | 0.91102000000001 | 1.63241999999999 | 6.49053000000003 |
| Si | 4.19754000000004 | 6.01659000000003 | 6.60348999999997 |
| Si | 11.46811000000002 | 10.87768999999998 | 7.25600000000010 |
| Si | 1.66166000000000 | 5.02239999999999 | 7.57692999999994 |
| Si | 9.28667000000001 | 2.38490999999995 | 7.58583999999997 |
| Si | 6.52179000000001 | 1.48441000000003 | 7.58890999999994 |
| Si | 8.81880999999998 | 5.29454000000002 | 7.88977999999997 |
| Si | 6.20759000000003 | 10.01218000000002 | 8.39279000000010 |
| Si | 1.16603000000001 | 7.67142999999996 | 8.54078000000003 |
| Si | 4.03345999999999 | 0.30045000000004 | 8.83076000000006 |
| Si | 9.68524000000000 | 12.14075000000000 | 8.96686000000000 |
| Si | 7.16647999999999 | 7.23733000000004 | 8.97783000000003 |
| Si | 1.87645999999999 | 2.56144000000005 | 9.05034999999997 |
| Si | 3.45236000000000 | 9.54198999999997 | 9.37580000000003 |
| Si | 4.61912000000000 | 5.94270000000003 | 9.80684999999994 |
| Si | 11.57529000000003 | 9.37814000000003 | 9.91657000000010 |
| Si | 9.84757999999999 | 1.68549000000002 | 10.33173000000010 |
| Si | 9.17938999999999 | 6.14520000000004 | 10.61420999999997 |
| Si | 2.02201000000000 | 11.36570000000004 | 11.01221000000010 |
| Si | 9.45776000000002 | 10.06017999999996 | 11.38731000000006 |
| Si | 6.89244999999997 | 10.75082000000000 | 11.41328999999997 |
| Si | 4.62983000000000 | 12.22691000000004 | 11.78878000000010 |
| Si | 4.77658000000003 | 3.93277999999998 | 11.86347000000000 |
| Si | 0.00000000000000 | 0.64099999999999 | 11.89328000000000 |
| Si | 4.04675000000000 | 7.84950000000002 | 11.90235000000006 |
| Si | 1.46158789785866 | 3.35797267733982 | 12.00747453234976 |
| Si | 7.66341674503518 | 2.74821289232046 | 12.36970356197638 |
| Si | 11.78913036539570 | 10.90363670682371 | 12.51309990022422 |
| Si | 11.72227146156940 | 5.89291337331779 | 12.53748053648086 |
| Si | 1.21428863609793 | 8.66117832025543 | 12.65478139572474 |
| Si | 6.63191756072861 | 6.29024383743056 | 12.69991556027299 |
| Si | 6.94837360896311 | 12.99431588337603 | 13.13338618526906 |
| Si | 8.94573657830325 | 5.19272388614532 | 14.14135545939258 |

193

II-P3

| | | | |
|---|---|---|---|
| H | 6.58656000 | 12.90919000 | 0.00000000 |
| O | 12.34643000 | 6.23675000 | 0.32400000 |
| O | 6.32035000 | 0.68143000 | 0.34265000 |
| O | 6.64696000 | 10.90101000 | 0.43734000 |
| O | 1.43306000 | 8.88837000 | 0.50018000 |
| H | 9.03195000 | 2.06583000 | 0.63119000 |
| O | 3.86303000 | 10.46476000 | 0.65412000 |
| O | 9.40856000 | 10.33665000 | 0.75816000 |
| O | 8.97135000 | 1.15729000 | 0.95693000 |
| Si | 0.33529000 | 7.71048000 | 1.04689000 |
| O | 11.41461000 | 3.24415000 | 1.12214000 |
| O | 6.05809000 | 6.39254000 | 1.24787000 |
| O | 11.35216000 | 8.50616000 | 1.29237000 |
| Si | 11.95804000 | 4.88343000 | 1.29616000 |
| O | 5.57676000 | 8.67061000 | 1.29731000 |
| Si | 5.36204000 | 10.27901000 | 1.31001000 |
| Si | 2.45532000 | 9.93546000 | 1.35284000 |
| Si | 7.97589000 | 10.75088000 | 1.45379000 |
| Si | 7.46857000 | 0.73103000 | 1.50179000 |
| O | 11.74039000 | 11.12450000 | 1.65148000 |

[illegible]

| | | | |
|---|---|---|---|
| O | 11.35215999999998 | 8.50616000000004 | 1.29236999999989 |
| O | 5.57675999999996 | 8.67061000000003 | 1.29730999999993 |
| O | 11.74038999999997 | 11.12449999999999 | 1.65147999999999 |
| O | 11.94920000000000 | 0.43491999999997 | 1.76809000000003 |
| O | 1.79838999999998 | 11.37620000000001 | 1.80135999999993 |
| O | 1.06378000000003 | 4.55670000000004 | 1.82871000000000 |
| O | 8.39548000000004 | 7.04101000000000 | 1.84762000000001 |
| O | 8.12957999999999 | 12.28788000000000 | 2.10634000000004 |
| O | 10.66671999999996 | 5.50028000000003 | 2.11340999999993 |
| O | 7.64250999999997 | 9.37059999999996 | 2.35450999999989 |
| O | 8.23978000000001 | 4.36766000000002 | 2.37969999999996 |
| O | 1.05523999999999 | 7.20221000000000 | 2.42764000000011 |
| O | 6.96701000000000 | 1.98259000000004 | 2.49582000000009 |
| O | 2.94267000000003 | 9.03534000000004 | 2.65423000000010 |
| O | 5.57684999999999 | 11.01483000000002 | 2.79376000000002 |
| O | 5.55013999999998 | 4.38570000000005 | 2.83242999999993 |
| O | 3.63872000000002 | 6.39621999999996 | 2.88391000000002 |
| O | 0.55106000000004 | 2.29398999999997 | 3.02639999999997 |
| O | 10.16689999999996 | 9.96554999999999 | 3.28935000000000 |
| O | 2.92355999999999 | 3.04286999999996 | 3.38899999999990 |
| O | 10.43409999999997 | 1.98437000000002 | 3.43561000000000 |
| O | 6.09823999999998 | 7.13813000000000 | 3.49108000000003 |
| O | 0.48070999999996 | 11.99164999999995 | 4.02326999999990 |
| O | 9.16071000000002 | 6.09578000000000 | 4.18111999999997 |
| O | 0.97666999999999 | 4.54704999999998 | 4.38852999999994 |
| O | 3.08859000000003 | 0.48030000000004 | 4.59870000000003 |
| O | 7.12127999999996 | 3.54350999999997 | 4.68117000000006 |
| O | 12.16213999999997 | 9.04504000000001 | 4.71644999999990 |
| O | 2.01498000000001 | 7.68645999999997 | 4.73946000000000 |
| O | 6.00409999999997 | 9.36258000000003 | 4.82024000000000 |
| O | 7.69723999999998 | 11.39130999999996 | 4.91129999999997 |
| O | 4.79955000000004 | 2.37986000000001 | 5.01078000000006 |
| O | 10.79898000000004 | 4.10323000000004 | 5.01979000000006 |
| O | 11.97162999999998 | 6.52033999999996 | 5.16815999999994 |
| O | 5.27730999999998 | 11.86749000000002 | 5.19491999999990 |
| O | 9.61395000000002 | 8.45341999999998 | 5.37118000000010 |
| O | 2.78230000000004 | 11.22875999999996 | 5.39252000000010 |
| O | 12.12775999999999 | 2.03320999999998 | 5.54722999999990 |
| O | 7.36934000000001 | 7.19216000000004 | 5.73076999999990 |
| O | 5.07310000000001 | 4.93269999999997 | 5.77144999999997 |
| O | 4.69389000000002 | 7.40688999999997 | 5.89178999999990 |
| O | 10.51296999999998 | 10.77216000000004 | 5.91766000000006 |
| O | 2.24134999999998 | 2.44178999999999 | 5.95153000000003 |
| O | 9.68518000000001 | 1.77421999999996 | 6.10530999999990 |
| O | 0.99593000000004 | 0.04055999999996 | 6.30419000000006 |
| O | 6.62583000000004 | 0.44990999999996 | 6.32625000000006 |
| O | 2.60175000000003 | 5.71772000000001 | 6.42522000000010 |
| O | 0.52427000000003 | 10.85443999999999 | 6.61728999999990 |
| O | 9.59361000000001 | 6.33155999999997 | 6.88212999999997 |
| O | 0.24088999999996 | 4.58203999999997 | 6.89382999999997 |
| O | 0.63894000000003 | 7.98388000000003 | 7.03464000000010 |
| O | 6.09523999999999 | 2.98598000000003 | 7.14604000000010 |
| O | 8.85395999999999 | 3.87869000000004 | 7.14824999999997 |
| O | 6.56018999999997 | 11.02720000000000 | 7.22127999999997 |
| O | 4.04473999999998 | 12.85904000000001 | 7.33218000000010 |
| O | 8.76444999999999 | 12.06824999999998 | 7.62043999999990 |
| O | 7.29605000000001 | 5.87779999999998 | 8.06985000000010 |
| O | 0.72450999999998 | 1.93019000000004 | 8.08120000000010 |
| O | 6.89742999999996 | 8.56527999999996 | 8.09660000000010 |
| O | 11.16503000000000 | 12.17027999999997 | 8.21368000000006 |
| O | 4.36473999999996 | 6.02265999999997 | 8.21442000000003 |
| O | 8.00193000000000 | 1.62576999999997 | 8.21556000000010 |
| O | 2.52144999999999 | 3.83120999999999 | 8.25784999999997 |
| O | 4.60298000000000 | 9.87453999999997 | 8.29783999999994 |
| O | 11.23293999999997 | 9.65737000000002 | 8.33763000000000 |
| O | 2.62367000000001 | 8.34195000000002 | 8.68505000000003 |
| O | 5.53447000000000 | 0.91970000000001 | 8.72556000000010 |
| O | 1.28774000000001 | 6.06899000000004 | 8.77017000000000 |
| O | 10.31381000000000 | 2.22343999999997 | 8.86801999999987 |
| O | 3.02209000000000 | 1.46592000000000 | 9.34606000000006 |
| O | 9.41126000000003 | 5.06335000000000 | 9.39370000000010 |
| O | 2.55275999999996 | 10.91413000000001 | 9.53045999999994 |
| O | 8.59502000000002 | 7.45155999999996 | 9.76356000000010 |
| O | 0.25734999999999 | 8.20553000000001 | 9.78758999999990 |
| O | 6.67756999999997 | 10.56639999999996 | 9.82734000000006 |
| O | 9.35747000000002 | 10.86630999999996 | 9.95108000000003 |
| O | 5.93269000000002 | 6.91789000000004 | 9.99594000000003 |
| O | 9.20339000000000 | 0.22140000000002 | 10.04827999999997 |
| O | 4.65431999999998 | 12.45006999999997 | 10.15129000000003 |
| O | 4.77644999999996 | 4.38800999999999 | 10.27451999999994 |
| O | 1.23121999999997 | 3.17125999999997 | 10.40270000000010 |
| O | 1.22909999999998 | 12.81664000000003 | 10.76713999999994 |
| O | 4.13856999999997 | 9.06001000000002 | 10.78539000000000 |
| O | 12.00916000000002 | 10.72987999999997 | 10.81921000000010 |
| O | 10.42087000000003 | 8.80290999999998 | 10.93948000000000 |
| O | 3.59532999999998 | 6.56214999999997 | 10.94387000000006 |
| O | 8.69596000000002 | 2.57416000000001 | 11.09873999999990 |
| O | 10.68622999999996 | 6.10892000000003 | 11.26061000000003 |
| O | 11.16631000000004 | 1.47555999999997 | 11.27848999999990 |
| O | 1.69607000000000 | 9.81736000000004 | 11.56142999999997 |
| O | 7.83750999999998 | 5.87395000000004 | 11.56350999999994 |
| O | 7.58989999999996 | 12.19780000000004 | 11.79330000000003 |
| O | 7.96866000000004 | 9.60059000000000 | 11.88239999999997 |
| O | 6.05742999999996 | 2.87874000000000 | 11.92653999999994 |
| O | 5.45287764273664 | 10.81314227419446 | 12.15108270350266 |
| O | 3.09149425271346 | 11.86759202996854 | 12.17473546646851 |
| O | 3.15200244952389 | 3.56599500207070 | 12.27012990216896 |
| O | 0.42126377011195 | 4.61007891794859 | 12.45162220541648 |
| O | 0.91801868980867 | 1.86630693976115 | 12.58712465793930 |
| O | 5.55551222610413 | 7.61770201541472 | 12.58747188538563 |
| O | 5.36153260104991 | 0.21603736948834 | 12.73574652612387 |
| O | 0.24102513505015 | 7.27383065660790 | 12.65045763445574 |
| O | 10.19145293180495 | 10.73720289110222 | 12.69899983874189 |
| O | 5.42844982838957 | 5.14400347893260 | 12.78163830834979 |
| O | 12.09969530205358 | 12.48152621499555 | 12.87909968008397 |
| O | 2.76335192827817 | 8.07115853500250 | 12.96587491163245 |
| O | 7.72699705604702 | 1.32662243720455 | 13.25681095195501 |
| O | 8.35527571211895 | 3.86295817305816 | 13.39004857818099 |
| O | 0.16654848264306 | 9.75666712263528 | 13.42010682840282 |
| O | 10.53316248176981 | 5.57441810511413 | 13.68509201227606 |
| O | 7.67621243429740 | 6.36246172018367 | 13.91919755740355 |
| O | 8.97678770394992 | 5.16707310706382 | 15.79356892389389 |
| Pt | 7.97404776254336 | 9.15424578998991 | 16.34046567655379 |
| Pt | 5.70732837196331 | 8.06483691476713 | 17.69890514153462 |
| Pt | 7.72663856058189 | 6.51942678750651 | 16.60922621328746 |
| Pt | 7.28412209716937 | 11.46899009753772 | 14.93547654728899 |
| Si | 0.33529000000003 | 7.71048000000000 | 1.04689000000008 |
| Si | 11.95804000000004 | 4.88342999999995 | 1.29616000000010 |
| Si | 5.36203999999997 | 10.27900999999999 | 1.31000999999992 |
| Si | 2.45532000000003 | 9.93546000000002 | 1.35284000000002 |
| Si | 7.97589000000000 | 10.75088000000000 | 1.45379000000003 |
| Si | 7.46856999999998 | 0.73103000000003 | 1.50179000000003 |
| Si | 10.67104000000004 | 9.93228999999998 | 1.72913999999992 |
| Si | 6.89163999999996 | 7.82098000000001 | 1.90039999999999 |
| Si | 11.59966999999997 | 1.97594000000002 | 2.23462999999992 |
| Si | 0.41182999999997 | 12.06493000000001 | 2.35560000000010 |
| Si | 5.19710000000004 | 5.95740000000003 | 2.53515999999990 |
| Si | 9.11035999999999 | 5.76089000000002 | 2.59096999999997 |
| Si | 6.98767999999998 | 3.52843000000003 | 3.03292000000010 |
| Si | 1.42896000000001 | 3.64158999999998 | 3.13314999999989 |
| Si | 2.46711000000000 | 7.55467000000003 | 3.17378000000008 |
| Si | 6.19584000000002 | 10.88913999999998 | 4.31980000000000 |
| Si | 3.24937000000000 | 2.09392999999998 | 4.69687000000000 |
| Si | 10.60485000000002 | 9.57516000000005 | 4.79298999999990 |
| Si | 10.68432999999996 | 2.45871999999998 | 4.99262000000000 |
| Si | 6.04216999999995 | 7.79614000000002 | 5.06574000000000 |
| Si | 12.17232000000003 | 4.92029000000004 | 5.36653000000000 |
| Si | 0.52618000000003 | 7.80403999999996 | 5.42080000000010 |
| Si | 1.24491000000002 | 11.77372000000004 | 5.47459000000003 |
| Si | 8.97683999999999 | 6.98952999999997 | 5.51034000000003 |
| Si | 5.79442000000005 | 3.48931000000001 | 5.63946000000010 |
| Si | 3.88077999999997 | 12.41462000000000 | 5.72508999999990 |
| Si | 7.59978000000002 | 12.00861999999996 | 6.46257999999990 |
| Si | 0.91102000000001 | 1.63241999999999 | 6.49053000000003 |
| Si | 4.19754000000004 | 6.01659000000003 | 6.60348999999997 |
| Si | 11.46811000000002 | 10.87768999999998 | 7.25600000000010 |
| Si | 1.66166000000000 | 5.02239999999999 | 7.57692999999994 |
| Si | 9.28667000000001 | 2.38490999999995 | 7.58583999999997 |
| Si | 6.52179000000001 | 1.48441000000003 | 7.58890999999994 |
| Si | 8.81880999999998 | 5.29454000000002 | 7.88977999999997 |
| Si | 6.20759000000003 | 10.01218000000002 | 8.39279000000010 |
| Si | 1.16603000000001 | 7.67142999999996 | 8.54078000000003 |
| Si | 4.03345999999999 | 0.30045000000004 | 8.83076000000006 |
| Si | 9.68524000000000 | 12.14075000000000 | 8.96686000000000 |
| Si | 7.16647999999999 | 7.23733000000004 | 8.97783000000003 |
| Si | 1.87645999999999 | 2.56144000000005 | 9.05034999999997 |

[illegible]

| | | | |
|---|---|---|---|
| O | 5.57675999999996 | 8.67061000000003 | 1.29730999999993 |
| O | 11.74038999999997 | 11.12449999999999 | 1.65147999999999 |
| O | 11.94920000000000 | 0.43491999999997 | 1.76809000000003 |
| O | 1.79838999999998 | 11.37620000000001 | 1.80135999999993 |
| O | 1.06378000000003 | 4.55670000000004 | 1.82871000000000 |
| O | 8.39548000000004 | 7.04101000000000 | 1.84762000000001 |
| O | 8.12957999999999 | 12.28788000000000 | 2.10634000000004 |
| O | 10.66671999999996 | 5.50028000000003 | 2.11340999999993 |
| O | 7.64250999999997 | 9.37059999999996 | 2.35450999999989 |
| O | 8.23978000000001 | 4.36766000000002 | 2.37969999999996 |
| O | 1.05523999999999 | 7.20221000000000 | 2.42764000000011 |
| O | 6.96701000000000 | 1.98259000000004 | 2.49582000000009 |
| O | 2.94267000000003 | 9.03534000000004 | 2.65423000000010 |
| O | 5.57684999999999 | 11.01483000000002 | 2.79376000000002 |
| O | 5.55013999999998 | 4.38570000000005 | 2.83242999999993 |
| O | 3.63872000000002 | 6.39621999999996 | 2.88391000000002 |
| O | 0.55106000000004 | 2.29398999999997 | 3.02639999999997 |
| O | 10.16689999999996 | 9.96554999999999 | 3.28935000000000 |
| O | 2.92355999999999 | 3.04286999999996 | 3.38899999999990 |
| O | 10.43409999999997 | 1.98437000000002 | 3.43561000000000 |
| O | 6.09823999999998 | 7.13813000000000 | 3.49108000000000 |
| O | 0.48070999999996 | 11.99164999999995 | 4.02326999999990 |
| O | 9.16071000000002 | 6.09578000000000 | 4.18111999999997 |
| O | 0.97666999999999 | 4.54704999999998 | 4.38852999999994 |
| O | 3.08859000000003 | 0.48030000000000 | 4.59870000000000 |
| O | 7.12127999999999 | 3.54350999999997 | 4.68117000000006 |
| O | 12.16213999999997 | 9.04504000000001 | 4.71644999999990 |
| O | 2.01498000000001 | 7.68645999999997 | 4.73946000000000 |
| O | 6.00409999999997 | 9.36258000000003 | 4.82024000000000 |
| O | 7.69723999999998 | 11.39130999999999 | 4.91129999999997 |
| O | 4.79955000000004 | 2.37986000000001 | 5.01078000000006 |
| O | 10.79898000000004 | 4.10323000000004 | 5.01979000000006 |
| O | 11.97162999999998 | 6.52033999999996 | 5.16815999999994 |
| O | 5.27730999999998 | 11.86749000000002 | 5.19491999999990 |
| O | 9.61395000000002 | 8.45241999999998 | 5.37118000000010 |
| O | 2.78230000000004 | 11.22875999999996 | 5.39252000000010 |
| O | 12.12775999999999 | 2.03320999999998 | 5.54722999999990 |
| O | 7.36934000000001 | 7.19216000000004 | 5.73076999999990 |
| O | 5.07310000000001 | 4.93269999999997 | 5.77144999999997 |
| O | 4.69389000000001 | 7.40688999999997 | 5.89178999999990 |
| O | 10.51296999999998 | 10.77216000000004 | 5.91766000000006 |
| O | 2.24134999999998 | 2.44178999999999 | 5.95153000000003 |
| O | 9.68518000000001 | 1.77421999999996 | 6.10530999999990 |
| O | 0.99593000000004 | 0.04055999999999 | 6.30419000000006 |
| O | 6.62583000000004 | 0.44990999999996 | 6.32625000000006 |
| O | 2.60175000000003 | 5.71772000000001 | 6.42522000000010 |
| O | 0.52427000000003 | 10.85443999999999 | 6.61728999999990 |
| O | 9.59361000000001 | 6.33155999999997 | 6.88212999999997 |
| O | 0.24088999999996 | 4.58203999999997 | 6.89382999999997 |
| O | 0.63894000000003 | 7.98388000000003 | 7.03464000000010 |
| O | 6.09523999999998 | 2.98598000000003 | 7.14604000000010 |
| O | 8.85395999999999 | 3.87869000000004 | 7.14824999999997 |
| O | 6.56018999999997 | 11.02720000000000 | 7.22127999999997 |
| O | 4.04473999999998 | 12.85904000000001 | 7.33218000000010 |
| O | 8.76444999999999 | 12.06824999999998 | 7.62043999999990 |
| O | 7.29605000000001 | 5.87779999999998 | 8.06985000000010 |
| O | 0.72450999999998 | 1.93019000000004 | 8.08120000000010 |
| O | 6.89742999999996 | 8.56527999999996 | 8.09660000000010 |
| O | 11.16503000000000 | 12.17027999999997 | 8.21368000000006 |
| O | 4.36473999999996 | 6.02265999999997 | 8.21442000000003 |
| O | 8.00193000000000 | 1.62576999999997 | 8.21556000000010 |
| O | 2.52144999999999 | 3.83120999999999 | 8.25784999999997 |
| O | 4.60298000000000 | 9.87453999999997 | 8.29783999999994 |
| O | 11.23293999999997 | 9.65737000000002 | 8.33763000000000 |
| O | 2.62367000000001 | 8.34195000000002 | 8.68505000000003 |
| O | 5.53447000000000 | 0.91970000000001 | 8.72556000000010 |
| O | 1.28774000000001 | 6.06899000000004 | 8.77017000000000 |
| O | 10.31381000000000 | 2.32343999999997 | 8.86801999999987 |
| O | 3.02209000000000 | 1.46592000000000 | 9.34606000000006 |
| O | 9.41126000000003 | 5.06335000000000 | 9.39370000000010 |
| O | 2.55275999999996 | 10.91413000000001 | 9.53045999999994 |
| O | 8.59502000000002 | 7.45155999999996 | 9.76356000000010 |
| O | 0.25734999999999 | 8.20553000000001 | 9.78758999999990 |
| O | 6.67756999999997 | 10.56639999999996 | 9.82734000000006 |
| O | 9.35747000000002 | 10.86630999999996 | 9.95108000000003 |
| O | 5.93269000000002 | 6.91789000000004 | 9.99594000000003 |
| O | 9.20339000000000 | 0.22140000000002 | 10.04827999999997 |
| O | 4.65431999999998 | 12.45006999999997 | 10.15129000000003 |
| O | 4.77644999999996 | 4.38800999999999 | 10.27451999999994 |
| O | 1.23121999999997 | 3.17125999999997 | 10.40270000000010 |
| O | 1.22909999999998 | 12.81664000000003 | 10.76713999999994 |
| O | 4.13856999999997 | 9.06001000000002 | 10.78539000000000 |
| O | 12.00916000000002 | 10.72987999999997 | 10.81921000000010 |
| O | 10.42087000000003 | 8.80290999999998 | 10.93948000000000 |
| O | 3.59532999999998 | 6.56214999999997 | 10.94387000000006 |
| O | 8.69596000000002 | 2.57416000000001 | 11.09873999999990 |
| O | 10.68622999999996 | 6.10892000000003 | 11.26061000000003 |
| O | 11.16631000000004 | 1.47555999999997 | 11.27848999999990 |
| O | 5.44489194 | 5.12763982 | 12.79420498 |
| O | 12.09996461 | 12.48581677 | 12.87794695 |
| O | 2.76373503 | 8.07532337 | 12.96342849 |
| Si | 6.96286307 | 13.03024721 | 13.11413114 |
| O | 7.74174279 | 1.34909054 | 13.21002730 |
| O | 8.35239699 | 3.87220338 | 13.31916489 |
| O | 0.15775848 | 9.76271263 | 13.41509604 |
| O | 10.52355950 | 5.61393355 | 13.68831130 |
| O | 7.69687692 | 6.31236077 | 13.98738387 |
| Si | 8.94865544 | 5.17163574 | 14.15711112 |
| Pt | 7.25492365 | 12.02042905 | 15.10447282 |
| O | 9.11424811 | 4.81374362 | 15.72772855 |
| Pt | 7.99435908 | 1.07264944 | 15.92865284 |
| Pt | 8.62109793 | 3.25825891 | 16.80287634 |
| H | 7.88849666 | 2.69548975 | 18.16154661 |
| Pt | 8.74947534 | 1.05275329 | 18.34191528 |
| H | 10.80318678 | 0.01652259 | 19.78948714 |
| C | 9.73077701 | 12.89747051 | 19.79207493 |
| H | 9.50091406 | 11.82247063 | 19.81595505 |
| C | 8.78522020 | 0.72787993 | 20.18782372 |
| H | 8.13232319 | 0.96576633 | 21.02965288 |
| 193 | | | |
| III-R | | | |
| H | 6.58656000 | 12.90919000 | 0.00000000 |
| O | 12.34643000 | 6.23675000 | 0.32400000 |
| O | 6.32035000 | 0.68143000 | 0.34265000 |
| O | 6.64696000 | 10.90101000 | 0.43734000 |
| O | 1.43306000 | 8.88837000 | 0.50018000 |
| H | 9.03195000 | 2.06583000 | 0.63119000 |
| O | 3.86303000 | 10.46476000 | 0.65412000 |
| O | 9.40856000 | 10.33665000 | 0.75816000 |
| O | 8.97135000 | 1.15729000 | 0.95693000 |
| Si | 0.33529000 | 7.71048000 | 1.04689000 |
| O | 11.41461000 | 3.24415000 | 1.12214000 |
| O | 6.05809000 | 6.39254000 | 1.24787000 |
| O | 11.35216000 | 8.50616000 | 1.29237000 |
| Si | 11.95804000 | 4.88343000 | 1.29616000 |
| O | 5.57676000 | 8.67061000 | 1.29731000 |
| Si | 5.36204000 | 10.27901000 | 1.31001000 |
| Si | 2.45532000 | 9.93546000 | 1.35284000 |
| Si | 7.97589000 | 10.75088000 | 1.45379000 |
| Si | 7.46857000 | 0.73103000 | 1.50179000 |
| O | 11.74039000 | 11.12450000 | 1.65148000 |
| Si | 10.67104000 | 9.93229000 | 1.72914000 |
| O | 11.94920000 | 0.43492000 | 1.76809000 |
| O | 1.79839000 | 11.37620000 | 1.80136000 |
| [illegible] | | | |
| Pt | 6.83511256301393 | 1.00679594584956 | 16.42201049446295 |
| Pt | 9.29541138199115 | 0.27301890171112 | 16.63026481862154 |
| Si | 0.33529000000003 | 7.71048000000000 | 1.04689000000008 |
| Si | 11.95804000000004 | 4.88342999999995 | 1.29616000000010 |
| Si | 5.36203999999997 | 10.27900999999999 | 1.31000999999992 |
| Si | 2.45532000000003 | 9.93546000000002 | 1.35284000000002 |
| Si | 7.97589000000000 | 10.75088000000000 | 1.45379000000003 |
| Si | 7.46856999999998 | 0.73103000000003 | 1.50179000000003 |
| Si | 10.67104000000004 | 9.93228999999998 | 1.72913999999992 |
| Si | 6.89163999999996 | 7.82098000000001 | 1.90039999999999 |
| Si | 11.59966999999997 | 1.97594000000002 | 2.23462999999992 |
| Si | 0.41182999999997 | 12.06493000000001 | 2.35560000000010 |
| Si | 5.19710000000004 | 5.95740000000003 | 2.53515999999990 |
| Si | 9.11035999999999 | 5.76089000000002 | 2.59096999999997 |
| Si | 6.98767999999998 | 3.52843000000003 | 3.03292000000010 |
| Si | 1.42896000000001 | 3.64158999999998 | 3.13314999999989 |
| Si | 2.46711000000000 | 7.55467000000003 | 3.17378000000008 |
| Si | 6.19584000000002 | 10.88913999999998 | 4.31980000000000 |
| Si | 3.24937000000000 | 2.09392999999998 | 4.69687000000000 |
| Si | 10.60485000000002 | 9.57516000000005 | 4.79298999999990 |
| Si | 10.68432999999996 | 2.45871999999998 | 4.99262000000000 |
| Si | 6.04216999999995 | 7.79614000000002 | 5.06574000000000 |
| Si | 12.17232000000003 | 4.92029000000004 | 5.36653000000000 |
| Si | 0.52618000000003 | 7.80403999999996 | 5.42080000000010 |
| Si | 1.24491000000002 | 11.77372000000004 | 5.47459000000003 |
| Si | 8.97683999999999 | 6.98952999999997 | 5.51034000000003 |
| Si | 5.79442000000005 | 3.48931000000001 | 5.63946000000010 |
| Si | 3.88077999999997 | 12.41462000000000 | 5.72508999999990 |
| Si | 7.59978000000002 | 12.00861999999996 | 6.46257999999990 |
| Si | 0.91102000000001 | 1.63241999999999 | 6.49053000000003 |
| Si | 4.19754000000004 | 6.01659000000003 | 6.60348999999997 |
| Si | 11.46811000000002 | 10.87768999999998 | 7.25600000000010 |
| Si | 1.66166000000000 | 5.02239999999999 | 7.57692999999994 |
| Si | 9.28667000000001 | 2.38490999999995 | 7.58583999999997 |
| Si | 6.52179000000001 | 1.48441000000003 | 7.58890999999994 |
| Si | 8.81880999999998 | 5.29454000000002 | 7.88977999999997 |
| Si | 6.20759000000003 | 10.01218000000002 | 8.39279000000010 |
| Si | 1.16603000000001 | 7.67142999999996 | 8.54078000000003 |
| Si | 4.03345999999999 | 0.30045000000004 | 8.83076000000006 |
| Si | 9.68524000000000 | 12.14075000000000 | 8.96686000000000 |
| Si | 7.16647999999999 | 7.23733000000004 | 8.97783000000003 |
| Si | 1.87645999999999 | 2.56144000000005 | 9.05034999999997 |
| Si | 3.45236000000000 | 9.54198999999997 | 9.37580000000003 |
| Si | 4.61912000000000 | 5.94270000000003 | 9.80684999999994 |
| Si | 11.57529000000003 | 9.37814000000003 | 9.91657000000010 |
| Si | 9.84757999999999 | 1.68549000000002 | 10.33173000000010 |
| Si | 9.17938999999999 | 6.14520000000004 | 10.61420999999997 |
| Si | 2.02201000000000 | 11.36570000000004 | 11.01221000000010 |
| Si | 9.45776000000002 | 10.06017999999996 | 11.38731000000006 |
| Si | 6.89244999999997 | 10.75082000000000 | 11.41328999999997 |
| Si | 4.62983000000000 | 12.22691000000004 | 11.78878000000010 |
| Si | 4.77658000000003 | 3.93277999999998 | 11.86347000000000 |
| Si | 0.00000000000000 | 0.64099999999999 | 11.89328000000000 |
| Si | 4.04675000000000 | 7.84950000000003 | 11.90235000000006 |
| Si | 1.45850066242897 | 3.36062834508158 | 12.00802428201974 |
| Si | 7.66066265631796 | 2.73810987012403 | 12.34999277911197 |
| Si | 11.79159263113343 | 10.90323747866182 | 12.50820256308954 |
| Si | 11.70798146458250 | 5.88880071232543 | 12.55054612429558 |
| Si | 1.21146484558761 | 8.65635454099976 | 12.65346887883219 |
| Si | 6.64014804589031 | 6.27842357632658 | 12.71221893827738 |
| Si | 6.97301054461803 | 13.04698741636656 | 13.12194100527699 |
| Si | 8.93797632718284 | 5.14738250719025 | 14.16324175440944 |

193

III-P1

| | | | |
|---|---|---|---|
| H | 6.58656000 | 12.90919000 | 0.00000000 |
| O | 12.34643000 | 6.23675000 | 0.32400000 |
| O | 6.32035000 | 0.68143000 | 0.34265000 |
| O | 6.64696000 | 10.90101000 | 0.43734000 |
| O | 1.43306000 | 8.88837000 | 0.50018000 |
| H | 9.03195000 | 2.06583000 | 0.63119000 |
| O | 3.86303000 | 10.46476000 | 0.65412000 |
| O | 9.40856000 | 10.33665000 | 0.75816000 |
| O | 8.97135000 | 1.15729000 | 0.95693000 |
| Si | 0.33529000 | 7.71048000 | 1.04689000 |
| O | 11.41461000 | 3.24415000 | 1.12214000 |
| O | 6.05809000 | 6.39254000 | 1.24787000 |
| O | 11.35216000 | 8.50616000 | 1.29237000 |
| Si | 11.95804000 | 4.88343000 | 1.29616000 |
| O | 5.57676000 | 8.67061000 | 1.29731000 |
| Si | 5.36204000 | 10.27901000 | 1.31001000 |
| Si | 2.45532000 | 9.93546000 | 1.35284000 |
| Si | 7.97589000 | 10.75088000 | 1.45379000 |
| Si | 7.46857000 | 0.73103000 | 1.50179000 |

[illegible]

| | | | |
|---|---|---|---|
| O | 0.72450999999998 | 1.93019000000004 | 8.08120000000010 |
| O | 6.89742999999996 | 8.56527999999996 | 8.09660000000010 |
| O | 11.16503000000000 | 12.17027999999997 | 8.21368000000006 |
| O | 4.36473999999996 | 6.02265999999997 | 8.21442000000003 |
| O | 8.00193000000000 | 1.62576999999997 | 8.21556000000010 |
| O | 2.52144999999999 | 3.83120999999999 | 8.25784999999997 |
| O | 4.60298000000000 | 9.87453999999997 | 8.29783999999994 |
| O | 11.23293999999997 | 9.65737000000002 | 8.33763000000000 |
| O | 2.62367000000001 | 8.34195000000002 | 8.68505000000003 |
| O | 5.53447000000000 | 0.91970000000001 | 8.72556000000010 |
| O | 1.28774000000001 | 6.06899000000004 | 8.77017000000000 |
| O | 10.31381000000000 | 2.32343999999997 | 8.86801999999987 |
| O | 3.02209000000000 | 1.46592000000000 | 9.34606000000006 |
| O | 9.41126000000003 | 5.06335000000000 | 9.39370000000010 |
| O | 2.55275999999996 | 10.91413000000001 | 9.53045999999994 |
| O | 8.59502000000002 | 7.45155999999996 | 9.76356000000010 |
| O | 0.25734999999999 | 8.20553000000001 | 9.78758999999990 |
| O | 6.67756999999997 | 10.56639999999996 | 9.82734000000006 |
| O | 9.35747000000002 | 10.86630999999996 | 9.95108000000003 |
| O | 5.93269000000002 | 6.91789000000004 | 9.99594000000003 |
| O | 9.20339000000000 | 0.22140000000002 | 10.04827999999997 |
| O | 4.65431999999998 | 12.45006999999997 | 10.15129000000003 |
| O | 4.77644999999996 | 4.38800999999999 | 10.27451999999994 |
| O | 1.23121999999997 | 3.17125999999997 | 10.40270000000010 |
| O | 1.22909999999998 | 12.81664000000003 | 10.76713999999994 |
| O | 4.13856999999997 | 9.06001000000002 | 10.78539000000000 |
| O | 12.00916000000002 | 10.72987999999997 | 10.81921000000010 |
| O | 10.42087000000003 | 8.80290999999998 | 10.93948000000000 |
| O | 3.59532999999998 | 6.56214999999997 | 10.94387000000006 |
| O | 8.69596000000002 | 2.57416000000001 | 11.09873999999990 |
| O | 10.68622999999996 | 6.10892000000003 | 11.26061000000003 |
| O | 11.16631000000004 | 1.47555999999997 | 11.27848999999990 |
| O | 1.69607000000000 | 9.81736000000004 | 11.56142999999997 |
| O | 7.83750999999998 | 5.87395000000004 | 11.56350999999994 |
| O | 7.58989999999996 | 12.19780000000004 | 11.79330000000003 |
| O | 7.96866000000004 | 9.60059000000000 | 11.88239999999997 |
| O | 6.05742999999996 | 2.87874000000000 | 11.92653999999994 |
| O | 5.45479139919221 | 10.80931917827034 | 12.15668528201651 |
| O | 3.09308375680173 | 11.86562944003293 | 12.17473720295379 |
| O | 3.15222973444915 | 3.56654446335293 | 12.26976144717174 |
| O | 0.42315972083183 | 4.61233742627839 | 12.45118990002355 |
| O | 0.91760180928079 | 1.86782147842688 | 12.58746172684390 |
| O | 5.55849374860514 | 7.61798992983047 | 12.58921613585926 |
| O | 5.36922701364575 | 0.22064528939018 | 12.72003344769997 |
| O | 0.24512857080875 | 7.27570750268266 | 12.64747031950218 |
| O | 10.17789581597291 | 10.74289465652358 | 12.69115373819501 |

[illegible]

| | | | | | | | |
|---|---|---|---|---|---|---|---|
| O | 0.25734999999999 | 8.20553000000001 | 9.78758999999990 | O | 8.23978000000001 | 4.36766000000002 | 2.37968999999996 |
| O | 6.67756999999997 | 10.56639999999996 | 9.82734000000006 | O | 1.05523999999999 | 7.20221000000000 | 2.42764000000011 |
| O | 9.35747000000002 | 10.86630999999996 | 9.95108000000003 | O | 6.96701000000000 | 1.98259000000004 | 2.49582000000009 |
| O | 5.93269000000002 | 6.91789000000004 | 9.99594000000003 | O | 2.94267000000003 | 9.03534000000004 | 2.65423000000010 |
| O | 9.20339000000000 | 0.22140000000002 | 10.04827999999997 | O | 5.57684999999999 | 11.01483000000002 | 2.79376000000002 |
| O | 4.65431999999998 | 12.45006999999997 | 10.15129000000003 | O | 5.55013999999998 | 4.28570000000005 | 2.83242999999993 |
| O | 4.77644999999996 | 4.38800999999999 | 10.27451999999994 | O | 3.63872000000002 | 6.39621999999996 | 2.88391000000002 |
| O | 1.23121999999997 | 3.17125999999997 | 10.40270000000010 | O | 0.55106000000004 | 2.29398999999997 | 3.02639999999997 |
| O | 1.22909999999998 | 12.81664000000003 | 10.76713999999994 | O | 10.16689999999996 | 9.96554999999999 | 3.28935000000000 |
| O | 4.13856999999997 | 9.06001000000002 | 10.78539000000000 | O | 2.92355999999999 | 3.04286999999996 | 3.38899999999990 |
| O | 12.00916000000002 | 10.72987999999997 | 10.81921000000010 | O | 10.43409999999997 | 1.98437000000002 | 3.43561000000000 |
| O | 10.42087000000003 | 8.80290999999998 | 10.93948000000000 | O | 6.09823999999998 | 7.13813000000000 | 3.49108000000000 |
| O | 3.59532999999998 | 6.56214999999997 | 10.94387000000006 | O | 0.48070999999996 | 11.99164999999995 | 4.02326999999990 |
| O | 8.69596000000002 | 2.57416000000001 | 11.09873999999990 | O | 9.16071000000002 | 6.09578000000000 | 4.18111999999997 |
| O | 10.68622999999996 | 6.10892000000003 | 11.26061000000003 | O | 0.97666999999999 | 4.54704999999998 | 4.38852999999994 |
| O | 11.16631000000004 | 1.47555999999997 | 11.27848999999990 | O | 3.08859000000003 | 0.48030000000000 | 4.59870000000000 |
| O | 1.69607000000000 | 9.81736000000004 | 11.56142999999997 | O | 7.12127999999999 | 3.54350999999997 | 4.68117000000006 |
| O | 7.83750999999998 | 5.87395000000004 | 11.56350999999994 | O | 12.16213999999997 | 9.04504000000001 | 4.71644999999990 |
| O | 7.58989999999996 | 12.19780000000004 | 11.79330000000003 | O | 2.01498000000001 | 7.68645999999997 | 4.73946000000000 |
| O | 7.96866000000004 | 9.60059000000000 | 11.88239999999997 | O | 6.00409999999997 | 9.36258000000003 | 4.82024000000000 |
| O | 6.05742999999996 | 2.87874000000000 | 11.92653999999994 | O | 7.69723999999998 | 11.39130999999999 | 4.91129999999997 |
| O | 5.46529499391609 | 10.82759236474735 | 12.17282034026016 | O | 4.79955000000004 | 2.37986000000001 | 5.01078000000006 |
| O | 3.09496142541488 | 11.87089396661192 | 12.17653963192333 | O | 10.79898000000004 | 4.10323000000004 | 5.01979000000006 |
| O | 3.15060774337583 | 3.56766265650240 | 12.27072905158819 | O | 11.97162999999998 | 6.52033999999996 | 5.16815999999994 |
| O | 0.40851916363474 | 4.61296534079271 | 12.45842408878730 | O | 5.27730999999998 | 11.86749000000002 | 5.19491999999990 |
| O | 0.91563899901873 | 1.86982416504735 | 12.58882525430294 | O | 9.61395000000002 | 8.45341999999998 | 5.37118000000010 |
| O | 5.56217006417843 | 7.59903661053143 | 12.59892078882672 | O | 2.78230000000004 | 11.22875999999996 | 5.39252000000010 |
| O | 5.36166474839464 | 0.23516930228289 | 12.71988706795331 | O | 12.12775999999999 | 2.03320999999998 | 5.54722999999990 |
| O | 0.23733812483827 | 7.27945971753025 | 12.65870381290960 | O | 7.36934000000001 | 7.19216000000004 | 5.73076999999990 |
| O | 10.17458256530220 | 10.76220782813628 | 12.68242465109165 | O | 5.07310000000001 | 4.93269999999997 | 5.77144999999997 |
| O | 5.44206706013680 | 5.12876047307669 | 12.79473882409754 | O | 4.69389000000001 | 7.40688999999997 | 5.89178999999990 |
| O | 12.10210536612318 | 12.48471743114566 | 12.88309824139789 | O | 10.51296999999998 | 10.77216000000004 | 5.91766000000006 |
| O | 2.76118724420873 | 8.07364626593943 | 12.96300633412944 | O | 2.24134999999998 | 2.44178999999999 | 5.95153000000003 |
| O | 7.76481293805846 | 1.34304858569134 | 13.18111899022710 | O | 9.68518000000001 | 1.77421999999996 | 6.10530999999990 |
| O | 8.35342309563725 | 3.88616551028224 | 13.27867819969946 | O | 0.99593000000004 | 0.04055999999999 | 6.30419000000006 |
| O | 0.15949943846631 | 9.76663507153991 | 13.41754359841888 | O | 6.62583000000004 | 0.44990999999996 | 6.32625000000006 |
| O | 10.51850795766675 | 5.60983030477693 | 13.70692008581091 | O | 2.60175000000003 | 5.71772000000001 | 6.42522000000010 |
| O | 7.69789441405324 | 6.31150876109935 | 13.98898321917818 | O | 0.52427000000003 | 10.85443999999999 | 6.61728999999990 |
| O | 9.01834471795466 | 4.72447673825990 | 15.72589841003741 | O | 9.59361000000001 | 6.33155999999997 | 6.88212999999997 |
| Pt | 7.24191830632468 | 12.14874922277778 | 15.17436475942829 | O | 0.24088999999996 | 4.58203999999997 | 6.89382999999997 |
| Pt | 8.36926576754383 | 2.99417424853042 | 16.36347943991472 | O | 0.63894000000003 | 7.98388000000003 | 7.03464000000010 |
| Pt | 6.75371851734712 | 1.20216800930863 | 16.50408632612902 | O | 6.09523999999999 | 2.98598000000003 | 7.14604000000010 |
| Pt | 9.25751418358002 | 0.59630712204170 | 16.51318957565399 | O | 8.85395999999999 | 3.87869000000004 | 7.14824999999997 |
| Si | 0.33529000000003 | 7.71048000000000 | 1.04689000000008 | O | 6.56018999999997 | 11.02720000000000 | 7.22127999999997 |
| Si | 11.95804000000004 | 4.88342999999995 | 1.29616000000010 | O | 4.04473999999998 | 12.85904000000001 | 7.33218000000010 |
| Si | 5.26203999999997 | 10.27900999999999 | 1.31000999999992 | O | 8.76444999999999 | 12.06824999999998 | 7.62043999999990 |

[illegible]

O 10.79898000000004 4.10323000000004 5.01979000000006
O 11.97162999999998 6.52033999999996 5.16815999999994
O 5.27730999999998 11.86749000000002 5.19491999999990
O 9.61395000000002 8.45241999999998 5.37118000000010
O 2.78230000000004 11.22875999999996 5.39252000000010
O 12.12775999999999 2.03320999999998 5.54722999999990
O 7.36934000000001 7.19216000000004 5.73076999999990
O 5.07310000000001 4.93269999999997 5.77144999999997
O 4.69389000000001 7.40688999999997 5.89178999999990
O 10.51296999999998 10.77216000000004 5.91766000000006
O 2.24134999999998 2.44178999999999 5.95153000000003
O 9.68518000000001 1.77421999999996 6.10530999999990
O 0.99593000000004 0.04055999999999 6.30419000000006
O 6.62583000000004 0.44990999999996 6.32625000000006
O 2.60175000000003 5.71772000000001 6.42522000000010
O 0.52427000000003 10.85443999999999 6.61728999999990
O 9.59361000000001 6.33155999999997 6.88212999999997
O 0.24088999999996 4.58203999999997 6.89382999999997
O 0.63894000000003 7.98388000000003 7.03464000000010
O 6.09523999999999 2.98598000000003 7.14604000000010
O 8.85395999999999 3.87869000000004 7.14824999999997
O 6.56018999999997 11.02720000000000 7.22127999999997
O 4.04473999999998 12.85904000000001 7.33218000000010
O 8.76444999999999 12.06824999999998 7.62043999999990
O 7.29605000000001 5.87779999999998 8.06985000000010
O 0.72450999999998 1.93019000000004 8.08120000000010
O 6.89742999999996 8.56527999999996 8.09660000000010
O 11.16503000000000 12.17027999999997 8.21368000000006
O 4.36473999999996 6.02265999999997 8.21442000000003
O 8.00193000000000 1.62576999999997 8.21556000000010
O 2.52144999999999 3.83120999999999 8.25784999999997
O 4.60298000000000 9.87453999999997 8.29783999999994
O 11.23293999999997 9.65737000000002 8.33763000000000
O 2.62367000000001 8.34195000000002 8.68505000000003
O 5.53447000000000 0.91970000000001 8.72556000000010
O 1.28774000000001 6.06899000000004 8.77017000000000
O 10.31381000000000 2.32343999999997 8.86801999999987
O 3.02209000000000 1.46592000000000 9.34606000000006
O 9.41126000000003 5.06335000000000 9.39370000000010
O 2.55275999999996 10.91413000000001 9.53045999999994
O 8.59502000000002 7.45155999999996 9.76356000000010
O 0.25734999999999 8.20553000000001 9.78758999999990
O 6.67756999999997 10.56639999999996 9.82734000000006
O 9.35747000000002 10.86630999999996 9.95108000000003
O 5.93269000000002 6.91789000000004 9.99594000000003
O 9.20339000000000 0.22140000000002 10.04827999999997
O 4.65431999999998 12.45006999999997 10.15129000000003
O 4.77644999999996 4.38802999999999 10.27451999999994
O 1.23121999999997 3.17125999999997 10.40270000000010
O 1.22909999999998 12.81664000000003 10.76713999999994
O 4.13856999999997 9.06001000000002 10.78539000000000
O 12.00916000000002 10.72987999999997 10.81921000000010
O 10.42087000000003 8.80290999999998 10.93948000000000
O 3.59532999999998 6.56214999999997 10.94387000000006
O 8.69596000000002 2.57416000000001 11.09873999999990
O 10.68622999999996 6.10892000000003 11.26061000000003
O 11.16631000000004 1.47555999999997 11.27848999999990
O 1.69607000000000 9.81736000000004 11.56142999999997
O 7.83750999999998 5.87395000000004 11.56350999999994
O 7.58889999999996 12.19780000000004 11.79330000000003
O 7.96866000000004 9.60059000000002 11.88239999999997
O 6.05742999999996 2.87874000000000 11.92653999999994
O 5.46659010802339 10.82712678443547 12.17655109652128
O 3.09534268545151 11.86988912013077 12.17626643952531
O 3.15057685768304 3.56772871185518 12.27064602637075
O 0.40831554635097 4.61213161980390 12.45819755865386
O 0.91452064842207 1.87036079114096 12.59008244187430
O 5.56295716416495 7.59719925262170 12.59973993683821
O 5.36446962386744 0.23999400683905 12.70930294594269
O 0.23733341031688 7.27837157121326 12.65928503886902
O 10.17385614693824 10.76112374045714 12.68262507818576
O 5.44222171973992 5.12709134743016 12.79643558233078
O 12.10214377853163 12.48485806573047 12.88125357405123
O 2.76105459381263 8.07294622240875 12.96249970685357
O 7.77150901704403 1.33929492570026 13.17560740168189
O 8.35321316674179 3.88789273908509 13.26858417011792
O 0.15844764144220 9.76499417206048 13.41641995638112
O 10.51916332643496 5.60523820292268 13.70754976607690
O 7.69842180933300 6.30575669187515 13.99347707435731
O 9.02314749986621 4.70233349660820 15.72510893723869
Pt 7.17144468155320 12.11090015440430 15.16269001981117
Pt 8.41487574678361 2.93406514053779 16.25036563073446
Pt 6.66603654203981 1.19697564876995 16.27685713181472

Si 2.0220 11.3657 11.0122
Si 9.4578 10.0602 11.3873
Si 6.8925 10.7508 11.4133
Si 4.6298 12.2269 11.7888
Si 4.7766 3.9328 11.8635
Si 0.0000 0.6410 11.8933
Si 4.0468 7.8495 11.9023
Si 1.4656 3.3548 12.0065
Si 7.6751 2.7159 12.4155
Si 11.7985 10.9001 12.5098
Si 11.7209 5.8791 12.5341
Si 1.2144 8.6537 12.6488
Si 6.6056 6.2868 12.6561
Si 6.9823 12.9682 13.1951
Si 8.9347 5.2381 14.1847
Sn 8.7888 11.6945 17.0560
Sn 8.2589 8.4249 18.4536
Sn 8.2330 9.1514 15.0529
196
III-TS3
C 5.11599999999990 10.20647166666663 19.24522166666611
C 4.97177499999991 9.13434333333310 18.39452333333293
H 6.58656000000001 12.90918999999996 0.00000000000000
H 9.03195000000001 2.06583000000001 0.63119000000006
H 4.76465333333342 11.20255999999986 18.97423833333347
H 5.51170166666661 10.13531666666688 20.26828666666688
H 5.44451833333345 10.63893499999988 16.22055499999943
H 5.03679166666657 8.09909166666684 18.73342666666691
O 12.34642999999997 6.23675000000002 0.32400000000007
O 6.32035000000003 0.68143000000001 0.34265000000005
O 6.64696000000003 10.90101000000003 0.43733999999995
O 1.43306000000000 8.88837000000004 0.50018000000000
O 3.86302999999998 10.46476000000001 0.65411999999992
O 9.40855999999997 10.33665000000001 0.75816000000009
O 8.97135000000001 1.15729000000001 0.95693000000006
O 11.41461000000000 3.24415000000001 1.12213999999995
O 6.05808999999996 6.39254000000003 1.24786999999992
O 11.35215999999998 8.50616000000004 1.29236999999989
O 5.57675999999996 8.67061000000003 1.29730999999993
O 11.74038999999997 11.12449999999999 1.65147999999999
O 11.94920000000000 0.43491999999997 1.76809000000003
O 1.79838999999998 11.37620000000001 1.80135999999993
O 1.06378000000003 4.55670000000004 1.82871000000000
O 8.39548000000004 7.04101000000000 1.84762000000001
O 8.12957999999999 12.28788000000000 2.10634000000004
O 10.66671999999996 5.50028000000003 2.11340999999993
O 7.64250999999997 9.37059999999996 2.35450999999989
O 8.23978000000001 4.36766000000002 2.37969999999996
O 1.05523999999999 7.20221000000000 2.42764000000011
O 6.96701000000000 1.98259000000004 2.49582000000009
O 2.94267000000003 9.03534000000004 2.65423000000010
O 5.57684999999999 11.01483000000002 2.79376000000002
O 5.55013999999998 4.38570000000005 2.83242999999993
O 3.63872000000002 6.39621999999996 2.88391000000002
O 0.55106000000004 2.29389999999997 3.02639999999997
O 10.16689999999996 9.96554999999999 3.28935000000000
O 2.92355999999999 3.04286999999996 3.38899999999990
O 10.43409999999997 1.98437000000002 3.43561000000000
O 6.09823999999996 7.13813000000000 3.49108000000000
O 0.48070999999996 11.99164999999995 4.02326999999990
O 9.16071000000002 6.09578000000000 4.18111999999997
O 0.97666999999999 4.54704999999998 4.38852999999994
O 3.08859000000003 0.48030000000000 4.59870000000000
O 7.12127999999999 3.54350999999997 4.68117000000006
O 12.16213999999997 9.04504000000001 4.71644999999990
O 2.01498000000001 7.68645999999997 4.73946000000000
O 6.00409999999997 9.36258000000003 4.82024000000000
O 7.69723999999996 11.39130999999999 4.91129999999997
O 4.79955000000004 2.37986000000001 5.01078000000006
O 10.79898000000004 4.10323000000004 5.01979000000006
O 11.97162999999998 6.52033999999996 5.16815999999994
O 5.27730999999998 11.86749000000002 5.19491999999990
O 9.61395000000002 8.45241999999998 5.37118000000010
O 2.78230000000004 11.22875999999996 5.39252000000010
O 12.12775999999999 2.03320999999998 5.54722999999990
O 7.36934000000001 7.19216000000004 5.73076999999990
O 5.07310000000001 4.93269999999997 5.77144999999997
O 4.69389000000001 7.40688999999997 5.89178999999990
O 10.51296999999998 10.77216000000004 5.91766000000006
O 2.24134999999998 2.44178999999999 5.95153000000003
O 9.68518000000001 1.77421999999996 6.10530999999990
O 0.99593000000004 0.04055999999999 6.30419000000006
O 6.62583000000004 0.44990999999996 6.32625000000006

O 12.34642999999997 6.23675000000002 0.32400000000007
O 6.32035000000003 0.68143000000001 0.34265000000005
O 6.64696000000003 10.90101000000003 0.43733999999995
O 1.43306000000000 8.88837000000004 0.50018000000000
O 3.86302999999998 10.46476000000001 0.65411999999992
O 9.40855999999997 10.33665000000001 0.75816000000009
O 8.97135000000001 1.15729000000001 0.95693000000006
O 11.41461000000000 3.24415000000001 1.12213999999995
O 6.05808999999996 6.39254000000003 1.24786999999992
O 11.35215999999998 8.50616000000004 1.29236999999989
O 5.57675999999996 8.67061000000003 1.29730999999993
O 11.74038999999997 11.12449999999999 1.65147999999999
O 11.94920000000000 0.43491999999997 1.76809000000003
O 1.79838999999998 11.37620000000001 1.80135999999993
O 1.06378000000003 4.55670000000004 1.82871000000000
O 8.39548000000004 7.04101000000000 1.84762000000001
O 8.12957999999999 12.28788000000000 2.10634000000004
O 10.66671999999996 5.50028000000003 2.11340999999993
O 7.64250999999997 9.37059999999996 2.35450999999989
O 8.23978000000001 4.36766000000002 2.37969999999996
O 1.05523999999999 7.20221000000000 2.42764000000011
O 6.96701000000000 1.98259000000004 2.49582000000009
O 2.94267000000003 9.03534000000004 2.65423000000010
O 5.57684999999999 11.01483000000002 2.79376000000002
O 5.55013999999998 4.38570000000005 2.83242999999993
O 3.63872000000002 6.39621999999996 2.88391000000002
O 0.55106000000004 2.29398999999997 3.02639999999997
O 10.16689999999996 9.96554999999999 3.28935000000000
O 2.92355999999999 3.04286999999996 3.38899999999990
O 10.43409999999997 1.98437000000002 3.43561000000000
O 6.09823999999998 7.13813000000000 3.49108000000000
O 0.48070999999996 11.99164999999995 4.02326999999990
O 9.16071000000002 6.09578000000000 4.18111999999997
O 0.97666999999999 4.54704999999998 4.38852999999994
O 3.08859000000003 0.48030000000000 4.59870000000000
O 7.12127999999999 3.54350999999997 4.68117000000006
O 12.16213999999997 9.04504000000001 4.71644999999990
O 2.01498000000001 7.68645999999997 4.73946000000000
O 6.00409999999997 9.36258000000003 4.82024000000000
O 7.69723999999998 11.39130999999999 4.91129999999997
O 4.79955000000004 2.37986000000001 5.01078000000006
O 10.79898000000004 4.10323000000004 5.01979000000006
O 11.97162999999998 6.52033999999996 5.16815999999994
O 5.27730999999998 11.86749000000002 5.19491999999990
O 9.61395000000002 8.45241999999998 5.37118000000010
O 2.78230000000004 11.22875999999996 5.39252000000010
O 12.12775999999999 2.03320999999998 5.54722999999990
O 7.36934000000001 7.19216000000004 5.73076999999990
O 5.07310000000001 4.93269999999997 5.77144999999997
O 4.69389000000001 7.40688999999997 5.89178999999990
O 10.51296999999998 10.77216000000004 5.91766000000006
O 2.24134999999998 2.44178999999999 5.95153000000003
O 9.68518000000001 1.77421999999996 6.10530999999990
O 0.99593000000004 0.04055999999999 6.30419000000006
O 6.62583000000004 0.44990999999996 6.32625000000006
O 2.60175000000003 5.71772000000001 6.42522000000010
O 0.52427000000003 10.85443999999999 6.61728999999990
O 9.59361000000001 6.33155999999997 6.88212999999997
O 0.24088999999996 4.58203999999997 6.89382999999997
O 0.63894000000003 7.98388000000003 7.03464000000010
O 6.09523999999999 2.98598000000003 7.14604000000010
O 8.85395999999999 3.87869000000004 7.14824999999997
O 6.56018999999997 11.02720000000000 7.22127999999997
O 4.04473999999998 12.85904000000001 7.33218000000010
O 8.76444999999999 12.06824999999998 7.62043999999990
O 7.29605000000001 5.87779999999998 8.06985000000010
O 0.72450999999998 1.93019000000004 8.08120000000010
O 6.89742999999996 8.56527999999996 8.09660000000010
O 11.16503000000000 12.17027999999997 8.21368000000006
O 4.36473999999996 6.02265999999997 8.21442000000003
O 8.00193000000000 1.62576999999997 8.21556000000010
O 2.52144999999999 3.83120999999999 8.25784999999997
O 4.60298000000000 9.87453999999997 8.29783999999994
O 11.23293999999997 9.65737000000002 8.33763000000000
O 2.62367000000001 8.34195000000002 8.68505000000003
O 5.53447000000000 0.91970000000001 8.72556000000010
O 1.28774000000001 6.06899000000004 8.77017000000000
O 10.31381000000000 2.23343999999997 8.86801999999987
O 3.02209000000000 1.46592000000000 9.34606000000006
O 9.41126000000003 5.06335000000000 9.39370000000010
O 2.55275999999996 10.91413000000001 9.53045999999994
O 8.59502000000002 7.45155999999996 9.76356000000010
O 0.25734999999999 8.20553000000001 9.78758999999990
O 6.67756999999997 10.56639999999996 9.82734000000006
O 9.35747000000002 10.86630999999996 9.95108000000003
O 5.93269000000002 6.91789000000004 9.99594000000003
O 9.20339000000000 0.22140000000002 10.04827999999997
O 4.65431999999998 12.45006999999997 10.15129000000003
O 4.77644999999996 4.38800999999999 10.27451999999994
O 1.23121999999997 3.17125999999997 10.40270000000010
O 1.22909999999998 12.81664000000003 10.76713999999994
O 4.13856999999997 9.06001000000002 10.78539000000000
O 12.00916000000002 10.72987999999997 10.81921000000010
O 10.42087000000003 8.80290999999998 10.93948000000000
O 3.59532999999998 6.56214999999997 10.94387000000006
O 8.69596000000002 2.57416000000001 11.09873999999990
O 10.68622999999996 6.10892000000003 11.26061000000003
O 11.16631000000004 1.47555999999997 11.27848999999990
O 1.69607000000000 9.81736000000004 11.56142999999997
O 7.83750999999998 5.87395000000004 11.56350999999994
O 7.58989999999996 12.19780000000004 11.79330000000003
O 7.96866000000004 9.60059000000000 11.88239999999997
O 6.05742999999996 2.87874000000000 11.92653999999994
O 5.46768166666661 10.82957166666682 12.17836999999962
O 3.09435499999991 11.86772166666688 12.17590500000029
O 3.14949166666659 3.56867666666671 12.27092166666694
O 0.40602166666650 4.61066333333346 12.45900500000061
O 0.91541000000008 1.86936166666671 12.59096333333331
O 5.56104999999986 7.59577499999983 12.59934166666653
O 5.35832000000008 0.25481666666678 12.69004166666694
O 0.23439333333327 7.27743666666648 12.66440666666675
O 10.17376500000010 10.76201999999995 12.68267999999990
O 5.43932166666691 5.12762333333351 12.79946833333360
O 12.10184666666678 12.48515166666676 12.88330333333306
O 2.75783499999990 8.06993500000019 12.96126000000003
O 7.77518333333347 1.33237333333350 13.16564666666726
O 8.34181999999982 3.87323166666678 13.28597166666691
O 0.15811833333313 9.76461500000008 13.41664166666691
O 10.51045333333341 5.60255166666648 13.69929833333389
O 7.69604666666659 6.30008499999989 13.99355333333325
O 9.04142166666691 4.72704333333308 15.73146166666698
Pt 7.09138499999976 12.20783666666685 15.21293333333302
Pt 8.72395166666662 2.92146166666659 16.34813333333318
Pt 6.83583333333318 1.36688166666645 16.23966166666627
Pt 9.14931666666680 0.39759833333350 16.61250333333306
Si 0.33529000000003 7.71048000000000 1.04689000000008
Si 11.95804000000004 4.88342999999995 1.29616000000010
Si 5.36203999999997 10.27900999999999 1.31000999999992
Si 2.45532000000003 9.93546000000002 1.35284000000002
Si 7.97589000000000 10.75088000000000 1.45379000000003
Si 7.46856999999998 0.73103000000003 1.50179000000003
Si 10.67104000000004 9.93228999999998 1.72913999999992
Si 6.89163999999996 7.82098000000001 1.90039999999999
Si 11.59966999999997 1.97594000000002 2.23462999999992
Si 0.41182999999997 12.06493000000001 2.35560000000010
Si 5.19710000000004 5.95740000000003 2.53515999999990
Si 9.11035999999999 5.76089000000002 2.59096999999997
Si 6.98767999999998 3.52843000000003 3.03292000000010
Si 1.42896000000001 3.64158999999998 3.13314999999989
Si 2.46711000000000 7.55467000000003 3.17378000000008
Si 6.19584000000002 10.88913999999998 4.31980000000000
Si 3.24937000000000 2.09392999999998 4.69687000000000
Si 10.60485000000002 9.57516000000005 4.79298999999990
Si 10.68432999999996 2.45871999999998 4.99262000000000
Si 6.04216999999995 7.79614000000002 5.06574000000000
Si 12.17232000000003 4.92029000000004 5.36653000000000
Si 0.52618000000003 7.80403999999996 5.42080000000010
Si 1.24491000000002 11.77372000000004 5.47459000000003
Si 8.97683999999999 6.98952999999997 5.51034000000003
Si 5.79442000000005 3.48931000000001 5.63946000000010
Si 3.88077999999997 12.41462000000000 5.72508999999990
Si 7.59978000000002 12.00861999999996 6.46257999999990
Si 0.91102000000001 1.63241999999999 6.49053000000003
Si 4.19754000000004 6.01659000000003 6.60348999999997
Si 11.46811000000002 10.87768999999998 7.25600000000010
Si 1.66166000000000 5.02239999999999 7.57692999999994
Si 9.28667000000001 2.38490999999995 7.58583999999997
Si 6.52179000000001 1.48441000000003 7.58890999999994
Si 8.81880999999998 5.29454000000002 7.88977999999997
Si 6.20759000000003 10.01218000000002 8.39279000000010
Si 1.16603000000001 7.67142999999996 8.54078000000003
Si 4.03345999999999 0.30045000000004 8.83076000000006
Si 9.68524000000000 12.14075000000000 8.96686000000000
Si 7.16647999999999 7.23733000000004 8.97783000000003
Si 1.87645999999999 2.56144000000005 9.05034999999997
Si 3.45236000000000 9.54198999999997 9.37580000000003
Si 4.61912000000000 5.94270000000003 9.80684999999994
Si 11.57529000000003 9.37814000000003 9.91657000000010
Si 9.84757999999999 1.68549000000002 10.33173000000010
Si 9.17938999999999 6.14520000000004 10.61420999999997
Si 2.02201000000000 11.36570000000004 11.01221000000010
Si 9.45776000000002 10.06017999999996 11.38731000000006
Si 6.89244999999997 10.75082000000000 11.41328999999997
Si 4.62983000000000 12.22691000000004 11.78878000000010
Si 4.77658000000003 3.93277999999998 11.86347000000000
Si 0.00000000000000 0.64099999999999 11.89328000000000
Si 4.04675000000000 7.84950000000003 11.90235000000006
Si 1.45879166666670 3.35983833333323 12.00739833333363
Si 7.66115166666688 2.72741666666670 12.35445666666646
Si 11.79085666666678 10.90110999999987 12.50806833333354
Si 11.70763833333356 5.88739333333330 12.54853333333312
Si 1.21029999999987 8.65571499999995 12.65204999999968
Si 6.63685166666651 6.27873499999988 12.71066833333315
Si 6.95129000000006 13.04737833333346 13.12106166666627
Si 8.93344333333322 5.14299666666679 14.15124833333334
193
III-P4
H 6.58660000 12.90920000 0.00000000
O 12.34640000 6.23680000 0.32420000
O 6.32040000 0.68140000 0.34270000
O 6.64700000 10.90100000 0.43740000
O 1.43310000 8.88830000 0.50020000
H 9.03200000 2.06580000 0.63100000
O 3.86310000 10.46470000 0.65410000
O 9.40850000 10.33660000 0.75810000
O 8.97140000 1.15730000 0.95680000
Si 0.33530000 7.71050000 1.04700000
O 11.41450000 3.24420000 1.12220000
O 6.05810000 6.39260000 1.24800000
O 11.35210000 8.50620000 1.29250000
Si 11.95800000 4.88350000 1.29630000
O 5.57680000 8.67070000 1.29730000
Si 5.36210000 10.27900000 1.31010000
Si 2.45530000 9.93550000 1.35300000
Si 7.97590000 10.75090000 1.45380000
Si 7.46850000 0.73110000 1.50180000
O 11.74040000 11.12450000 1.65150000
Si 10.67110000 9.93230000 1.72930000
O 11.94920000 0.43500000 1.76800000
O 1.79840000 11.37620000 1.80130000
O 1.06370000 4.55670000 1.82880000
O 8.39550000 7.04100000 1.84770000
Si 6.89170000 7.82100000 1.90050000
O 8.12960000 12.28790000 2.10620000
O 10.66670000 5.50030000 2.11330000
Si 11.59970000 1.97590000 2.23460000
O 7.64250000 9.37070000 2.35460000
Si 0.41180000 12.06500000 2.35550000
O 8.23970000 4.36770000 2.37980000
O 1.05530000 7.20220000 2.42750000
O 6.96700000 1.98260000 2.49570000
Si 5.19710000 5.95730000 2.53500000
Si 9.11040000 5.76090000 2.59100000
O 2.94270000 9.03540000 2.65410000
O 5.57680000 11.01480000 2.79390000
O 5.55020000 4.38570000 2.83230000
O 3.63870000 6.39630000 2.88380000
O 0.55100000 2.29390000 3.02660000
Si 6.98770000 3.52850000 3.03300000
Si 1.42900000 3.64160000 3.13310000
Si 2.46710000 7.55470000 3.17380000
O 10.16690000 9.96550000 3.28930000
O 2.92350000 3.04290000 3.38910000
O 10.43410000 1.98430000 3.43550000
O 6.09820000 7.13810000 3.49120000
O 0.48080000 11.99170000 4.02340000
O 9.16080000 6.09580000 4.18110000
Si 6.19580000 10.88910000 4.31970000
O 0.97670000 4.54710000 4.38850000
O 3.08860000 0.48030000 4.59870000
O 7.12130000 3.54350000 4.68130000
Si 3.24940000 2.09400000 4.69700000
O 12.16210000 9.04510000 4.71650000
O 2.01500000 7.68650000 4.73950000
Si 10.60480000 9.57510000 4.79300000
O 6.00410000 9.36250000 4.82020000
O 7.69720000 11.39130000 4.91140000
Si 10.68440000 2.45870000 4.99260000
O 4.79950000 2.37980000 5.01090000
O 10.79900000 4.10320000 5.01980000
Si 6.04210000 7.79610000 5.06560000
O 11.97170000 6.52030000 5.16830000
O 5.27730000 11.86750000 5.19490000
Si 12.17230000 4.92030000 5.36640000
O 9.61390000 8.45250000 5.37120000
O 2.78230000 11.22870000 5.39260000
Si 0.52620000 7.80400000 5.42080000
Si 1.24490000 11.77370000 5.47460000
Si 8.97690000 6.98950000 5.51040000
O 12.12780000 2.03330000 5.54720000
Si 5.79440000 3.48920000 5.63940000
Si 3.88070000 12.41460000 5.72510000
O 7.36930000 7.19210000 5.73090000
O 5.07310000 4.93270000 5.77150000
O 4.69400000 7.40690000 5.89180000
O 10.51290000 10.77210000 5.91780000
O 2.24140000 2.44170000 5.95170000
O 9.68520000 1.77430000 6.10530000
O 0.99600000 0.04050000 6.30430000
O 6.62590000 0.44990000 6.32640000
O 2.60180000 5.71780000 6.42530000
Si 7.59980000 12.00860000 6.46270000
Si 0.91110000 1.63240000 6.49060000
Si 4.19750000 6.01660000 6.60350000
O 0.52430000 10.85450000 6.61730000
O 9.59370000 6.33160000 6.88220000
O 0.24090000 4.58200000 6.89380000
O 0.63890000 7.98390000 7.03460000
O 6.09520000 2.98590000 7.14590000
O 8.85390000 3.87870000 7.14820000
O 6.56020000 11.02720000 7.22140000
Si 11.46810000 10.87770000 7.25600000
O 4.04470000 12.85900000 7.33220000
Si 1.66160000 5.02230000 7.57700000
Si 9.28670000 2.38500000 7.58590000
Si 6.52180000 1.48450000 7.58880000
O 8.76440000 12.06820000 7.62050000
Si 8.81890000 5.29460000 7.88990000
O 7.29610000 5.87770000 8.06980000
O 0.72450000 1.93020000 8.08130000
O 6.89740000 8.56520000 8.09660000
O 11.16500000 12.17030000 8.21380000
O 4.36470000 6.02260000 8.21440000
O 8.00190000 1.62580000 8.21570000
O 2.52150000 3.83120000 8.25790000
O 4.60290000 9.87450000 8.29790000
O 11.23290000 9.65730000 8.33760000
Si 6.20760000 10.01220000 8.39260000
Si 1.16610000 7.67140000 8.54080000
O 2.62370000 8.34190000 8.68510000
O 5.53450000 0.91970000 8.72540000
O 1.28780000 6.06890000 8.77020000
Si 4.03340000 0.30040000 8.83070000
O 10.31380000 2.23350000 8.86820000
Si 9.68530000 12.14080000 8.96670000
Si 7.16640000 7.23740000 8.97790000
Si 1.87650000 2.56150000 9.05020000
O 3.02210000 1.46600000 9.34590000
Si 3.45230000 9.54190000 9.37570000
O 9.41130000 5.06340000 9.39360000
O 2.55280000 10.91410000 9.53060000
O 8.59510000 7.45150000 9.76350000
O 0.25740000 8.20550000 9.78750000
Si 4.61910000 5.94270000 9.80670000
O 6.67760000 10.56640000 9.82720000
Si 11.57530000 9.37810000 9.91650000
O 9.35740000 10.86630000 9.95100000
O 5.93270000 6.91790000 9.99580000
O 9.20340000 0.22140000 10.04830000
O 4.65430000 12.45010000 10.15140000
O 4.77650000 4.38800000 10.27460000
Si 9.84750000 1.68550000 10.33180000
O 1.23120000 3.17120000 10.40260000
Si 9.17940000 6.14530000 10.61410000
O 1.22910000 12.81660000 10.76700000
O 4.13860000 9.06000000 10.78530000
O 12.00920000 10.72990000 10.81920000
O 10.42090000 8.80280000 10.93950000
O 3.59530000 6.56210000 10.94400000
Si 2.02200000 11.36580000 11.01220000
O 8.69590000 2.57420000 11.09890000
O 10.68620000 6.10890000 11.26050000
O 11.16630000 1.47550000 11.27840000
Si 9.45780000 10.06020000 11.38720000
Si 6.89240000 10.75090000 11.41340000
O 1.69610000 9.81730000 11.56130000
O 7.83750000 5.87390000 11.56350000
Si 4.62980000 12.22690000 11.78880000
O 7.58990000 12.19780000 11.79330000
Si 4.77660000 3.93270000 11.86340000
O 7.96870000 9.60050000 11.88220000
Si 0.00000000 0.64100000 11.89310000
Si 4.04670000 7.84950000 11.90240000
O 6.05740000 2.87870000 11.92640000
Si 1.45870000 3.35980000 12.00740000
O 3.09400000 11.86750000 12.17600000
O 5.46760000 10.82990000 12.17820000
O 3.14920000 3.56880000 12.27100000
Si 7.66100000 2.72620000 12.35520000

O 9.20339000 0.22140000 10.04828000
O 4.65432000 12.45007000 10.15129000
O 4.77645000 4.38801000 10.27452000
Si 9.84758000 1.68549000 10.33173000
O 1.23122000 3.17126000 10.40270000
Si 9.17939000 6.14520000 10.61421000
O 1.22910000 12.81664000 10.76714000
O 4.13857000 9.06001000 10.78539000
O 12.00916000 10.72988000 10.81921000
O 10.42087000 8.80291000 10.93948000
O 3.59533000 6.56215000 10.94387000
Si 2.02201000 11.36570000 11.01221000
O 8.69596000 2.57416000 11.09874000
O 10.68623000 6.10892000 11.26061000
O 11.16631000 1.47556000 11.27849000
Si 9.45776000 10.06018000 11.38731000
Si 6.89245000 10.75082000 11.41329000
O 1.69607000 9.81736000 11.56143000
O 7.83751000 5.87395000 11.56351000
Si 4.62983000 12.22691000 11.78878000
O 7.58990000 12.19780000 11.79330000
Si 4.77658000 3.93278000 11.86347000
O 7.96866000 9.60059000 11.88240000
Si 0.00000000 0.64100000 11.89328000
Si 4.04675000 7.84950000 11.90235000
O 6.05743000 2.87874000 11.92654000
Si 1.46511738 3.35471310 12.00700928
O 5.45546351 10.80848909 12.16185190
O 3.09400413 11.86474122 12.17471320
O 3.15306041 3.56727499 12.26956471
Si 7.67559544 2.71658056 12.41617713
O 0.42251608 4.61151655 12.45116097
Si 11.79468199 10.89993880 12.51067737
Si 11.72219298 5.88051646 12.53420063
O 0.91756512 1.86805763 12.58730605
O 5.55992743 7.61884338 12.58967016
Si 6.60295372 6.28485662 12.64762960
O 0.24562803 7.27583605 12.64793684
Si 1.21448558 8.65553782 12.65048779
O 10.17695885 10.74311628 12.69183882
O 5.36971200 0.22477868 12.71424944
O 5.41585290 5.14476098 12.78087529
O 12.09471866 12.48365783 12.87859428
O 2.76575167 8.07370652 12.96576284
Si 6.97451010 12.97634679 13.18749331
O 7.73968221 1.31969828 13.26187207
O 0.16062816 9.75999921 13.41761586
O 8.34358254 3.86571026 13.42634089
O 10.52589850 5.58594926 13.66680921
O 7.63993660 6.35669171 13.92197144
Si 8.92772853 5.23834527 14.18666360
O 9.04838153 5.26244451 15.80670558
Pt 6.70505592 11.50680036 15.04662262
Sn 7.85476312 9.02274470 15.13977236
H 5.57194142 10.66364412 16.05265503
Pt 9.22123427 7.18593867 16.45965394
Pt 9.85609283 9.61086867 17.01364767
Sn 8.18192353 11.67535516 17.14526246
Pt 6.22755230 9.84398387 17.42438450
Sn 8.23319668 8.23288901 18.63833553
H 5.00321137 8.14385714 19.06351976
C 5.56663302 9.08460578 19.18396381
C 5.44652260 9.73387535 20.35774823
H 5.99880181 10.64813292 20.59144119
H 4.76943757 9.37423014 21.14047859
196
III-TS4
C 5.09878000000002 10.32815000000012 19.13220333333357
C 4.87201666666656 9.32240666666652 18.21895333333351
H 6.58656000000001 12.90918999999996 0.00000000000000
H 9.03195000000001 2.06583000000001 0.63119000000006
H 4.48855666666666 11.23998999999995 19.07582000000026
H 5.76085333333327 10.21994999999998 19.99502000000006
H 4.00950666666658 9.05363666666664 17.64276333333328
H 6.18483666666684 8.48479000000005 17.22071666666695
O 12.34642999999997 6.23675000000002 0.32400000000007
O 6.32035000000003 0.68143000000001 0.34265000000005
O 6.64696000000003 10.90101000000003 0.43733999999995
O 1.43306000000000 8.88837000000004 0.50018000000000
O 3.86302999999998 10.46476000000001 0.65411999999992
O 9.40855999999997 10.33665000000001 0.75816000000009
O 8.97135000000001 1.15729000000001 0.95693000000006
O 11.41461000000000 3.24415000000001 1.12213999999995
O 6.05808999999996 6.39254000000003 1.24786999999992
O 11.35215999999998 8.50616000000004 1.29236999999989
O 5.57675999999996 8.67061000000003 1.29730999999993
O 11.74038999999997 11.12449999999999 1.65147999999999
O 11.94920000000000 0.43491999999997 1.76809000000003
O 1.79838999999998 11.37620000000001 1.80135999999993
O 1.06378000000003 4.55670000000004 1.82871000000000
O 8.39548000000004 7.04101000000000 1.84762000000001
O 8.12957999999999 12.28788000000000 2.10634000000004
O 10.66671999999996 5.50028000000003 2.11340999999993
O 7.64250999999997 9.37059999999996 2.35450999999989
O 8.23978000000001 4.36766000000002 2.37969999999996
O 1.05523999999999 7.20221000000000 2.42764000000011
O 6.96701000000000 1.98259000000004 2.49582000000009
O 2.94267000000003 9.03534000000004 2.65423000000010
O 5.57684999999999 11.01483000000002 2.79376000000002
O 5.55013999999998 4.38570000000005 2.83242999999993
O 3.63872000000002 6.39621999999996 2.88391000000002
O 0.55106000000004 2.29398999999997 3.02639999999997
O 10.16689999999996 9.96554999999999 3.28935000000000
O 2.92355999999999 3.04286999999996 3.38899999999990
O 10.43409999999997 1.98437000000002 3.43561000000000
O 6.09823999999998 7.13813000000000 3.49108000000000
O 0.48070999999996 11.99164999999995 4.02326999999990
O 9.16071000000002 6.09578000000000 4.18111999999997
O 0.97666999999999 4.54704999999998 4.38852999999994
O 3.08859000000003 0.48030000000000 4.59870000000000
O 7.12127999999999 3.54350999999997 4.68117000000006
O 12.16213999999997 9.04504000000001 4.71644999999990
O 2.01498000000001 7.68645999999997 4.73946000000000
O 6.00409999999997 9.36258000000003 4.82024000000000
O 7.69723999999998 11.39130999999999 4.91129999999997
O 4.79955000000004 2.37986000000001 5.01078000000006
O 10.79898000000004 4.10323000000004 5.01979000000006
O 11.97162999999998 6.52033999999996 5.16815999999994
O 5.27730999999998 11.86749000000002 5.19491999999990
O 9.61395000000002 8.45241999999998 5.37118000000010
O 2.78230000000004 11.22875999999996 5.39252000000010
O 12.12775999999999 2.03320999999998 5.54722999999990
O 7.36934000000001 7.19216000000004 5.73076999999990
O 5.07310000000001 4.93269999999997 5.77144999999997
O 4.69389000000001 7.40688999999997 5.89178999999990
O 10.51296999999998 10.77216000000004 5.91766000000006
O 2.24134999999998 2.44178999999999 5.95153000000003
O 9.68518000000001 1.77421999999996 6.10530999999990
O 0.99593000000004 0.04055999999999 6.30419000000006
O 6.62583000000004 0.44990999999996 6.32625000000006
O 2.60175000000003 5.71772000000001 6.42522000000010
O 0.52427000000003 10.85443999999999 6.61728999999990
O 9.59361000000001 6.33155999999997 6.88212999999997
O 0.24088999999996 4.58203999999997 6.89382999999997
O 0.63894000000003 7.98388000000003 7.03464000000010
O 6.09523999999999 2.98598000000003 7.14604000000010
O 8.85395999999999 3.87869000000004 7.14824999999997
O 6.56018999999997 11.02720000000000 7.22127999999997
O 4.04473999999998 12.85904000000001 7.33218000000010
O 8.76444999999999 12.06824999999998 7.62043999999990
O 7.29605000000001 5.87779999999998 8.06985000000010
O 0.72450999999998 1.93019000000004 8.08120000000010
O 6.89742999999996 8.56527999999996 8.09660000000010
O 11.16503000000000 12.17027999999997 8.21368000000006
O 4.36473999999996 6.02265999999997 8.21442000000003
O 8.00193000000000 1.62576999999997 8.21556000000010
O 2.52144999999999 3.83120999999999 8.25784999999997
O 4.60298000000000 9.87453999999997 8.29783999999994
O 11.23293999999997 9.65737000000002 8.33763000000000
O 2.62367000000001 8.34195000000002 8.68505000000003
O 5.53447000000000 0.91970000000001 8.72556000000010
O 1.28774000000001 6.06899000000004 8.77017000000000
O 10.31381000000000 2.23343999999997 8.86801999999987
O 3.02209000000000 1.46592000000000 9.34606000000006
O 9.41126000000003 5.06335000000000 9.39370000000010
O 2.55275999999996 10.91413000000001 9.53045999999994
O 8.59502000000002 7.45155999999996 9.76356000000010
O 0.25734999999999 8.20553000000001 9.78758999999990
O 6.67756999999997 10.56639999999996 9.82734000000006
O 9.35747000000002 10.86630999999996 9.95108000000003
O 5.93269000000002 6.91789000000004 9.99594000000003
O 9.20339000000000 0.22140000000002 10.04827999999997
O 4.65431999999998 12.45006999999997 10.15129000000003
O 4.77644999999996 4.38800999999999 10.27451999999994
O 1.23121999999997 3.17125999999997 10.40270000000010
O 1.22909999999998 12.81664000000003 10.76713999999994
O 4.13856999999997 9.06001000000002 10.78539000000000
O 12.00916000000002 10.72987999999997 10.81921000000010
O 10.42087000000003 8.80290999999998 10.93948000000000
O 3.59532999999998 6.56214999999997 10.94387000000006
O 8.69596000000002 2.57416000000001 11.09873999999990
O 10.68622999999996 6.10892000000003 11.26061000000003
O 11.16631000000004 1.47555999999997 11.27848999999990
O 1.69607000000000 9.81736000000004 11.56142999999997
O 7.83750999999998 5.87395000000004 11.56350999999994
O 7.58989999999996 12.19780000000004 11.79330000000003
O 7.96866000000004 9.60059000000000 11.88239999999997
O 6.05742999999996 2.87874000000000 11.92653999999994
O 5.45464333333336 10.80667333333332 12.15309333333357
O 3.09276000000011 11.86735666666678 12.17383999999994
O 3.15248666666676 3.56579333333313 12.26954333333334
O 0.42345999999993 4.61197000000002 12.45090000000003
O 0.91746666666663 1.86714666666646 12.58743666666714
O 5.55882666666656 7.61856666666688 12.58926333333283
O 5.37474333333335 0.20354666666651 12.73957333333350
O 0.24447666666659 7.27587666666651 12.64767999999958
O 10.18026999999990 10.74252999999996 12.69224000000010
O 5.41651333333330 5.14363333333332 12.78091333333350
O 12.09566333333340 12.48370999999983 12.87920666666694
O 2.76496000000001 8.07332333333337 12.96638000000029
O 7.73894666666674 1.31889000000011 13.25825000000010
O 8.34196999999991 3.86528666666679 13.42339333333347
O 0.16151333333315 9.75981666666661 13.41844333333325
O 10.52590666666665 5.58583666666657 13.66684000000010
O 7.64125999999996 6.36004999999987 13.93021333333354
O 9.05207000000012 5.24959333333331 15.80666000000019
Pt 7.08125333333338 11.62638999999989 15.15443333333360
Pt 9.35379999999994 7.15183666666657 16.47664666666675
Pt 10.19600333333348 9.51100999999997 17.00543000000016
Pt 6.54248333333347 10.04408333333329 17.31547999999997
Si 0.33529000000003 7.71048000000000 1.04689000000008
Si 11.95804000000004 4.88342999999995 1.29616000000010
Si 5.36203999999997 10.27900999999999 1.31000999999992
Si 2.45532000000003 9.93546000000002 1.35284000000002
Si 7.97589000000000 10.75088000000000 1.45379000000003
Si 7.46856999999998 0.73103000000003 1.50179000000003
Si 10.67104000000004 9.93228999999998 1.72913999999992
Si 6.89163999999996 7.82098000000001 1.90039999999999
Si 11.59966999999997 1.97594000000002 2.23462999999992
Si 0.41182999999997 12.06493000000001 2.35560000000010
Si 5.19710000000004 5.95740000000003 2.53515999999990
Si 9.11035999999999 5.76089000000002 2.59096999999997
Si 6.98767999999998 3.52843000000003 3.03292000000010
Si 1.42896000000001 3.64158999999998 3.13314999999989
Si 2.46711000000000 7.55467000000003 3.17378000000008
Si 6.19584000000002 10.88913999999998 4.31980000000000
Si 3.24937000000000 2.09392999999998 4.69687000000000
Si 10.60485000000002 9.57516000000005 4.79298999999990
Si 10.68432999999996 2.45871999999998 4.99262000000000
Si 6.04216999999995 7.79614000000002 5.06574000000000
Si 12.17232000000003 4.92029000000004 5.36653000000000
Si 0.52618000000003 7.80403999999996 5.42080000000010
Si 1.24491000000002 11.77372000000004 5.47459000000003
Si 8.97683999999999 6.98952999999997 5.51034000000003
Si 5.79442000000005 3.48931000000001 5.63946000000010
Si 3.88077999999997 12.41462000000000 5.72508999999990
Si 7.59978000000002 12.00861999999996 6.46257999999990
Si 0.91102000000001 1.63241999999999 6.49053000000003
Si 4.19754000000004 6.01659000000003 6.60348999999997
Si 11.46811000000002 10.87768999999998 7.25600000000010
Si 1.66166000000000 5.02239999999999 7.57692999999994
Si 9.28667000000001 2.38490999999995 7.58583999999997
Si 6.52179000000001 1.48441000000003 7.58890999999994
Si 8.81880999999998 5.29454000000002 7.88977999999997
Si 6.20759000000003 10.01218000000002 8.39279000000010
Si 1.16603000000001 7.67142999999996 8.54078000000003
Si 4.03345999999999 0.30045000000004 8.83076000000006
Si 9.68524000000000 12.14075000000000 8.96686000000000
Si 7.16649000000000 7.23733000000004 8.97783000000003
Si 1.87645999999999 2.56144000000005 9.05034999999997
Si 3.45236000000000 9.54198999999997 9.37580000000003
Si 4.61912000000000 5.94270000000003 9.80684999999994
Si 11.57529000000003 9.37814000000003 9.91657000000010
Si 9.84757999999999 1.68549000000002 10.33173000000010
Si 9.17938999999999 6.14520000000004 10.61420999999997
Si 2.02201000000000 11.36570000000004 11.01221000000010
Si 9.45776000000002 10.06017999999996 11.38731000000006
Si 6.89244999999997 10.75082000000000 11.41328999999997
Si 4.62983000000000 12.22691000000004 11.78878000000010
Si 4.77658000000003 3.93277999999998 11.86347000000000
Si 0.00000000000000 0.64099999999999 11.89328000000000
Si 4.04675000000000 7.84950000000003 11.90235000000006
Si 1.46536000000002 3.35497666666659 12.00680333333344
Si 7.67599000000000 2.72136333333345 12.41423999999984
Si 11.79966666666652 10.90008666666660 12.50928333333286
Si 11.72137333333341 5.87902000000022 12.53449999999974
Si 1.21473666666655 8.65287666666669 12.64897999999949
Si 6.60527999999994 6.28676666666673 12.65482999999994
Si 6.98539999999988 12.96733666666671 13.19742666666662
Si 8.92425333333328 5.23720666666662 14.18948999999974
Sn 8.79209333333350 11.72673666666675 17.02980333333334
Sn 8.31795333333342 8.35660666666672 18.52492000000007
Sn 8.35452666666666 9.15032999999997 14.91735999999990
196
III-P4
H 6.58656000 12.90919000 0.00000000
O 12.34643000 6.23675000 0.32400000
O 6.32035000 0.68143000 0.34265000
O 6.64696000 10.90101000 0.43734000
O 1.43306000 8.88837000 0.50018000
H 9.03195000 2.06583000 0.63119000
O 3.86303000 10.46476000 0.65412000
O 9.40856000 10.33665000 0.75816000
O 8.97135000 1.15729000 0.95693000
Si 0.33529000 7.71048000 1.04689000
O 11.41461000 3.24415000 1.12214000
O 6.05809000 6.39254000 1.24787000
O 11.35216000 8.50616000 1.29237000
Si 11.95804000 4.88343000 1.29616000
O 5.57676000 8.67061000 1.29731000
Si 5.36204000 10.27901000 1.31001000
Si 2.45532000 9.93546000 1.35284000
Si 7.97589000 10.75088000 1.45379000
Si 7.46857000 0.73103000 1.50179000
O 11.74039000 11.12450000 1.65148000
Si 10.67104000 9.93229000 1.72914000
O 11.94920000 0.43492000 1.76809000
O 1.79839000 11.37620000 1.80136000
O 1.06378000 4.55670000 1.82871000
O 8.39548000 7.04101000 1.84762000
Si 6.89164000 7.82098000 1.90040000
O 8.12958000 12.28788000 2.10634000
O 10.66672000 5.50028000 2.11341000
Si 11.59967000 1.97594000 2.23463000
O 7.64251000 9.37060000 2.35451000
Si 0.41183000 12.06493000 2.35560000
O 8.23978000 4.36766000 2.37970000
O 1.05524000 7.20221000 2.42764000
O 6.96701000 1.98259000 2.49582000
Si 5.19710000 5.95740000 2.53516000
Si 9.11036000 5.76089000 2.59097000
O 2.94267000 9.03534000 2.65423000
O 5.57685000 11.01483000 2.79376000
O 5.55014000 4.38570000 2.83243000
O 3.63872000 6.39622000 2.88391000
O 0.55106000 2.29399000 3.02640000
Si 6.98768000 3.52843000 3.03292000
Si 1.42896000 3.64159000 3.13315000
Si 2.46711000 7.55467000 3.17378000
O 10.16690000 9.96555000 3.28935000
O 2.92356000 3.04287000 3.38900000
O 10.43410000 1.98437000 3.43561000
O 6.09824000 7.13813000 3.49108000
O 0.48071000 11.99165000 4.02327000
O 9.16071000 6.09578000 4.18112000
Si 6.19584000 10.88914000 4.31980000
O 0.97667000 4.54705000 4.38853000
O 3.08859000 0.48030000 4.59870000
O 7.12128000 3.54351000 4.68117000
Si 3.24937000 2.09393000 4.69687000
O 12.16214000 9.04504000 4.71645000
O 2.01498000 7.68646000 4.73946000
Si 10.60485000 9.57516000 4.79299000
O 6.00410000 9.36258000 4.82024000
O 7.69724000 11.39131000 4.91130000
Si 10.68433000 2.45872000 4.99262000
O 4.79955000 2.37986000 5.01078000
O 10.79898000 4.10323000 5.01979000
Si 6.04217000 7.79614000 5.06574000
O 11.97163000 6.52034000 5.16816000
O 5.27731000 11.86749000 5.19492000
Si 12.17232000 4.92029000 5.36653000
O 9.61395000 8.45242000 5.37118000
O 2.78230000 11.22876000 5.39252000
Si 0.52618000 7.80404000 5.42080000
Si 1.24491000 11.77372000 5.47459000
Si 8.97684000 6.98953000 5.51034000
O 12.12776000 2.03321000 5.54723000
Si 5.79442000 3.48931000 5.63946000
Si 3.88078000 12.41462000 5.72509000
O 7.36934000 7.19216000 5.73077000
O 5.07310000 4.93270000 5.77145000
O 4.69389000 7.40689000 5.89179000
O 10.51297000 10.77216000 5.91766000
O 2.24135000 2.44179000 5.95153000
O 9.68518000 1.77422000 6.10531000
O 0.99593000 0.04056000 6.30419000
O 6.62583000 0.44991000 6.32625000
O 2.60175000 5.71772000 6.42522000

| | | | |
|---|---|---|---|
| O | 0.40580000 | 4.61050000 | 12.45920000 |
| Si | 11.79090000 | 10.90090000 | 12.50780000 |
| Si | 11.70720000 | 5.88720000 | 12.54850000 |
| O | 0.91540000 | 1.86920000 | 12.59140000 |
| O | 5.56060000 | 7.59550000 | 12.59940000 |
| Si | 1.21010000 | 8.65560000 | 12.65180000 |
| O | 0.23390000 | 7.27750000 | 12.66530000 |
| O | 10.17390000 | 10.76230000 | 12.68260000 |
| O | 5.35740000 | 0.25720000 | 12.68700000 |
| Si | 6.63670000 | 6.27880000 | 12.71070000 |
| O | 5.43870000 | 5.12780000 | 12.80000000 |
| O | 12.10180000 | 12.48510000 | 12.88380000 |
| O | 2.75720000 | 8.06940000 | 12.96100000 |
| Si | 6.94860000 | 13.04650000 | 13.12190000 |
| O | 7.77500000 | 1.33170000 | 13.16580000 |
| O | 8.34010000 | 3.86930000 | 13.29150000 |
| O | 0.15830000 | 9.76470000 | 13.41700000 |
| O | 10.50900000 | 5.60240000 | 13.69700000 |
| O | 7.69540000 | 6.29920000 | 13.99330000 |
| Si | 8.93300000 | 5.14190000 | 14.15140000 |
| Pt | 7.08690000 | 12.22650000 | 15.22210000 |
| H | 11.92630000 | 0.03580000 | 15.48380000 |
| O | 9.05040000 | 4.73370000 | 15.73180000 |
| Pt | 6.85200000 | 1.38960000 | 16.24100000 |
| H | 11.06850000 | 11.71400000 | 16.28740000 |
| C | 11.28310000 | 12.78870000 | 16.29820000 |
| Pt | 8.74760000 | 2.94220000 | 16.39870000 |
| Pt | 9.15700000 | 0.39980000 | 16.63040000 |
| C | 10.95990000 | 0.46400000 | 17.38880000 |
| H | 8.50240000 | 12.61650000 | 17.74750000 |
| H | 11.42010000 | 1.21680000 | 18.02590000 |
| Si | 7.59978000 | 12.00862000 | 6.46258000 |
| Si | 0.91102000 | 1.63242000 | 6.49053000 |
| Si | 4.19754000 | 6.01659000 | 6.60349000 |
| O | 0.52427000 | 10.85444000 | 6.61729000 |
| O | 9.59361000 | 6.33156000 | 6.88213000 |
| O | 0.24089000 | 4.58204000 | 6.89383000 |
| O | 0.63894000 | 7.98388000 | 7.03464000 |
| O | 6.09524000 | 2.98598000 | 7.14604000 |
| O | 8.85396000 | 3.87869000 | 7.14825000 |
| O | 6.56019000 | 11.02720000 | 7.22128000 |
| Si | 11.46811000 | 10.87769000 | 7.25600000 |
| O | 4.04474000 | 12.85904000 | 7.33218000 |
| Si | 1.66166000 | 5.02240000 | 7.57693000 |
| Si | 9.28667000 | 2.38491000 | 7.58584000 |
| Si | 6.52179000 | 1.48441000 | 7.58891000 |
| O | 8.76445000 | 12.06825000 | 7.62044000 |
| Si | 8.81881000 | 5.29454000 | 7.88978000 |
| O | 7.29605000 | 5.87780000 | 8.06985000 |
| O | 0.72451000 | 1.93019000 | 8.08120000 |
| O | 6.89743000 | 8.56528000 | 8.09660000 |
| O | 11.16503000 | 12.17028000 | 8.21368000 |
| O | 4.36474000 | 6.02266000 | 8.21442000 |
| O | 8.00193000 | 1.62577000 | 8.21556000 |
| O | 2.52145000 | 3.83121000 | 8.25785000 |
| O | 4.60298000 | 9.87454000 | 8.29784000 |
| O | 11.23294000 | 9.65737000 | 8.33763000 |
| Si | 6.20759000 | 10.01218000 | 8.39279000 |
| Si | 1.16603000 | 7.67143000 | 8.54078000 |
| O | 2.62367000 | 8.34195000 | 8.68505000 |
| O | 5.53447000 | 0.91970000 | 8.72556000 |
| O | 1.28774000 | 6.06899000 | 8.77017000 |
| Si | 4.03346000 | 0.30045000 | 8.83076000 |
| O | 10.31381000 | 2.32344000 | 8.86802000 |
| Si | 9.68524000 | 12.14075000 | 8.96686000 |
| Si | 7.16649000 | 7.23733000 | 8.97783000 |
| Si | 1.87646000 | 2.56144000 | 9.05035000 |
| O | 3.02209000 | 1.46592000 | 9.34606000 |
| Si | 3.45236000 | 9.54199000 | 9.37580000 |
| O | 9.41126000 | 5.06335000 | 9.39370000 |
| O | 2.55276000 | 10.91413000 | 9.53046000 |
| O | 8.59502000 | 7.45156000 | 9.76356000 |
| O | 0.25735000 | 8.20553000 | 9.78759000 |
| Si | 4.61912000 | 5.94270000 | 9.80685000 |
| O | 6.67757000 | 10.56640000 | 9.82734000 |
| Si | 11.57529000 | 9.37814000 | 9.91657000 |
| O | 9.35747000 | 10.86631000 | 9.95108000 |
| O | 5.93269000 | 6.91789000 | 9.99594000 |
| O | 9.20339000 | 0.22140000 | 10.04828000 |
| O | 4.65432000 | 12.45007000 | 10.15129000 |
| O | 4.77645000 | 4.38801000 | 10.27452000 |
| Si | 9.84758000 | 1.68549000 | 10.33173000 |
| O | 1.23122000 | 3.17126000 | 10.40270000 |
| Si | 9.17939000 | 6.14520000 | 10.61421000 |
| O | 1.22910000 | 12.81664000 | 10.76714000 |
| O | 4.13857000 | 9.06001000 | 10.78539000 |
| O | 12.00916000 | 10.72988000 | 10.81921000 |
| O | 10.42087000 | 8.80291000 | 10.93948000 |
| O | 3.59533000 | 6.56215000 | 10.94387000 |
| Si | 2.02201000 | 11.36570000 | 11.01221000 |
| O | 8.69596000 | 2.57416000 | 11.09874000 |
| O | 10.68623000 | 6.10892000 | 11.26061000 |
| O | 11.16631000 | 1.47556000 | 11.27849000 |
| Si | 9.45776000 | 10.06018000 | 11.38731000 |
| Si | 6.89245000 | 10.75082000 | 11.41329000 |
| O | 1.69607000 | 9.81736000 | 11.56143000 |
| O | 7.83751000 | 5.87395000 | 11.56351000 |
| Si | 4.62983000 | 12.22691000 | 11.78878000 |
| O | 7.58990000 | 12.19780000 | 11.79330000 |
| Si | 4.77658000 | 3.93278000 | 11.86347000 |
| O | 7.96866000 | 9.60059000 | 11.88240000 |
| Si | 0.00000000 | 0.64100000 | 11.89328000 |
| Si | 4.04675000 | 7.84950000 | 11.90235000 |
| O | 6.05743000 | 2.87874000 | 11.92654000 |
| Si | 1.46717431 | 3.35273334 | 12.00685341 |
| O | 5.45745220 | 10.80863042 | 12.15901207 |
| O | 3.09341044 | 11.86387276 | 12.17382096 |
| O | 3.15508982 | 3.56410224 | 12.26859305 |
| Si | 7.67261811 | 2.70698441 | 12.41454140 |
| O | 0.42501806 | 4.61031239 | 12.45051863 |
| Si | 11.79957520 | 10.90060465 | 12.50868868 |
| Si | 11.72192499 | 5.87513065 | 12.53553603 |
| O | 0.91699775 | 1.86649930 | 12.58658665 |
| O | 5.55901168 | 7.61551948 | 12.58909749 |
| Si | 1.21491856 | 8.65189060 | 12.64841338 |
| O | 0.24341169 | 7.27382080 | 12.64847551 |
| Si | 6.60954299 | 6.28719809 | 12.65675353 |
| O | 10.18064680 | 10.74481342 | 12.69033416 |
| O | 5.37437251 | 0.20392119 | 12.73496366 |
| O | 5.42102218 | 5.14335791 | 12.78310544 |
| O | 12.09837308 | 12.48282335 | 12.87969399 |
| O | 2.76467306 | 8.07162850 | 12.96437481 |
| Si | 6.98057609 | 12.95861852 | 13.19661803 |
| O | 7.72988438 | 1.31309712 | 13.26487451 |
| O | 8.34206902 | 3.85839075 | 13.41597341 |
| O | 0.15981613 | 9.75620425 | 13.41670732 |
| O | 10.52822511 | 5.57535420 | 13.67264344 |
| O | 7.65233058 | 6.36245744 | 13.92953958 |
| Si | 8.92391041 | 5.22253792 | 14.18004967 |
| O | 9.03608964 | 5.19575801 | 15.80212459 |
| Sn | 8.42516622 | 9.10812691 | 15.04623743 |
| Pt | 7.05380549 | 11.45792822 | 15.09159044 |
| Pt | 9.17445225 | 7.05847343 | 16.57614439 |
| Pt | 6.15836609 | 9.85213151 | 17.00482824 |
| Sn | 8.45585981 | 11.53018494 | 17.24662616 |
| H | 5.52534409 | 8.43034502 | 17.31725406 |
| Pt | 9.93459132 | 9.36606758 | 17.36932957 |
| H | 3.61860260 | 9.99527060 | 17.67629584 |
| C | 4.56699857 | 10.02551064 | 18.22616415 |
| Sn | 7.77831531 | 8.16757860 | 18.47237828 |
| C | 4.60397004 | 10.32364029 | 19.53004310 |
| H | 3.69415814 | 10.66644920 | 20.04291381 |
| H | 5.51064410 | 10.28350966 | 20.13826974 |